\documentclass[a4paper,11pt]{article}
\pdfoutput=1 

\usepackage{jheppub} 

\usepackage[utf8]{inputenc}
\usepackage{graphicx} 
\usepackage{grffile}
\usepackage{subcaption}
\usepackage{dcolumn}  
\usepackage{makecell}
\usepackage{colordvi}
\usepackage{color}
\usepackage{epstopdf}
\usepackage{amssymb}
\usepackage{amsmath}
\usepackage{bm}
\usepackage{slashed}
\usepackage{enumerate}
\usepackage{url}
\usepackage{morefloats}

\usepackage{siunitx}

\usepackage{natbib}

\usepackage[colorlinks=true,
urlcolor=blue,
linkcolor=blue,
citecolor=blue,
linktocpage=true,
pdfproducer=medialab,
pdfa=true,
anchorcolor=blue]{hyperref}

\usepackage{tabularx,array}

\newcolumntype{C}{>{\centering\arraybackslash}X}
\usepackage{multirow}

\usepackage{lineno}

\usepackage{float}
\usepackage{xcolor}

\usepackage[title]{appendix}

\renewcommand{\thesection}{\arabic{section}}
\renewcommand{\thesubsection}{\thesection.\arabic{subsection}}
\renewcommand{\thesubsubsection}{\thesubsection.\arabic{subsubsection}}

\usepackage{minted}
\usepackage[ruled,linesnumbered]{algorithm2e}

\usepackage{changepage}
\usepackage{booktabs}

\input belle2sym.tex

\def\ee{\ensuremath{\Pep\Pem}\xspace}
\def\qq{\ensuremath{\Pq\Paq}\xspace}
\def\WW{\ensuremath{\PWp\PWm}\xspace}

\def\ntrkoff{\ensuremath{{\rm N}_{\rm Trk}^{\rm Offline}}\xspace}

\def\dphi{\ensuremath{\Delta\phi}\xspace}

\newcommand{\resetmarkpar}{\color{black}}          

\usepackage{hepnames}
\usepackage{xpatch}

\makeatletter
\xpatchcmd\@HepConStyle
 {\edef\@upcode{\updefault}}
 {\ifdefined\shapedefault\edef\@upcode{\shapedefault}\else\edef\@upcode{\updefault}\fi}
 {}{}
\makeatother

\preprint{}

\begin{document}

\title{Analysis note: Long-range near-side correlation in \boldmath \ee with  W-boson-pair events at 183--209 GeV with ALEPH archived data}

\author[a]{Tzu-An Sheng,}
\author[a]{Yu-Chen Chen,}
\author[b]{Yi Chen,}
\author[h]{Kyle Sun,}
\author[g]{Jason Chiu,}
\author[a]{Michael Peters,}
\author[c]{Marcello Maggi,}
\author[d]{Austin Baty,}
\author[e]{Anthony Badea,}
\author[a]{Chris McGinn,}
\author[a]{Jesse Thaler,}
\author[a]{Gian Michele Innocenti,}
\author[b]{Jingyu Zhang,}
\author[f]{Paoti Chang}
\author[a]{Yen-Jie Lee,}

\affiliation[a]{Massachusetts Institute of Technology, Cambridge, MA, USA}
\affiliation[b]{Vanderbilt University, Nashville, TN, USA}
\affiliation[c]{INFN Sezione di Bari, Bari, Italy}
\affiliation[d]{University of Illinois Chicago, Chicago, IL, USA}
\affiliation[e]{University of Chicago, Chicago, IL, USA}
\affiliation[f]{National Taiwan University, Taipei, Taiwan}%
\affiliation[g]{University of Illinois Urbana-Champaign, Urbana, IL, USA}
\affiliation[h]{University of California, Berkeley, Berkeley, CA, USA}

\emailAdd{janice\_c@mit.edu}
\emailAdd{yenjie@mit.edu}
\emailAdd{chenyi@mit.edu}
\emailAdd{pchang@phys.ntu.edu.tw}


\date{\today}

\abstract{






Events characterized by a high multiplicity of charged particles have been
a central focus in the study of
collective behavior across both large and small collision systems.
A previous measurement of two-particle angular correlations in \ee collisions at
center-of-mass energies up to $\sqrt{s} = 209\,\mathrm{GeV}$, using LEP2 data,
revealed discrepancies with Monte Carlo (MC) predictions at high multiplicity, suggesting
the possible emergence of long-range near-side correlations even in the simplest collision
system. Unlike at lower energies, where quark--antiquark production dominates, \WW
processes become increasingly important at multiplicities above 30.
On the one hand,
the observed excess in long-range correlations may reflect the more complex
color-string configurations arising from \WW production.
On the other hand,
it can simply arise from the higher final-state multiplicity made possible by the increased collision energy, independent of the underlying production mechanism.

To discriminate between these competing interpretations,
we present a measurement of two-particle angular correlations
in \ee collisions at $\sqrt{s} = 183\text{--}209\,\mathrm{GeV}$, with a focus on enhancing
the contribution from \WW processes. The analysis uses data collected by the ALEPH
detector during the LEP2 program. Correlation functions are evaluated across a broad
range of pseudorapidities and full azimuth, in bins of charged-particle multiplicity.
A ridge-like modulation is seen for multiplicity above 50, deviating from the MC reference.
In addition, the correlation functions are further decomposed into a Fourier series,
and the resulting harmonic coefficients $v_n$ are compared with predictions from the archived Monte Carlo sample.
For multiplicity starting from 30, the signed \(v_2\)-like proxy goes from negative to positive,
also deviating from the MC baseline.
}



\maketitle
\flushbottom


\section{Introduction}
\label{sec:Introduction}

Two-particle angular correlations~\cite{STAR:2005ryu,STAR:2009ngv,PHOBOS:2009sau,Chatrchyan:2012wg,Aamodt:2011by,Adam:2019woz} have been widely used in heavy-ion collisions to study the properties of the Quark-Gluon Plasma (QGP), as reviewed in Ref.~\cite{Busza:2018rrf}.
These studies revealed a \emph{long-range} \emph{near-side} correlation in azimuthal angle, commonly referred to as the \emph{ridge structure}~\cite{STAR:2009ngv,PHOBOS:2009sau},
which has been observed across a wide range of collision systems and energies. 
In two-particle correlation measurements, a ridge denotes an enhancement of pair yield at small relative azimuth, $|\Delta\phi|\approx 0$, that remains visible when the two particles are separated by a large pseudorapidity interval $|\Delta\eta|$.
In heavy-ion collisions, the ridge is commonly explained by the collective expansion of the produced medium, with elliptic flow ($v_2$) in non-central collisions driven primarily by the almond-shaped nuclear overlap geometry; higher harmonics are sensitive to finer-scale structure in the initial density profile~\cite{Alver:2010gr}.
However, following the start of LHC operations, a similar ridge was first reported by the CMS Collaboration in high-multiplicity proton-proton (pp) collisions~\cite{Khachatryan:2010gv}
and subsequently confirmed in other collision systems, including intermediate-multiplicity pp~\cite{Aad:2015gqa}, proton-ion (pA)~\cite{CMS:2012qk,ALICE:2012eyl,ALICE:2013snk,ATLAS:2012cix,Aaij:2015qcq},
and deuteron-ion (dA)~\cite{PHENIX:2013ktj} collisions. More recently, measurements in light-ion OO and NeNe collisions at the LHC
have revealed significant azimuthal anisotropies, including elliptic and
triangular flow harmonics, using two- and multi-particle correlations at
$\sqrt{s_{\rm NN}}=5.36$ TeV~\cite{ATLAS:2025OOFlow,ALICE:2025OOFlow,CMS:2025OOFlow}.
The interpretation of similar long-range correlations in small collision systems remains under active debate~\cite{Dumitru:2010iy,Dusling:2013qoz,Bozek:2011if,He:2015hfa,Nagle:2018nvi}.

Over the last few years, experimentalists have extended two-particle correlation studies to even smaller systems, such as photonuclear PbPb~\cite{ATLAS:2021jhn,CMS:2022doq},
$e$p~\cite{ZEUS:2019jya}, and $e^+e^-$~\cite{Badea:2019vey,Belle:2022fvl,The:2022lun} collisions.
Dedicated ridge searches in these smaller systems have not reported a long-range near-side excess of the kind seen in high-multiplicity pp or AA collisions, although measurements in ultraperipheral collisions (UPCs) do observe non-zero $v_2$ after non-flow subtraction~\cite{ATLAS:2021jhn}. These results provide important constraints on possible mechanisms responsible for the ridge~\cite{Bierlich:2019wld,Bierlich:2020naj,Castorina:2020iia,Agostini:2021xca,Larkoski:2021hee,Baty:2021ugw}.
Within the same archived-ALEPH \ee program, complementary measurements and analysis notes have also studied jet spectra and substructure, energy-energy correlators, and event-shape observables at the \(Z\) pole~\cite{Chen:2021uws,Chen:2021uwsNote,Bossi:2025eecNote,Badea:2025thrustNote}.
These studies provide related constraints on perturbative showering, hadronization, and event reconstruction in the clean \ee environment; the present note instead focuses on the LEP2 high-multiplicity regime and tests whether the event composition and two-dipole-enriched diboson topology modify long-range two-particle correlations.

A recent review~\cite{Grosse-Oetringhaus:2024bwr} summarizes ongoing experimental and theoretical efforts to study collective behavior in small systems.
It concludes with the observation that there is no clear onset of new physics above a minimal characteristic scale.
Remarkably, many phenomena associated with collectivity in large AA collisions have also been seen in smaller pp and pPb collisions.
Nevertheless, how the ridge-like correlations emerge as the system size and final-state multiplicity increase remains an open question.

To address this question, recent measurements have pursued new experimental strategies to search for the onset of ridge-like collectivity by going to smaller and smaller collision systems or lower and lower multiplicity events.
The ALICE collaboration has shown that the ridge-like structure in high-multiplicity pp collisions persists down to low charged-particle multiplicities, $8 \lesssim \langle N_{\rm ch} \rangle \lesssim 24$~\cite{ALICE:2023ulm}. Another approach, performed using CMS pp data, measures two-particle correlation within high-multiplicity jets~\cite{cms:2023iam}. The study reports that the long-range elliptic anisotropic coefficient, denoted $v_2^*$, increases with in-jet charged-particle multiplicity. This measurement probes whether ridge-like correlation can arise within a nominally vacuum-like parton-shower and fragmentation environment, as final-state multiplicity within the jet increases. Complementary measurements in ultraperipheral PbPb collisions also report non-zero $v_2$ after non-flow subtraction, extending the small-system collectivity discussion to the UPC environment~\cite{ATLAS:2021jhn}.

\begin{figure}[ht]
\centering
    \includegraphics[width=0.45\textwidth]{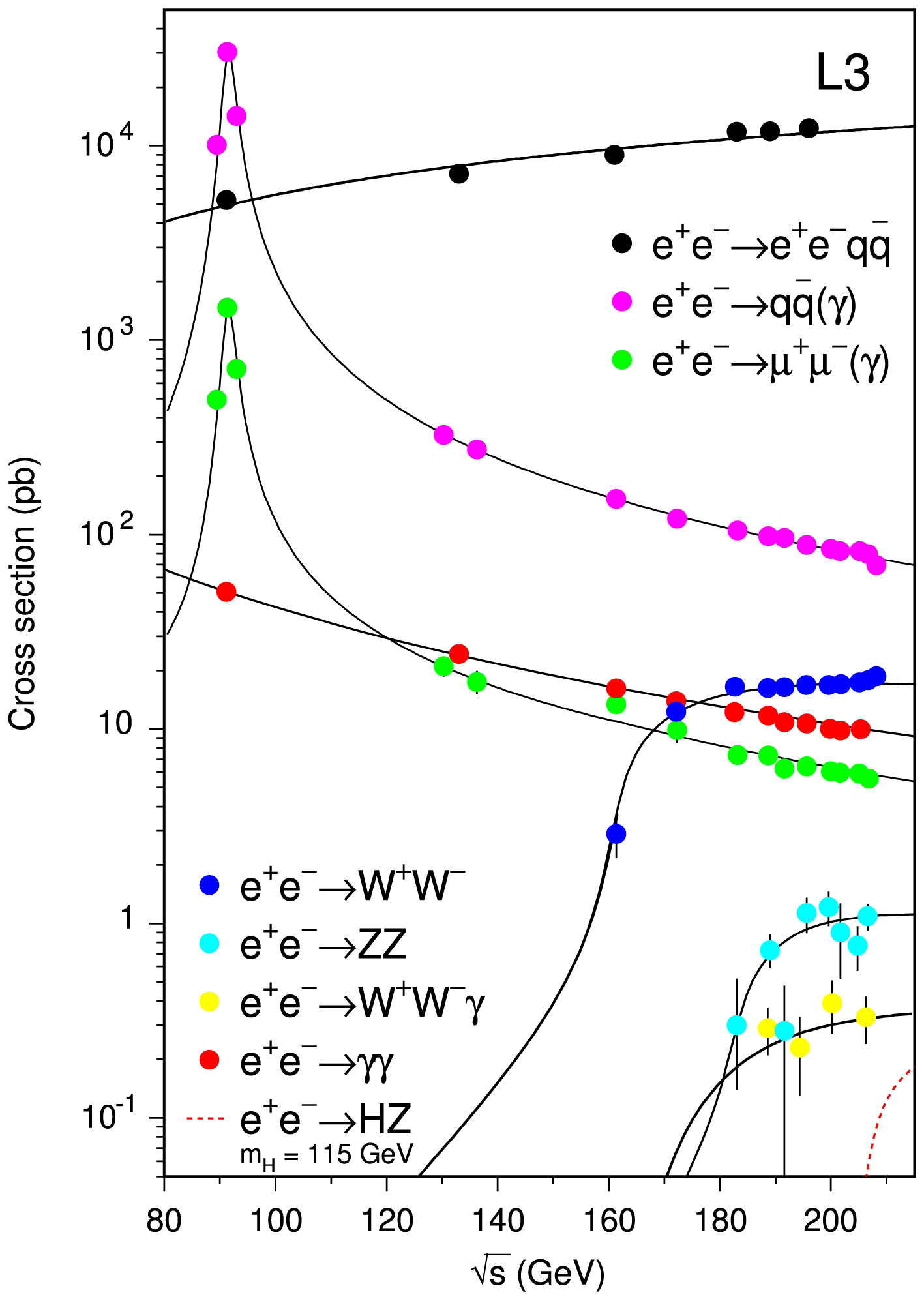}
    \caption{Cross-sections of electroweak SM processes at LEP energies. Courtesy of Ref.~\cite{2013119}}
    \label{fig:qed_xsection}
\end{figure}

In this note, we study the high-energy \ee collision data taken by ALEPH experiment at
LEP2.
This dataset offers several unique advantages including:
\begin{enumerate}
    \item The \ee collisions are point-like, and free from initial-state color correlations;
    \item The final-state evolution occurs entirely in vacuum;
    \item The collision energy lies above the diboson production thresholds, so final states such as \(\PWp\PWm\), \(\PZ\PZ\), and \(\PW\Pe\Pnu\) are open alongside \(\Pep\Pem \to \Pq\Paq\), leading to a richer mixture of partonic topologies than at LEP1.
      In contrast, the events at the lower-energy runs at LEP1 are dominated by
      \(\Pep\Pem \to \Pq\Paq\) and
      \(\Pep\Pem \to \Pq\Paq(\Pgamma)\), as shown in Fig.~\ref{fig:qed_xsection}.
\end{enumerate}

In a previous study using inclusive hadronic events, a signal that may indicate collectivity was reported in the high-multiplicity region~\cite{Chen:2023njr,Chen:2023nsi}.
A long-range, near-side enhancement in the two-particle correlation function was identified as a deviation from the smooth Monte Carlo (MC) baseline.
Moreover, the Fourier decomposition of the correlation revealed that the deviation between data and MC increases with the transverse momentum of the associated charged particles,
suggesting a potential $p_{\mathrm{T}}$-dependent effect that suggests collectivity in high-multiplicity $e^+e^-$ collisions.

At LEP2 energies, the high-multiplicity region is particularly interesting because the event composition shifts from quark-antiquark (\qq) production toward a larger \WW contribution.
In hadronic \WW decays, each \PW boson decays into two quarks at the parton level, resulting in a more complicated picture than \qq process.
At early times following the collision---before fragmentation and hadronization take
over---each hadronic $\PW$ decay gives rise to a color string connecting the outgoing
quark pair.
With two such decays present, the event contains two color-singlet hadronic systems whose associated string or dipole configurations may occupy overlapping space-time regions during hadronization.
This provides a unique opportunity to test whether the event topology and nonperturbative inter-string effects, such as color reconnection, modify final-state angular correlations.
These hadronization-level effects should be distinguished from partonic rescattering mechanisms; both are discussed below only as qualitative motivations for studying the multiplicity dependence of the correlation observables.

From the phenomenological perspective, Nagle~et~al.~\cite{Nagle:2017sjv} explored the
effect of initial-stage complexity using a modified \emph{A Multi Phase Transport} (AMPT) model,
initialized with either a single or multiple color-string configuration.
It is demonstrated that in the single-string case, varying partonic and hadronic interactions has only a marginal effect on the correlation function.
For the two-string configuration, Ref.~\cite{Nagle:2017sjv} finds that long-range enhancements and non-zero $v_2$ arise primarily from parton-level rescattering; turning off hadronic rescattering leaves $v_2$ nearly unchanged, so the dominant sensitivity is to partonic scattering.
In this context, hadronic \(\PWp\PWm\) decays in \ee collisions provide an experimentally accessible two-dipole-like topology, while not being a direct realization of the AMPT partonic-rescattering scenario.

\begin{figure}[ht]
\centering
    \includegraphics[width=0.95\textwidth]{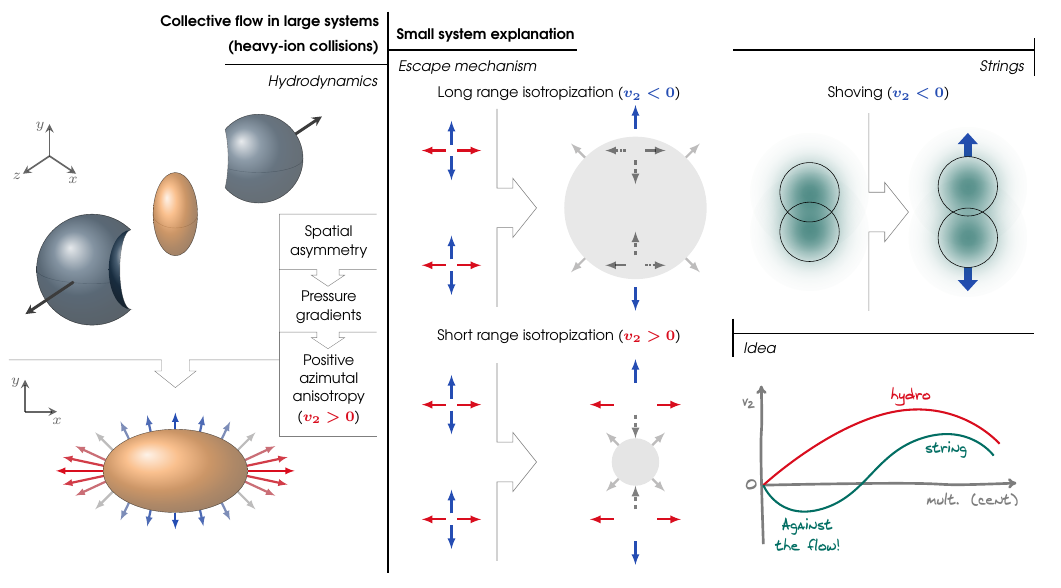}
    \caption{Illustration of how the sign of the elliptic flow, \(v_2\), varies for
      different systems and the hydrodynamic vs the string-shoving model.
      Courtesy of Ref.~\cite{Bierlich:2024lmb}.}
    \label{fig:string-shoving}
\end{figure}

More recently, other microscopic models have also been proposed for small systems.
The string shoving model~\cite{Bierlich:2024lmb} is implemented in an MC hadronization picture where confining strings can repel each other, modifying particle production relative to independent fragmentation. It predicts a transition of the $v_2$ coefficient from negative to positive values with increasing multiplicity due to an escape mechanism.
As illustrated in Fig.~\ref{fig:string-shoving}, this behavior is a distinct signature from positive elliptic flow in hydrodynamic models. 

These two frameworks emphasize mechanisms operating at different stages—partonic rescattering versus string-level interactions. While the present analysis does not aim to distinguish between them quantitatively, we compare with archived LEP2 Monte Carlo samples and examine the multiplicity dependence of $v_2$, thereby providing insight into the underlying mechanism.

Building on the inclusive measurement with \ee collisions at LEP2~\cite{Chen:2023njr,Chen:2023nsi}, this measurement has the following aims:
\begin{enumerate}
    \item To enhance the fraction of hadronic diboson events (predominantly hadronic \WW, with a two-dipole-like topology) using a multivariate method, thereby optimizing the sensitivity and significance of potential collectivity-like signals;
    \item To investigate the multiplicity dependence of the $v_2$ coefficient, providing insight into the underlying mechanisms of collectivity-like behavior in small systems;
    \item To improve the statistical treatment of the harmonic extraction by introducing a bootstrap estimate of the bin-to-bin covariance of the long-range azimuthal differential yield, and a Bayesian propagation of this covariance into the Fourier fit. This procedure captures event-sampling fluctuations that are correlated across $\Delta\phi$ bins, which is an effect that is not addressed when bins are treated as statistically independent.
\end{enumerate}

The analysis note is structured as follows: Section~\ref{sec:Sample} presents the data and
Monte Carlo sample. In Section~\ref{sec:Selection}, we revisit the event and track
selections used in the \ee two-particle correlation analysis, and the Boosted Decision
Tree (BDT)-based selections to enhance the \WW signal purity. The two-particle correlation results with the \WW-enriched selection are presented in Section~\ref{sec:TwoParticleCorrelationFunction}. Section~\ref{sec:flow} presents the extracted two-particle Fourier coefficients and signed square-root proxies. The note concludes with a brief summary in Section~\ref{sec:summary}.

\resetmarkpar
\section{Data and Monte Carlo sample}
\label{sec:Sample}


The ALEPH detector concept and its performance at LEP are documented in Refs.~\cite{Decamp:1990jra,ALEPH:1994ayc}.
For LEP2 the silicon vertex detector (VDET~II) was upgraded with increased length along the beam axis, reduced passive material in the tracking volume, a radiation-hard preamplifier chip, and finer strip pitch to increase granularity~\cite{Creanza:1998nim}.
Very briefly, the central part of the detector is designed for the efficient reconstruction of charged particles. Their trajectories are measured by a two-layer silicon strip vertex detector, a cylindrical drift chamber, and a large time projection chamber (TPC). These tracking detectors are situated inside a 1.5 T axial magnetic field generated by a superconducting solenoidal coil. The transverse momenta of the charged particles are reconstructed with a resolution of \(\delta p_{\rm T}/p_{\rm T} = 6 \times 10^{-4} p_{\rm T} \oplus 0.005\) (GeV/$c$).

Electrons and photons are identified in the electromagnetic calorimeter (ECAL), which is located between the TPC and the superconducting coil. The ECAL is a sampling calorimeter composed of lead plates and proportional wire chambers segmented into \(0.9^\circ \times 0.9^\circ\) projective towers. These are read out in three depth sections and have a total thickness of approximately 22 radiation lengths. Isolated photons are reconstructed with a relative energy resolution of \(0.18/\sqrt{E} + 0.009\) GeV.

The iron return yoke, constructed with 23 layers of streamer tubes, also serves as the hadron calorimeter (HCAL) for the detection of charged and neutral hadrons. The relative energy resolution for hadrons is \(0.85/\sqrt{E}\). Muons are identified based on their patterns in the HCAL and by the muon chambers, which are made of two double layers of streamer tubes located outside the HCAL.

The information from the trackers and calorimeters is integrated using an energy-flow algorithm~\cite{ALEPH:1994ayc}. This algorithm generates a set of charged and neutral particles, called energy-flow objects. The charged particles reconstructed in the energy-flow algorithm are used in the correlation function analysis.


We use the same data sample as what is studied in the previous LEP2 two-particle correlation measurement~\cite{Chen:2023njr,Chen:2023nsi}. The dataset was collected by ALEPH between 1997 and 2000, corresponding to \ee collision energies of $\sqrt{s} = 183$--209~GeV.

To derive the correction to the reconstruction effects and provide a baseline for comparison with data, we utilize Monte Carlo (MC) samples produced by the ALEPH collaboration.
In the ALEPH archived LEP2 MC, different MC generators are used to simulate different physics processes: $\ee \to q\bar{q} (\gamma)$ events are generated with \textsc{kk} version~4.14~\cite{Jadach:1999vf}; $WW$-like ($4f$) final states are generated with \textsc{koralw} version~1.51~\cite{Jadach:2001mp}; and $ZZ$-like final states are generated with \textsc{pythia}, with double counting of contributions from $WW$-like compatible final states being avoided. The hadronization of quarks into final states is modeled by \textsc{jetset 7.4}~\cite{Sjostrand:1993yb} or \textsc{pythia 6.1}~\cite{Sjostrand:2000wi}. In the archived MC samples, the reconstruction and detector simulation are performed under detector conditions corresponding to each data-taking period of the ALEPH experiment. Details of the ALEPH MC simulation at LEP2 energies are discussed in Ref.~\cite{ALEPH:2006cdc}.

We consider the cross-section normalizations of different physics processes calculated by the event generators, and the integrated luminosities corresponding to different data-taking periods.
Table~\ref{tab:physics_processes} summarizes the integrated luminosities for each subsample, collected at collision energies of
$\sqrt{s}=183$, 187, 192, 196, 200, 202, 205, and 207~GeV, along with the cross-section normalizations associated with different MC processes at each collision energy.

\begin{table}[htbp]
\centering
\scriptsize
\caption{List of physics processes with cross-section normalizations (in units of picobarn (pb)) calculated from event generators, at different LEP2 collision energies. The corresponding integrated luminosities (pb$^{-1}$) of each subsample are also shown.}
\begin{tabular}{r|cccccccc}
\hline
year & 1997 & 1998 & \multicolumn{4}{c}{1999} & \multicolumn{2}{c}{2000} \\
energy [GeV] & 182.65 & 188.63 & 191.58 & 195.52 & 199.52 & 201.62 & 204.86 & 206.53 \\
luminosity [pb$^{-1}$] & 56.81 & 174.21 & 28.93 & 79.86 & 86.28 & 41.89 & 81.41 & 133.21 \\
\hline
\multicolumn{9}{c}{\( \gamma \gamma \rightarrow \text{hadrons} \), cross section (pb)} \\
\hline
\texttt{[GGUD]} \( \gamma\gamma \rightarrow uu/dd \)
& 1687.05 & 1716.50 & 1727.30 & 1750.76 & 1770.35 & 1773.91 & 1786.82 & 1795.31 \\
\texttt{[GGSS]} \( \gamma\gamma \rightarrow ss \)
& 82.99 & 84.18 & 85.97 & 86.11 & 87.05 & 87.15 & 88.26 & 88.49 \\
\texttt{[GGCC]} \( \gamma\gamma \rightarrow cc \)
& 251.79 & 257.81 & 261.51 & 263.56 & 265.94 & 267.62 & 270.23 & 271.91 \\
\texttt{[GGBB]} \( \gamma\gamma \rightarrow bb \)
& 1.30 & 1.29 & 1.29 & 1.34 & 1.37 & 1.36 & 1.38 & 1.41 \\
\hline
\multicolumn{9}{c}{\( e^+e^- \rightarrow \text{hadrons} \), cross section (pb)}\\
\hline
\texttt{[KQQ]} \( e^+e^- \rightarrow qq \)
& 110.53 & 101.54 & 97.54 & 92.66 & 88.14 & 85.93 & 82.55 & 81.01 \\
\texttt{[TT]} \( e^+e^- \rightarrow \tau^+\tau^- \)
& 8.88 & 8.27 & 7.96 & 7.59 & 7.25 & 7.09 & 6.84 & 6.74 \\
\hline
\multicolumn{9}{c}{\( e^+e^- \rightarrow 4f \), cross section (pb)}\\
\hline
\texttt{[KWW4F]} \(\WW \)
& 16.60 & 17.70 & 18.08 & 18.45 & 18.71 & 18.80 & 18.93 & 18.96 \\
\texttt{[KWENU]} \( W e \nu \)
& 0.53 & 0.58 & 0.61 & 0.66 & 0.70 & 0.72 & 0.75 & 0.77 \\
\texttt{[PZEE]} \( Z e e \)
& 7.87 & 8.03 & 8.11 & 8.20 & 8.32 & 8.36 & 8.43 & 8.47 \\
\texttt{[PZZ]} \( ZZ \)
& 1.89 & 2.13 & 2.26 & 2.32 & 2.35 & 2.35 & 2.35 & 2.34 \\
\texttt{[ZNN]} \( Z \nu\nu \)
& 0.01 & 0.01 & 0.01 & 0.01 & 0.02 & 0.02 & 0.02 & 0.02 \\
\hline
\end{tabular}
\label{tab:physics_processes}
\end{table}

The numerical values in Table~\ref{tab:physics_processes} are taken from the
archived ALEPH LEP2 sample bookkeeping used to normalize the corresponding MC
subsamples. The generator setup is documented in Ref.~\cite{ALEPH:2006cdc} and
the generator references cited above, while the LEP2 electroweak cross-section
context is summarized in Ref.~\cite{2013119}. The values are listed explicitly
here because they define the MC mixture used below.

For the analysis, we construct a stratified MC sample that reproduces the realistic
composition of the data, by sampling events from different MC subsamples according to
their cross-section normalizations and integrated luminosities, as listed in
Table~\ref{tab:physics_processes}.
This stratified sampling is different from the method used in the previous LEP2
measurement~\cite{Chen:2023njr,Chen:2023nsi}, where all events in the MC samples were combined via a weighted sum based on the relative abundance of each process.

\newpage

\resetmarkpar
\section{Event selections and \WW identification with BDT}
\label{sec:Selection}



As discussed in the previous LEP-II two-particle correlation measurement~\cite{Chen:2023njr,Chen:2023nsi}, we suppress events with a large QED initial-state radiation (ISR) component in the $q\bar q(\gamma)$ sample by requirements on the visible dijet invariant mass $M_{\rm vis}$ and on the reduced squared center-of-mass energy $s'$, following the ALEPH prescription of Ref.~\cite{ALEPH:2003obs}.
In this analysis, $M_{\rm vis}$ and $s'$ are obtained by clustering the energy-flow inputs to exactly two jets with the exclusive $\ee$ $k_\mathrm{T}$ scheme (FastJet \texttt{ee\_kt\_algorithm}~\cite{Cacciari:2011ma}).
The quantities are obtained by clustering each event into two jets, where
\begin{equation}
s' = \frac{\sin \theta_1 + \sin \theta_2 - | \sin(\theta_1 + \theta_2) |}{ \sin \theta_1 + \sin \theta_2 + | \sin(\theta_1 + \theta_2) | } \times s,
\end{equation}
where $\theta_{1,2}$ denote the polar angles of the two jets with respect to the beam ($z$) axis.
The visible mass $M_{\rm vis}$ is the invariant mass of the same two-jet system; events are accepted if $M_{\rm vis}/\sqrt{s} > 0.7$ or $\sqrt{s'/s} > 0.9$~\cite{ALEPH:2003obs}.

To select hadronic events, we require the polar angle of the event sphericity axis to be from $7\pi/36$ to $29\pi/36$. Events that have fewer than five tracks or a total reconstruction charged-particle energy smaller than $15$~GeV are rejected.

In the two-particle correlation analysis, we consider charged particles from the primary collision, by restricting the displacement between tracks and the interaction point. The tracks should fall within the geometrical acceptance of the tracking detector. Additional quality requirements on the track transverse momentum and the number of TPC hits are applied to ensure a high-purity track sample. All event and track selections follow the criteria established in the previous measurement~\cite{Chen:2023njr,Chen:2023nsi}, and are summarized in Table~\ref{tab:SelectionSummary}. Throughout the analysis, we study the correlation function as a function of offline multiplicity, denoted as \({\rm N}_{\rm Trk}^{\rm Offline}\), which represents the number of selected tracks after applying the above criteria.

\begin{table}[ht]
\caption{Summary table for particle selections and event selections.  Neutral particles are selected mainly for the event thrust calculation.}
\begin{center}
\begin{tabularx}{\textwidth}{l|l}
\hline\hline
\multicolumn{2}{l}{Event selection}  \\
\hline
ISR                     & $\sqrt{s'} \ge 0.9 \sqrt{s}$ \\
                        & $M_{\rm vis} \ge 0.7 \sqrt{s}$ \\
Hadronic events         & at least five good tracks \\
                        & total reconstructed charged-particle energy $\ge 15$~GeV \\
Acceptance              & $7\pi/36 \le \theta_{\rm sphericity} \le 29\pi/36$\\
\hline
\multicolumn{2}{l}{Charged particles}  \\
\hline
Acceptance              & $|\cos\theta|<0.94$ \\
High quality tracks     & $p_{\rm T} \ge 0.2$ GeV\\
                        & at least 4 TPC hits \\
Impact parameter        & $d_0<2$~cm, $z_0<10$~cm \\
\hline
\multicolumn{2}{l}{Neutral particles (for thrust calculation)}  \\
\hline
Acceptance              & $|\cos\theta|<0.98$ \\
Energy cut              & $E>0.4$~GeV \\
\hline\hline
\end{tabularx} 
\label{tab:SelectionSummary}
\end{center}
\end{table}

\begin{figure}[htbp]
\centering
    \begin{subfigure}[b]{0.38\textwidth}
    \centering
        \includegraphics[width=\textwidth]{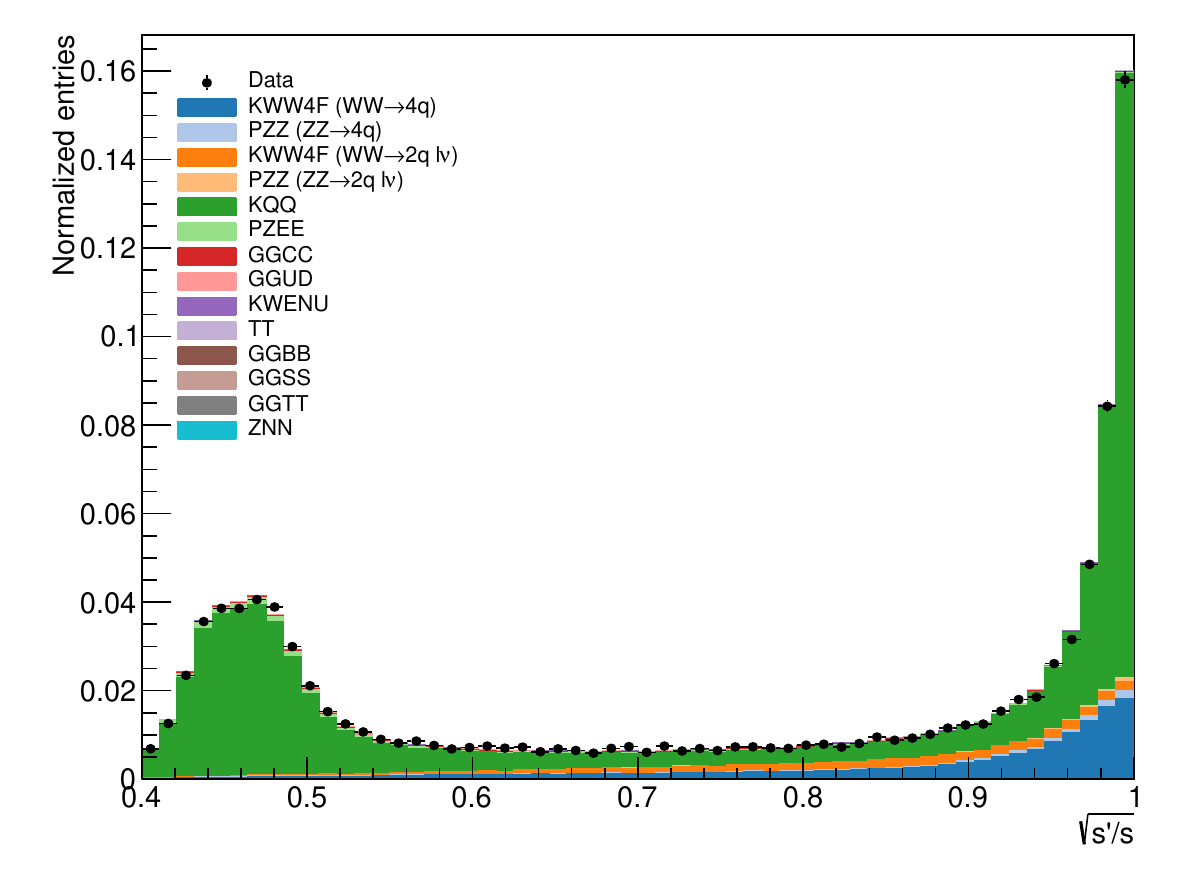}
        \caption{\(\sqrt{s'/s}\)}
    \end{subfigure}
    \begin{subfigure}[b]{0.38\textwidth}
    \centering
        \includegraphics[width=\textwidth]{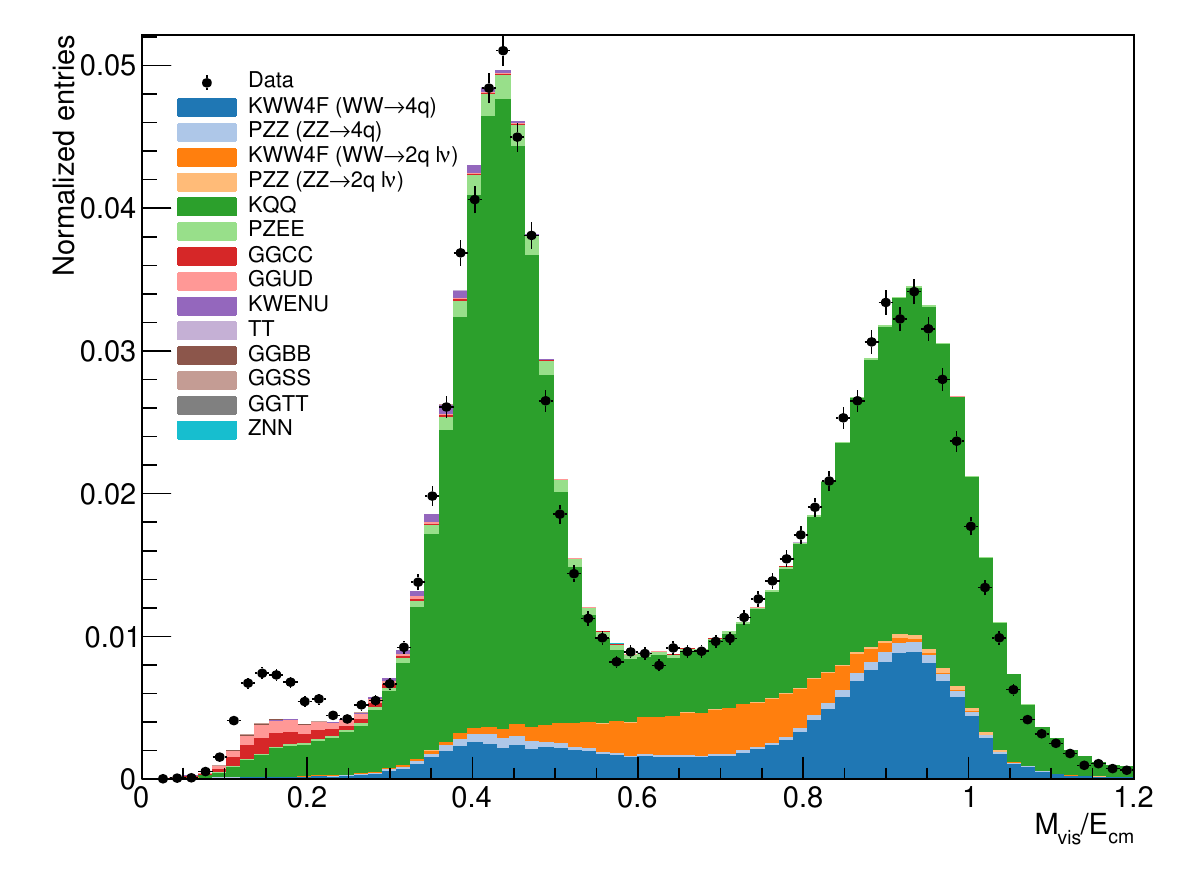}
        \caption{\(M_{\rm vis}/E_{\rm cm}\)}
    \end{subfigure}
\caption{Distributions of \(\sqrt{s'/s}\) and \(M_{\rm vis}/E_{\rm cm}\) without ISR selection. Hadronic-event selection is applied.
The low-value tails in \(\sqrt{s'/s}\) and \(M_{\rm vis}/E_{\rm cm}\) illustrate the radiative component.
Data are shown as black markers; the stacked histograms correspond to the merged LEP2 combined MC sample.
All distributions are normalized to unit area.}
\label{fig:SelectionQA_evtSel}
\end{figure}

\begin{figure}[htbp]
\centering
    \begin{subfigure}[b]{0.38\textwidth}
    \centering
        \includegraphics[width=\textwidth]{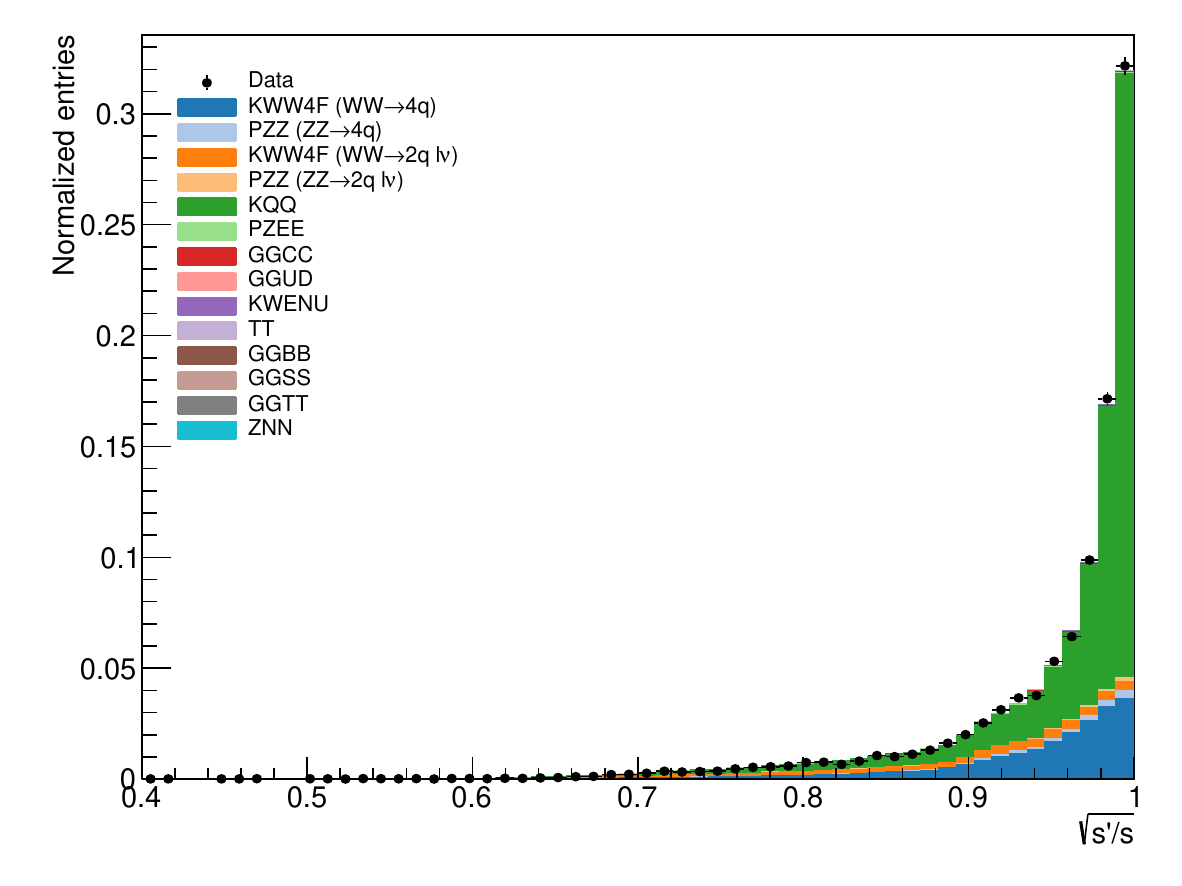}
        \caption{\(\sqrt{s'/s}\)}
    \end{subfigure}
    \begin{subfigure}[b]{0.38\textwidth}
    \centering
        \includegraphics[width=\textwidth]{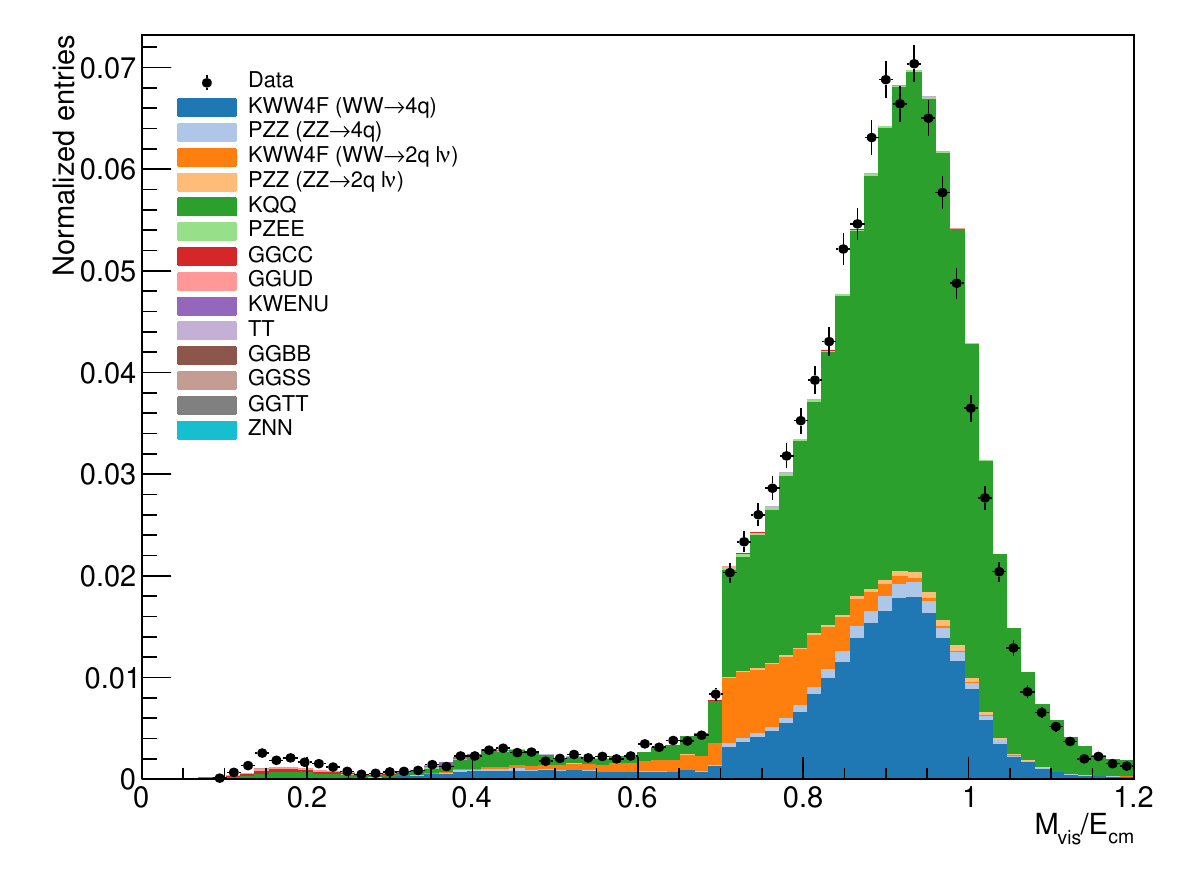}
        \caption{\(M_{\rm vis}/E_{\rm cm}\)}
    \end{subfigure}
    \begin{subfigure}[b]{0.38\textwidth}
    \centering
        \includegraphics[width=\textwidth]{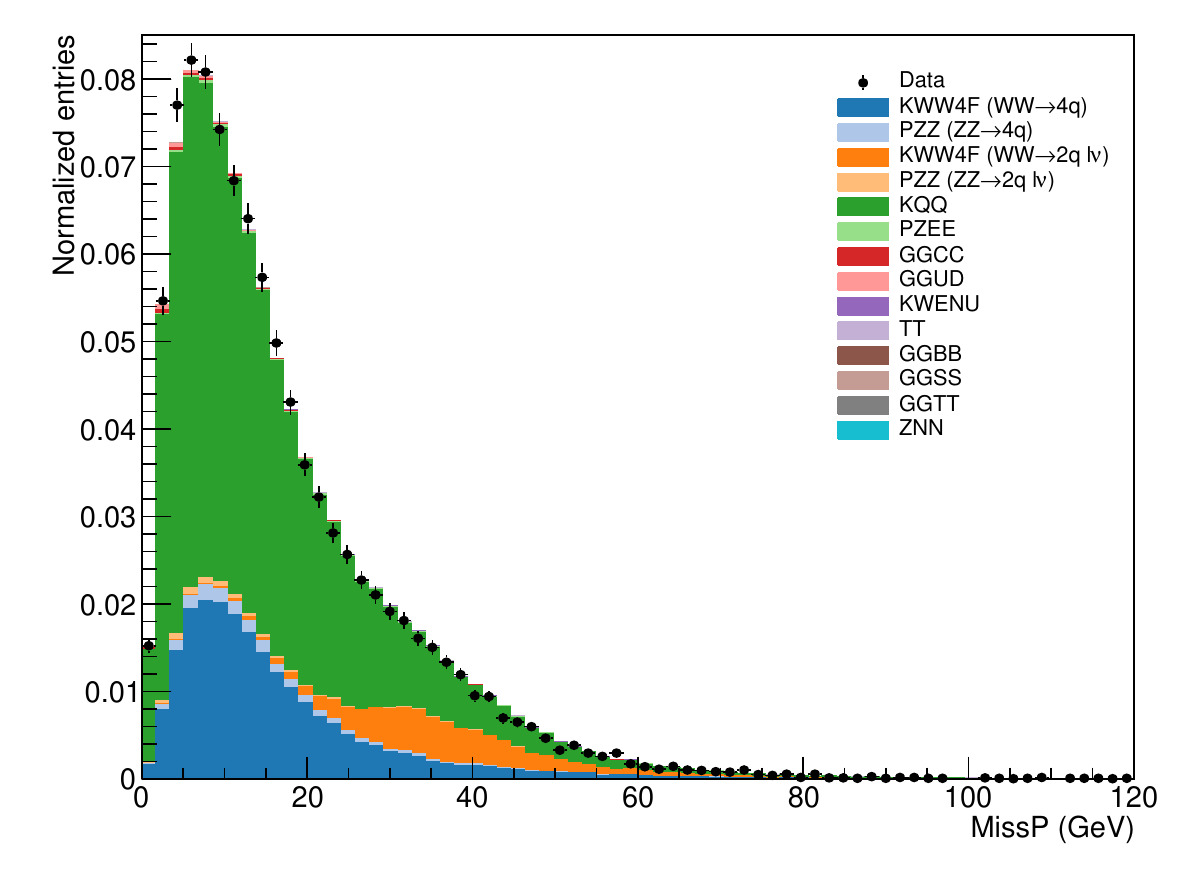}
        \caption{\(\Vec{p}_{\rm miss}\)}
    \end{subfigure}
    \begin{subfigure}[b]{0.38\textwidth}
    \centering
        \includegraphics[width=\textwidth]{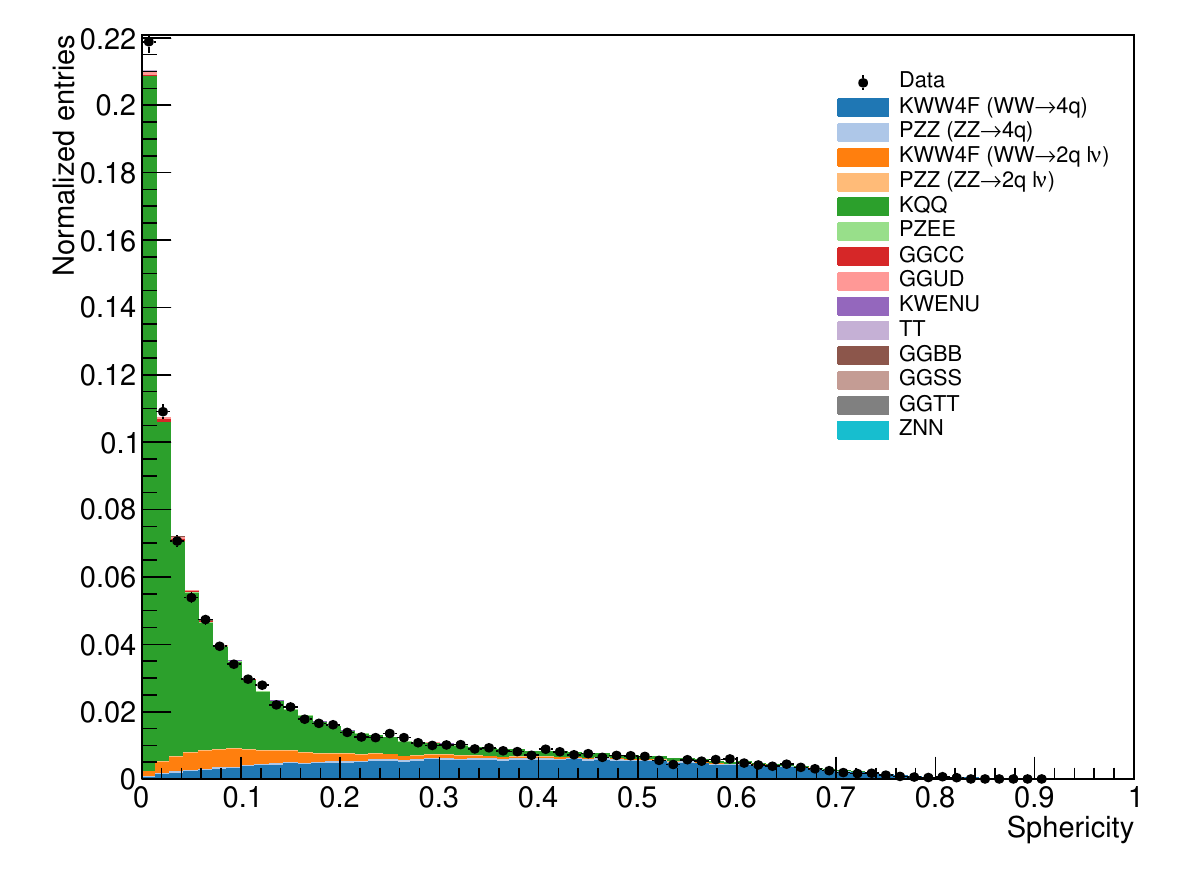}
        \caption{Sphericity}
    \end{subfigure}
\caption{Selection variable distributions after the full event selections (including ISR selection).
The \(\sqrt{s'/s}\) and \(M_{\rm vis}/E_{\rm cm}\) distributions are shifted toward larger values, consistent with the requirement
\(M_{\rm vis}/\sqrt{s} > 0.7\) or \(s'/s > 0.81\).
The retained sample is also reasonably described by the combined MC in the missing-momentum and sphericity observables.
All distributions are normalized to unit area.}
\label{fig:SelectionQA_evtISR_shape1}
\end{figure}

To make the effect of the event selections explicit, Figure~\ref{fig:SelectionQA_evtSel}
show normalized selection variable distributions for the merged LEP2 high-energy sample
\((\sqrt{s} = 183\text{--}207~\mathrm{GeV})\), comparing data to the stacked combined MC sample.
In Figure~\ref{fig:SelectionQA_evtSel}, before the ISR selection, the \(\sqrt{s'/s}\) and \(M_{\rm vis}/E_{\rm cm}\) distributions retain a visible low-value tail,
as expected from radiative-return-like configurations.
After imposing the ISR selection, these tails are strongly reduced and the accepted sample is concentrated in the large-\(\sqrt{s'}\),
large-\(M_{\rm vis}\) region, as shown in Figure~\ref{fig:SelectionQA_evtISR_shape1}.
After event selection, these event-level variables show reasonable 
agreement between data and the combined MC.

\clearpage

\subsection{High-multiplicity categorization with \texorpdfstring{$\ntrkoff \ge 30$}{Ntrk >= 30}}
\label{sec:ntrkskim30}

For the intention of searching collectivity-like signals in \ee collisions, which are expected to be more prominent in the high-multiplicity phase space, we introduce a simplified event categorization based on a dedicated skim requiring $\ntrkoff \ge 30$.
The purpose is to focus the analysis on the multiplicity region where the collectivity-like deviation is most visible, while keeping the MC composition and the multivariate classification easy to interpret.
In this regime, the available hadronic MC statistics can be used efficiently without being limited by the low-multiplicity production of the $\gamma\gamma$ samples that was relevant for the multiplicity-inclusive stratified cocktail (simplified as ``inclusive cocktail'' in the context below).

\begin{figure}[htbp]
  \centering
  \begin{minipage}[t]{0.48\textwidth}
    \centering
    \includegraphics[width=\linewidth]{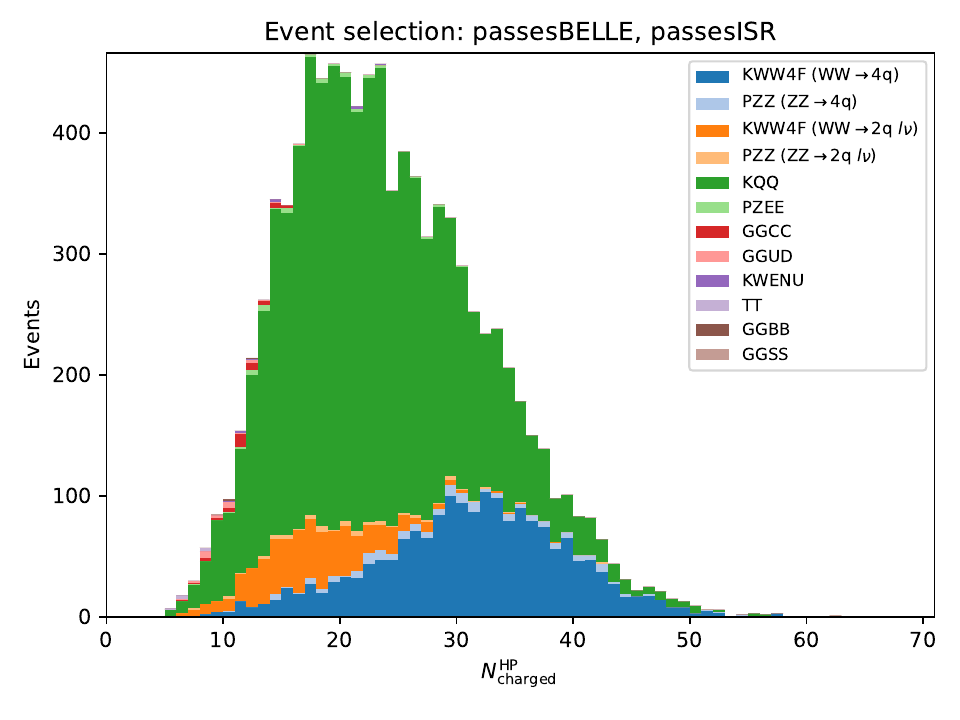}
  \end{minipage}\hfill
  \begin{minipage}[t]{0.48\textwidth}
    \centering
    \includegraphics[width=\linewidth]{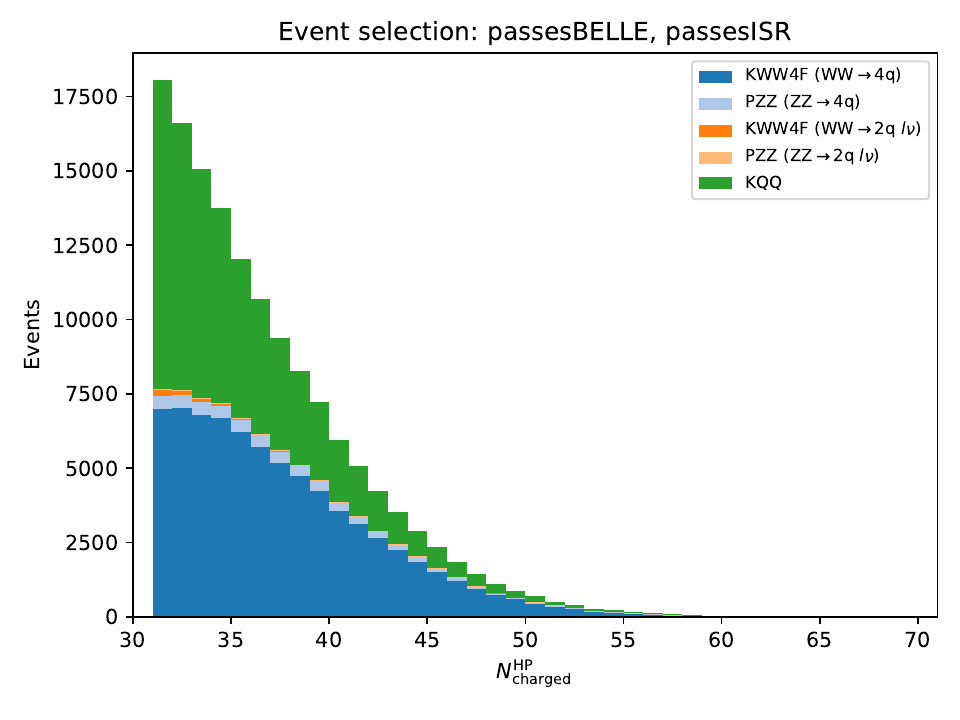}
  \end{minipage}
  \caption{Distribution of \ntrkoff\ at $\sqrt{s} = 192$~GeV, decomposed by MC sources: (left) stratified inclusive cocktail; (right) after $\ntrkoff \ge 30$ skim requirement with three-source cocktail (\qq, \WW, $ZZ$).}
  \label{fig:nChargedParticleHP-stratified-vs-ntrkskim30}
\end{figure}

With the $\ntrkoff \ge 30$ skim applied, the MC cocktail is reduced to the dominant ALEPH simulated components:
\qq production (\texttt{KQQ}), $WW$ production (\texttt{KWW4F}), and $ZZ$ production (\texttt{PZZ}).
After the high-multiplicity requirement, the remaining electroweak and two-photon contributions become negligible for the present purpose, so the cocktail can be described to good approximation by these three sources alone.
This simplification is useful for defining a clearer signal-versus-background problem for the BDT.

Figure~\ref{fig:nChargedParticleHP-stratified-vs-ntrkskim30} compares the \ntrkoff distributions for two configurations: the standard inclusive cocktail, and the dedicated high-multiplicity $\ntrkoff \ge 30$ skim. The selected events have passed through the baseline event selections as mentioned in Section~\ref{sec:Selection}.
The histograms are shown as a stacked decomposition by MC sources.

In the left panel, the multiplicity spectrum covers the full kinematically allowed range.
The bulk of the events accumulate at moderate charged multiplicity, where the \qq\ component dominates as expected.
The diboson contributions are subleading in the inclusive sample but grow in relative importance toward larger \ntrkoff.

The right panel illustrates the same observable with the three-source cocktail in the $\ntrkoff > 30$ skim.
The requirement removes the low-multiplicity tail and restricts the sample to the regime where the analysis is intended to be sensitive.
In particular, the fractional contribution from diboson hadronic decays events becomes more visible than in the inclusive stratified spectrum.

\subsection{Signal event definition}
\label{sec:orgdd765bf}

The primary focus of this analysis is to enhance the hadronic diboson component in the data sample, corresponding primarily to two color-singlet hadronic boson decays, to obtain more differential information on the generic \ee\ collision system, and to gain deeper insight into the collectivity-like excess observed in the previous inclusive analysis~\cite{Chen:2023njr}.

Our \emph{signal} is defined as hadronic diboson events, \(\PWp\PWm \to 4\Pq\), with a subdominant contribution from \(\PZ\PZ \to 4\Pq\). These events provide a sample enriched in two color-singlet hadronic decays, or equivalently a two-dipole-like topology. Under this definition, the dominant background is \(\Pep\Pem \to \Pq\APquark(\Pgamma)\), including events with additional perturbative radiation.
Additional backgrounds include semileptonic \(\PWp\PWm \to \ell \Pnu \Pq \APquark\) decays and \(\PZ\PZ\) final states with one hadronic \PZ decay and one invisible or charged-lepton \PZ decay, such as \(\PZ\PZ \to \Pq\Paq\nu\bar{\nu}\) or \(\PZ\PZ \to \Pq\Paq\ell^+\ell^-\).
Furthermore, to enhance the hadronic diboson component in the data samples, we select the subset of hadronic events compatible with the four-jet topology expected from hadronic diboson
decays. This is achieved with a multivariate BDT method and detailed in Section~\ref{sec:org9f991ca}.

\subsection{Discriminating variables for \WW identification}
\label{sec:org9f991ca}

We first set up a boosted decision tree (BDT) trained on variables with discriminating power between signal and background, to enhance the \WW component.
The current framework is trained using a minimal set of input variables to avoid introducing a significant event-shape bias to the analysis. We perform a data-MC agreement check to ensure there are no obvious data-MC discrepancies on the BDT training inputs and score, which is discussed later.
The set of input variables considered is as follows:

\begin{itemize}
\item Missing momentum ($\Vec{p}_{\rm miss}$)\\
The invisible objects, such as neutrinos, and non-reconstructed particles at the reconstruction level, are accounted for as the missing momentum, given by
\begin{equation}
\Vec{p}_{\rm miss} = -\sum\nolimits_{\rm neu, chg} \Vec{p}.
\label{eqn:MissP}
\end{equation}

\item Jet-pairing mass compatibility ($d^2$)\\
By clustering an event into a four-jet topology, we can derive a useful quantity to select hadronic \WW signals. For example, in Ref.~\cite{ALEPH:2003obs}, $d^2$ computes the compatibility of two-jet pairing masses with the $W$ mass. This is done by first clustering the event into four jets, and then rescaling the jet energies such that energy and momentum are conserved ($E_{\rm total} = E_{\rm cm}$, and $\vec{p}_{\rm total} = 0$). By combining two jets out of the four jets, with jet indices \(i,j,k,l\), one can compute the compatibility of the two-jet mass with the W mass,
\begin{equation}
d^2 = {\rm min} [\frac{(m_{ij}-M_W)^2+(m_{\rm kl}-M_W)^2}{M_W^2}],
\end{equation}
where $M_W = 80.4~\rm GeV$.

\item Visible mass ($M_{\rm vis}$) \\
By clustering an event into two jets, this quantity is the invariant mass of the two jets (as discussed in Sec.~\ref{sec:Selection}).

\end{itemize}

\begin{figure}[ht]
    \begin{center}
    \begin{subfigure}[b]{0.45\textwidth}
        \includegraphics[width=\textwidth]{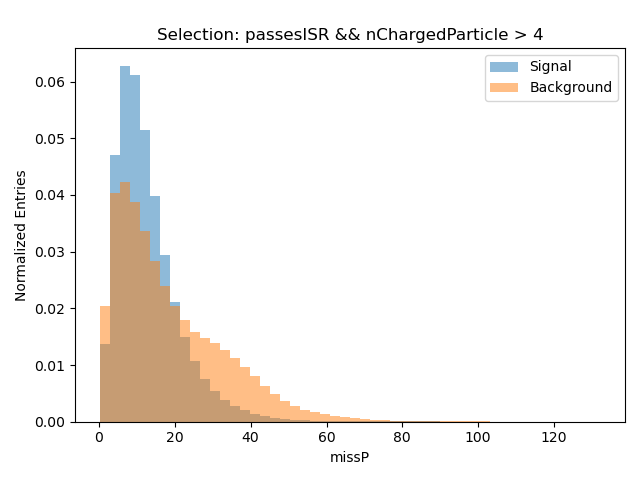}
        \caption{$\Vec{p}_{\rm miss}$}
    \end{subfigure}
    \begin{subfigure}[b]{0.45\textwidth}
        \includegraphics[width=\textwidth]{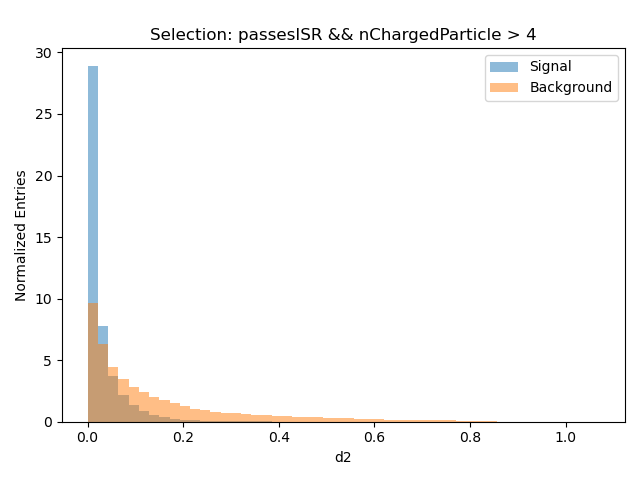}
        \caption{$d^2$}
    \end{subfigure}
    \begin{subfigure}[b]{0.45\textwidth}
        \includegraphics[width=\textwidth]{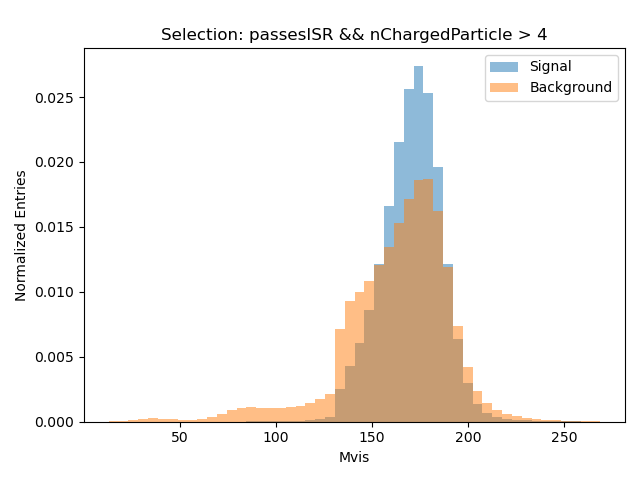}
        \caption{$M_{\rm vis}$}
    \end{subfigure}
    \end{center}
  \caption{Identification variables that are used in BDT training. \emph{Signal} refers to simulated \(\PWp\PWm \to 4\Pq\) (and a small \(\PZ\PZ \to 4\Pq\) contribution); \emph{background} refers primarily to \(\Pep\Pem \to \Pq\Paq(\Pgamma)\), semileptonic \(\PWp\PWm\) decays, and non-fully-hadronic \(\PZ\PZ\) decays, as defined in Sec.~\ref{sec:orgdd765bf}. The signal distribution is the blue histogram, and the background one is the orange histogram.}
  \label{fig:BDTVariables}
\end{figure}

The signal and background distributions of the BDT input variables are shown in Fig.~\ref{fig:BDTVariables}.

\begin{figure}[ht]
    \begin{center}
        \begin{subfigure}{1.0\textwidth}
            \includegraphics[width=1.0\textwidth]{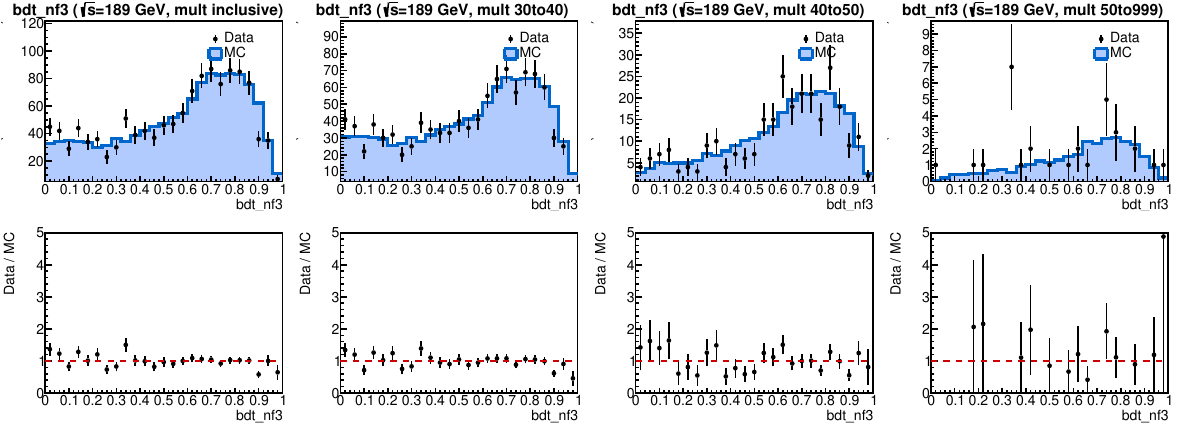}
            \vspace{-20pt}
            \caption{bdt\_nf3 with $\sqrt{s}=189$ GeV}
            \vspace{7pt}
        \end{subfigure}
        
        \begin{subfigure}{1.0\textwidth}
            \includegraphics[width=1.0\textwidth]{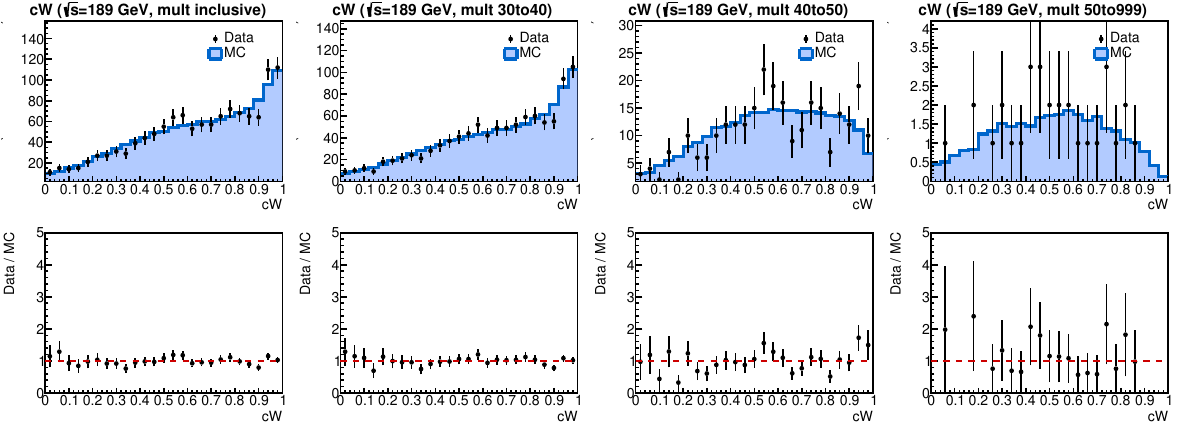}
            \vspace{-20pt}
            \caption{$c_W$ with $\sqrt{s}=189$ GeV}
            \vspace{7pt}
        \end{subfigure}
        
        \begin{subfigure}{1.0\textwidth}
            \includegraphics[width=1.0\textwidth]{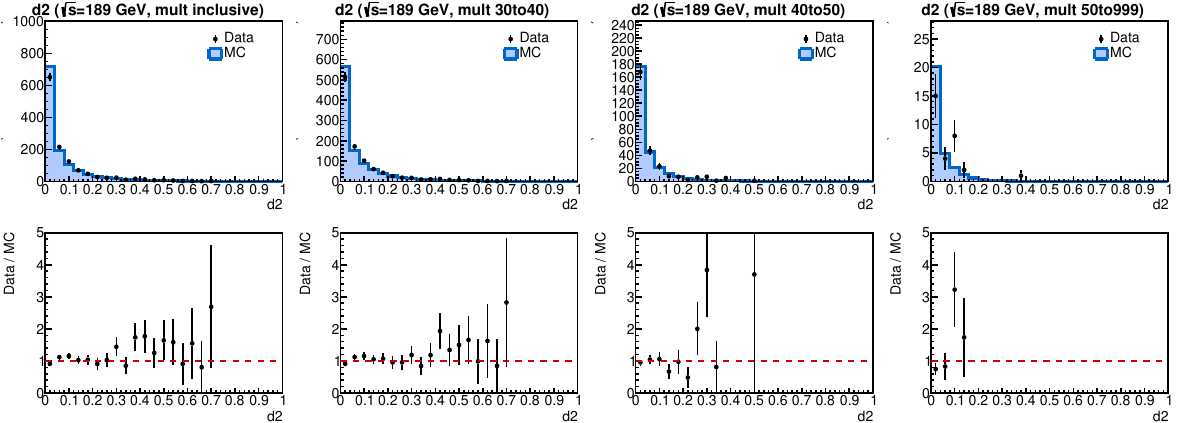}
            \vspace{-20pt}
            \caption{$d^2$ with $\sqrt{s}=189$ GeV}
        \end{subfigure}
    \end{center}
    \caption{Data and normalized Monte Carlo (MC) histograms \& Data/MC ratio plots for $\sqrt{s}=189$ GeV. Variables compared: bdt\_nf3, $c_W$, and $d^2$. For each set of histograms, the data (black markers) are compared to MC (blue histogram) across charged-particle multiplicity bins: inclusive, 30-40, 40-50, 50+, left to right. Below each histogram is a Data/MC ratio plot, where the red dashed line corresponds to unity.}
  \label{fig:DataVsMCVariables189A}
\end{figure}

\begin{figure}[h]
    \begin{center}
        \begin{subfigure}{1.0\textwidth}
            \includegraphics[width=1.0\textwidth]{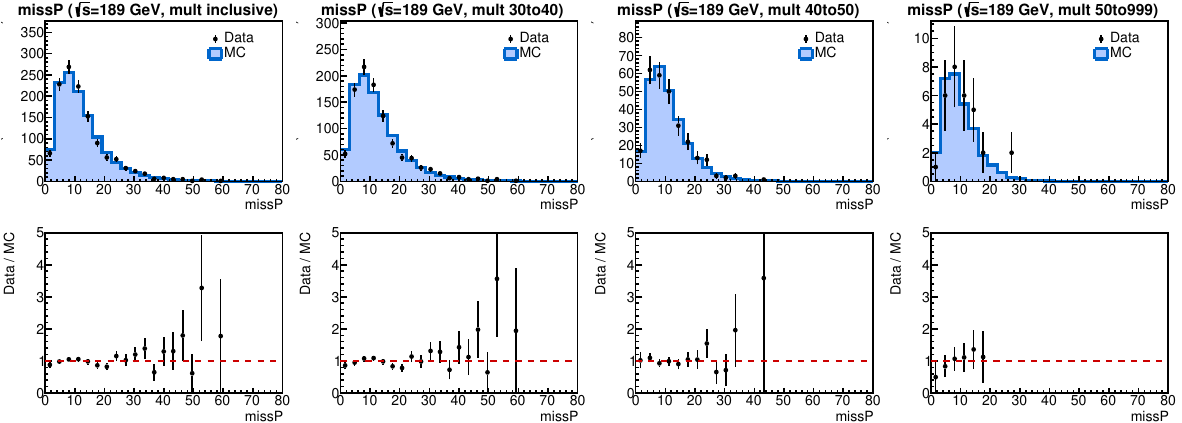}
            \vspace{-20pt}
            \caption{$\Vec{p}_{\rm miss}$ with $\sqrt{s}=189$ GeV}
            \vspace{7pt}
        \end{subfigure}
        \begin{subfigure}{1.0\textwidth}
            \includegraphics[width=1.0\textwidth]{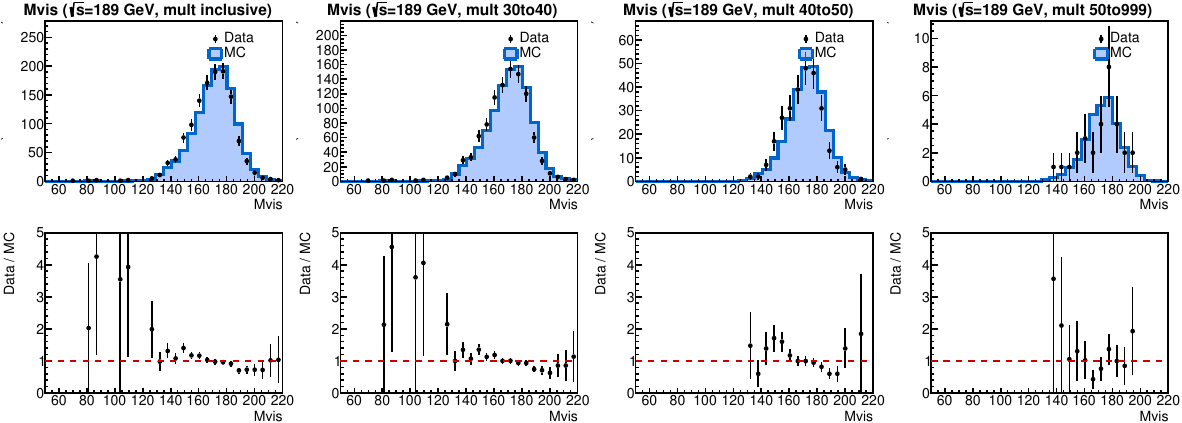}
            \vspace{-20pt}
            \caption{$M_{\rm vis}$ with $\sqrt{s}=189$ GeV}
            \vspace{7pt}
        \end{subfigure}
        \begin{subfigure}{1.0\textwidth}
            \includegraphics[width=1.0\textwidth]{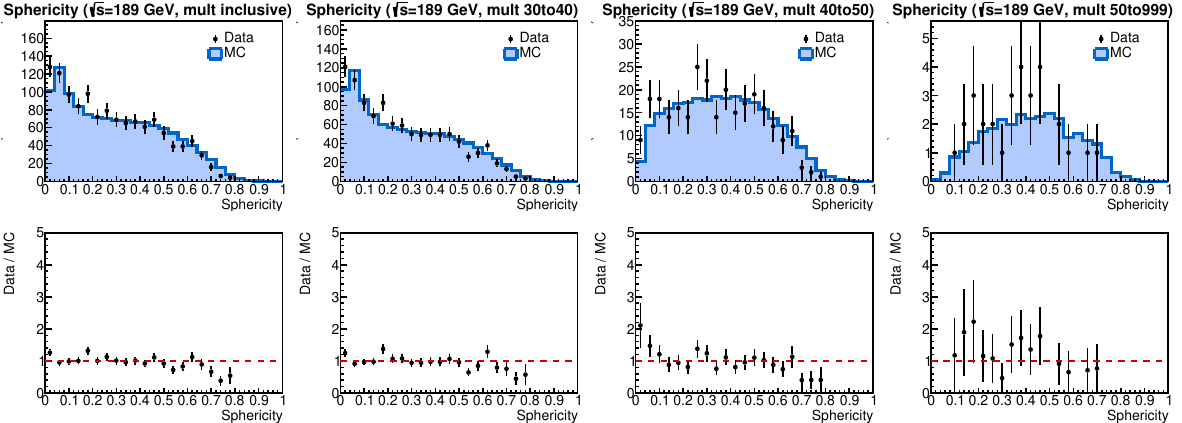}
            \vspace{-20pt}
            \caption{Sphericity with $\sqrt{s}=189$ GeV}
        \end{subfigure}
    \end{center}
    \caption{Same as Fig.~\ref{fig:DataVsMCVariables189A}, but with variables compared: $\Vec{p}_{\rm miss}$, $M_{\rm vis}$, and Sphericity.}
  \label{fig:DataVsMCVariables189B}
\end{figure}

Fig.~\ref{fig:DataVsMCVariables189A} and Fig.~\ref{fig:DataVsMCVariables189B} show the data and MC histogram and ratio comparisons for $\sqrt{s}=189$ GeV, searching for potential deviations and injected biases. We consider the three input variables, along with three new variables.
\begin{itemize}

\item Sphericity \\
This is an event-shape variable used as a quantifier of final-state momentum in an event.
Following Ref.~\cite{ALEPH:2003obs}, we form the normalized quadratic momentum tensor
\begin{equation}
  M^{\alpha\beta} = \frac{\sum_i p_i^\alpha p_i^\beta}{\sum_i |\vec{p}_i|^2},
  \qquad \alpha,\beta = 1,2,3,
\end{equation}
where the sum runs over reconstructed final-state particles.
Diagonalizing $M$ gives eigenvalues $\lambda_1 \ge \lambda_2 \ge \lambda_3$ with $\lambda_1+\lambda_2+\lambda_3=1$, and the sphericity is defined as
\begin{equation}
  S = \frac{3}{2}(\lambda_2 + \lambda_3).
\end{equation}

Sphericity characterizes to what degree events are isotropic versus jet-like, with a value of $0$ corresponding to pencil-like, two-jet events and a value of $1$ corresponding to isotropic event topologies. 

\item $c_W$ \\
This variable is used in the ALEPH LEP2 hadronic selection to suppress four-fermion ($WW$, $ZZ$, $Z\gamma^\ast$) background~\cite{ALEPH:2003obs}, following the same four-jet clustering procedure as discussed in the $d^2$ variable calculation, the cosine of interjet angle is defined as
\begin{equation}
  c_W = \cos(\text{smallest interjet angle}).
\end{equation}

\item Boosted Decision Tree score (denoted as \texttt{bdt\_nf3}) \\
This variable is the output of a boosted decision tree (BDT) classifier. Scores range from $0$ to $1$ and describe how signal-like an event is. A score close to $0$ indicates background-like events, while scores close to $1$ describe signal-like events.
\end{itemize}

Inspecting the ratio (Data/MC) in the lower set of plots, we see reasonable agreement between data and MC, with most ratios lying within one standard error of unity. This agreement implies that BDT training did not inject significant event-shape bias.
Histograms and plots for $\sqrt{s}=183, 192, 205$ GeV are in the appendix (Fig.~\ref{fig:DataVsMCVariables183A}, \ref{fig:DataVsMCVariables183B}, \ref{fig:DataVsMCVariables192A}, \ref{fig:DataVsMCVariables192B}, \ref{fig:DataVsMCVariables205A}, and \ref{fig:DataVsMCVariables205B}).

\begin{figure}[ht]
    \begin{center}
    \includegraphics[width=.6\linewidth]{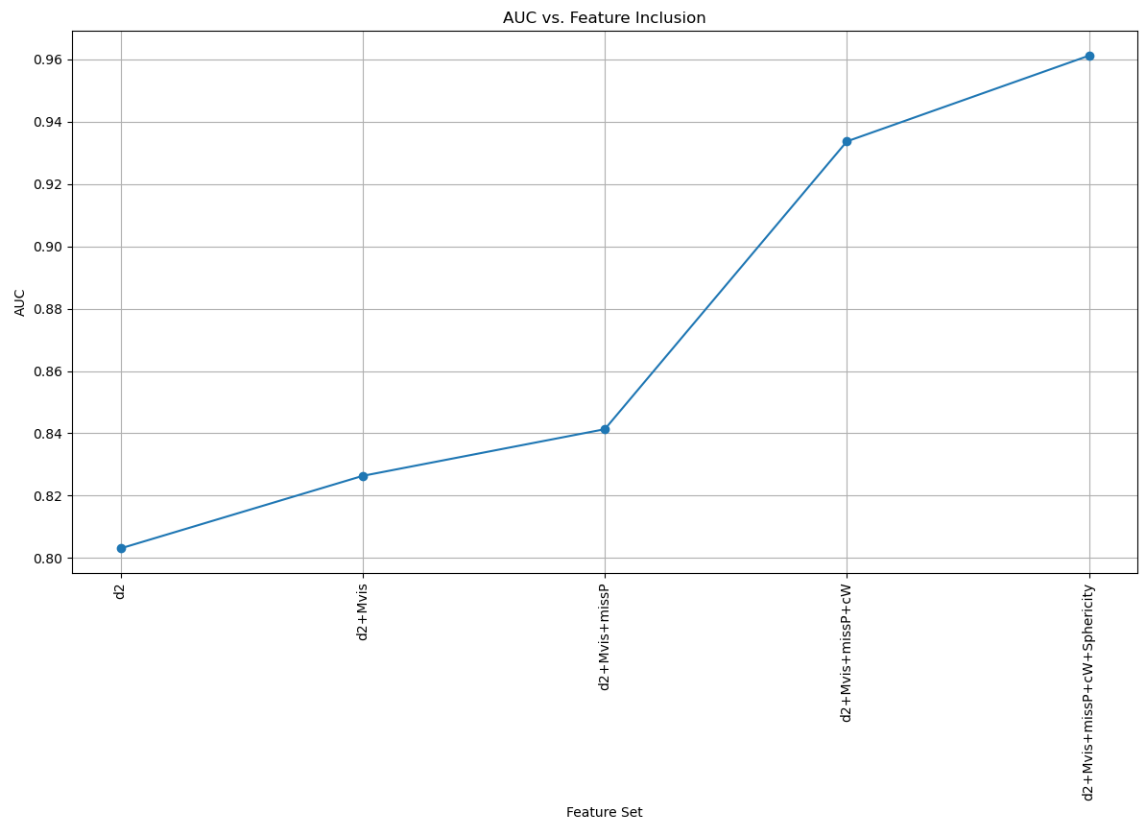}
    \end{center}
  \caption{The area under curve with the inclusion of different set of input features}
  \label{fig:BDT_auc}
\end{figure}

At the training stage, only the MC samples with \(\sqrt{s}=\SI{189}{\giga\eV}\) are selected as our training sample.
We adopted a 60-20-20 3-way split for training, testing and validation.
Event classification is performed with a gradient BDT using the XGBoost\cite{chen_xgboost_2016} classifier, set up for a binary classification task using logistic regression-style outputs.
The trees can grow up to depth 6.
The learning rate is set at 0.3, and we train 200 boosting rounds.
We’re using the ``hist'' tree method, which builds trees using histogram binning of
feature values instead of exact splits.
We also experiment with different sets of input variables using BDT, and Fig.~\ref{fig:BDT_auc} shows that the area under curve (AUC) reaches 0.85 when we include the 3 discriminating variables, which is the working setup we choose. The classifier performs better when either the sphericity or the cosine of the inter-jet angle\footnote{This refers to the smallest inter-jet angle among the four jets obtained when the event is clustered into four jets, as used in the jet clustering procedure for calculating $d^2$.} is included, but we decide to leave them out to avoid biasing the event shape further.

\begin{figure}[ht]
    \begin{center}
    \includegraphics[width=.32\linewidth]{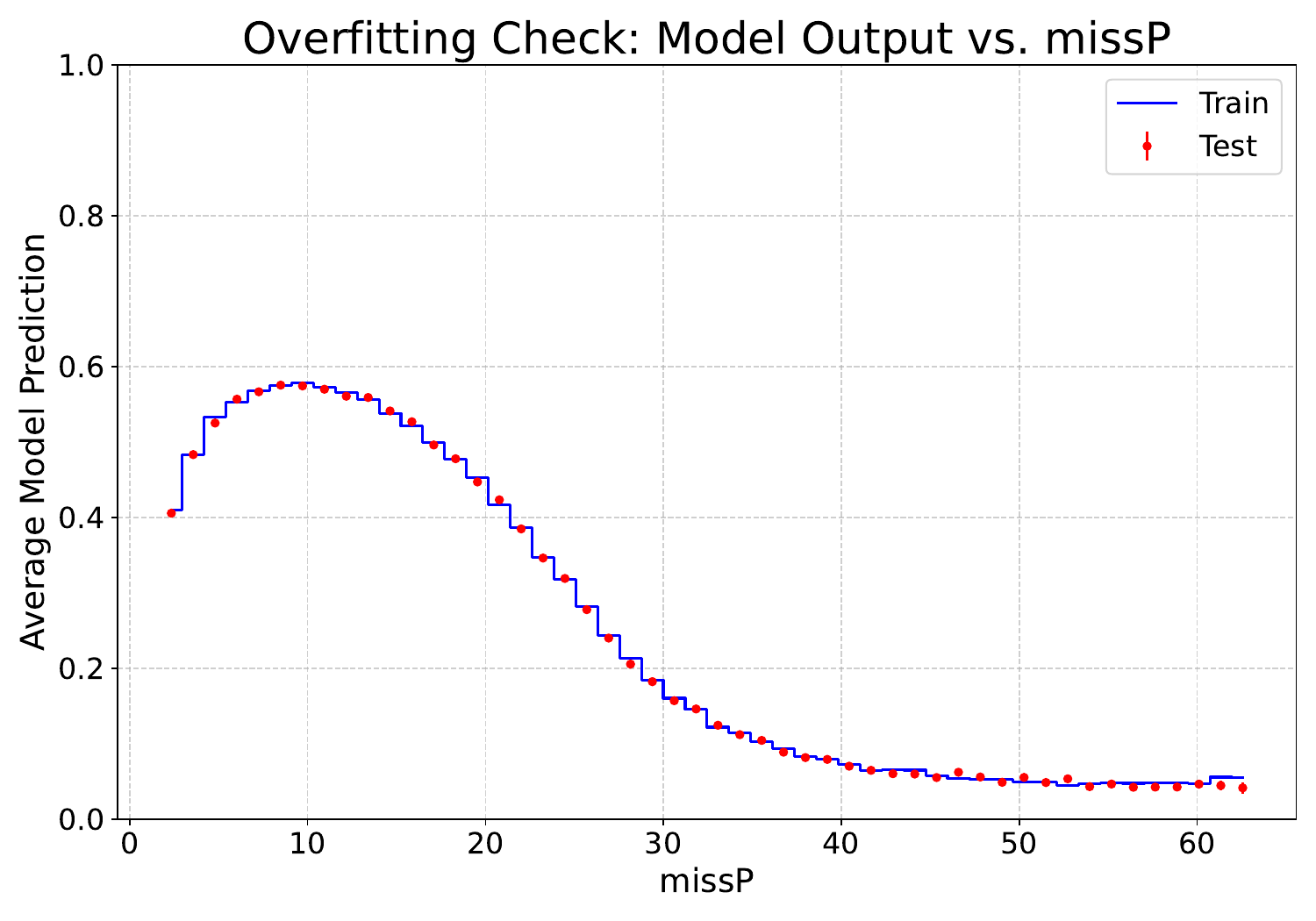}
    \includegraphics[width=.32\linewidth]{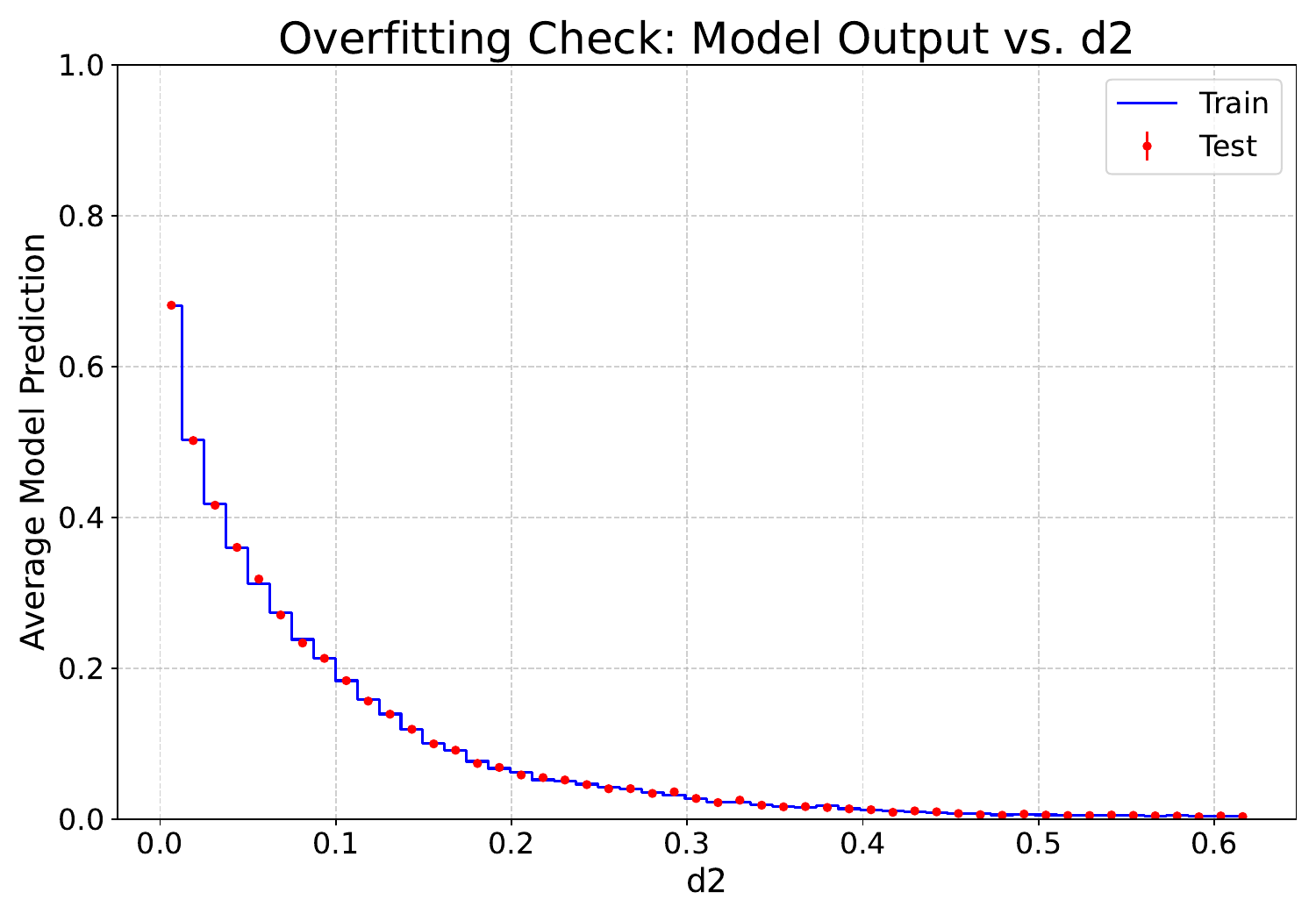}
    \includegraphics[width=.32\linewidth]{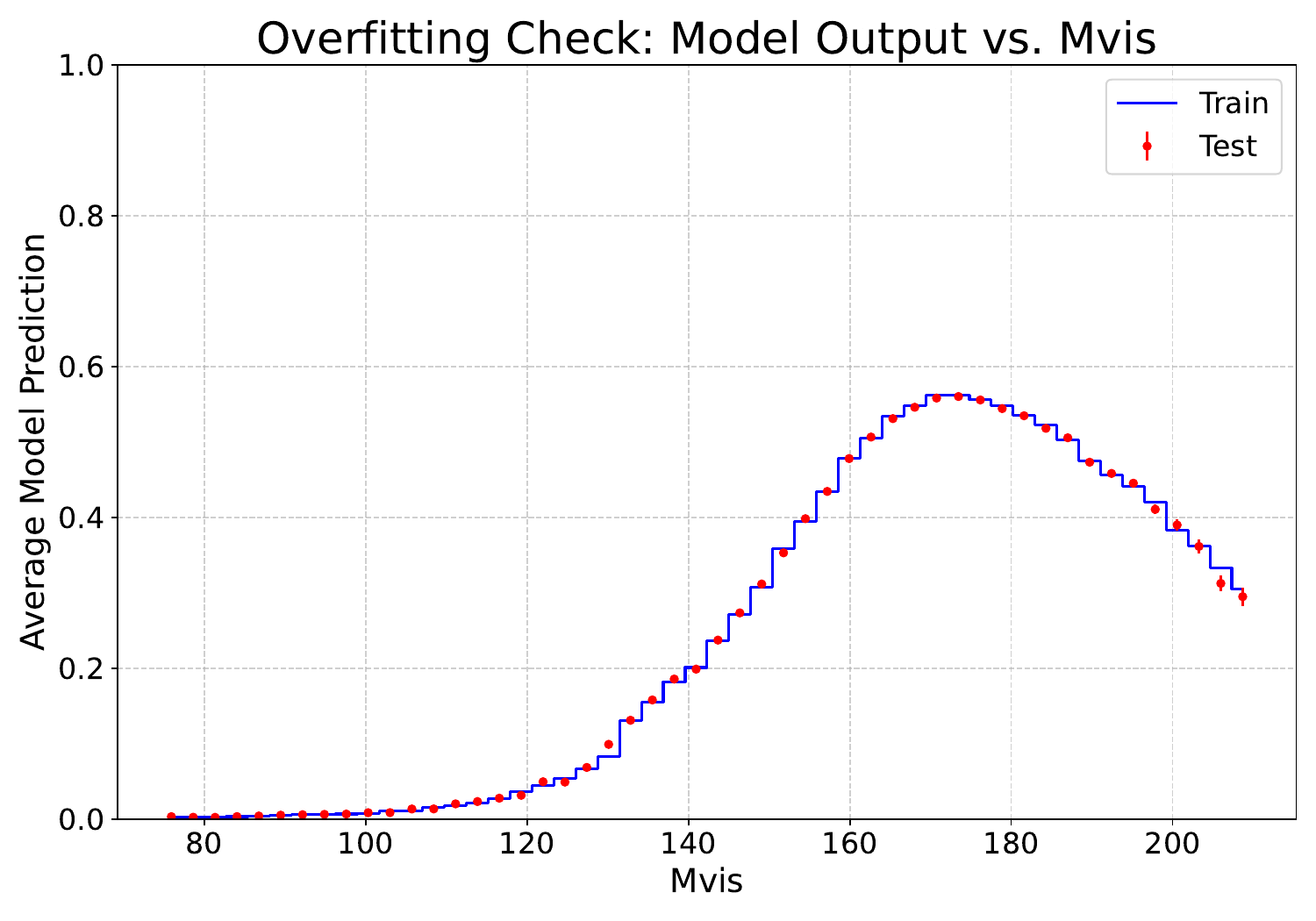}
    \end{center}
    \caption{Average BDT model output vs training features for the training dataset (blue
      lines) and test dataset (red dots). Left: missing momentum. Middle: \(d^2\). Right:
      missing mass. The high consistency between the training and the test dataset
      indicates that there is no overfitting.}
  \label{fig:BDT_overfit}
\end{figure}

With the training and testing sets, we also check for overfitting. In Fig.~\ref{fig:BDT_overfit}, the BDT output scores as a function of input features $\Vec{p}_{\rm miss}, d^2$ and $M_{\rm vis}$ for the training and the testing sets are overlaid. The compatible trend shows that no significant overfitting is detected.

\begin{figure}[ht]
    \begin{center}
    \includegraphics[width=.70\linewidth]{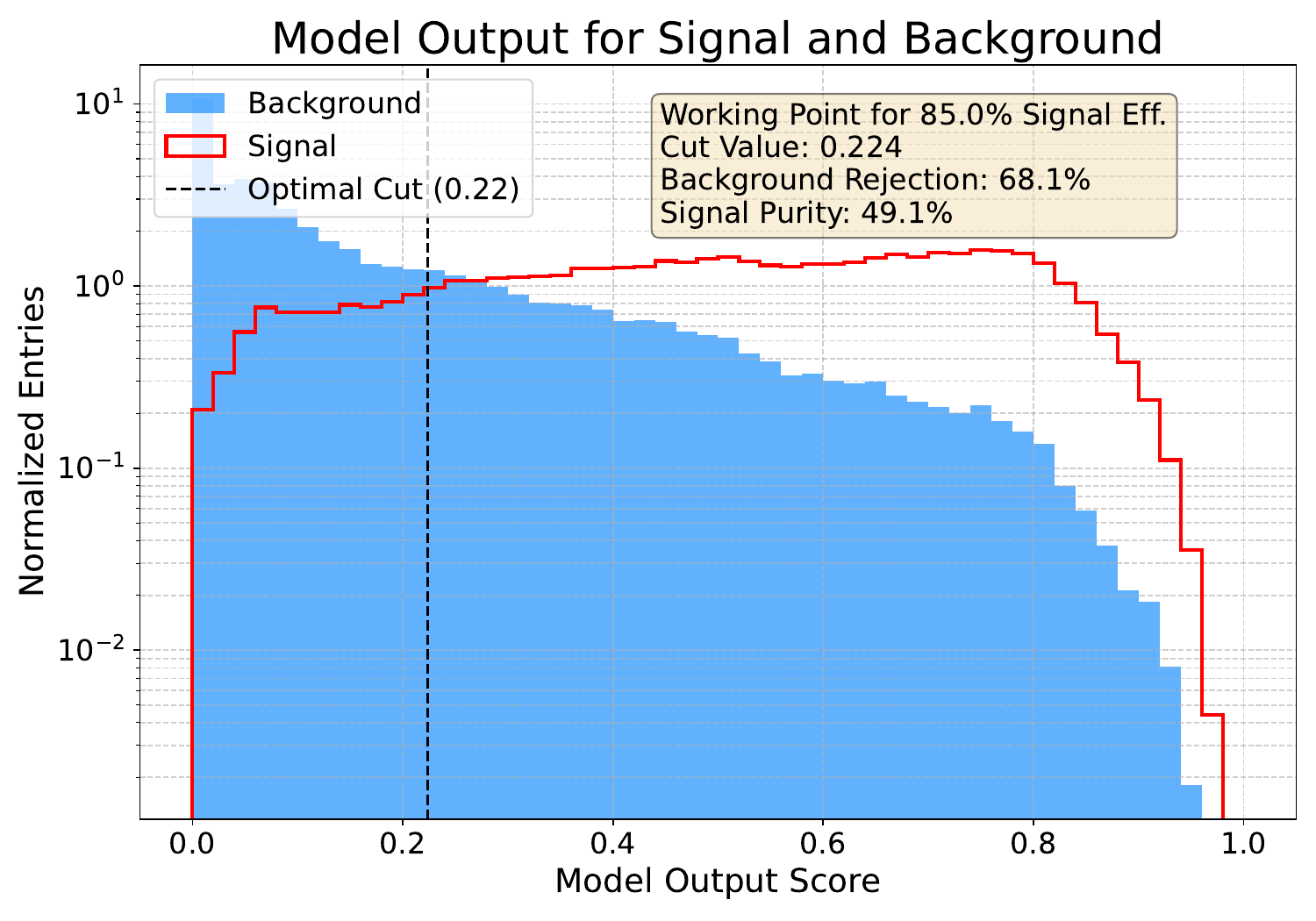}
    \end{center}
  \caption{Signal vs background model output score distribution.}
  \label{fig:BDT_output}
\end{figure}

In Fig.~\ref{fig:BDT_output}, we show the BDT output score distributions for the signal  (red curve) and background (blue histogram) events. A clear separation of the output BDT scores between the signal and background classes is seen.
In this work, we apply different signal-efficiency working points with the BDT classifier. The black line in Fig.~\ref{fig:BDT_output} indicates the chosen cut value, corresponding to a signal efficiency of $85\%$ in this example.

\subsection{Multiplicity-dependence of hadronic $WW$ signal fraction}
\label{sec:orgcf3e273}

\begin{figure}[ht]
\centering
    \begin{subfigure}[b]{0.47\textwidth}
    \centering
        \includegraphics[width=\textwidth]{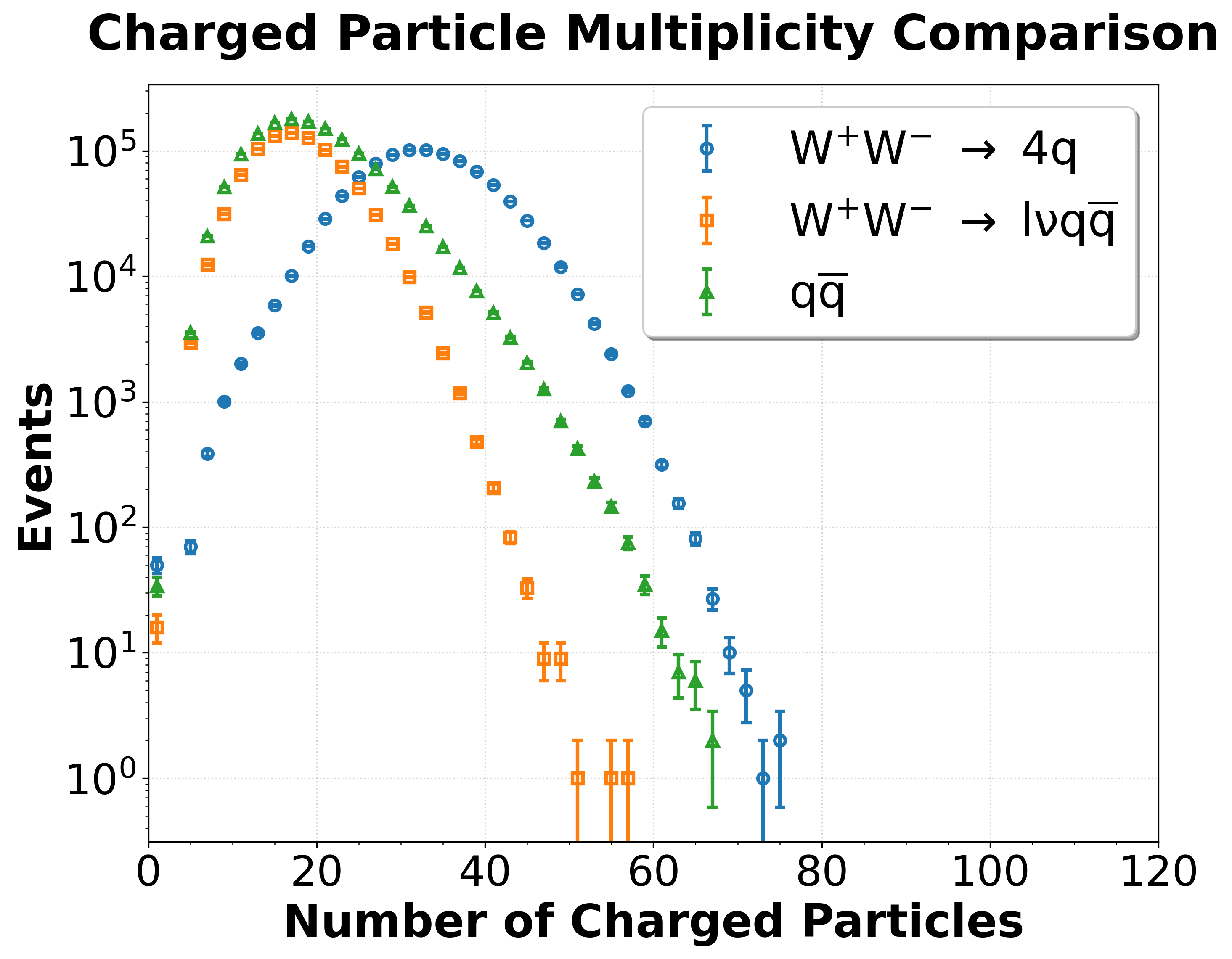}
        \caption{Normalized charged multiplicity distributions of \(\PWp\PWm \to 4\Pq\), \(\PWp\PWm \to \ell \Pnu \Pq \APquark\) and \(\Pep\Pem \to \Pq\APquark\).}
        \label{fig:ntrkoff_for_differentProcesses}
    \end{subfigure}
     \begin{subfigure}[b]{0.47\textwidth}
     \centering
        \includegraphics[width=\textwidth]{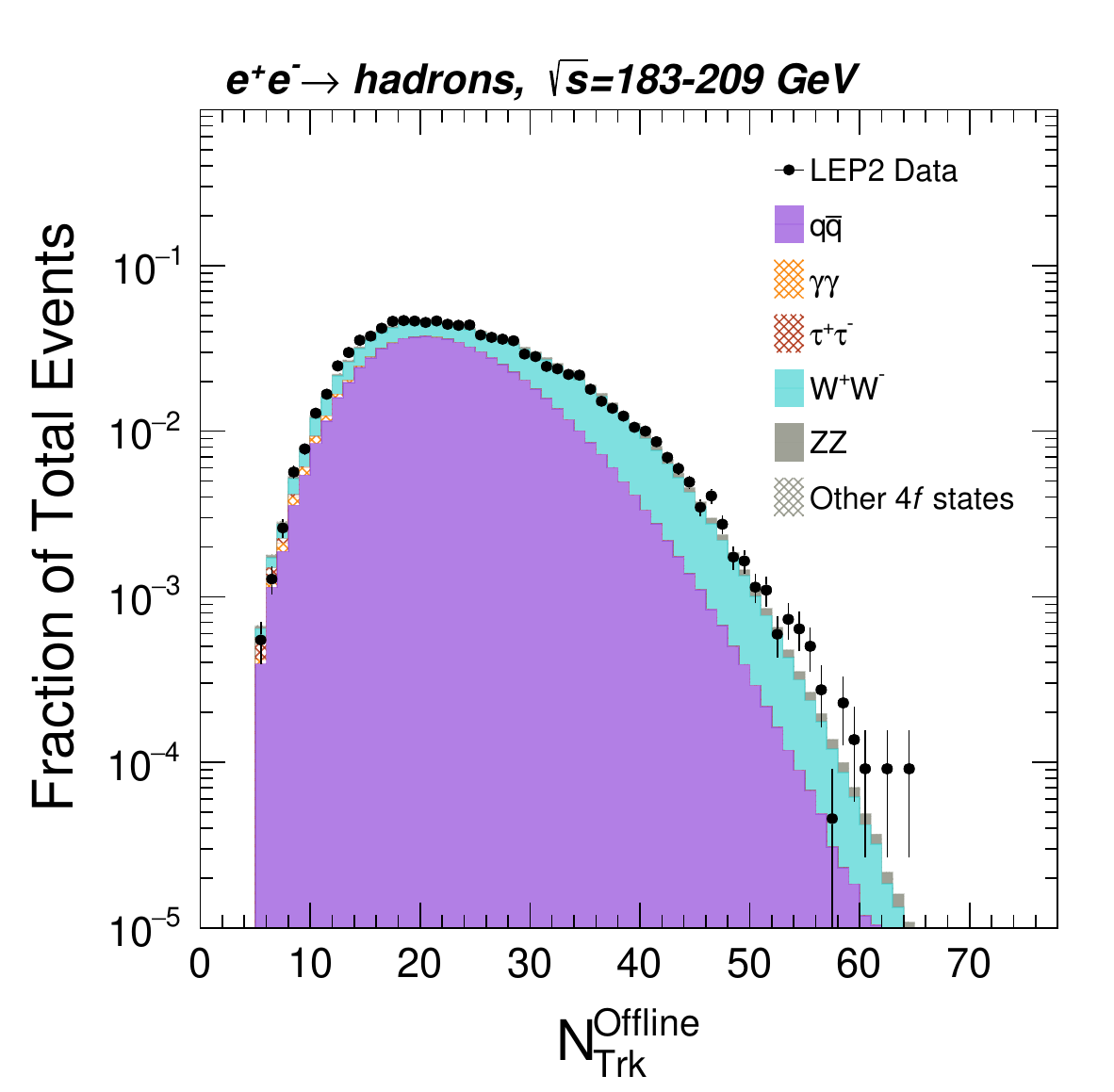}
        \caption{Total \ntrkoff}
        \label{fig:ntrkoff_merged}
    \end{subfigure}
\caption{The \ntrkoff multiplicity distributions of the sample at LEP-II energies. 
Figure (a) shows the normalized charged multiplicity distribution of \(\PWp\PWm \to 4\Pq\) (blue), \(\PWp\PWm \to \ell \Pnu \Pq \APquark\) (orange), and \(\Pep\Pem \to \Pq\APquark\) (green) MC events. Figure (b) shows the total \ntrkoff. Black error bars are data, overlaid with the stacked histogram in which each MC physics process is scaled by the expected cross section.}
\label{fig:HighEnergySampleBySources}
\end{figure}

The sample at LEP-II energies consists of a mixture of physics processes whose relative proportions vary strongly with charged-particle multiplicity. Figure~\ref{fig:ntrkoff_for_differentProcesses} shows the normalized $\ntrkoff$ distributions for the three dominant processes: $\PWp\PWm\to 4\Pq$ (blue), $\PWp\PWm\to \ell\Pnu\Pq\APquark$ (orange), and $\Pep\Pem\to \Pq\APquark$ (green). 
Because these three spectra peak at different positions, the fractional composition of the sample shifts continuously with $\ntrkoff$.
Figure~\ref{fig:ntrkoff_merged} shows the resulting stacked $\ntrkoff$ distribution overlaid on data, with each MC channel scaled by its production cross section: at low and intermediate multiplicity the sample is dominated by $\Pq\APquark$, while the hadronic $WW$ contribution grows toward higher $\ntrkoff$ and becomes the leading component in the high-multiplicity tail.

\begin{figure}[ht]
    \begin{center}
    \includegraphics[width=.8\linewidth]{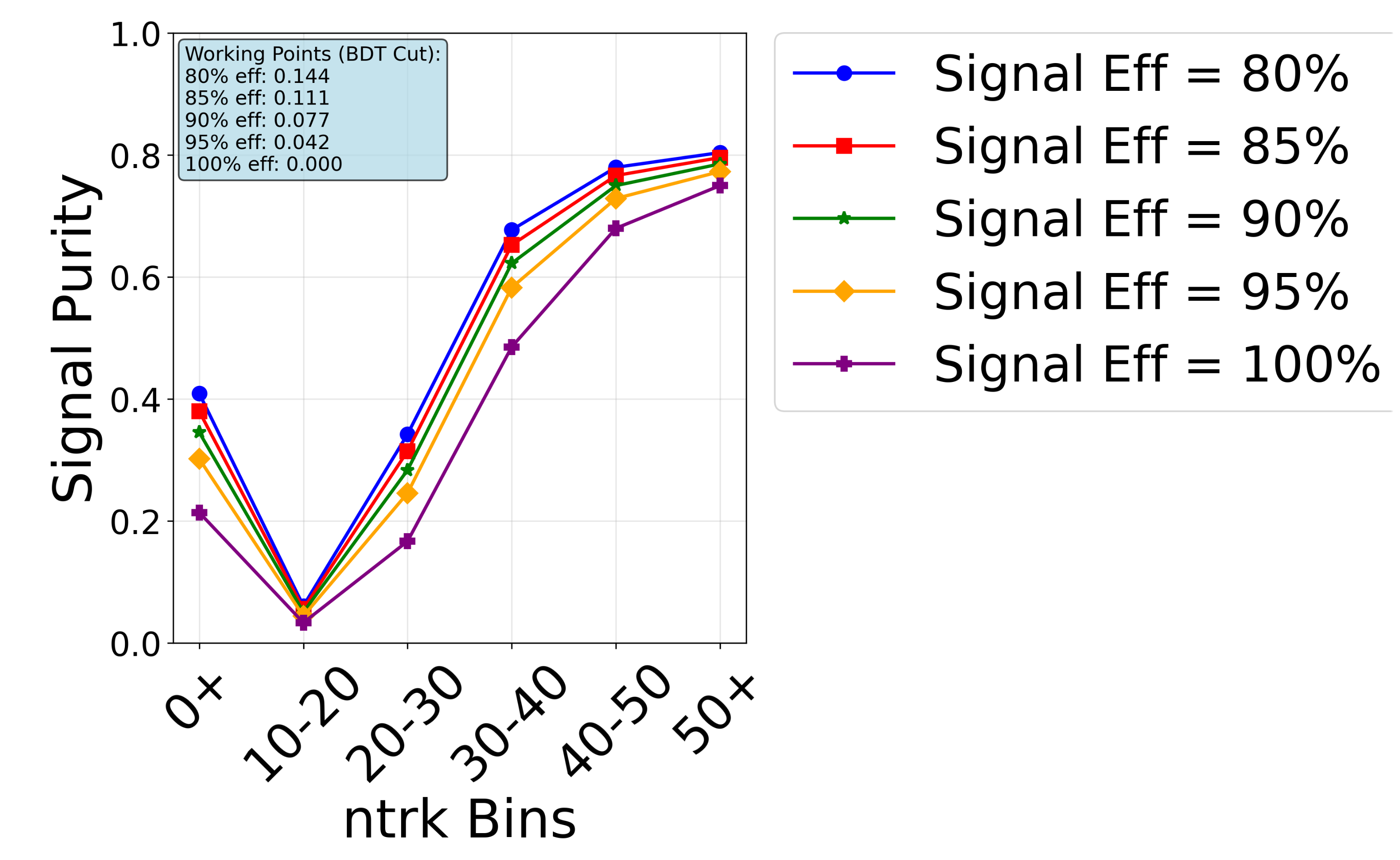}
    \end{center}
  \caption{Signal purity vs efficiency for different charged multiplicity classes.
    The leftmost column indicates the purity of the inclusive case.}
  \label{fig:BDT_purity}
\end{figure}

The system therefore exhibits a multiplicity-dependent signal fraction. We define the \emph{signal purity} as \begin{equation} \mathrm{purity} = \frac{N_{\text{had. diboson}}}{N_{\text{had. diboson}} + N_{\text{other}}}, \label{eqn:purity} \end{equation} i.e.\ the fraction of hadronic diboson events (hadronic $\PWp\PWm\to 4\Pq$ and $\PZ\PZ\to 4\Pq$) among the total signal-plus-background sample. Figure~\ref{fig:BDT_purity} shows the purity in bins of $\ntrkoff$ at several BDT signal-efficiency working points ($\epsilon_{\rm sig} = 80\%$, $85\%$, $90\%$, $95\%$, and $100\%$ corresponding to no BDT cut). For each curve, the leftmost point is the inclusive purity and the remaining points give the purity within the indicated multiplicity slice. Without any BDT cut (purple, $\epsilon_{\rm sig}=100\%$), the purity in $\ntrkoff \in [30,40)$ remains slightly below $50\%$, so this interval is not yet dominated by hadronic $WW$ events in the unselected sample. The purity then continues to climb with multiplicity, reaching $\sim\!75\%$ in the highest-multiplicity bin. Applying the BDT raises the purity in every multiplicity slice, but the size of the improvement depends on multiplicity. The largest gain appears at intermediate $\ntrkoff$: at the $\epsilon_{\rm sig}\approx 85\%$ working point, the purity in $[30,40)$ rises to $\sim\!65\%$, making this BDT-selected interval hadronic-$WW$ enriched. In the highest-multiplicity bin the gain shrinks to a few percent, because the sample is already hadronic-diboson dominated and there is little room for the BDT to push the purity higher.
In this analysis note, we focus on discussing the results from the $85\%$ signal efficiency working point. Other working points at signal efficiencies $95\%$, $90\%$, and $80\%$ are also attempted. The results evaluated at different working points are displayed in the Appendix.

\clearpage
\resetmarkpar
\section{Two-particle correlation function analysis}
\label{sec:TwoParticleCorrelationFunction}

We calculate the two-particle correlation function following the same formulation as in the previous measurement using the full LEP2 ALEPH data~\cite{Chen:2023njr, Chen:2023nsi}:
\begin{equation}
\begin{aligned}
\frac{1}{{\rm N}_{\rm trk}^{\rm corr}}\frac{d^2{\rm N}^{\rm pair}}{d\Delta\eta d\Delta\phi} &= C(\Delta \eta, \Delta \phi ) = B(0,0) \times \frac{S(\Delta\eta, \Delta\phi)}{B(\Delta\eta, \Delta\phi)},\\
\end{aligned}
\label{eqn:2PC}
\end{equation}

where ${\rm N}_{\rm trk}^{\rm corr}$ is the number of tracks after efficiency correction in the event, and ${\rm N}^{\rm pair}$ is the number of pairing yields associated with trigger particles.
The signal correlation function $S(\Delta\eta, \Delta\phi)$ counts the per-trigger-particle pairing yields within a single event and is expressed as
\begin{equation}
\begin{aligned}
S(\Delta\eta, \Delta\phi) &= \frac{1}{{\rm N}_{\rm trk}^{\rm corr}}\frac{d^2{\rm N}^{\rm same}} {d\Delta\eta d\Delta\phi}.
\end{aligned}
\label{eqn:2PCSig}
\end{equation}

The purpose of the background two-particle-correlation function \(B(\Delta\eta,
\Delta\phi)\) is to define the baseline -- the correlation without  collectivity under a certain event shape -- in a data-driven way, using the correlations formed by any two accepted tracks across all events.
In the literature, as well as the LEP2 analysis~\cite{Chen:2023njr, Chen:2023nsi},
the background two-particle correlation function
is exclusively constructed with the ``event-mixing'' technique.
With this method, the angular correlation is calculated between
the trigger particles in a signal event and the associated particles in another event in a
two-step process.
The first step is to form a ``mixed event'' by combining multiple events
drawn from a pool with similar properties, such as multiplicity, as the event containing the trigger track.
The random combinatorial background is then estimated by correlating the trigger track with all other tracks from the mixed event.
The background two-particle correlation function constructed in this fashion captures the
two-particle correlation distribution one would find if the pairs of particles are
independently drawn from the same underlying angular distribution.

This method, however, has technical limitations.
Event mixing is an operational approximation: the mixed-event pool contains a finite
number of candidate events, and the constructed background relies on the sampled
mixed events being statistically representative of the parent distribution within a
given event class.
This assumption becomes less robust in stringent bins, especially at high multiplicity,
where the number of eligible events can be small.
In the present analysis, a finite mixed-event sample may therefore underrepresent the
distribution of thrust-axis configurations in the parent sample, introducing additional
fluctuations or possible sampling bias in $B(\Delta\eta, \Delta\phi)$.
Constructing sufficiently large mixed-event ensembles is also computationally demanding
in both time and storage.

In this analysis, we therefore adopt an alternative construction of the background
correlation function.
This method employs the autocorrelation of the single-particle angular distribution
$S_{\text{1-part.}}(\eta,\phi)$, without relying on an explicit mixed-event pool.
In the continuous case, this is equivalent to taking the convolution

\begin{equation}
\begin{aligned}
B(\Delta\eta, \Delta\phi) = \frac{1}{{\rm N}_{\rm trk}^{\rm corr}} \int S_{\text{1-part.}}(\eta, \phi) \cdot S_{\text{1-part.}}(\eta + \Delta \eta, \phi + \Delta \phi)~d \eta d \phi.
\end{aligned}
\end{equation}

In practice, the single-particle spectrum $S_{\text{1-part.}} (\eta, \phi)$ is obtained
from the one-particle $(\eta, \phi)$ distribution in the given \ntrkoff multiplicity class.
The single-particle convolution implicitly includes each track paired with itself, which is subtracted via the same-event term so that the background retains only distinct-track combinations.
In the idealized limit in which all events in the mixing pool contribute to the
background estimate, this convolution construction is the deterministic analogue of
the same factorized baseline, but without the additional approximation associated with
drawing a finite mixed-event sample.
It therefore provides an empirical background function directly from the measured
single-particle spectrum, while reducing the extra fluctuations and possible
pool-sampling bias that can arise in small high-multiplicity event classes and
substantially improving computational efficiency.

\begin{figure}[ht]
    \begin{center}
    \begin{subfigure}[b]{0.32\textwidth}
        \includegraphics[width=\textwidth]{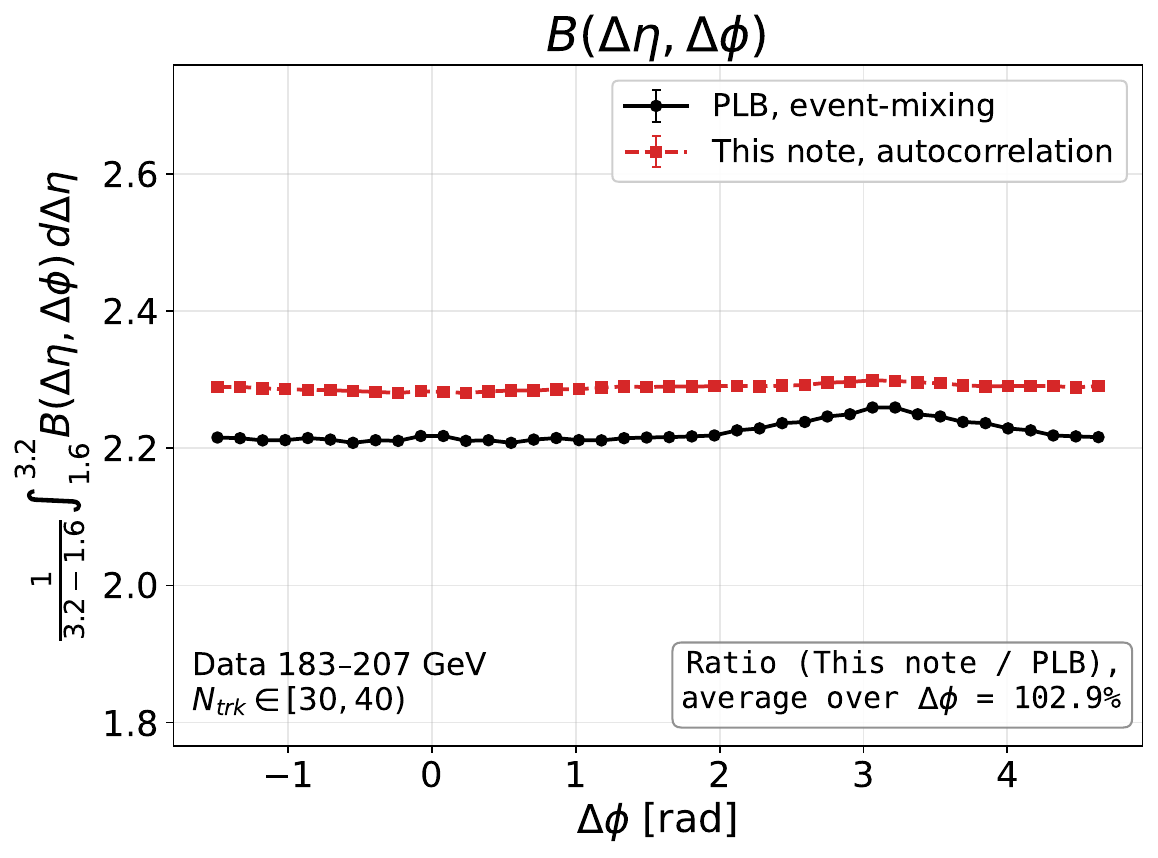}
        \caption{$N_{\mathrm{trk}}\in[30,40)$}
    \end{subfigure}
    \begin{subfigure}[b]{0.32\textwidth}
        \includegraphics[width=\textwidth]{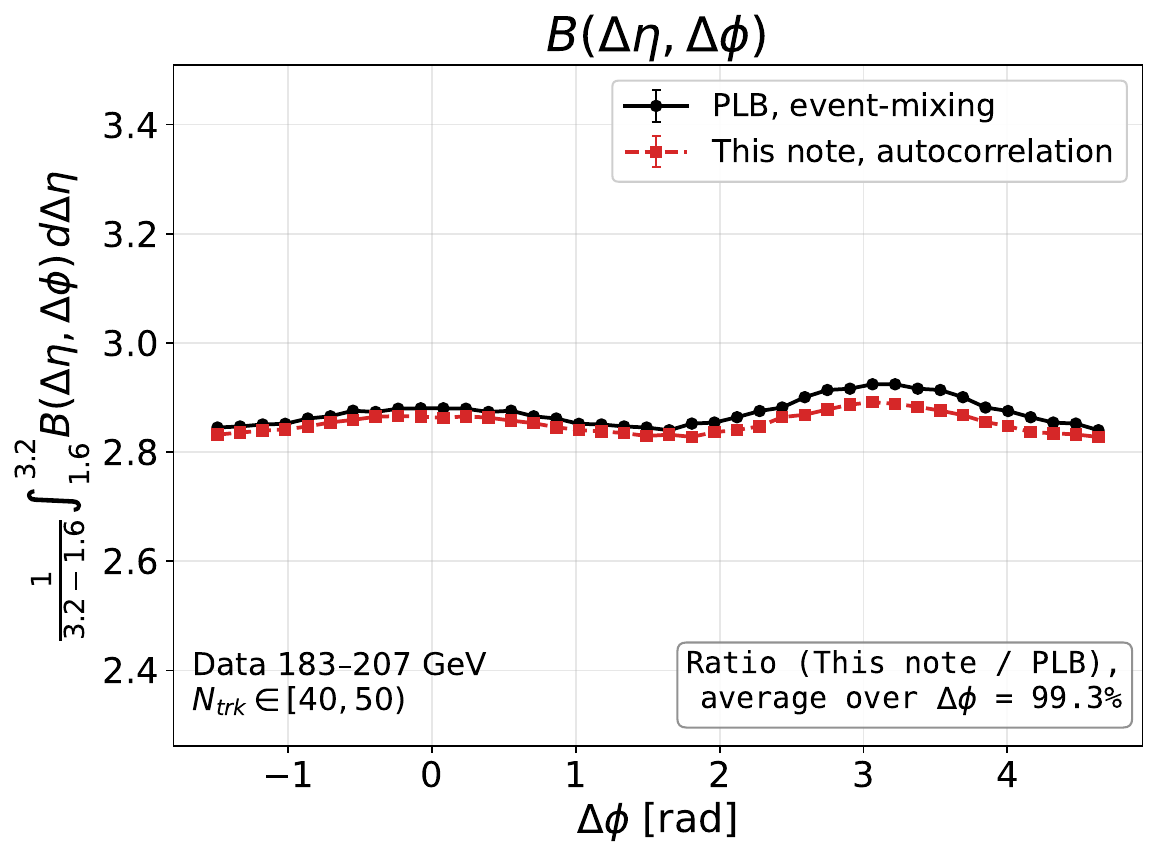}
        \caption{$N_{\mathrm{trk}}\in[40,50)$}
    \end{subfigure}
    \begin{subfigure}[b]{0.32\textwidth}
        \includegraphics[width=\textwidth]{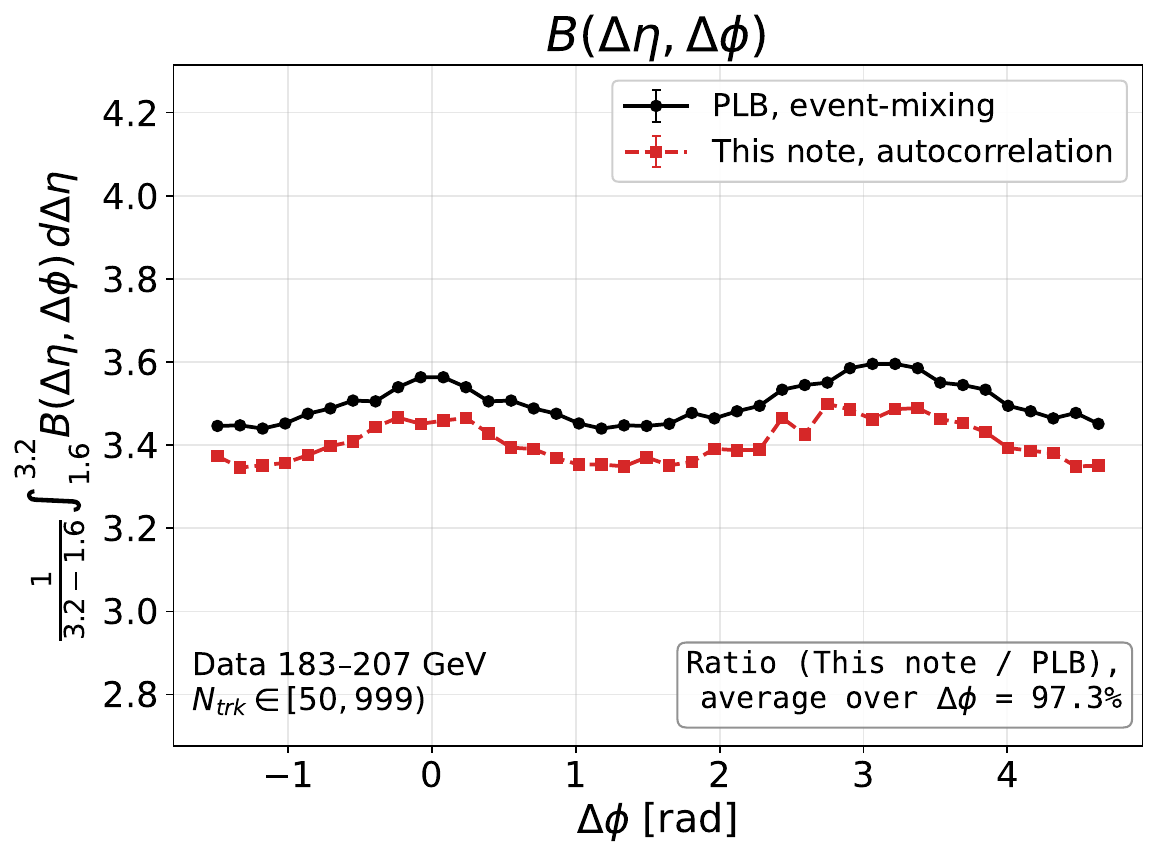}
        \caption{$N_{\mathrm{trk}}\in[50,999)$}
    \end{subfigure}
    \end{center}
    \caption{Closure test of the background $B(\Delta\eta,\Delta\phi)$: long-range
    ($1.6<|\Delta\eta|<3.2$) $\Delta\phi$ projection obtained with the autocorrelation
    construction of this note (red squares) and with the event-mixing baseline (black
    circles) in three multiplicity intervals.
    The $\Delta\phi$-averaged ratio (this note / baseline) is quoted in each panel.}
    \label{fig:val_bkgrnd}
\end{figure}

\begin{figure}[ht]
    \begin{center}
    \begin{subfigure}[b]{0.32\textwidth}
        \includegraphics[width=\textwidth]{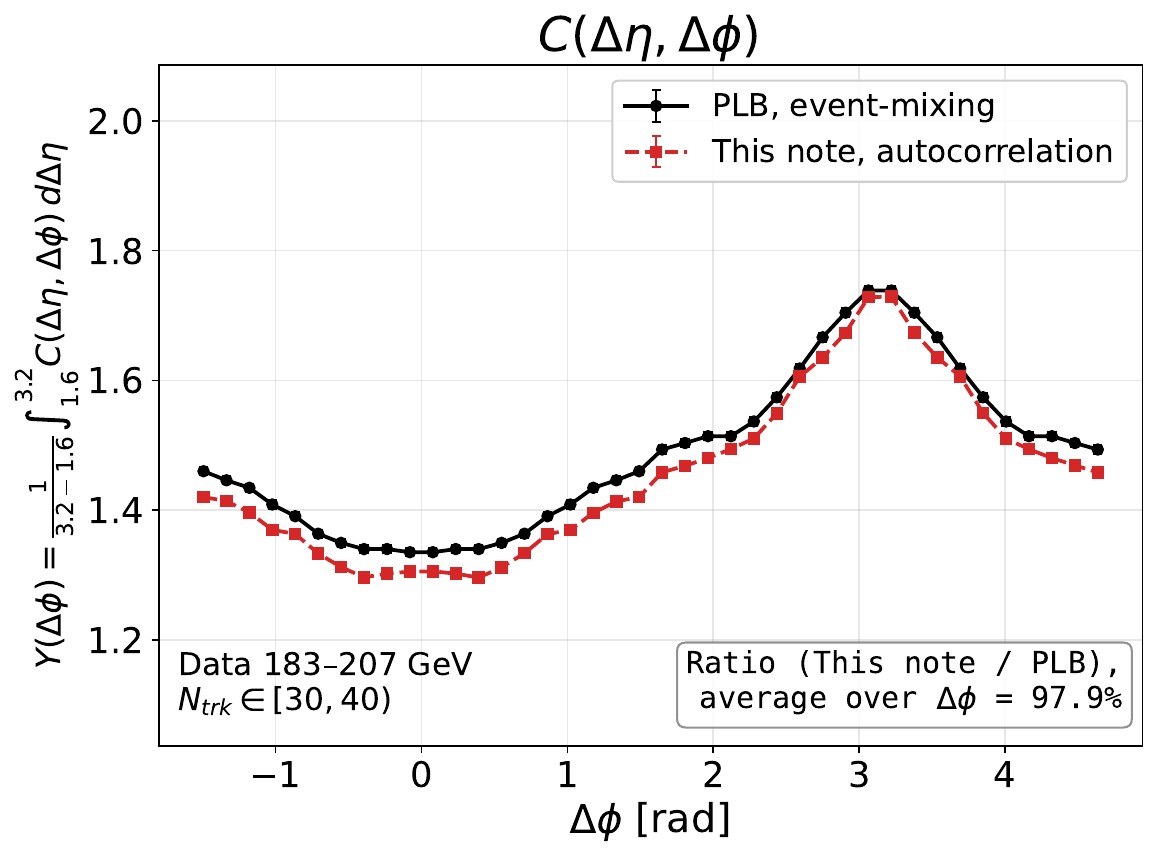}
        \caption{$N_{\mathrm{trk}}\in[30,40)$}
    \end{subfigure}
    \begin{subfigure}[b]{0.32\textwidth}
        \includegraphics[width=\textwidth]{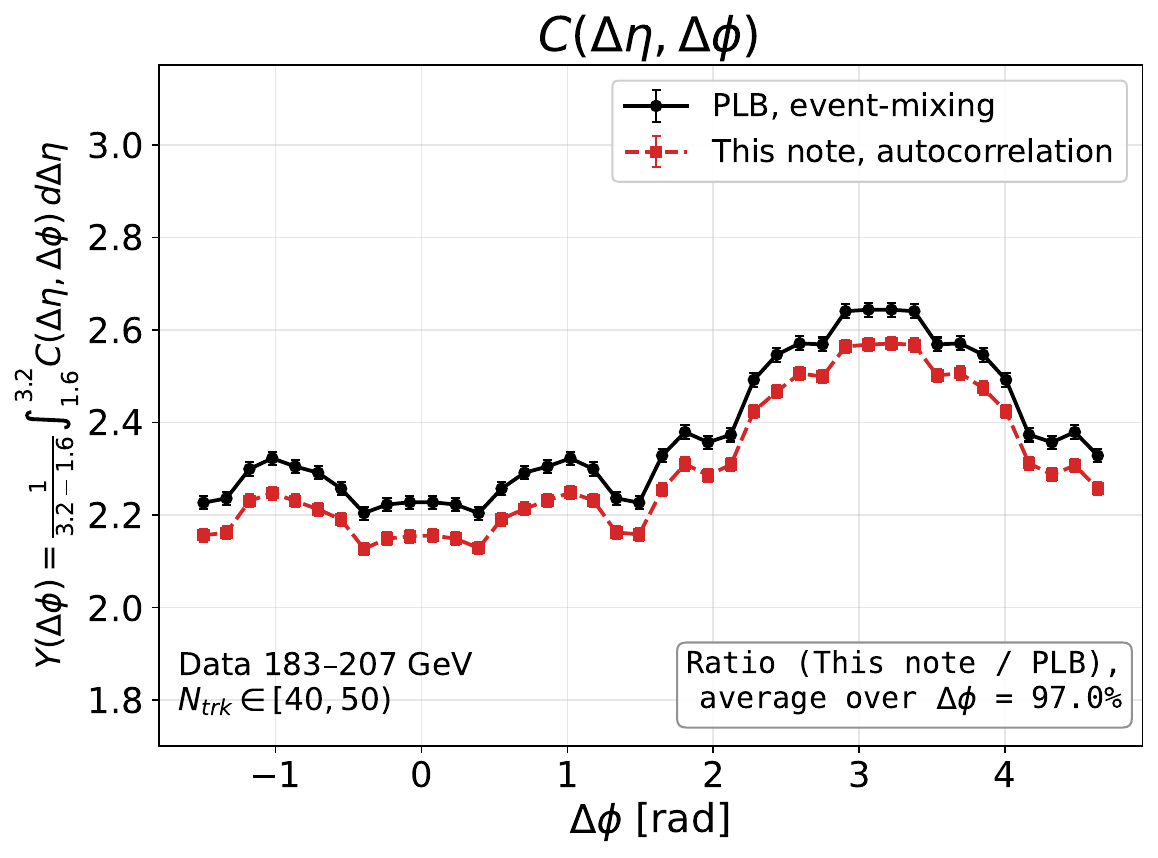}
        \caption{$N_{\mathrm{trk}}\in[40,50)$}
    \end{subfigure}
    \begin{subfigure}[b]{0.32\textwidth}
        \includegraphics[width=\textwidth]{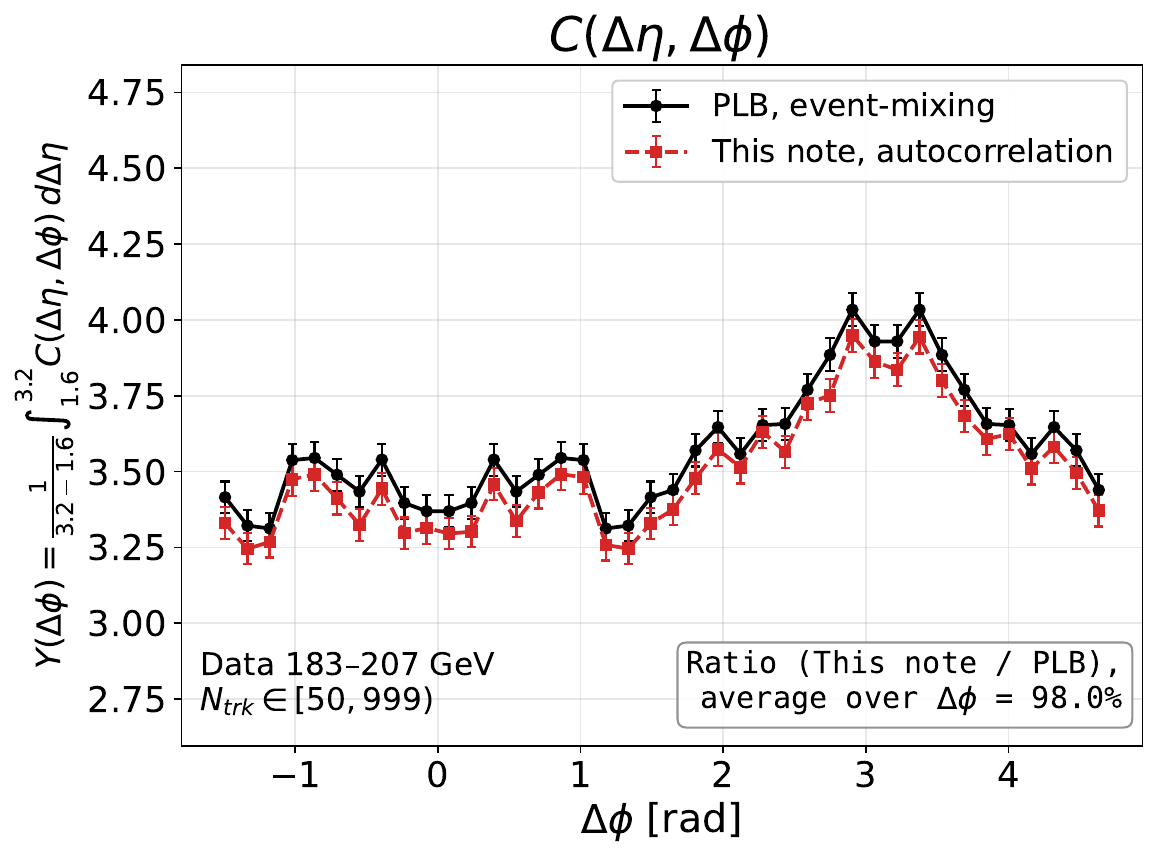}
        \caption{$N_{\mathrm{trk}}\in[50,999)$}
    \end{subfigure}
    \end{center}
    \caption{Closure test of the correlation function $C(\Delta\eta,\Delta\phi)$: long-range
    ($1.6<|\Delta\eta|<3.2$) $\Delta\phi$ projection obtained with the autocorrelation
    construction of this note (red squares) and with the event-mixing baseline (black
    circles) in three multiplicity intervals.
    The $\Delta\phi$-averaged ratio (this note / baseline) is quoted in each panel.}
    \label{fig:val_ratio}
\end{figure}

A quantitative closure test between the autocorrelation-based background construction used in this note and the event-mixing-based construction of the published baseline~\cite{Chen:2023njr, Chen:2023nsi} has been carried out on the same data sample at $\sqrt{s}=183$--$207$~GeV. Figures~\ref{fig:val_bkgrnd}~and~\ref{fig:val_ratio} overlay the long-range ($1.6<|\Delta\eta|<3.2$) $\Delta\phi$ projections of the background $B(\Delta\eta,\Delta\phi)$ and of the correlation function $C(\Delta\eta,\Delta\phi)$ obtained with the two methods, in three multiplicity intervals $N_{\mathrm{trk}}\in[30,40)$, $[40,50)$, and $[50,999)$. The two background estimates agree to within $3\%$ (averaged over $\Delta\phi$: $103\%$, $99\%$, and $97\%$ of the mixed-event reference, respectively), and the resulting correlation functions agree to within $2$--$3\%$ ($98\%$, $97\%$, and $98\%$), with the full $\Delta\phi$ shape---including the near-side and away-side structure---reproduced in all bins. The two constructions are not identical: event mixing is an operational approximation of the combinatorial background, whereas the autocorrelation method builds it from the single-particle spectra of the same events. The residual difference between them is estimated at the $\sim3\%$ level and is approximately flat in $\Delta\phi$, so it does not bias the extracted $\Delta\phi$ modulation. This confirms that the autocorrelation method reproduces the event-mixing baseline within its approximations and validates the methodology against the published result.

The detail is given below in Algorithm~\ref{algo}.
Finally, the normalization factor $B(0,0)$ in Eq.~\ref{eqn:2PC} accounts for the arbitrary normalization of the constructed background function $B(\Delta\eta, \Delta\phi)$.

\begin{algorithm}[H]
\SetAlgoLined
\KwData{$S$ (single-particle histogram), $N_\eta$, $N_\phi$}
\KwResult{$B$ (background 2-particle histogram)}

\For{$i_\eta := 0$ \KwTo $N_\eta-1$}{
  \For{$i_\phi := 0$ \KwTo $N_\phi-1$}{
    \For{$j_\eta := 0$ \KwTo $N_\eta-1$}{
      \For{$j_\phi := 0$ \KwTo $N_\phi-1$}{
        $\Delta\eta := j_\eta - i_\eta + \epsilon$\;
        $\Delta\eta' := i_\eta - j_\eta + \epsilon$\;
        $\Delta\phi := (j_\phi - i_\phi + \epsilon)\bmod N_\phi$\;
        $\Delta\phi' := (i_\phi - j_\phi + \epsilon)\bmod N_\phi$\;
        \eIf{$j_\eta = i_\eta$ \textbf{and} $j_\phi = i_\phi$}{
        $C := S[i_\eta][i_\phi] \times$ Average($S[j_\eta - 2\,\KwTo\, j_\eta + 2][j_\phi - 2 \,\KwTo\, j_\phi + 2]$)\;
        }{
        $C := S[i_\eta][i_\phi] \times S[j_\eta][j_\phi]$\;
        }
        $B[\Delta\eta][\Delta\phi] := B[\Delta\eta][\Delta\phi] + C$\;
        $B[\Delta\eta'][\Delta\phi] := B[\Delta\eta'][\Delta\phi] + C$\;
        $B[\Delta\eta][\Delta\phi'] := B[\Delta\eta][\Delta\phi'] + C$\;
        $B[\Delta\eta'][\Delta\phi'] := B[\Delta\eta'][\Delta\phi'] + C$\;
      }
    }
  }
}
\caption{Construct the background 2-particle correlation histogram\\
\footnotesize{$\epsilon$ is a tiny positive offset (${\cal O}(10^{-6})$ in bin-index space) added when forming $\Delta\eta$ and $\Delta\phi$ bin indices.
It prevents ambiguities at bin boundaries and ensures that pairs with $\Delta\eta=0$ are unambiguously assigned to the correct background histogram cell when the difference $j-i$ is evaluated as a floating-point or wrapped index.
The same offset is applied consistently to the $(\Delta\eta,\Delta\phi)$ and $(-\Delta\eta,-\Delta\phi)$ mirror entries filled in lines 7--10 of the algorithm.}}
\label{algo}
\end{algorithm}

As discussed in the previous two-particle correlation analyses in the \ee collisions~\cite{Badea:2019vey,Belle:2022fvl,The:2022lun,Chen:2023njr, Chen:2023nsi}, we are particularly interested in the correlation function analyzed in the ``thrust-axis'' coordinate.
The event thrust~\cite{PhysRevLett.39.1587} is defined in the center-of-mass frame as
\begin{equation}
\begin{aligned}
T &\equiv \max_{\hat{n}} \frac{\sum\nolimits_i |\vec{p}_i \cdot \hat{n}|}{\sum\nolimits_i |\vec{p}_i|}, \\
\hat{n}_{T} &\equiv \operatorname*{arg\,max}_{\hat{n}} \frac{\sum\nolimits_i |\vec{p}_i \cdot \hat{n}|}{\sum\nolimits_i |\vec{p}_i|},
\end{aligned}
\label{eqn:Thrust}
\end{equation}
where \(T\) is the scalar thrust value, \(\hat{n}_{T}\) is the resulting thrust axis, and \(\vec{p}_i\) is the momentum of the \(i\)-th reconstructed visible energy-flow object. The sum includes reconstructed charged and neutral particles used in the event-shape calculation; the missing momentum \(\vec{p}_{\rm miss}\) is not included in the thrust maximization. The thrust axis captures the dominant visible outgoing energy-flow direction at the reconstruction level. In the thrust-axis analysis, the new reference $z$-axis is the thrust axis $\hat{n}_{T}$, and the choice of $\phi=0$ (or new reference $x$-axis) is assigned with $\hat{n}_{T}\times(\hat{n}_{T}\times\hat{z})$.
Particles' pseudorapidity and azimuth positions are then re-defined with respect to the new reference frame, and are used for calculating two-particle angular separations in the thrust-axis analysis.
The thrust axis coordinate is a reference frame more sensitive to the particle fragmentation activity surrounding the outgoing final-state direction at the parton level in the \ee system.
For the dominant \(\Pep\Pem \to \Pq\Paq\) topology in LEP1 dataset, this axis is closely related to the single color-string / two-jet event geometry.
For hadronic \(\PWp\PWm \to \Pq\Paq\Pq\Paq\) events, there is no unique axis associated with one individual color string, since two hadronic \(W\) decays contribute to the final state.
Nevertheless, the thrust axis remains a well-defined global event axis that summarizes the dominant energy flow of the full reconstructed final state.
This makes it a practical choice for the present study: it preserves continuity with the published LEP1 and LEP2 thrust-axis correlation measurement~\cite{Badea:2019vey,Chen:2023njr,Chen:2023nsi}, avoids introducing an additional event-by-event combinatorial choice into the correlation observable itself, and permits a direct comparison between the inclusive and \(\mathrm{WW}\)-enhanced selections.

Similarly to the previous \ee two-particle correlation measurements~\cite{Badea:2019vey,Belle:2022fvl,The:2022lun,Chen:2023njr, Chen:2023nsi},  we explore from the short-$\Delta\eta$-range to the long-$\Delta\eta$-range projection, azimuthal differential associated yields (one-dimensional correlations in \dphi)
\begin{equation}
Y(\Delta\phi) = \frac{1}{{\rm N}_{\rm trk}^{\rm corr}}\frac{d{\rm N}^{\rm pair}}{d\Delta\phi}= \frac{1}{\Delta\eta_{\rm max}-\Delta\eta_{\rm min}}\int\limits_{\Delta\eta_{\rm min}}^{\Delta\eta_{\rm max}} \frac{1}{{\rm N}_{\rm trk}^{\rm corr}}\frac{d^2{\rm N}^{\rm pair}}{d\Delta\eta d\Delta\phi}d\Delta\eta.
\label{eqn:DeltaPhiAssociatedYield}
\end{equation}
Three $\Delta\eta$ regions are studied: short range ($0 \le |\Delta\eta| < 1$), middle range ($1 \le |\Delta\eta| < 1.6$) and long range ($1.6 \le |\Delta\eta| < 3.2$). The long-range azimuthal differential associated yield is denoted as $Y_l(\Delta \phi)$ in this analysis note.

\subsection{Correction of Tracking Efficiency}
\label{sec:Corrections}

\begin{figure}[ht]
\centering
    \includegraphics[width=0.48\textwidth]{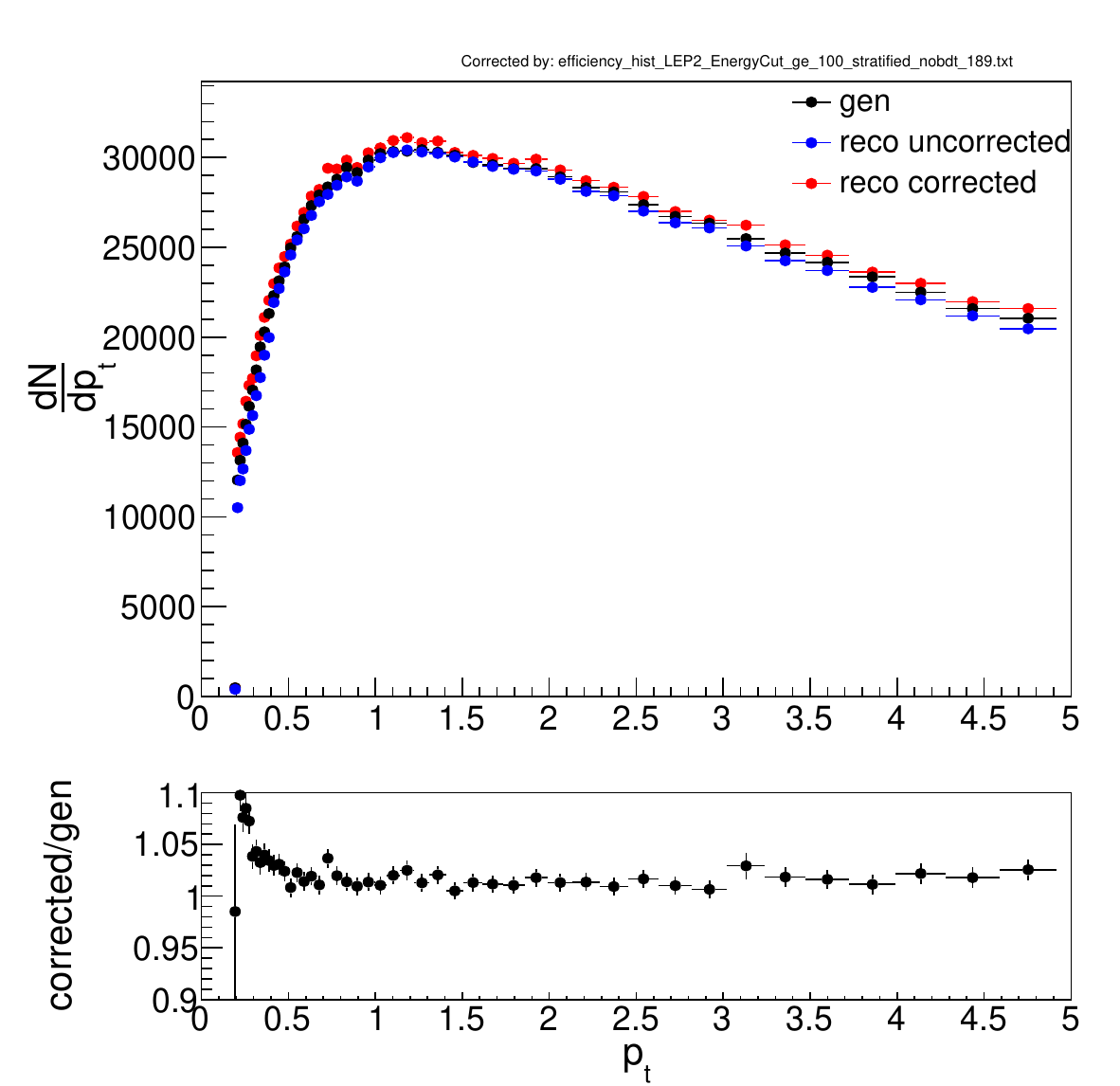}
    \includegraphics[width=0.48\textwidth]{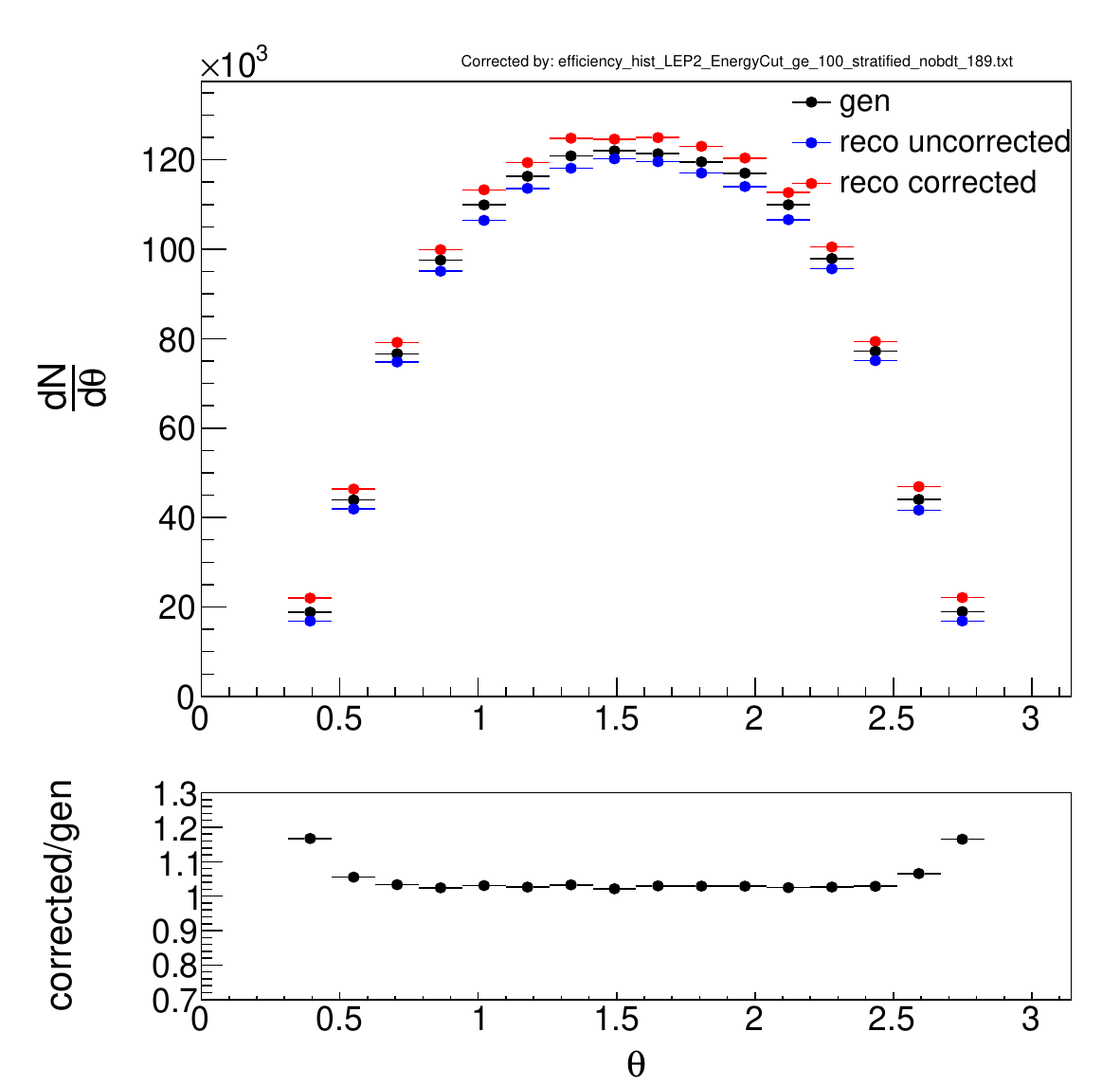}
    \includegraphics[width=0.48\textwidth]{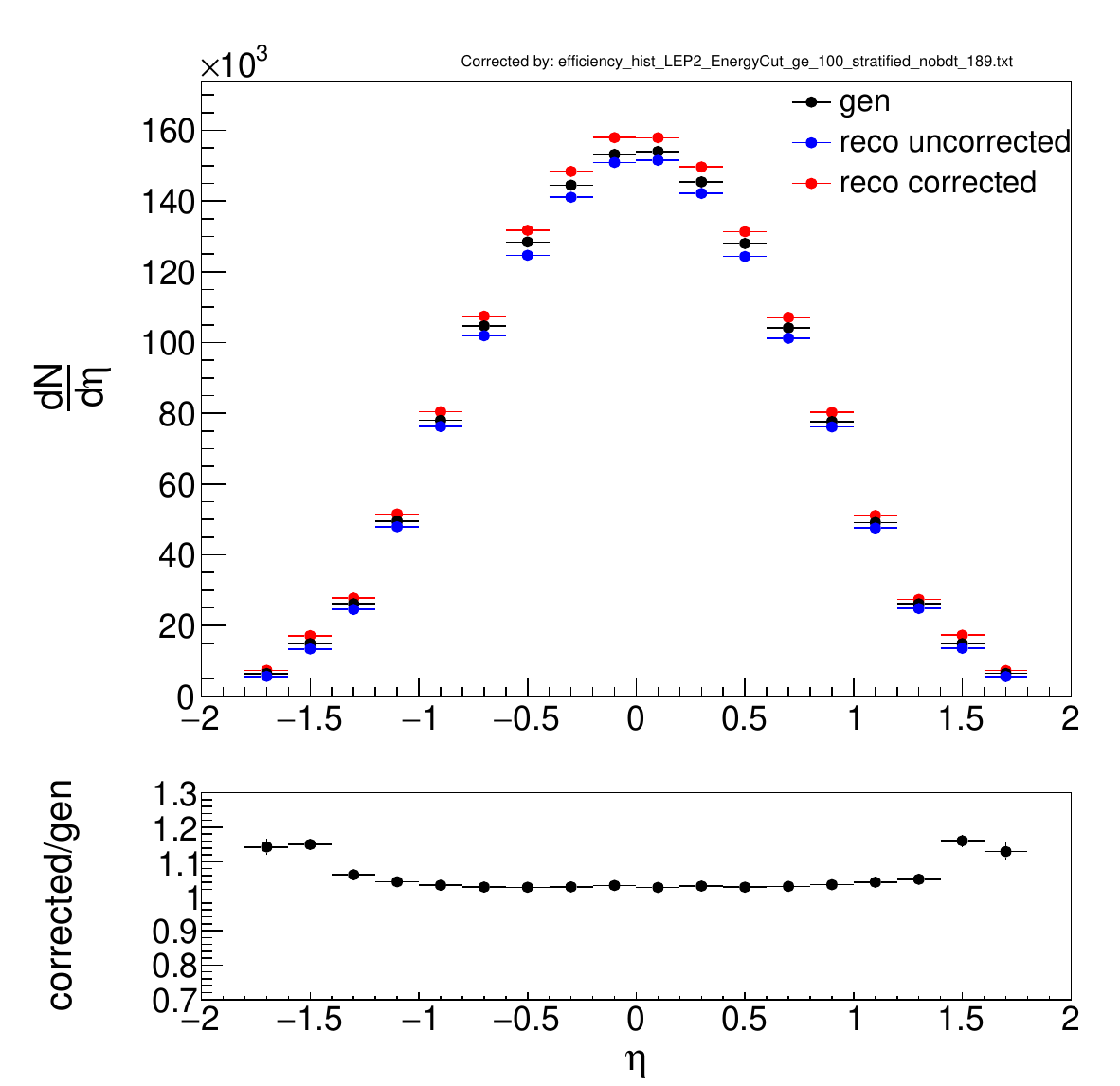}
    \includegraphics[width=0.48\textwidth]{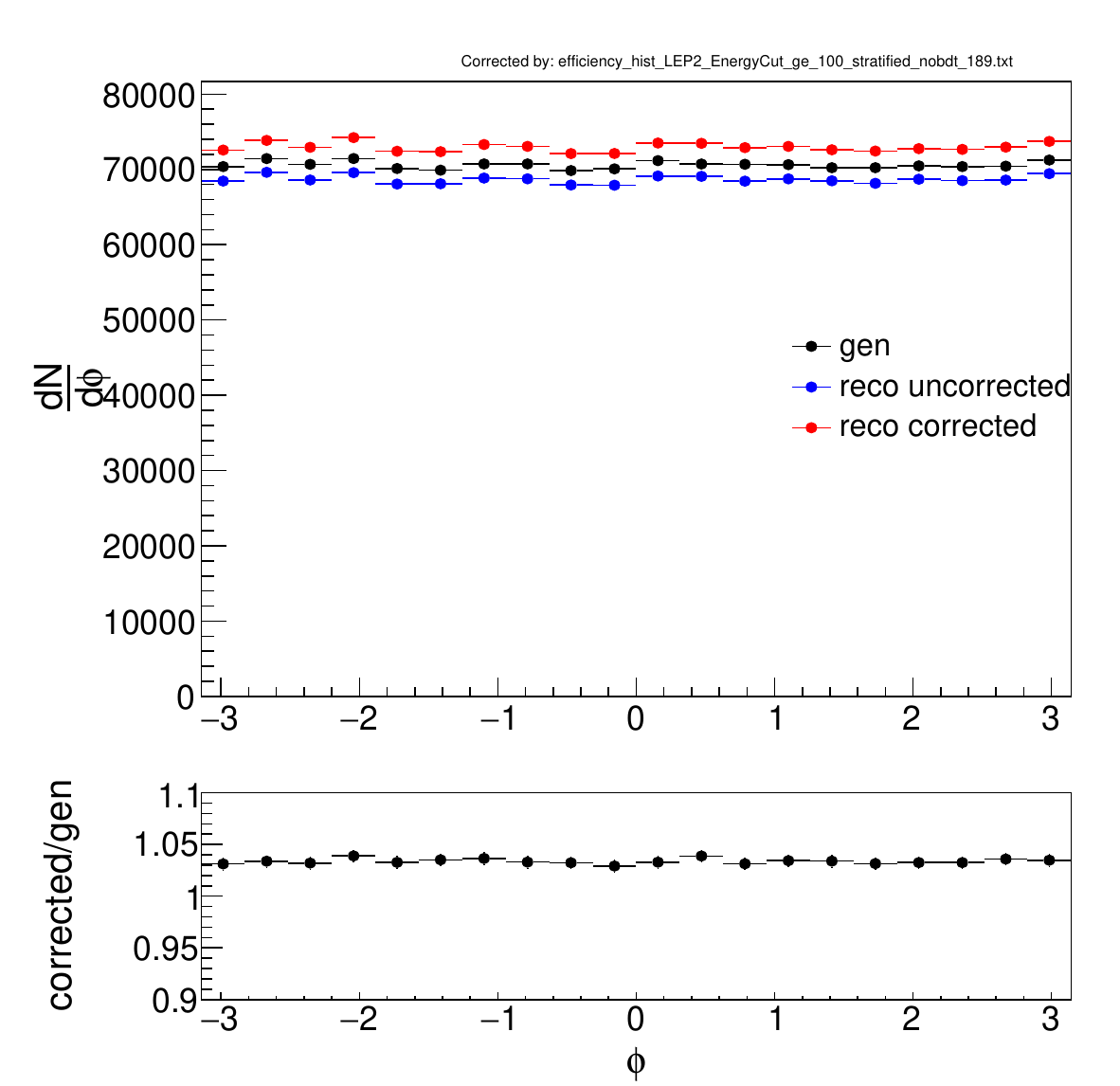}
    \caption{Tracking-efficiency closure check in the MC sample for $\sqrt{s} = 183$--$207~\mathrm{GeV}$. Generator-level spectra (black) are compared with the reconstruction-level spectra before (blue) and after (red) applying the efficiency correction, shown as projections in \(p_{\rm T}\) (top left), \(\theta\) (top right), \(\eta\) (bottom left), and \(\phi\) (bottom right). The corrected spectra recover the generator-level shapes more closely over the populated acceptance region.}
\label{fig:tracking_efficiency_closure}
\end{figure}

Similar to the previous LEP2 analysis~\cite{Chen:2023njr,Chen:2023nsi}, we build the tracking efficiency corrector with the MC-based efficiency (in Section~\ref{sec:Sample}) by
\begin{equation}
\varepsilon(p_{\rm T}, \theta, \phi, {\rm N}_{\rm Trk}^{\rm Offline}) = \left[\frac{d^3 {\rm N}^{\rm reco}}{dp_{\rm T} d\theta d\phi}/\frac{d^3 {\rm N}^{\rm gen}}{dp_{\rm T} d\theta d\phi}\right]_{{\rm N}_{\rm Trk}^{\rm Offline}},
\end{equation}
where \( {\rm N}^{\rm reco} \) denotes the number of charged particles at the reconstruction level, and \( {\rm N}^{\rm gen} \) denotes the same at the generator level.
In this extended analysis, the efficiency correction is derived separately for each sub-sample corresponding to different center-of-mass energies.
The efficiency correction factor establishes a correspondence between the reconstruction level and the generator level across the \( p_{\rm T}, \theta, \) and \( \phi \) spectra.
The reconstructed tracks are reweighted by the inverse of the tracking efficiency, which brings the spectra of the reconstructed tracks closer to reflecting the geometrical acceptance of the generator.
Figure~\ref{fig:tracking_efficiency_closure} compares the generator-level single-particle spectra with the reconstruction-level spectra before and after the efficiency correction in the MC sample for $\sqrt{s} = 183$--$207~\mathrm{GeV}$.
The corrected spectra more closely reproduce the generator-level shapes across the populated \(p_{\rm T}\), \(\theta\), \(\eta\), and \(\phi\) ranges, providing a closure check for the track-level correction used in this analysis.

\subsection{Results}
\label{sec:Rst}
We measure the thrust-axis two-particle correlation function in bins of \ntrkoff multiplicity classes.
The previous LEP2 publication~\cite{Chen:2023njr,Chen:2023nsi} targeted the high-multiplicity regime, specifically $\ntrkoff \ge 50$, to maximize sensitivity to the long-range near-side ridge, and reported tentative hints of such a structure in data.
In this note we improve the statistical treatment by incorporating bootstrap bin-to-bin covariances of the long-range azimuthal yield (Sec.~\ref{sec:flow_bootstrap}) and revisiting the significance of the modulation in the inclusive no-BDT configuration.
This correlated-error treatment inflates the overall statistical uncertainty by roughly a factor of two relative to an uncorrelated-bin estimate, and the data--MC difference in the long-range $\Delta\phi$ modulation is no longer statistically significant.
These findings underscore the importance of propagating bin-to-bin correlations when interpreting the long-range azimuthal structure.

\begin{figure}[ht]
\centering
    \includegraphics[width=0.45\textwidth]{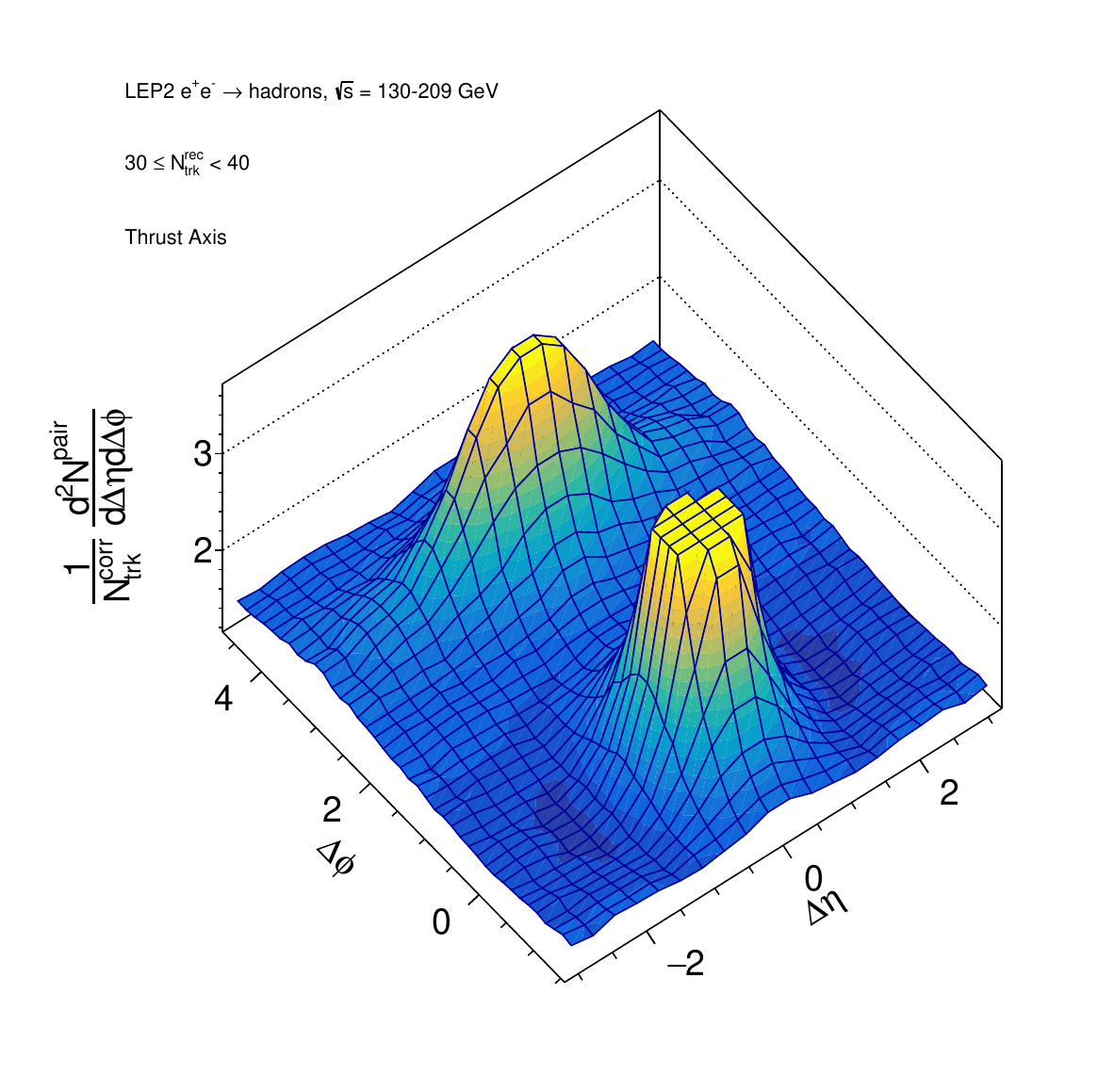}
    \includegraphics[width=0.45\textwidth]{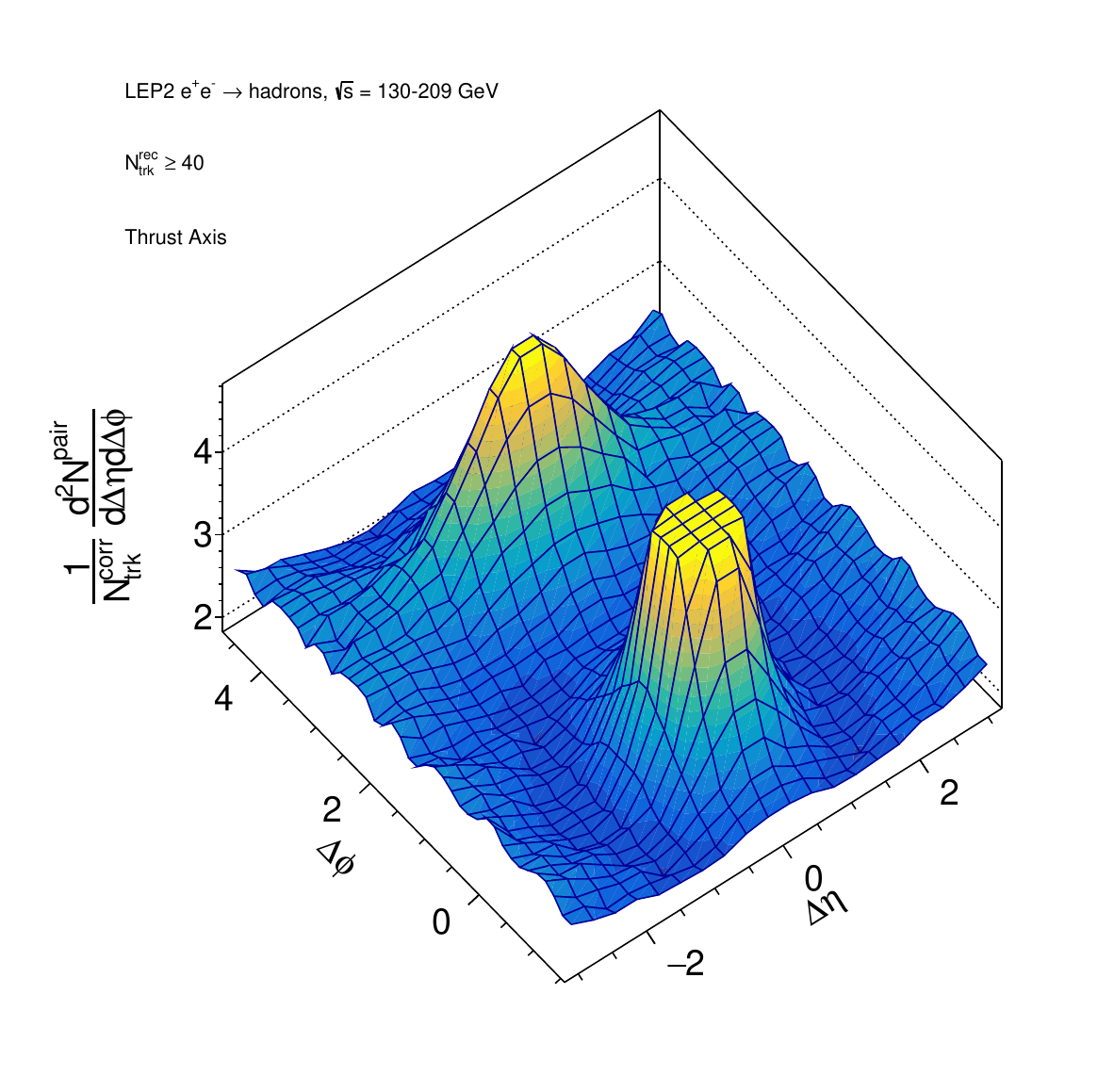}
    \caption{Two-particle correlation function with respect to the thrust axis,
      without the \WW-enhancing BDT selection, in the offline multiplicity
      intervals $[30,40)$ (left) and $[40,\infty)$ (right).}
\label{fig:feat_rst_nobdt_40phibins_0908_final_2PC}
\end{figure}

\begin{figure}[ht]
\centering
    \includegraphics[width=0.45\textwidth]{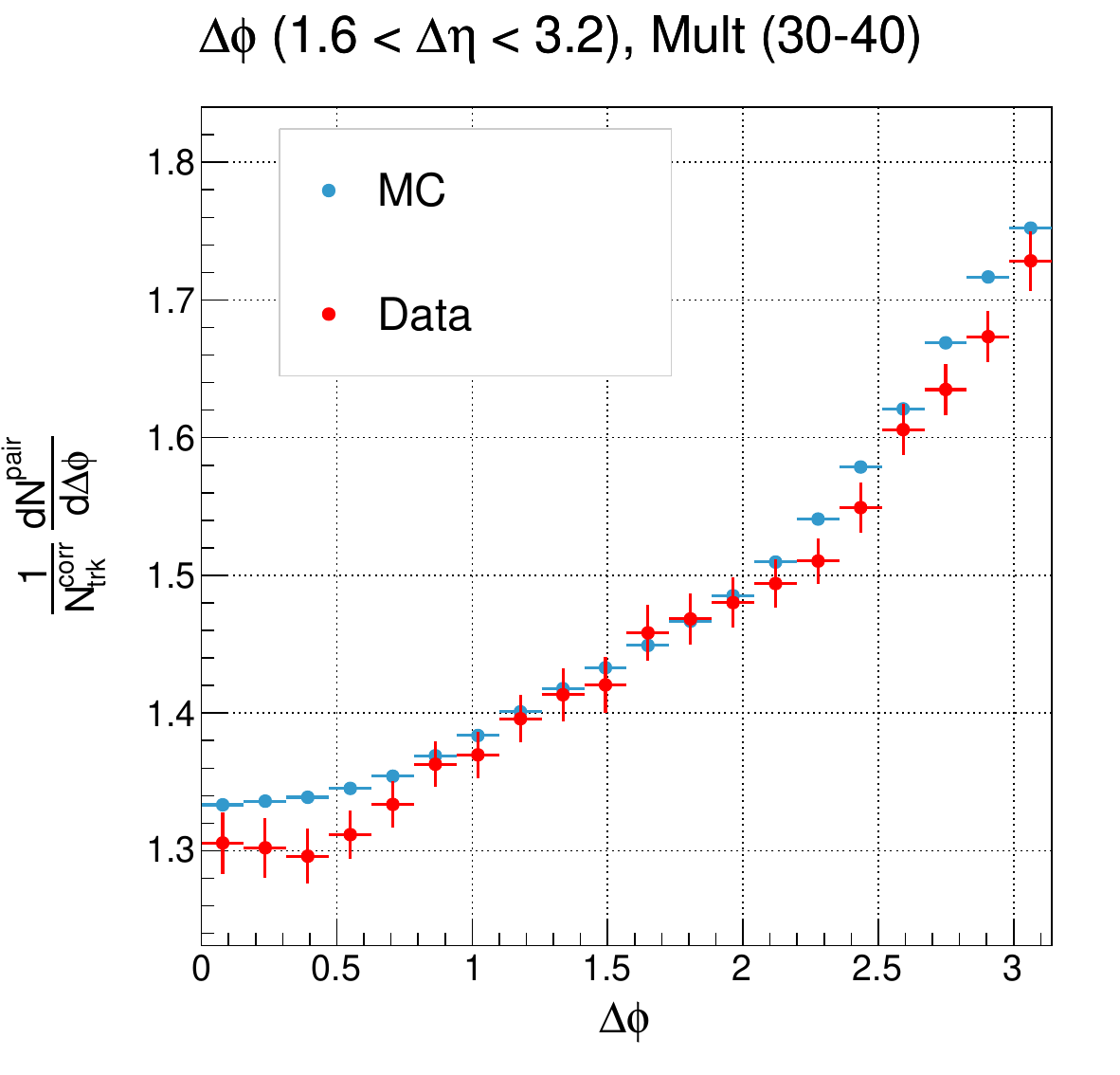}
    \includegraphics[width=0.45\textwidth]{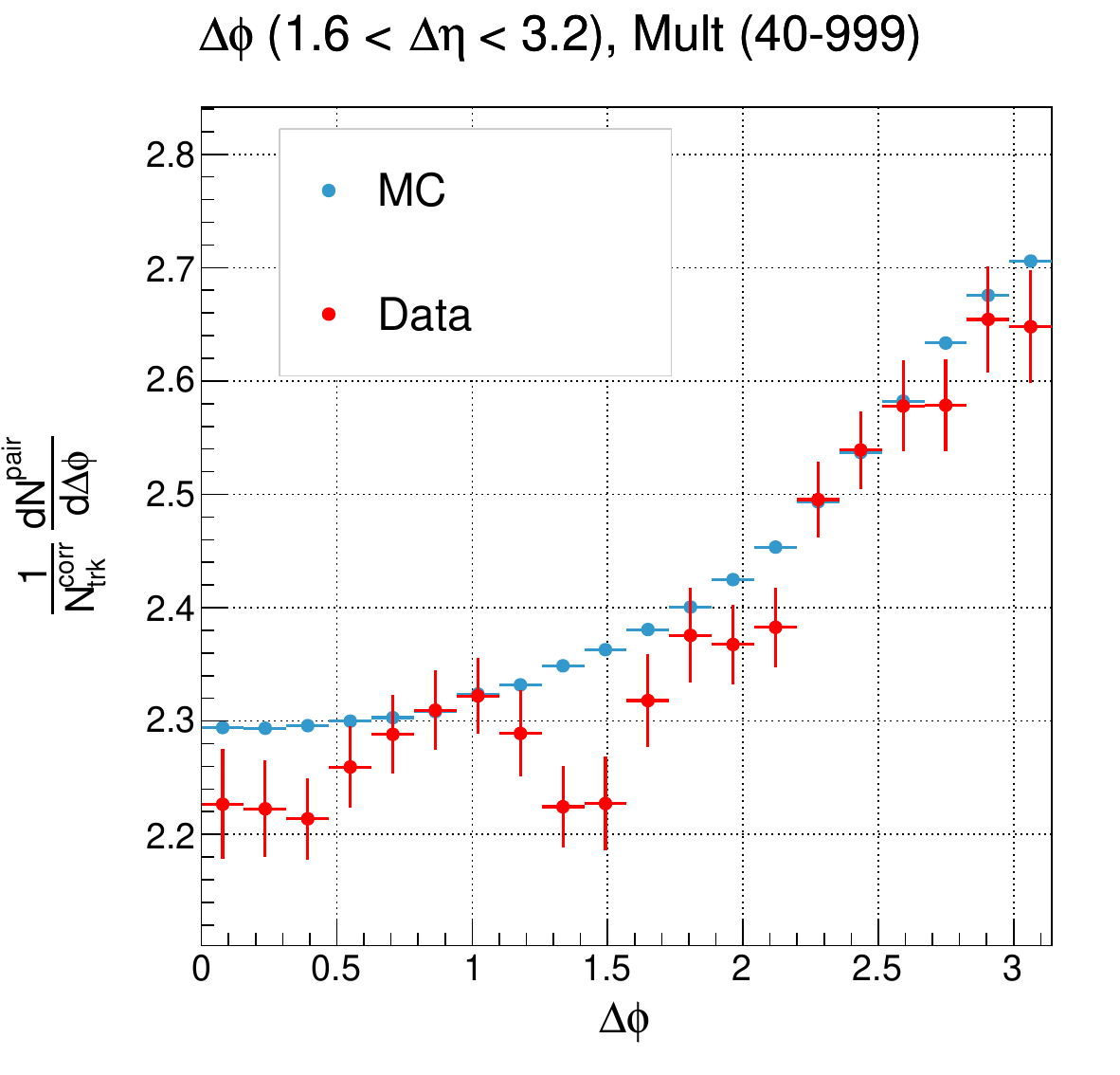}
    \caption{Long-range ($1.6<\Delta\eta<3.2$) azimuthal differential yield
      $Y_l(\Delta\phi)$ with respect to the thrust axis, without the
      \WW-enhancing BDT selection, in the offline multiplicity intervals
      $[30,40)$ (left) and $[40,\infty)$ (right).
      Data (red) are shown with bootstrap-derived statistical uncertainties
      (Sec.~\ref{sec:flow_bootstrap}); MC (blue) is shown with its
      statistical errors.}
\label{fig:feat_rst_nobdt_40phibins_0908_final_dNdphi}
\end{figure}

To establish a baseline, we first show the inclusive configuration \emph{without} BDT selection in Figures~\ref{fig:feat_rst_nobdt_40phibins_0908_final_2PC} and~\ref{fig:feat_rst_nobdt_40phibins_0908_final_dNdphi}; despite the updated stratified MC sample and the convolution-based background introduced above, the correlation shapes remain qualitatively compatible with Refs.~\cite{Chen:2023njr,Chen:2023nsi} and the published conclusions are unchanged. In Figure~\ref{fig:feat_rst_nobdt_40phibins_0908_final_dNdphi}, $Y_l(\Delta\phi)$ shows similar data and MC shapes in the $[30,40)$ multiplicity bin. The data--MC difference in $\Delta\phi$ modulation grows toward higher \ntrkoff, in line with the rising hadronic-\WW fraction discussed in Section~\ref{sec:Selection}.

\begin{figure}[ht]
\centering
    \includegraphics[width=0.45\textwidth]{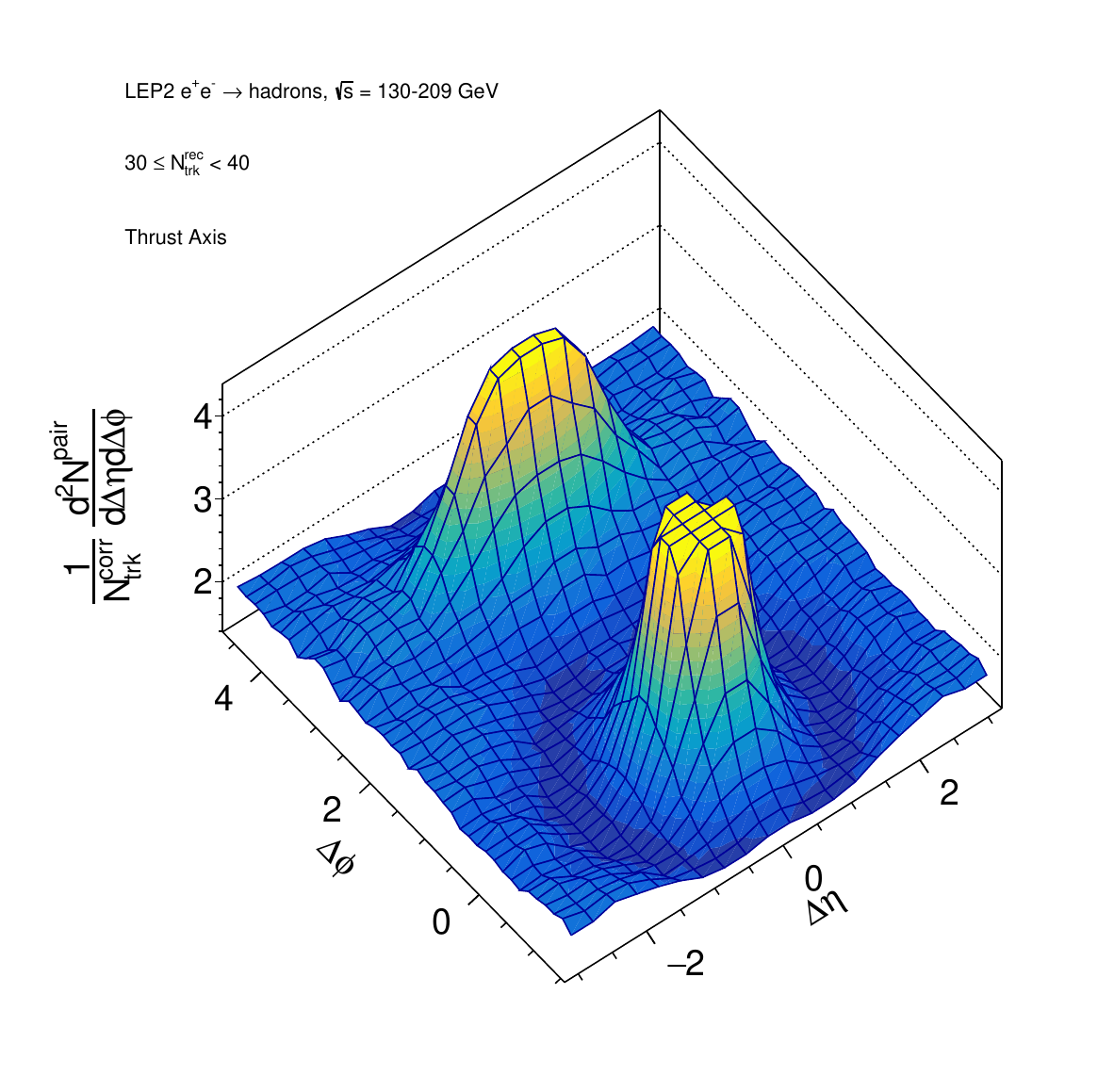}
    \includegraphics[width=0.45\textwidth]{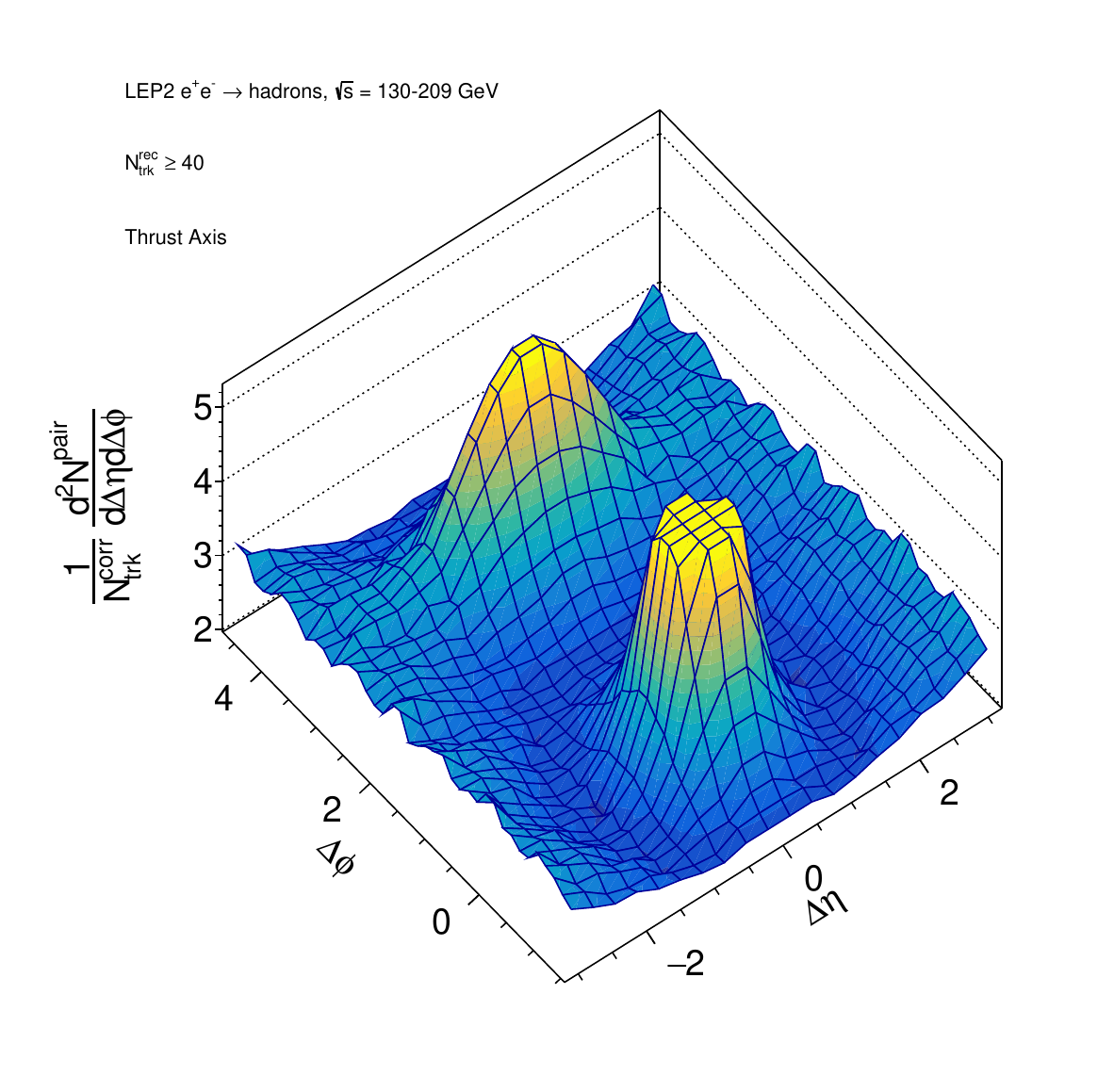}
\caption{Two-particle correlation function with respect to the thrust axis,
  with the \WW-enhanced BDT selection at $\epsilon_{\rm sig}\approx 85\%$,
  in the offline multiplicity intervals $[30,40)$ (left) and
  $[40,\infty)$ (right).}
\label{fig:feat_rst_85pbdt_40phibins_0908_final_2PC}
\end{figure}

\begin{figure}[ht]
\centering
    \includegraphics[width=0.45\textwidth]{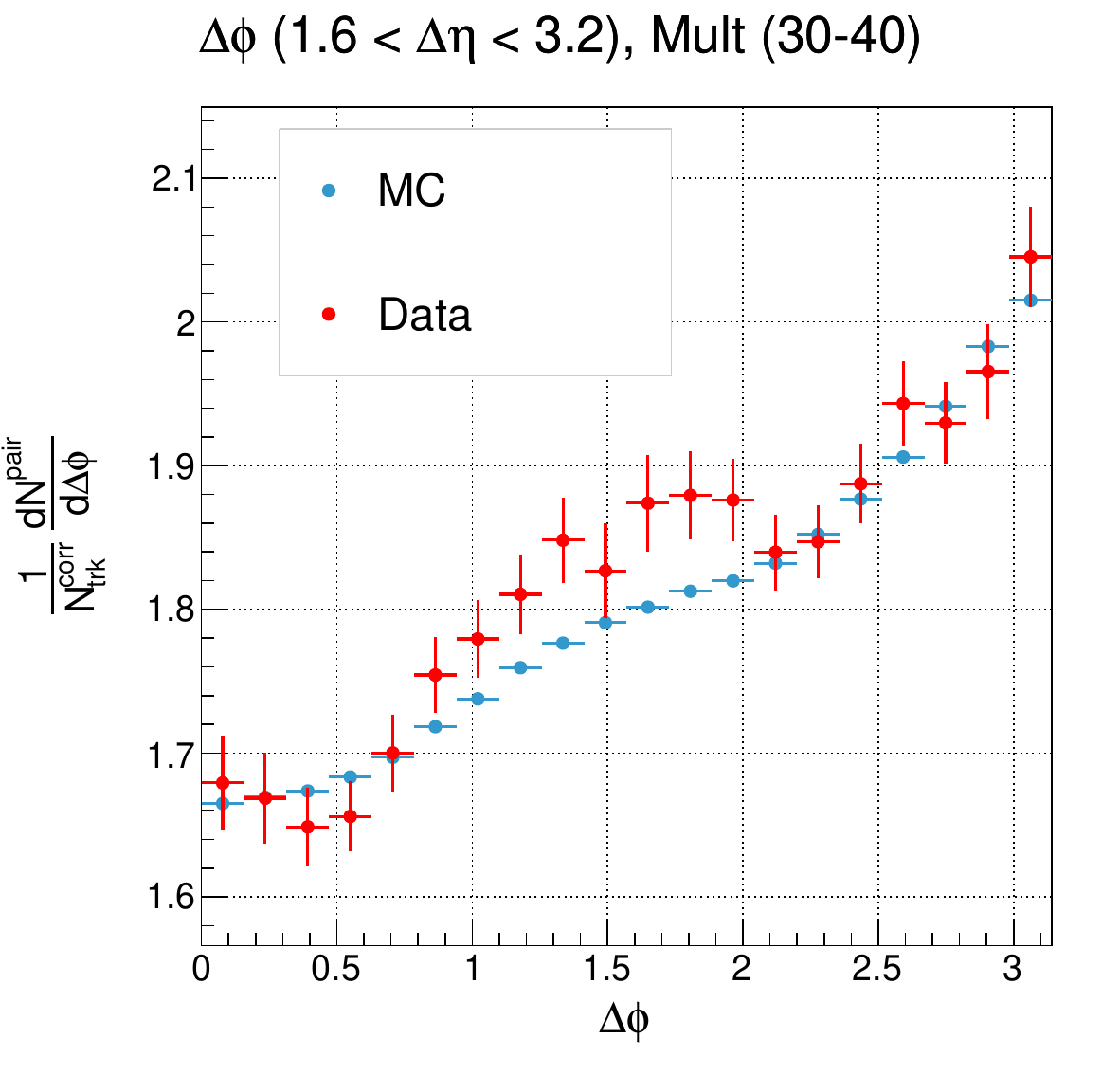}
    \includegraphics[width=0.45\textwidth]{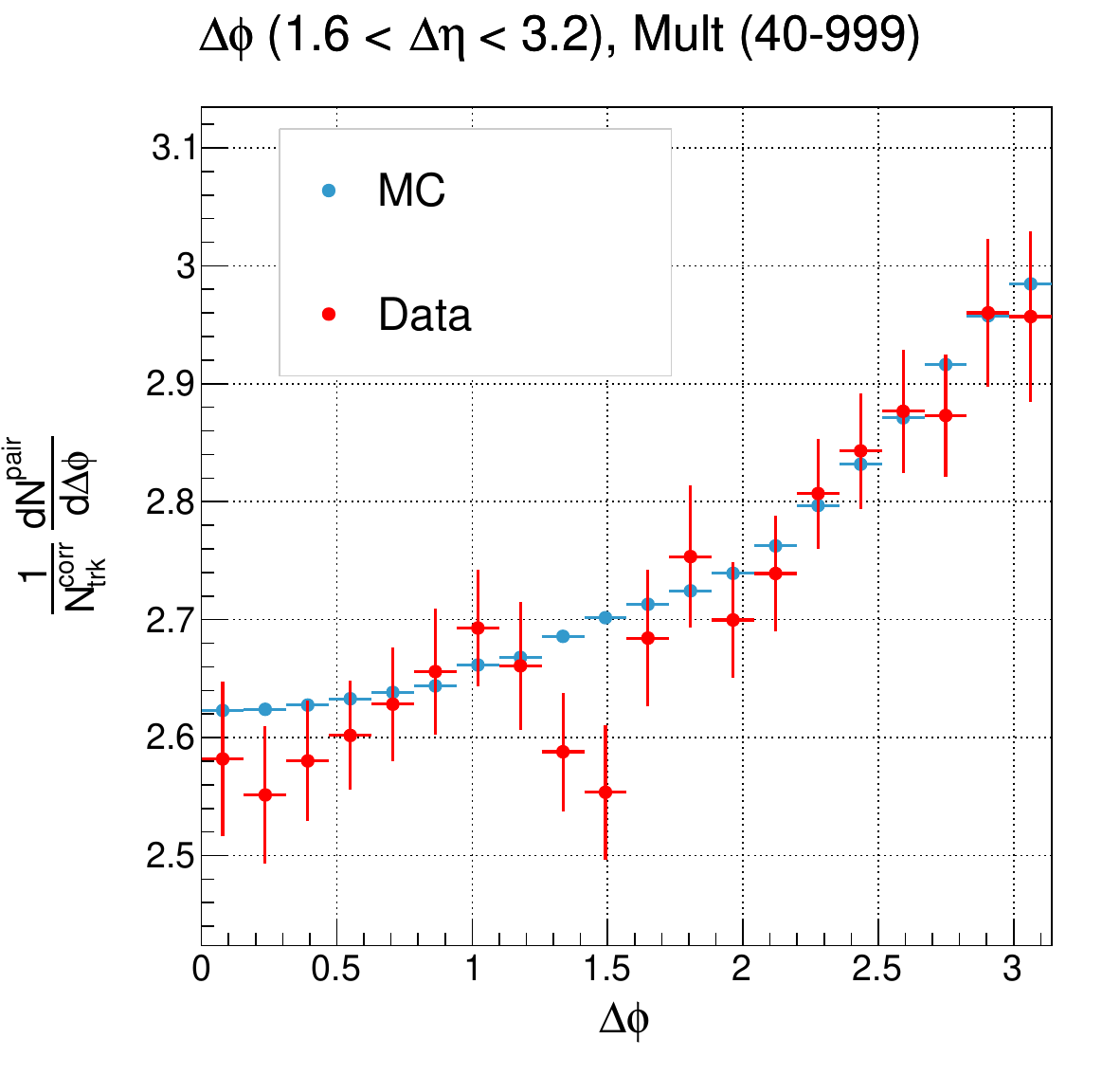}
\caption{Long-range ($1.6<\Delta\eta<3.2$) azimuthal differential yield
  $Y_l(\Delta\phi)$ with respect to the thrust axis, with the
  \WW-enhanced BDT selection at $\epsilon_{\rm sig}\approx 85\%$,
  in the offline multiplicity intervals $[30,40)$ (left) and
  $[40,\infty)$ (right).
  Data (red) are shown with bootstrap-derived statistical uncertainties
  (Sec.~\ref{sec:flow_bootstrap}); MC (blue) is shown with its
  statistical errors.}
\label{fig:feat_rst_85pbdt_40phibins_0908_final_dNdphi}
\end{figure}

Applying the BDT selection most strongly improves sensitivity in the intermediate-multiplicity regime: at the $\epsilon_{\rm sig}\approx 85\%$ working point, the \WW signal purity rises by roughly $30\%$ (relative) for $30 \le \ntrkoff < 40$, and by roughly $10\%$ for $\ntrkoff \ge 40$ (Fig.~\ref{fig:BDT_purity}).
The corresponding 2D and 1D correlation functions at the $85\%$ working
point are shown in
Figures~\ref{fig:feat_rst_85pbdt_40phibins_0908_final_2PC}
and~\ref{fig:feat_rst_85pbdt_40phibins_0908_final_dNdphi}; the full set
across all multiplicities and five BDT configurations (no-BDT + BDT at different working points) is collected
in the appendix.
With the BDT applied, the data exhibit a more pronounced $\Delta\phi$
modulation than MC --- most clearly in the $30 \le \ntrkoff < 40$
bin --- and a generally larger data--MC discrepancy across multiplicities.

\clearpage


To clarify the origin of the structure in $1 \lesssim \Delta\phi \lesssim 2$, we decompose the generator-level MC long-range yield $Y_l(\Delta\phi)$ into its two dominant hadronic components --- $\WW/ZZ \to 4q$ (pure diboson) and $q\bar{q}$ --- and compare them with the nominal stratified MC mixture. To remove the trivial differences in absolute scale between the samples, each curve is offset by its own ZYAM (zero-yield-at-minimum) constant $C_{\rm ZYAM}$, obtained from a low-order Fourier fit over $0 < \Delta\phi < \pi/2$ to assess the minimal correlation yield values across $\Delta \phi$. The subtracted constants are quoted in the legends so that only the \emph{shape} of $Y_l(\Delta\phi)$ is compared.

\begin{figure}[ht]
\centering
    \includegraphics[width=0.45\textwidth]{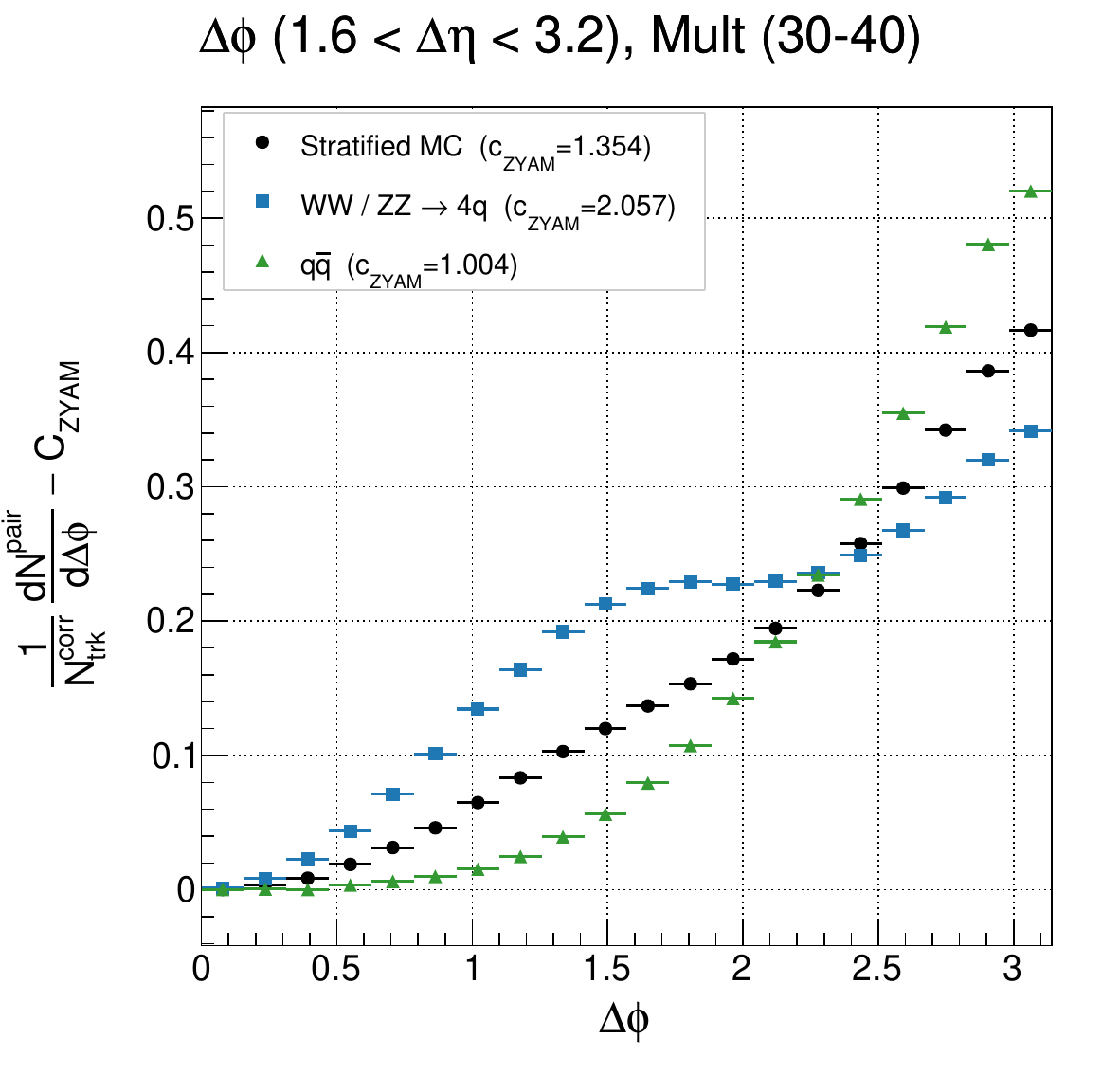}
    \includegraphics[width=0.45\textwidth]{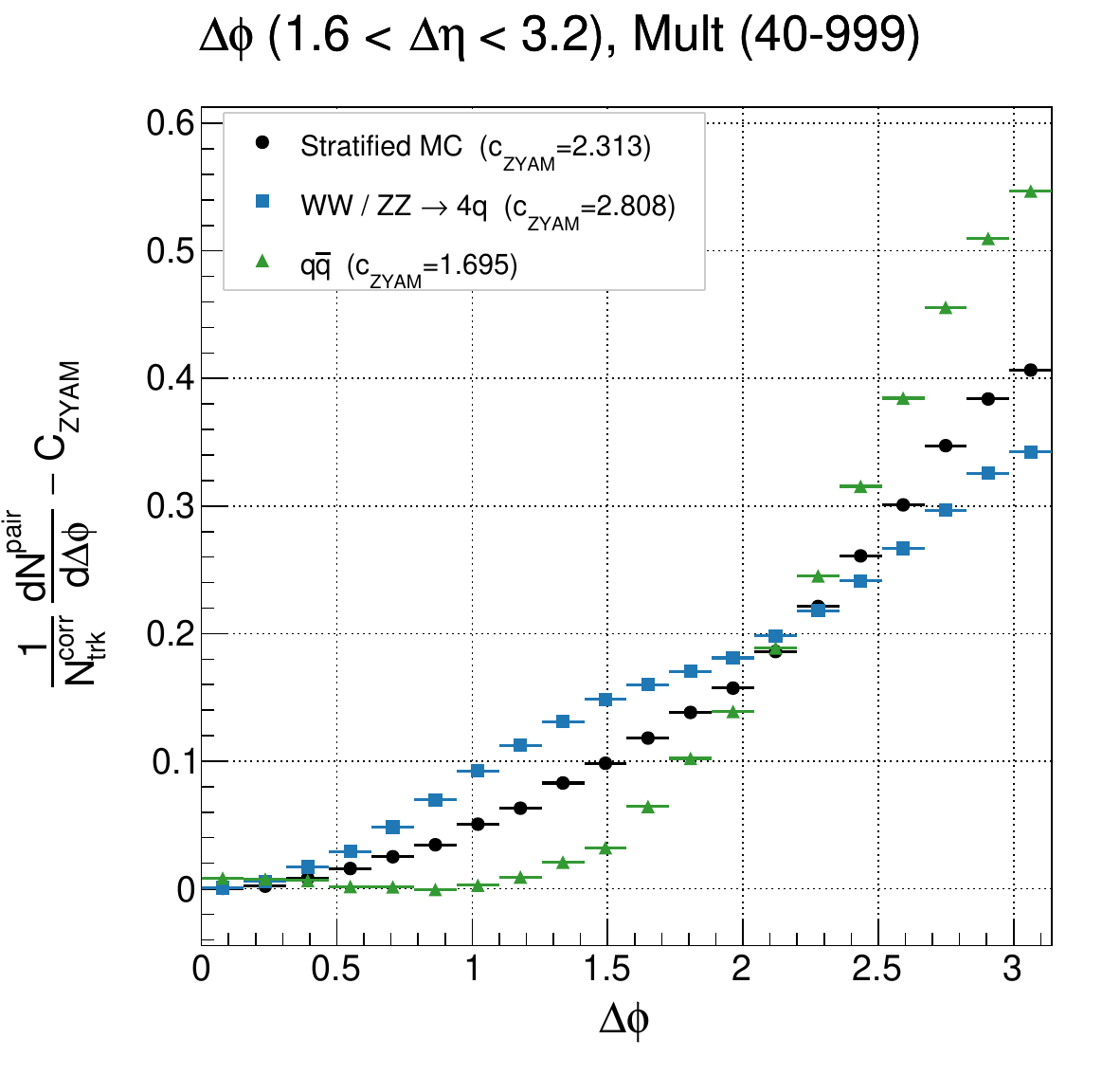}
\caption{Generator-level MC long-range ($1.6<\Delta\eta<3.2$) azimuthal yield
  $Y_l(\Delta\phi)$ with respect to the thrust axis, \emph{without} BDT
  selection, decomposed into the $\WW/ZZ\to 4q$ (blue) and $q\bar{q}$ (green)
  components and compared with the stratified MC (black), in the offline
  multiplicity intervals $[30,40)$ (left) and $[40,\infty)$ (right). Each curve
  is offset by its ZYAM constant $C_{\rm ZYAM}$ (quoted in the legend) so that
  only the shape is compared.}
\label{fig:dndphi_sample_overlay_nobdt}
\end{figure}

\begin{figure}[ht]
\centering
    \includegraphics[width=0.45\textwidth]{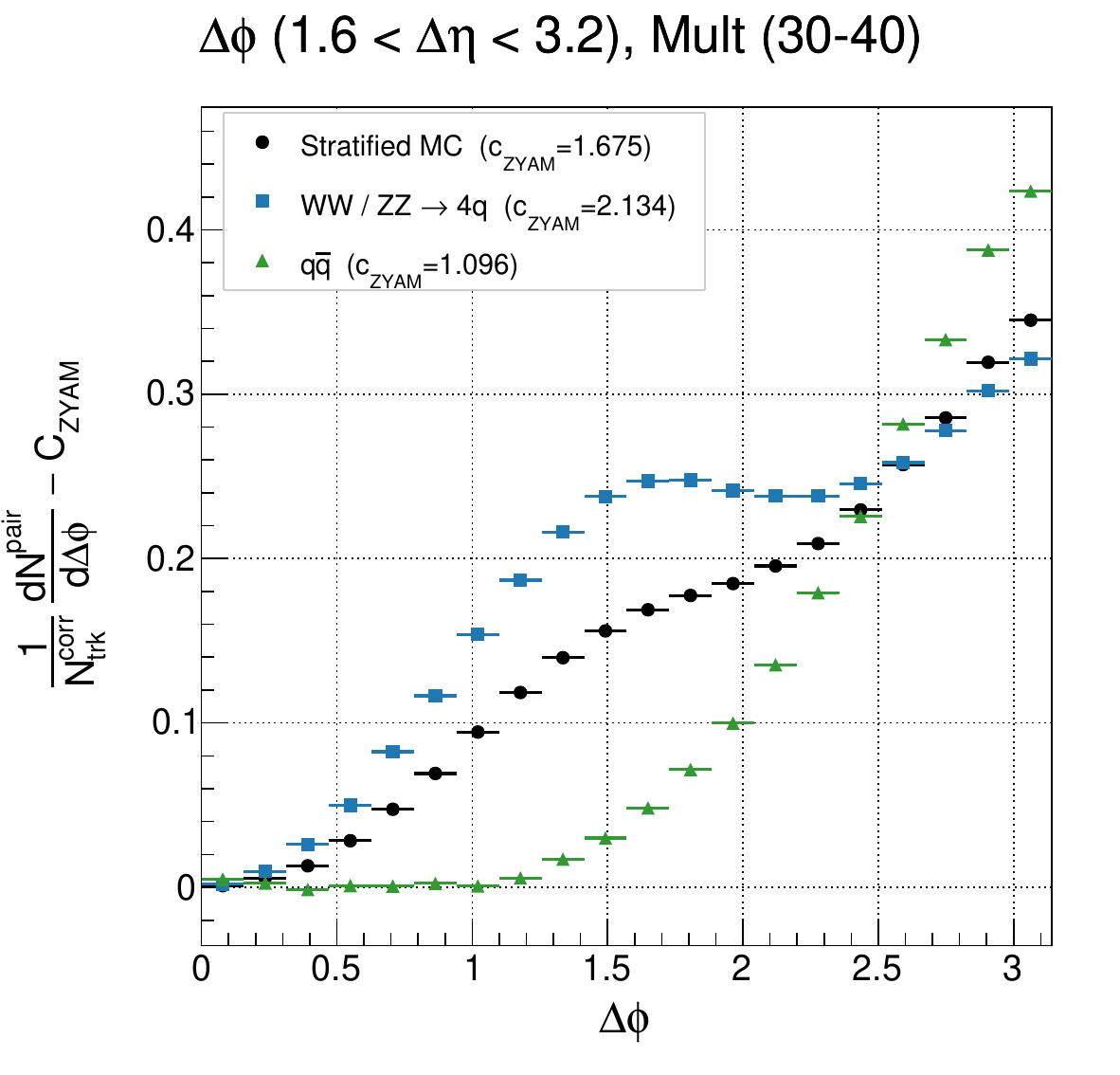}
    \includegraphics[width=0.45\textwidth]{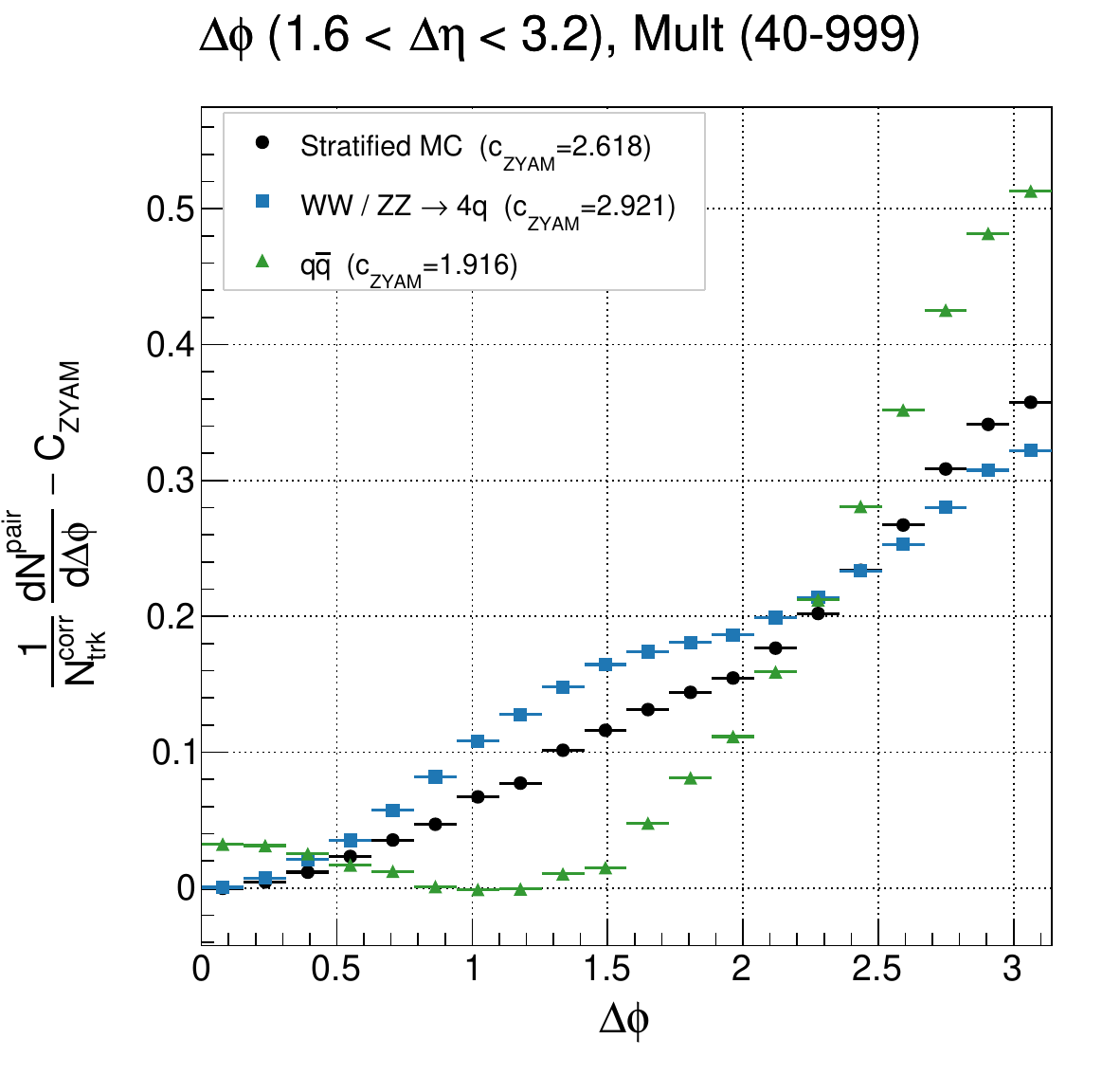}
\caption{Same decomposition as Fig.~\ref{fig:dndphi_sample_overlay_nobdt}, with
  the \WW-enhanced BDT selection at $\epsilon_{\rm sig}\approx 85\%$, in the
  offline multiplicity intervals $[30,40)$ (left) and $[40,\infty)$ (right).}
\label{fig:dndphi_sample_overlay_85pbdt}
\end{figure}

The $\WW/ZZ \to 4q$ component shows a distinct shoulder in the intermediate region $1 \lesssim \Delta\phi \lesssim 2$, whereas the $q\bar{q}$ component remains suppressed at small $\Delta\phi$ and rises monotonically toward the away side ($\Delta\phi \to \pi$), as expected for a predominantly back-to-back two-jet topology.
The effect is most visible in the $30 \le \ntrkoff < 40$ interval and persists, though less pronounced, for $\ntrkoff \ge 40$ (Figs.~\ref{fig:dndphi_sample_overlay_nobdt} and~\ref{fig:dndphi_sample_overlay_85pbdt}), consistent at the no-BDT and $\epsilon_{\rm sig}\approx 85\%$ working points.

Studying the two-particle correlation observable along with this extended reach down to $\ntrkoff \ge 30$ enables a differential study of how collective-like
long-range structures evolve with multiplicity---a proxy for effective
system size in \ee.
Rather than being restricted to the sparse $\ntrkoff \ge 50$ tail, the
Fourier analysis in Section~\ref{sec:flow} extracts the azimuthal
anisotropy $v_2$ across the broadened range $\ntrkoff \ge 30$.
This observable is motivated by models such as string
shoving~\cite{Bierlich:2024lmb}, which predict a characteristic
multiplicity dependence of $v_2$ in small systems.

\clearpage

\section{Flow analysis}
\label{sec:flow}

Following the previous LEP2 measurement~\cite{Chen:2023njr,Chen:2023nsi}, we further quantify the azimuthal modulation of the long-range differential yields $Y_l(\Delta\phi)$ (Eq.~\eqref{eqn:DeltaPhiAssociatedYield}) through Fourier coefficients of the two-particle correlation.
The possible short-range non-flow contribution is suppressed in the large $|\Delta \eta|$ region. The remaining long-range azimuthal modulations are characterized by a Fourier decomposition analysis~\cite{CMS:2011cqy,ALICE:2011svq,ATLAS:2012at}.

In this work, the extraction of the two-particle coefficients \(V_{n\Delta}\) and their signed square-root proxies is carried out in three stages.
First, a bootstrap ensemble of the efficiency-corrected long-range yield
is used to estimate the bin-to-bin covariance matrix of $Y_l(\Delta\phi)$
at the reconstruction level (Section~\ref{sec:flow_bootstrap}).
Second, the nominal long-range yield is corrected for residual detector and
reconstruction effects with a multiplicity-dependent bin-by-bin MC correction
(Section~\ref{sec:flow_binbybin}).
Third, the bootstrap covariance and the corrected nominal yield are propagated
through a Bayesian toy procedure that refits the Fourier template and summarizes
the resulting $V_{n\Delta}$ and signed square-root proxy distributions (Section~\ref{sec:flow_bayesian}).

The corrected long-range yield is described with a Fourier series up to the sixth harmonic,
\begin{equation}
\begin{aligned}
Y_l(\Delta\phi) = \frac{1}{{\rm N}_{\rm trk}^{\rm corr}}\frac{d{\rm N}^{\rm pair}}{d\Delta\phi} = \frac{{\rm N}^{\rm assoc}}{2\pi} \bigg( 1 + \sum_{n} 2 V_{n\Delta} \cos(n\Delta\phi) \bigg),
\end{aligned}
\label{eqn:fourier}
\end{equation}
where ${\rm N}^{\rm assoc}$ is the number of associated track pairs in the $|\Delta \eta|$ region of interest ($1.6 \le |\Delta \eta| < 3.2$) and within the full $\Delta \phi$ range for a particular multiplicity.

The Fourier coefficients $V_{n\Delta}$ are the amplitudes of the cosine terms in Eq.~\eqref{eqn:fourier}, where the modulation can take either sign.
For identical trigger and associated particle collections, the usual flow notation would motivate a factorized form $V_{n\Delta} = v_n^{\rm trig} \times v_n^{\rm assoc}$ if the correlation arose from a positive-definite single-particle anisotropy.
In the present measurement, however, \(V_{n\Delta}\) can be negative; therefore we report the signed square-root proxy
\begin{equation}
\begin{aligned}
v_n^{\rm sgn} = \mathrm{sign}(V_{n\Delta})\,\sqrt{|V_{n\Delta}|},
\end{aligned}
\label{eqn:vn}
\end{equation}
which preserves the sign information of the fitted two-particle Fourier coefficient. For compactness, the figures label this quantity as \(v_n\). A negative value of this signed proxy indicates a negative \(V_{n\Delta}\) modulation; it should not be interpreted as a conventional negative single-particle elliptic or triangular flow coefficient.

\begin{figure}
    \centering
    \includegraphics[width=0.49\textwidth]{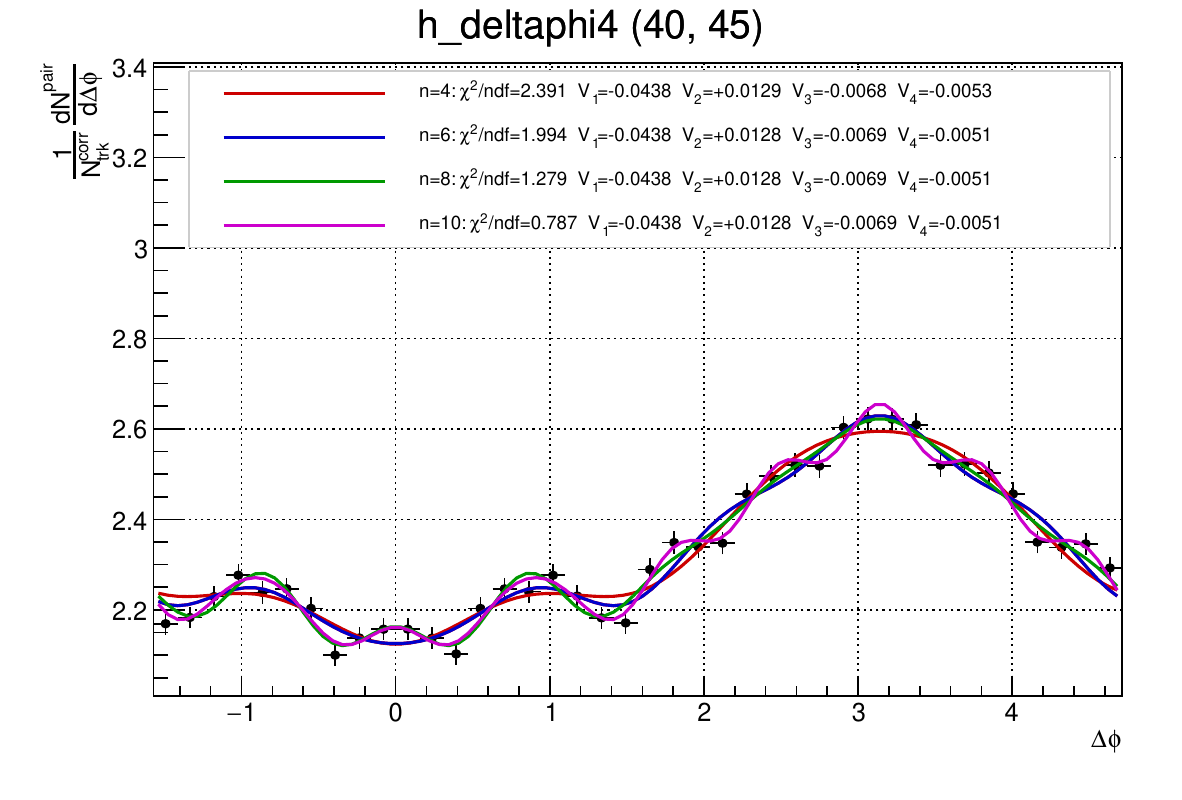}
    \includegraphics[width=0.5\textwidth]{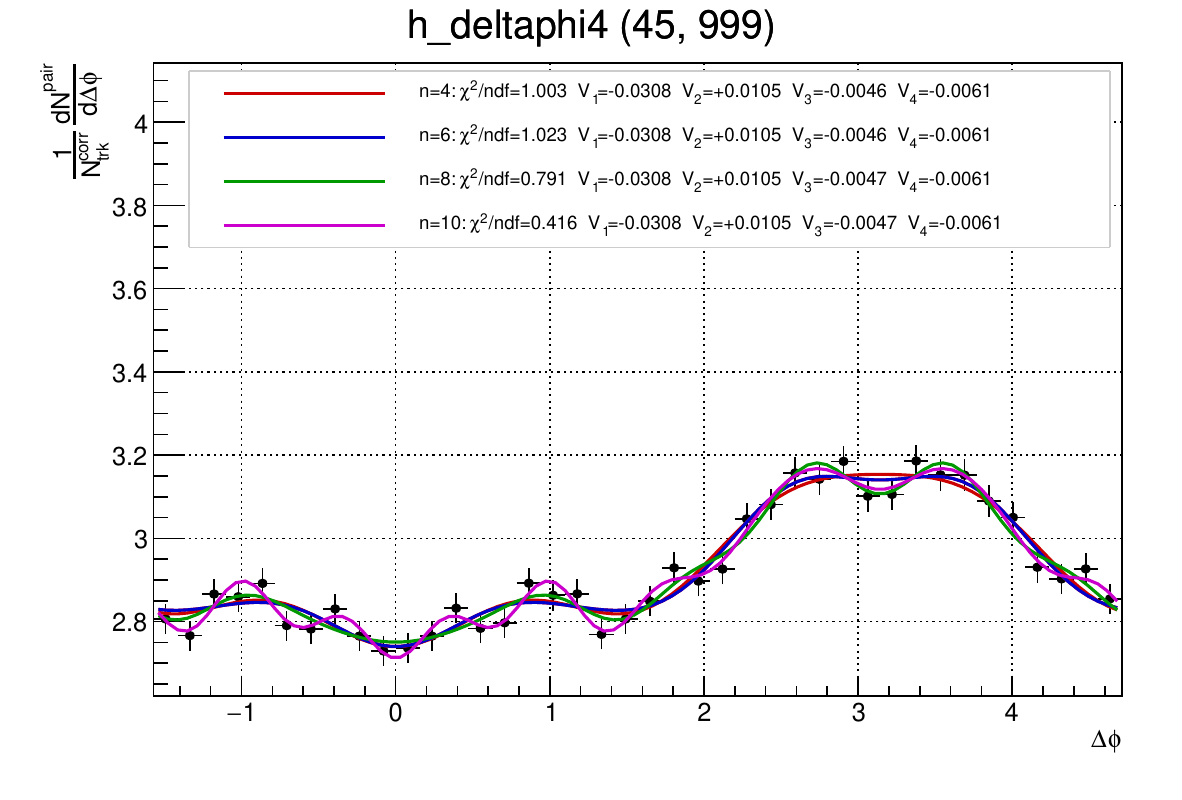}
    \caption{Fourier template with harmonic orders $n=4,6,8,10$ fitted to long-range yield distribution in azimuthal angle (data in black). Results shown for multiplicity intervals $[40, 45)$ (left) and $[45,999)$ (right).}
    \label{fig:23}
\end{figure}

\begin{figure}
    \centering
    \includegraphics[width=0.7\textwidth]{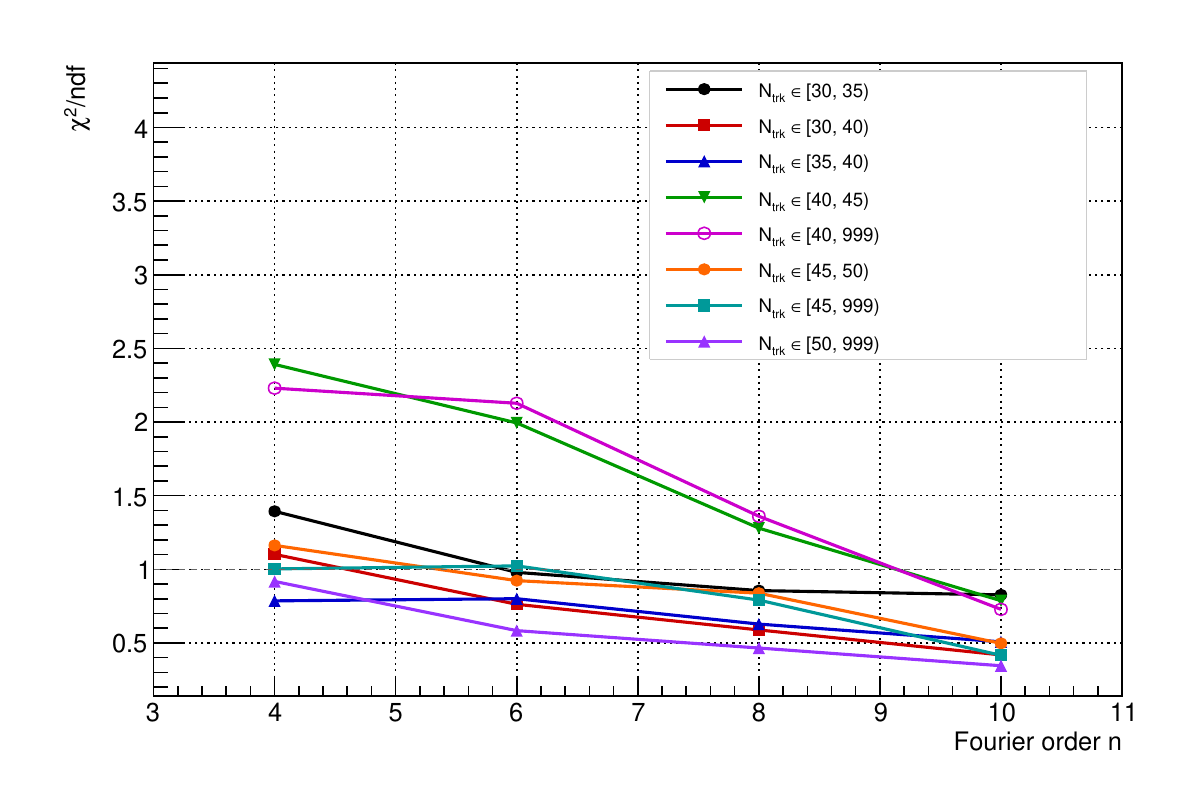}
    \caption{$\chi^2/$ndf for said values of $n$ with no BDT selection in various $N_{\rm trk}$ multiplicity bins.} 
    \label{fig:24}
\end{figure}

\begin{figure}
    \centering
    \includegraphics[width=\textwidth]{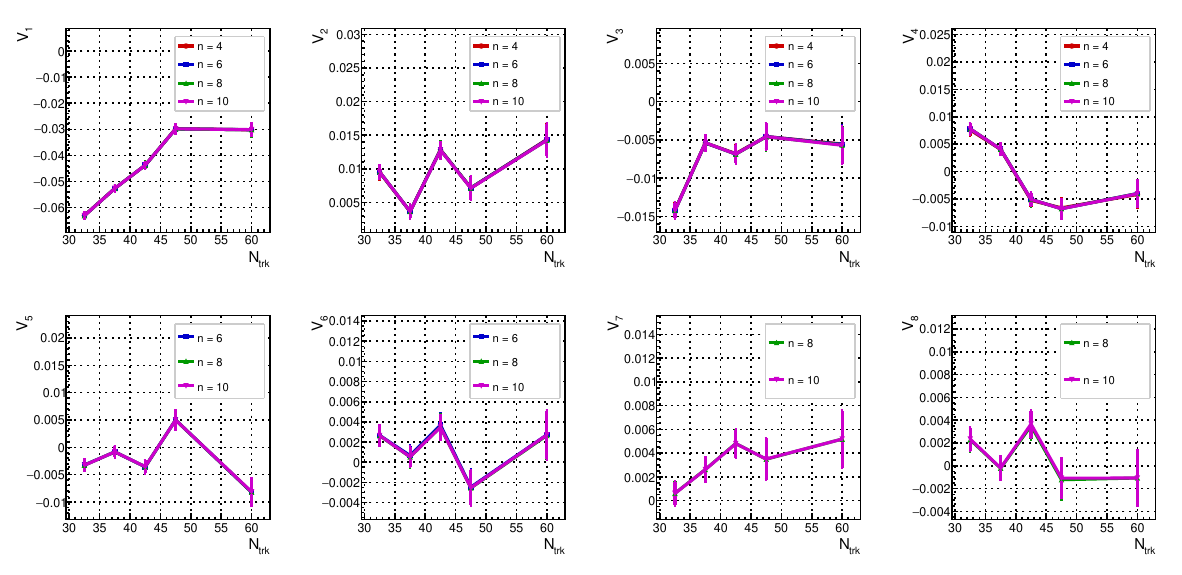}
    \caption{Fourier coefficients $V_{n\Delta}$ for each multiplicity bins, and graphed for said values of $n$ when applicable. Overlap of values show that higher order fits preserve lower order coefficients.} 
    \label{fig:25}
\end{figure}

We also evaluate the fit quality with choices of different degrees of freedom of the Fourier template. We extract the Fourier coefficients $V_{n\Delta}$ by attempting fit templates for harmonics of order $n=4, 6, 8, 10$ (Fig.~\ref{fig:23}), where the results show reasonable description on capturing data points.
We evaluated the reduced $\chi^2$ across 8 different multiplicity bins (Fig.~\ref{fig:24}). The $n=4$ and $n=10$ templates suggest underfitting and overfitting, respectively. The eight-harmonic is most ideal with statistics in all the explored bins, as its $\chi^2/$ndf values are closest to $1$, and the six-harmonic is also a reasonable configuration amongst all multiplicity bins. In Fig.~\ref{fig:25}, all the extracted $V_{n\Delta}$ show consistent results with different $n$-choices, suggesting the truncation of Fourier orders doesn't affect lower Fourier coefficients extraction.


\subsection{Bootstrap covariance of the long-range yield}
\label{sec:flow_bootstrap}

The long-range yield $Y_l(\Delta\phi)$ (Eq.~\eqref{eqn:DeltaPhiAssociatedYield})
is extracted from a finite event sample; event-sampling fluctuations therefore
induce correlated variations across $\Delta\phi$ bins that are not captured by
treating each bin independently.
A Fourier fit that ignores these bin-to-bin correlations can underestimate the
true uncertainty on $V_{n\Delta}$, particularly when pair statistics are limited.
To quantify this effect, we estimate the bin-to-bin covariance matrix of
$Y_l(\Delta\phi)$ directly from the data using a bootstrap procedure,
which is introduced for the first time in this analysis.

\paragraph{Step 1: Bootstrap signal collection.}
$N_{\rm boot}$ ($= \mathcal{O}(10^2)$) replica datasets are generated
by resampling the original data events with replacement, stratified by
center-of-mass energy to preserve the luminosity composition of each replica.
Each replica $b$ ($b=1,\ldots,N_{\rm boot}$) yields a 2D signal histogram $S^{(b)}(\Delta\eta,\Delta\phi)$, following the same definition
as Eq.~\eqref{eqn:2PCSig}.

\paragraph{Step 2: Bootstrap 2D correlation function.}
For each replica, the two-particle correlation function is evaluated using the
same formula as Eq.~\eqref{eqn:2PC}, with only the signal replaced by its
replica counterpart:
\begin{equation}
C^{(b)}(\Delta\eta,\Delta\phi)
= B(0,0)\times\frac{S^{(b)}(\Delta\eta,\Delta\phi)}{B(\Delta\eta,\Delta\phi)},
\quad b=1,\ldots,N_{\rm boot}.
\label{eqn:Chat}
\end{equation}
The background $B(\Delta\eta,\Delta\phi)$ and normalisation $B(0,0)$ are
identical to those defined in Section~\ref{sec:TwoParticleCorrelationFunction}
and are held fixed across all replicas.
The background correlation function is constructed from the single-particle
$(\eta,\phi)$ spectrum averaged over a large event pool and is a smooth,
high-statistics estimator of the uncorrelated pair baseline. Its statistical
fluctuations under event resampling are negligible compared to those of the
signal $S^{(b)}$, which is sensitive to the correlated pairs that drive
the azimuthal modulations in the long-range yield.
The bootstrap statistics is therefore designed to characterize the event-sampling
uncertainty of the signal correlation function specifically.

\paragraph{Step 3: Projection to 1D long-range yield.}
Each $C^{(b)}$ is projected onto the long-range $|\Delta\eta|$ interval
$[1.6,\,3.2]$ following Eq.~\eqref{eqn:DeltaPhiAssociatedYield} to yield $Y_l^{(b)}(\Delta\phi)$, which is represented using a vector of $N_{\Delta\phi}=40$ values per replica.

\paragraph{Step 4: Sample covariance matrix.}
The $N_{\rm boot}$ replica vectors are stacked to form the unbiased
$N_{\Delta\phi}\times N_{\Delta\phi}$ sample covariance matrix
\begin{equation}
\mathbf{V}_{ij} = \frac{1}{N_{\rm boot}-1}
\sum_{b=1}^{N_{\rm boot}}
\bigl(Y_{l,i}^{(b)}-\bar{Y}_{l,i}\bigr)
\bigl(Y_{l,j}^{(b)}-\bar{Y}_{l,j}\bigr),
\quad
\bar{Y}_{l,i}=\frac{1}{N_{\rm boot}}\sum_{b=1}^{N_{\rm boot}}Y_{l,i}^{(b)},
\label{eqn:bootstrap_cov}
\end{equation}
where $i,j\in\{1,\ldots,N_{\Delta\phi}\}$ label $\Delta\phi$ bins.
Assuming Gaussian statistics, $\mathbf{V}$ is passed to the Bayesian stage
(Section~\ref{sec:flow_bayesian}) as the covariance of the input yield.

\begin{figure}[ht]
\centering
    \includegraphics[width=0.43\textwidth]{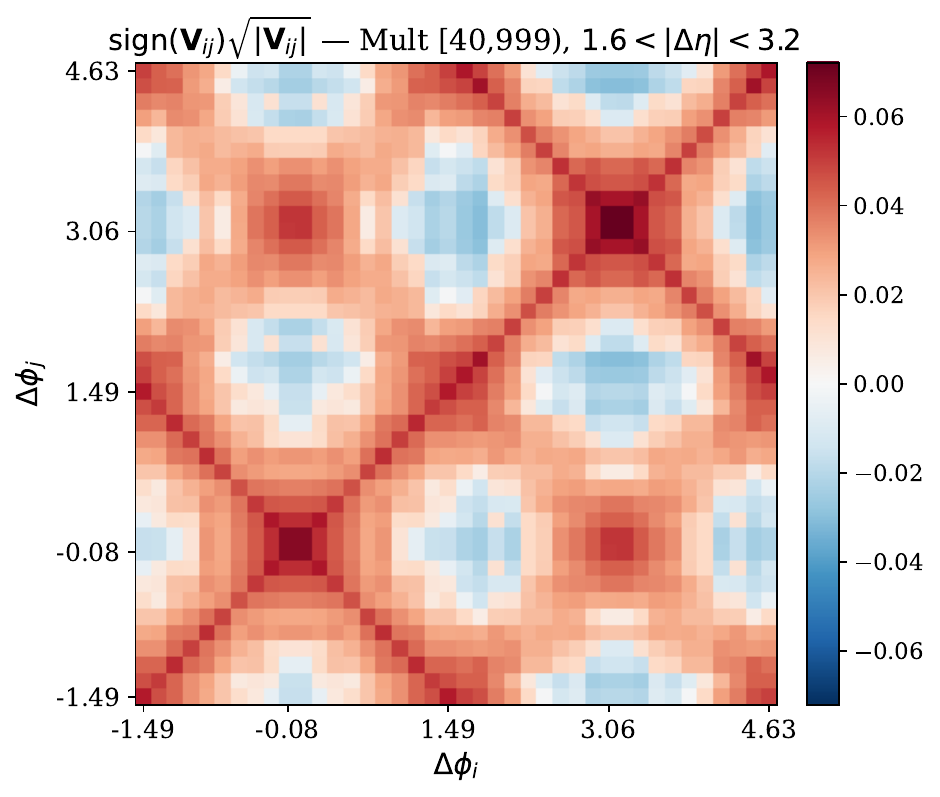}
    \includegraphics[width=0.47\textwidth]{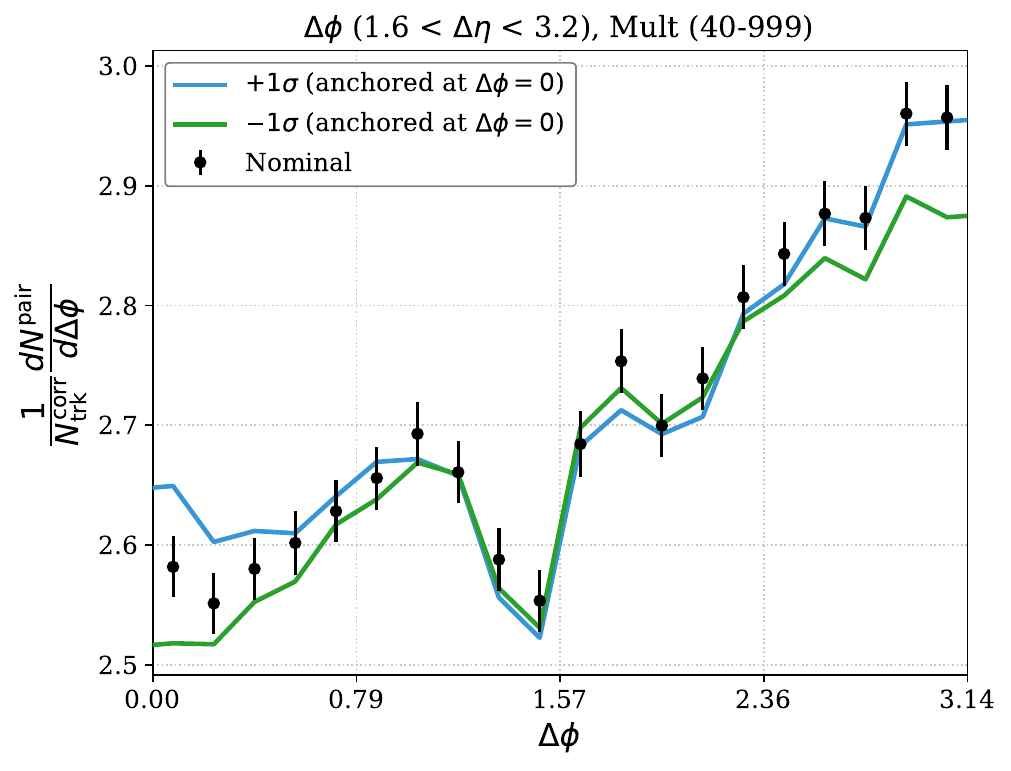}
\caption{Illustration of the bootstrap covariance in the multiplicity bin
  $[40,\infty)$ at long range ($1.6<\Delta\eta<3.2$), for the inclusive
  (no-BDT) selection at $\sqrt{s}=183$--$207\,$GeV.\\
  \textbf{Left:} the signed-square-root magnitude
  $\mathrm{sign}(\mathbf{V}_{ij})\sqrt{|\mathbf{V}_{ij}|}$ of the ratio
  covariance matrix; adjacent $\Delta\phi$ bins are positively correlated
  and bins separated by $\pi/2$ are anti-correlated.\\
  \textbf{Right:} the corresponding conditional $\pm 1\sigma$ propagation
  anchored at the $\Delta\phi=0$ bin (index $k_0$), with shifts
  $\delta_i=\pm\mathbf{V}_{i,k_0}/\sqrt{\mathbf{V}_{k_0,k_0}}$ applied coherently to all bins; the
  nominal $Y_l(\Delta\phi)$ is shown as black points.}
\label{fig:bootstrap_illustration}
\end{figure}

\subsection{Bin-by-bin residual MC correction}
\label{sec:flow_binbybin}

Before extracting the harmonic coefficients, we correct the measured correlation
observables for residual reconstruction effects that remain after the track-level
efficiency correction described in Section~\ref{sec:Corrections}.
This correction is derived independently in each multiplicity interval.
The procedure uses three inputs: the reconstructed data distribution, the
reconstructed MC reference, and the corresponding MC truth distribution.
The nominal correction factor is taken from the ratio of the truth-level and
reconstructed MC shapes,
\begin{equation}
Y_l(\Delta\phi)^{\rm corr} = Y_l(\Delta\phi)^{\rm data,reco}\times
\frac{Y_l(\Delta\phi)^{\rm MC,truth}}{Y_l(\Delta\phi)^{\rm MC,reco}},
\label{eqn:binbybin_corr}
\end{equation}
applied bin-by-bin.
The corrected $Y_l(\Delta\phi)$ is then passed to the Bayesian extraction
described in Section~\ref{sec:flow_bayesian}.

\subsection{Bayesian extraction of \(V_{n\Delta}\) and signed proxies}
\label{sec:flow_bayesian}

The harmonic coefficients are extracted from the corrected $Y_l(\Delta\phi)$
via a Bayesian toy analysis.
The corrected histogram $Y_l^{\rm corr}(\Delta\phi)$ defines the central
bin-content vector.
Each toy draw produces a correlated fluctuation of the bin contents by adding a
shift $\boldsymbol{\delta}$ drawn from the bootstrap covariance matrix $\mathbf{V}$
(Section~\ref{sec:flow_bootstrap}),
\begin{equation}
\tilde{Y}_l(\Delta\phi) = Y_l^{\rm corr}(\Delta\phi) + \boldsymbol{\delta},
\quad \boldsymbol{\delta}\sim\mathcal{N}(\mathbf{0},\,\mathbf{V}),
\label{eqn:toy_draw}
\end{equation}
so that the toy ensemble directly models the bin-to-bin correlated fluctuations
of the long-range yield estimated from the bootstrap.
Each toy histogram is refitted with the six-harmonic Fourier ansatz of
Eq.~\eqref{eqn:fourier}, and the fitted $V_{n\Delta}$ values are converted to
the signed square-root proxy \(v_n^{\rm sgn}\) through Eq.~\eqref{eqn:vn}.
The default configuration uses $N_{\rm toy}=1000$ toys per point.

The toy ensemble replaces a single best-fit estimate with a distribution of
solutions consistent with the correlated statistical uncertainties of
$Y_l(\Delta\phi)$.
The median of the toy \(v_n^{\rm sgn}\) distribution is taken as the central value, and
the 16th and 84th percentiles define the lower and upper statistical
uncertainties, naturally accommodating asymmetric intervals.

Several sources of systematic uncertainty are propagated through the same
Bayesian framework.
Alternative bin-by-bin correction factors, built from first- and second-order
Fourier variations of the residual MC correction, are substituted for the
nominal correction of Eq.~\eqref{eqn:binbybin_corr}; the resulting shifts in
\(v_n^{\rm sgn}\) are taken as the correction-factor systematic.
Robustness of the extraction is further assessed by reducing $N_{\rm toy}$
and by narrowing the $\Delta\phi$ fit range.

\begin{figure}[htbp]
  \centering
  \includegraphics[width=0.32\textwidth]{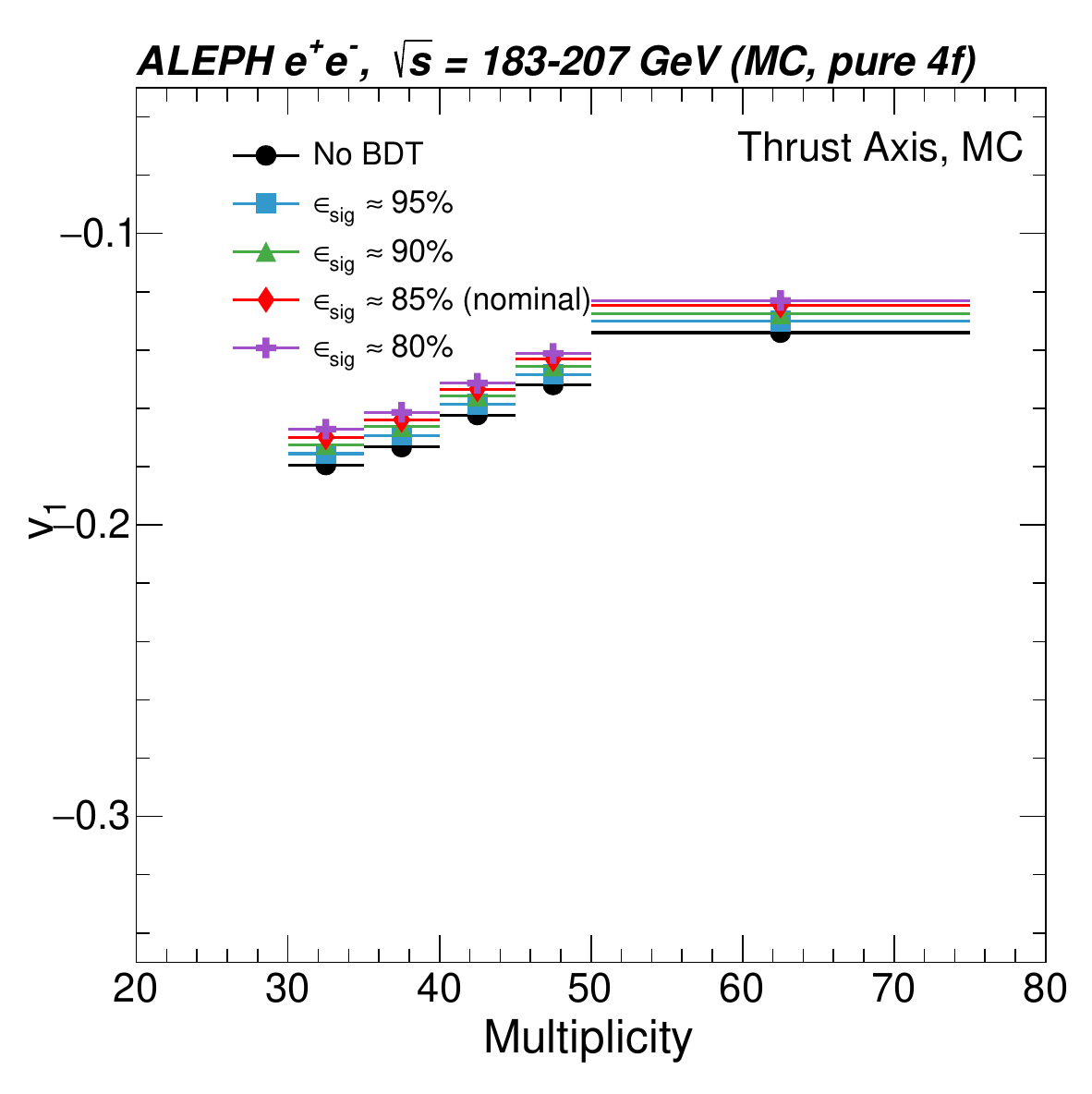}
  \includegraphics[width=0.32\textwidth]{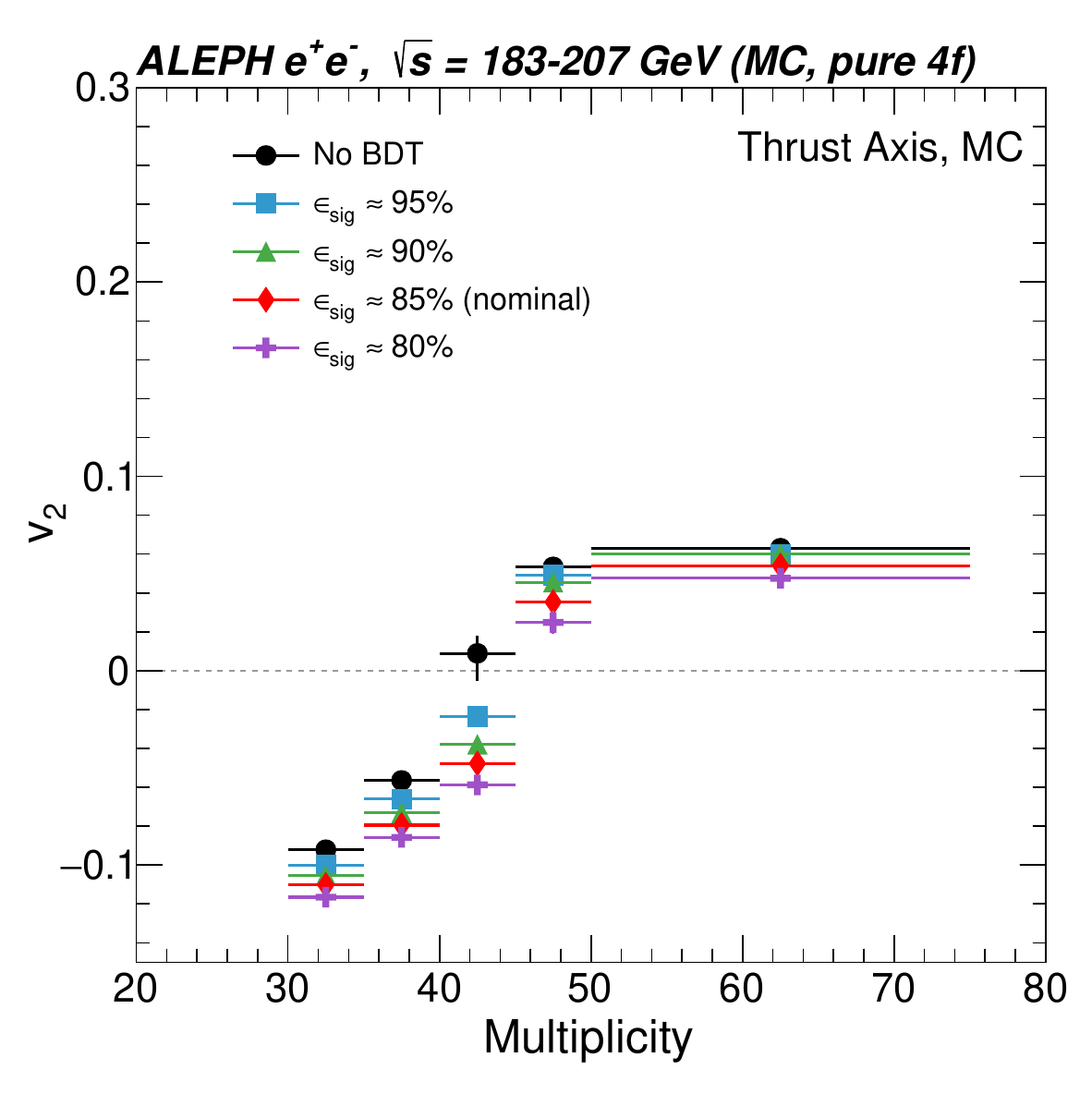}
  \includegraphics[width=0.32\textwidth]{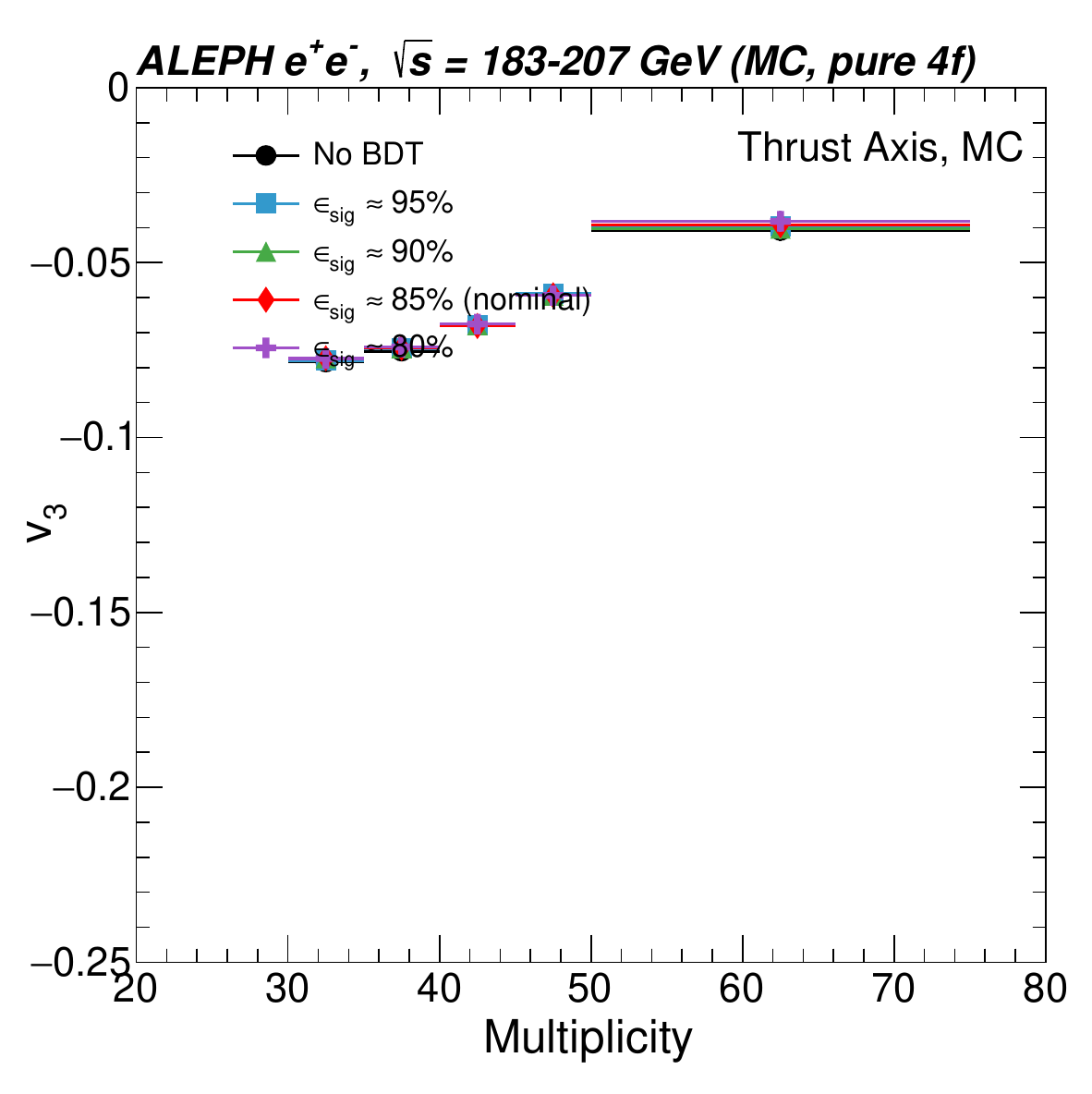}
  \caption{Signed square-root proxies, labeled $v_1$ (left), $v_2$ (center), and $v_3$ (right), extracted from the
    pure MC signal sample at four BDT working points
    ($\epsilon_{\rm sig} = 80\%$, $85\%$, $90\%$, $95\%$) as a function of
    charged-particle multiplicity.
    The spread across working points is used to define the BDT systematic
    (Eq.~\eqref{eqn:bdt_syst}).}
  \label{fig:vn_bdt_wp_overlay}
\end{figure}

An additional systematic uncertainty is assigned to account for the choice of
BDT working point.
It is evaluated on the pure MC signal sample (see Section~\ref{sec:orgdd765bf}), which is
free of single-string background contamination, by extracting \(v_n^{\rm sgn}\) at four
working points ($\epsilon_{\rm sig} = 80\%$, $85\%$, $90\%$, $95\%$).
The extracted signed proxies, labeled $v_1$, $v_2$, and $v_3$, versus multiplicity for all four working
points are shown in Fig.~\ref{fig:vn_bdt_wp_overlay}. The spread across working
points is more significant in the signed $v_2$ proxy, while almost robust for the signed $v_3$ proxy, across all multiplicity bins.
For each harmonic $n$ and each multiplicity bin, the BDT systematic is defined as
\begin{equation}
  \sigma_{\rm BDT}^{(n)} = \max_{\epsilon \in \{80\%,\,90\%,\,95\%\}}
  \left| \left(v_n^{\rm sgn}\right)^{\epsilon} - \left(v_n^{\rm sgn}\right)^{85\%} \right|,
  \label{eqn:bdt_syst}
\end{equation}
where the nominal working point is chosen to be $\epsilon_{\rm sig} = 85\%$.
This uncertainty is added in quadrature to the existing total systematic.
Amongst these sources, $\sigma_{\rm BDT}$ is found to be the dominant contribution to the
total systematic uncertainty.

\subsection{Results}

Using the thrust-axis two-particle correlations shown in Section~\ref{sec:TwoParticleCorrelationFunction}, we extract the two-particle Fourier coefficients and report their signed square-root proxies, labeled $v_1$, $v_2$, and $v_3$ in the figures for compactness.
The resulting signed proxies are presented as functions of the event multiplicity interval \ntrkoff.
All results in this section use the $\ntrkoff \ge 30$ skim (Section~\ref{sec:ntrkskim30}): events with $\ntrkoff < 30$ are excluded so that the focus is on the high-multiplicity regime where the hadronic \(\mathrm{WW}\) fraction rises, and where the BDT-selected samples become hadronic-\(\mathrm{WW}\) enriched, rather than on the lower-multiplicity region dominated by \(q\bar q\) production.

\begin{figure}[ht]
\centering
    \includegraphics[width=\textwidth]{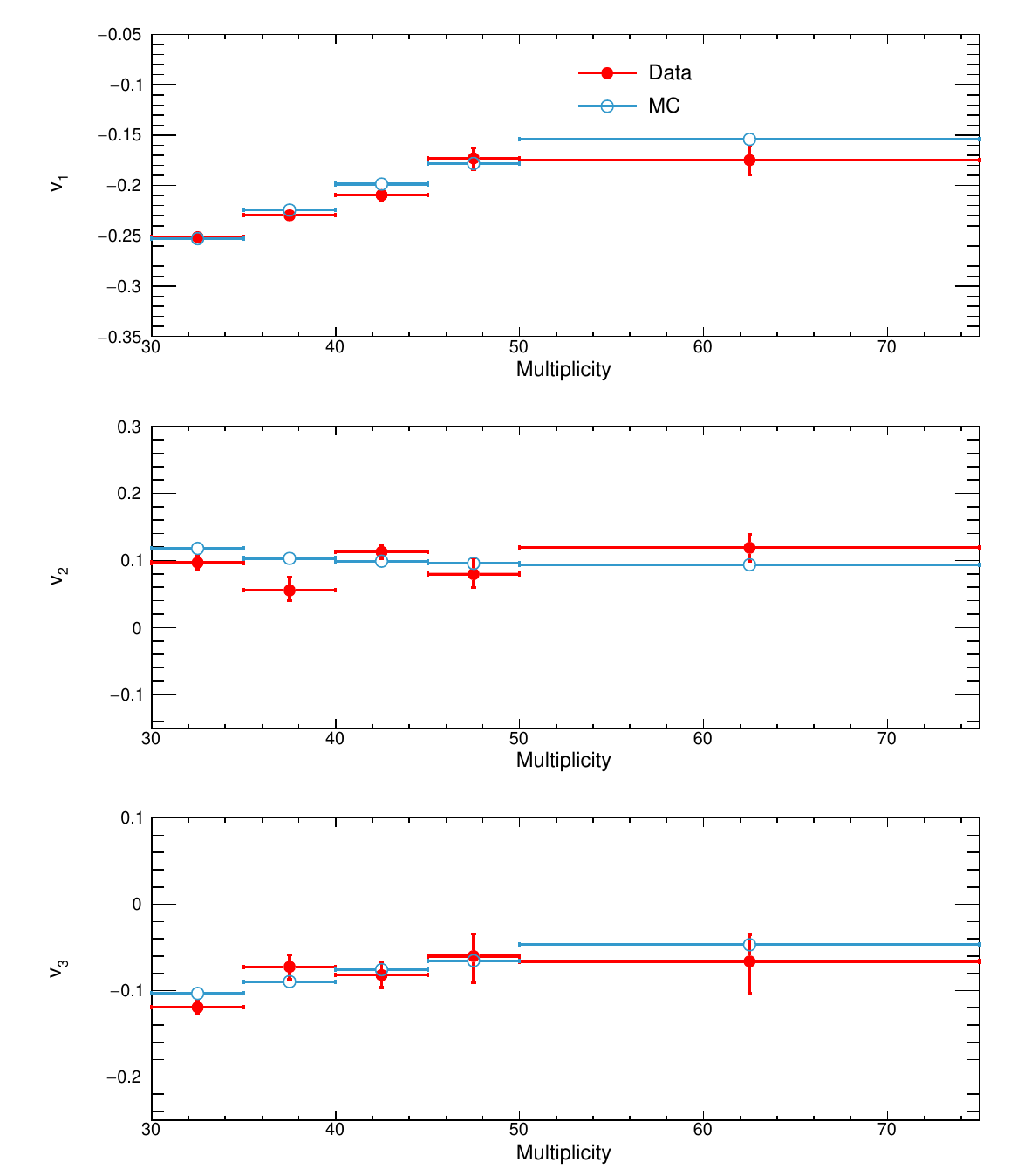}
\caption{The signed square-root proxies, labeled $v_1$, $v_2$ and $v_3$, as a function of offline multiplicity for the thrust-axis analysis of the LEP2 high-energy sample, without the \WW-enhancing BDT selection. Only events with $\ntrkoff \ge 30$ are included (Section~\ref{sec:ntrkskim30}). Data are shown in red, and MC in blue. The systematic uncertainties are shown as transparent boxes.}
\label{fig:vnVsMult_thrust_noBDT}
\end{figure}

Figure~\ref{fig:vnVsMult_thrust_noBDT} shows the signed \(v_n^{\rm sgn}\) proxies obtained from the long-range azimuthal differential yield $Y_l(\Delta \phi)$ for the LEP2 high-energy sample without \(\mathrm{WW}\) BDT enrichment, extracted from results of data (solid dots) and MC simulation (open dots) (in Figure~\ref{fig:rst_nobdt_40phibins_0908_final_dNdphi}).
Within the displayed bins, $|v_1|$ and $|v_3|$ generally decrease as \ntrkoff\ increases, with data and MC following similar slopes within the statistical uncertainties.
The signed \(v_2^{\rm sgn}\) proxy in this sample varies more slowly with multiplicity.
Overall, the data and MC display consistent trends in each signed proxy as a function of event multiplicity.

\begin{figure}[ht]
\centering
    \includegraphics[width=\textwidth]{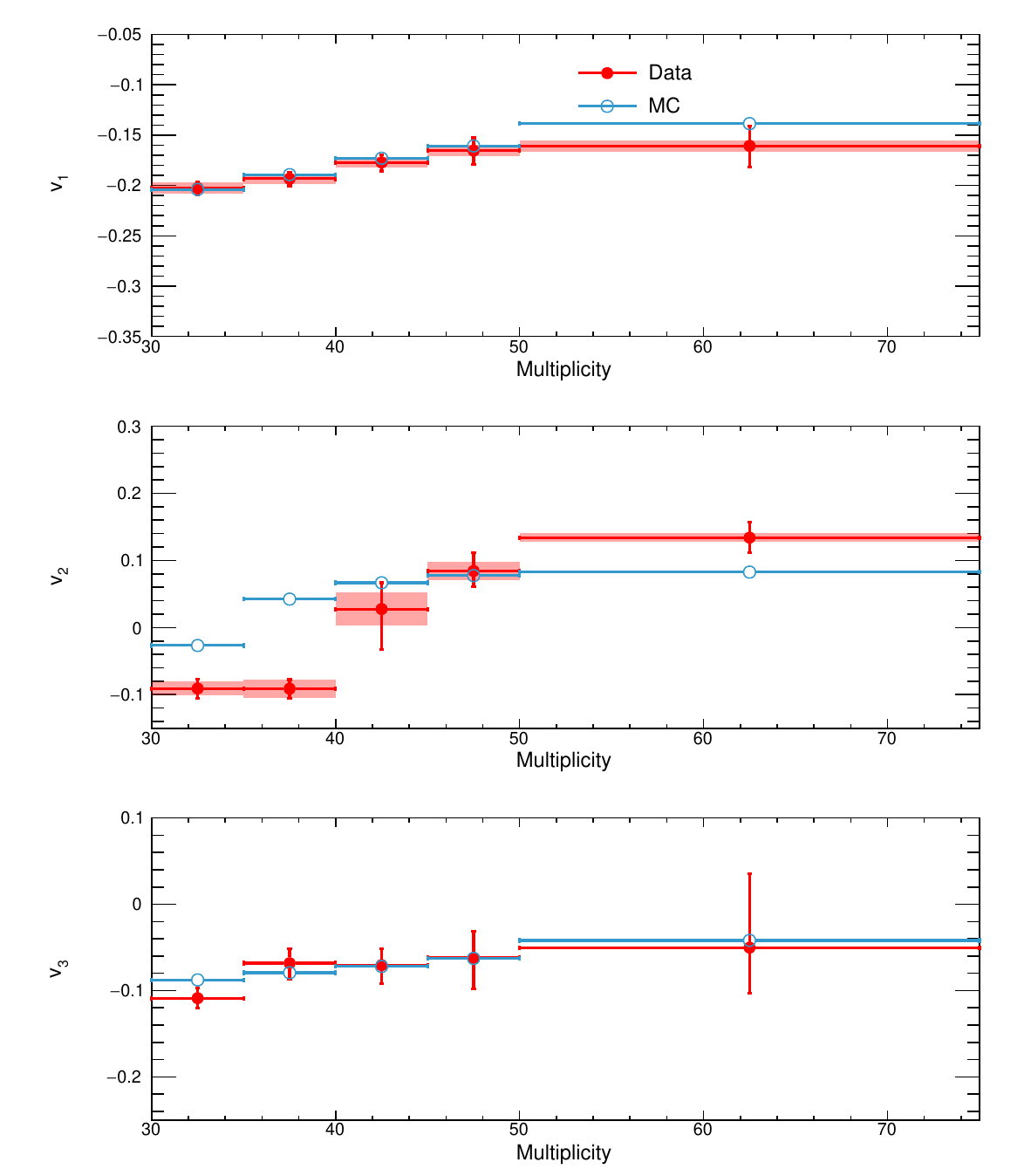}
\caption{Same as Figure~\ref{fig:vnVsMult_thrust_noBDT} but with the \WW-enhanced BDT selection at the 85\% signal-efficiency working point. Data are shown in red, and MC in blue. The systematic uncertainties are shown as transparent boxes.}
\label{fig:vnVsMult_thrust_yesBDT}
\end{figure}

The same exercise for the \WW-enriched sample (input $Y_l (\Delta\phi)$ in Figure~\ref{fig:rst_85pbdt_40phibins_0908_final_dNdphi}) is shown in Figure~\ref{fig:vnVsMult_thrust_yesBDT}.
Here, the data's signed \(v_2^{\rm sgn}\) proxy exhibits a clearer transition from negative toward positive values from $\ntrkoff \ge 30$, in contrast to the milder \(V_{2\Delta}\) multiplicity dependence without the BDT. The data--MC discrepancy also becomes more pronounced after applying the \WW-enhancing BDT selection.
This pattern is qualitatively discussed in the context of string-shoving expectations~\cite{Bierlich:2024lmb}.

\begin{figure}[ht]
\centering
    \begin{subfigure}[b]{0.45\textwidth}
    \includegraphics[width=\textwidth]{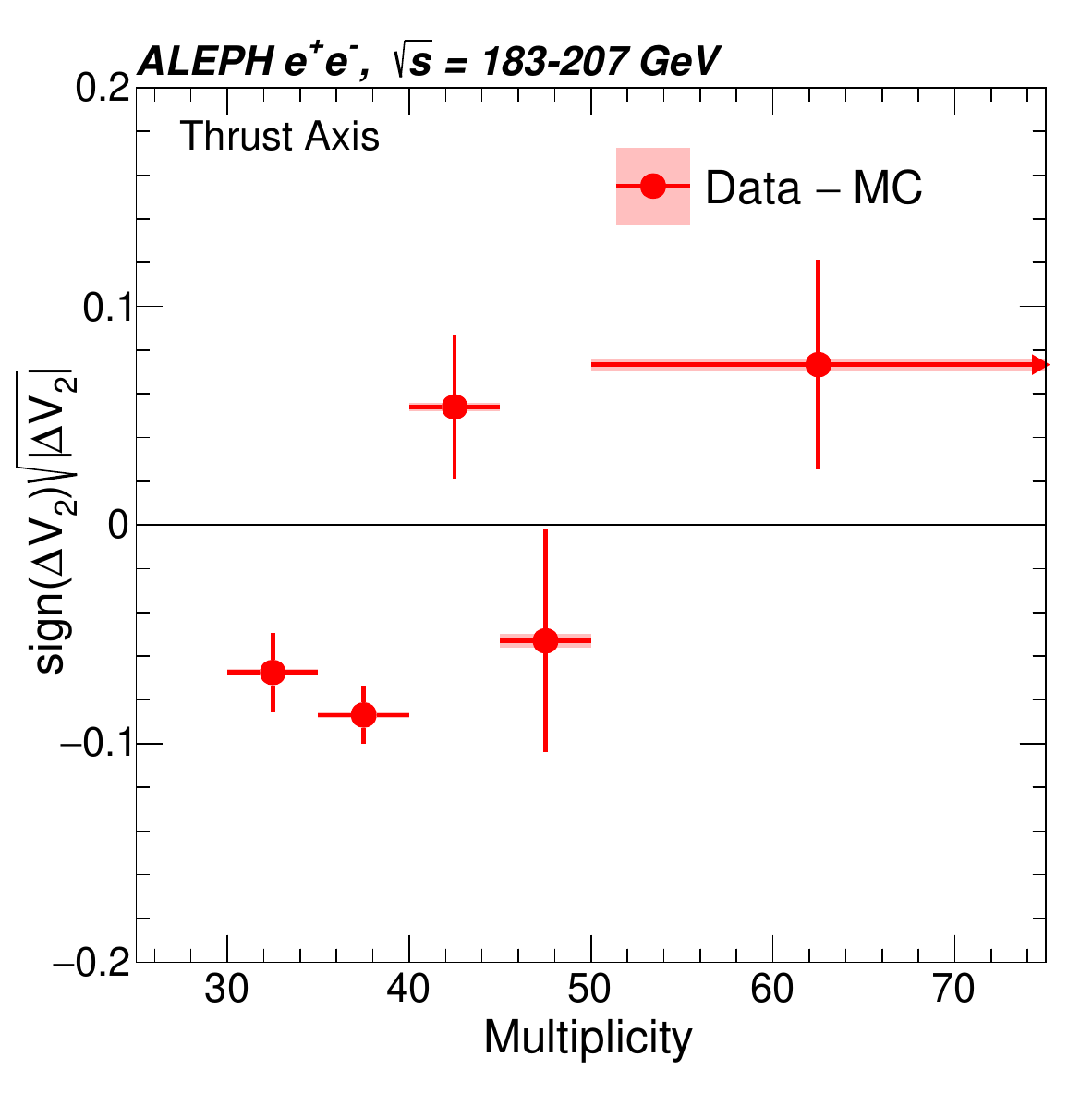}
    \caption{Inclusive (no BDT selection)}
        \label{fig:Dv2VsMult_thrust_noBDT}
    \end{subfigure}
    \begin{subfigure}[b]{0.45\textwidth}
    \includegraphics[width=\textwidth]{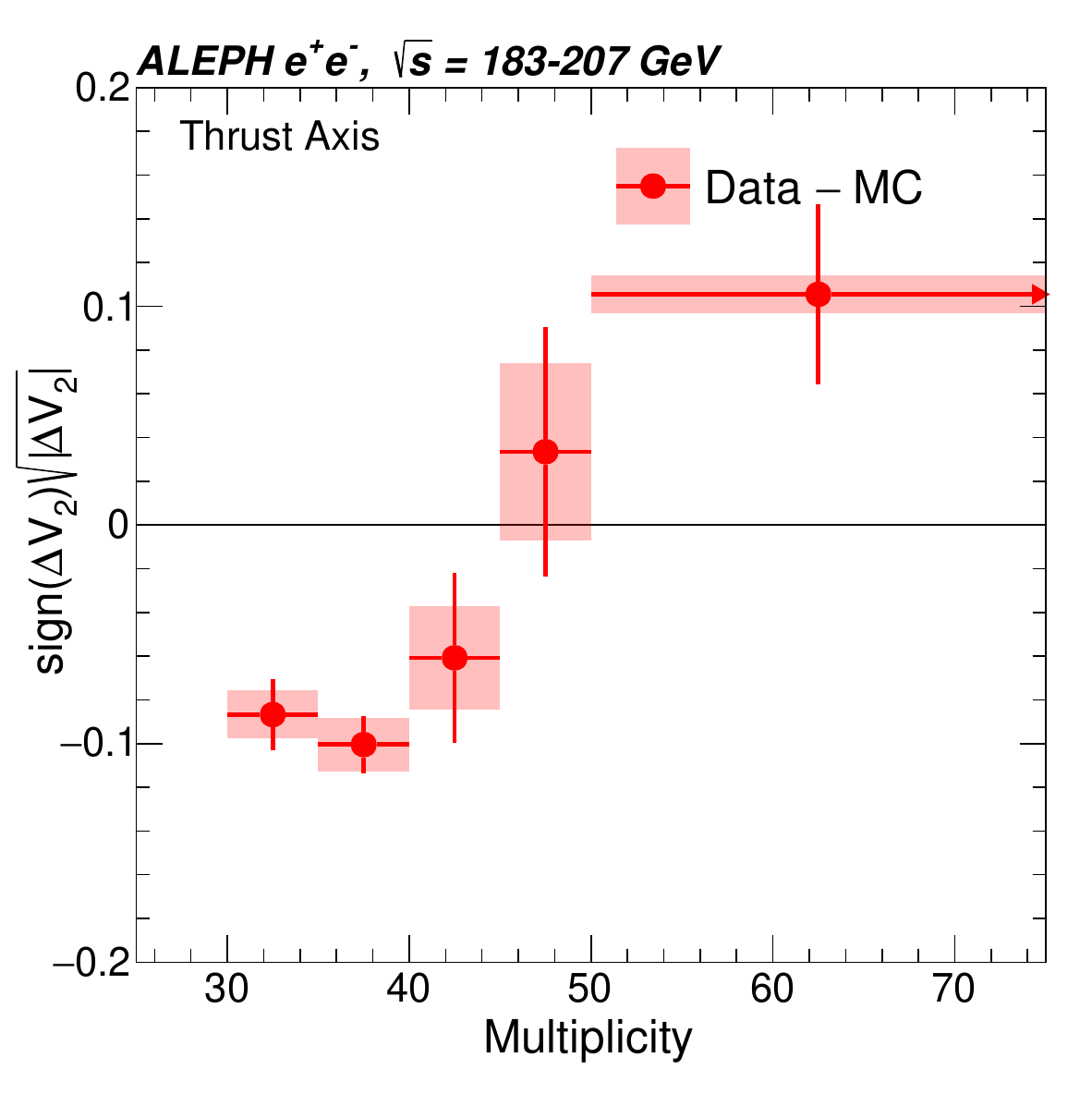}
    \caption{\WW-enhanced with BDT selection}
        \label{fig:Dv2VsMult_thrust_yesBDT}
    \end{subfigure}
\caption{$\mathrm{sign}(\Delta V_{2\Delta})\sqrt{|\Delta V_{2\Delta}|}$ as a function of the offline multiplicity for the thrust-axis analysis of the LEP2 high-energy sample, without (a) and with (b) the \WW-enhanced BDT selection ($\epsilon_{\rm sig}\approx 85\%$), where $\Delta V_{2\Delta} \equiv V_{2\Delta}^{\rm data}-V_{2\Delta}^{\rm MC}$. Vertical bars: bootstrap-propagated statistical uncertainty; shaded red boxes: systematic uncertainty.}
\label{fig:Dv2VsMult_thrust}
\end{figure}

To further highlight the modulation of data against MC, Figure~\ref{fig:Dv2VsMult_thrust} shows the transformed two-particle-harmonic difference
\(\mathrm{sign}(\Delta V_{2\Delta})\sqrt{|\Delta V_{2\Delta}|}\), where
\(\Delta V_{2\Delta} \equiv V_{2\Delta}^{\rm data} - V_{2\Delta}^{\rm MC}\).
This quantity is not a direct difference of the signed \(v_2^{\rm sgn}\) proxies; it is a signed square-root representation of the data--MC difference in the fitted two-particle Fourier coefficient.
Above $\ntrkoff = 30$, where the \WW signal purity is above 60\% for the BDT-selected sample as shown in Fig.~\ref{fig:BDT_purity}, this transformed difference moves from negative to positive in Figure~\ref{fig:Dv2VsMult_thrust_yesBDT}.

\clearpage
\resetmarkpar
\section{Summary}
\label{sec:summary}

Extending from the inclusive measurement of two-particle angular correlations for charged particles with hadronic \ee data at $\sqrt{s} = 183$--209~GeV~\cite{Chen:2023njr,Chen:2023nsi}, we further focus on the \WW-enriched subset using the BDT selection in order to enhance the contribution from multi-color-string configurations.

We follow similar approaches to calculate two-particle correlations, and to derive the two-particle Fourier coefficients and their signed square-root proxies, with four main extensions relative to the published inclusive study: (i)~a stratified combined MC sample matched to luminosity and generator cross-section normalizations; (ii)~a convolution-based background $B(\Delta\eta,\Delta\phi)$ in place of mixed events; (iii)~a multivariate BDT targeting hadronic \(\mathrm{WW}\) topologies; and (iv)~a bootstrap-derived bin-to-bin covariance of the long-range yield $Y_l(\Delta\phi)$, propagated into the Fourier fit through a Bayesian procedure to extract the signed proxies with correlated statistical uncertainties.
The correlation function measured with the inclusive hadronic data is qualitatively consistent with our previous publication when the BDT is not applied.
With the BDT, the long-range yield shows a stronger \(\Delta\phi\) modulation in some multiplicity intervals, including \(30 \le N_{\rm Trk}^{\rm Offline} < 40\), where \(\mathrm{WW}\) purity is rising.
The extracted signed \(v_2^{\rm sgn}\) proxy as a function of event multiplicity shows a dependence on \(\mathrm{WW}\) enrichment that differs from the archived MC in places, and suggests a sign change versus multiplicity in the same sense as discussed for the string shoving scenario~\cite{Bierlich:2024lmb}; quantitative agreement with that model is not claimed here.

The results presented in this note represent the current status of the \WW-enriched \ee two-particle correlation study. The autocorrelation-vs-mixed-event closure check and the systematic-uncertainty evaluation are included in Sections~\ref{sec:TwoParticleCorrelationFunction} and~\ref{sec:flow_bayesian}, respectively. Further MC-based cross-checks of the BDT-induced selection bias would be useful for a future revision, but the present results already include the closure and systematic studies described above.

\clearpage

\appendix
\renewcommand{\thesubsection}{\Alph{section}.\arabic{subsection}}
\renewcommand{\thesubsubsection}{\thesubsection.\arabic{subsubsection}}
\setcounter{subsection}{0}

\section{Details (other BDT working points)}

\begin{figure}[ht]
\centering
    \includegraphics[width=0.45\textwidth]{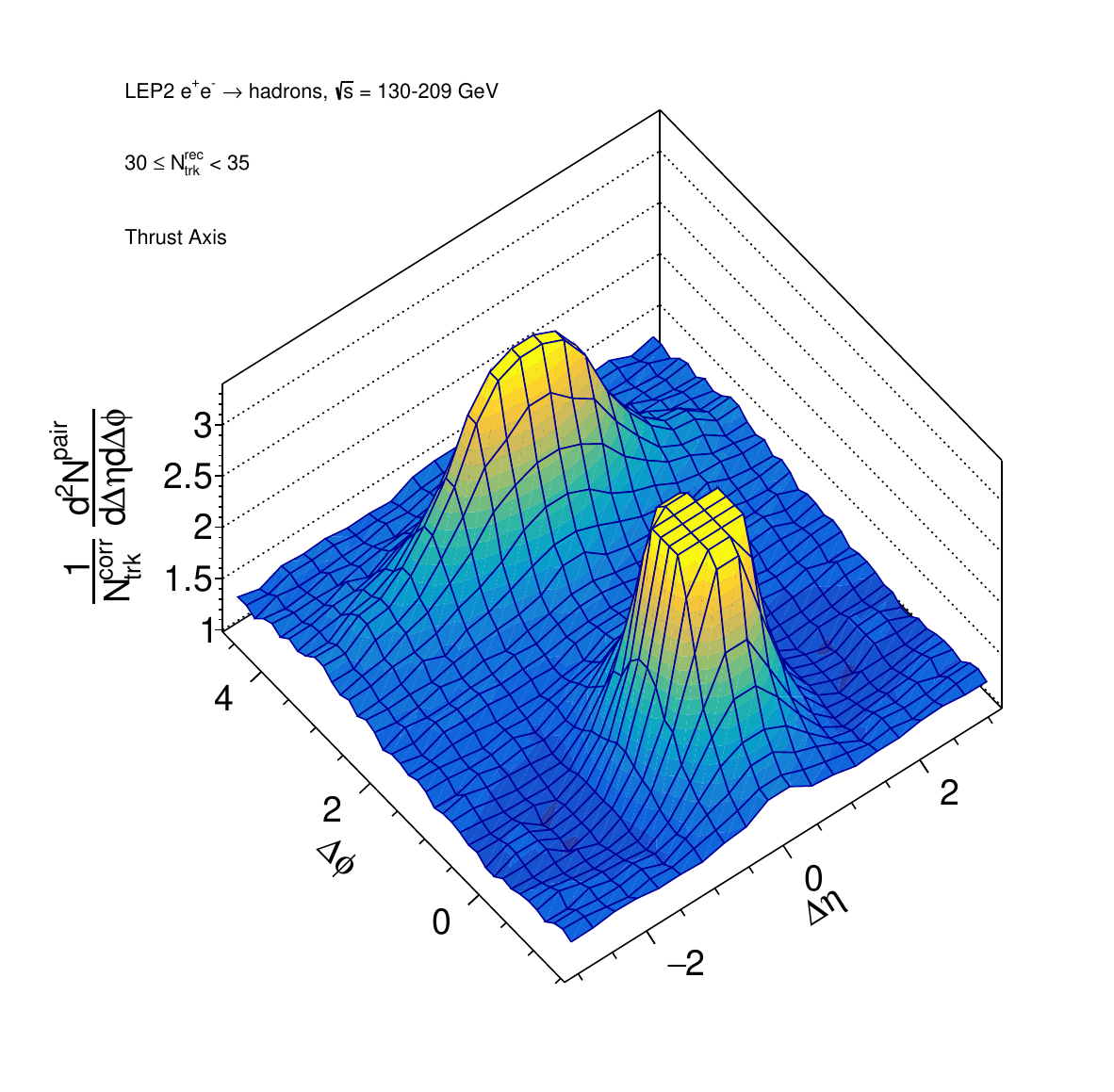}
    \includegraphics[width=0.45\textwidth]{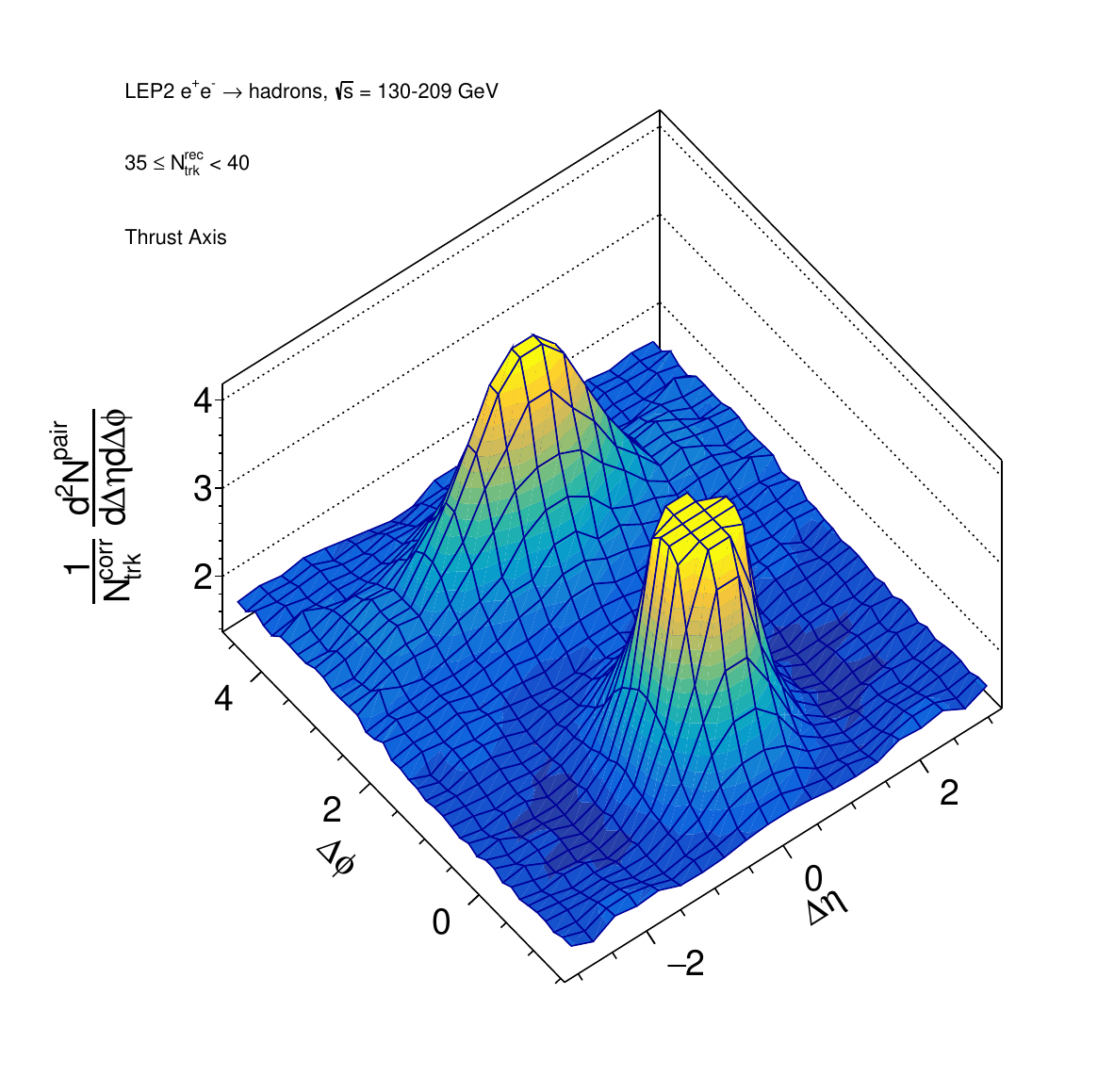}
    \includegraphics[width=0.45\textwidth]{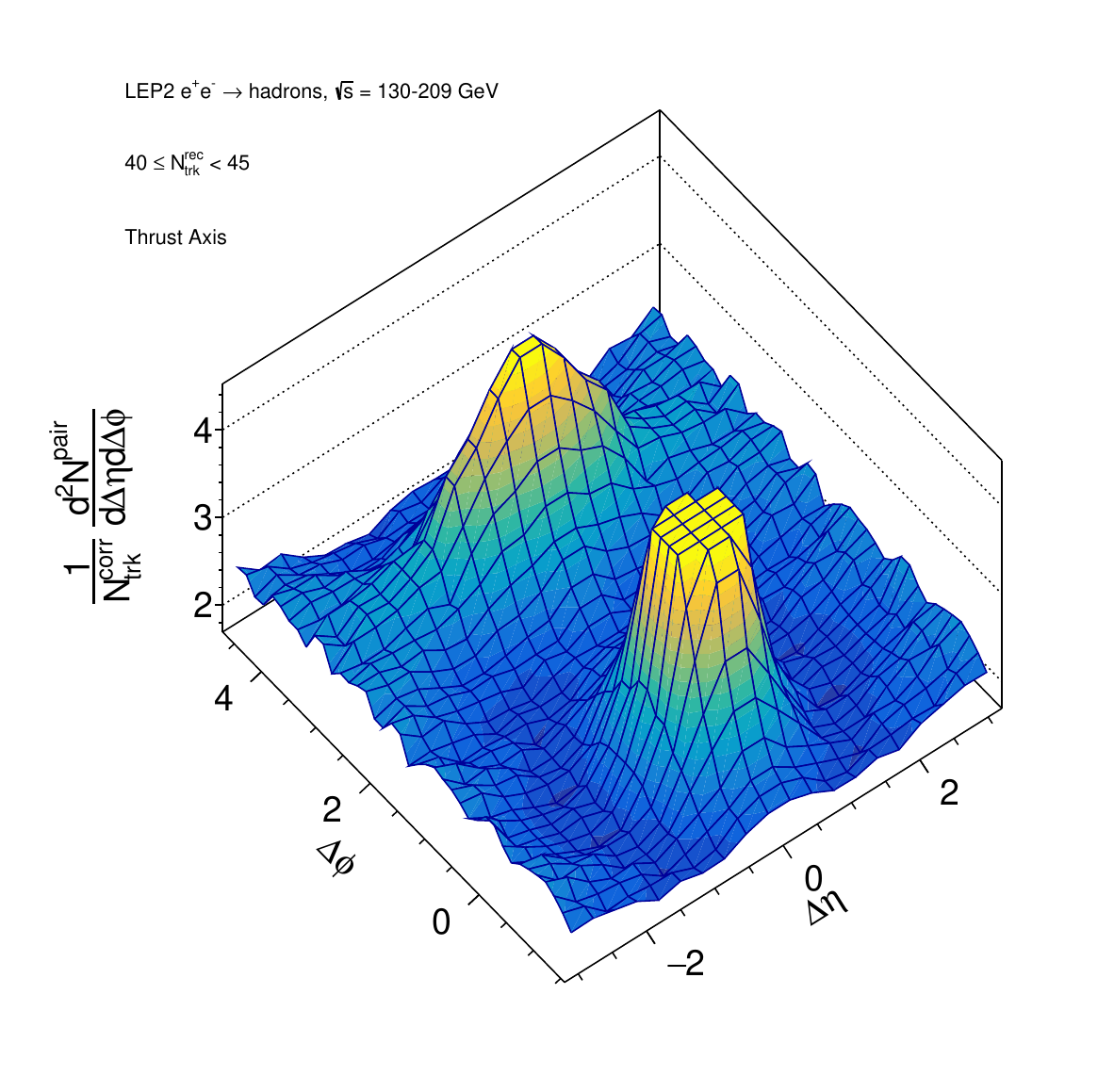}
    \includegraphics[width=0.45\textwidth]{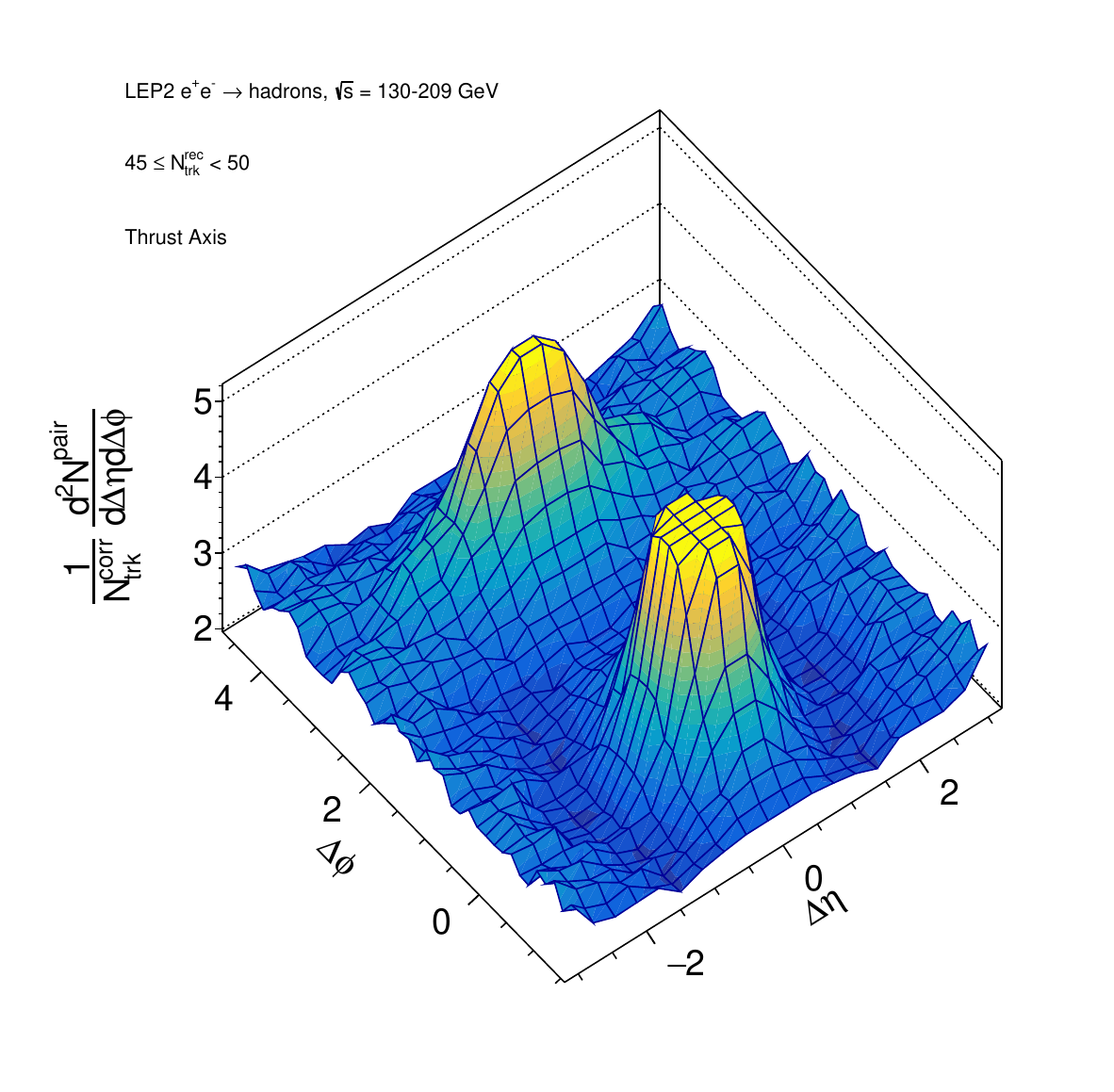}
    \includegraphics[width=0.45\textwidth]{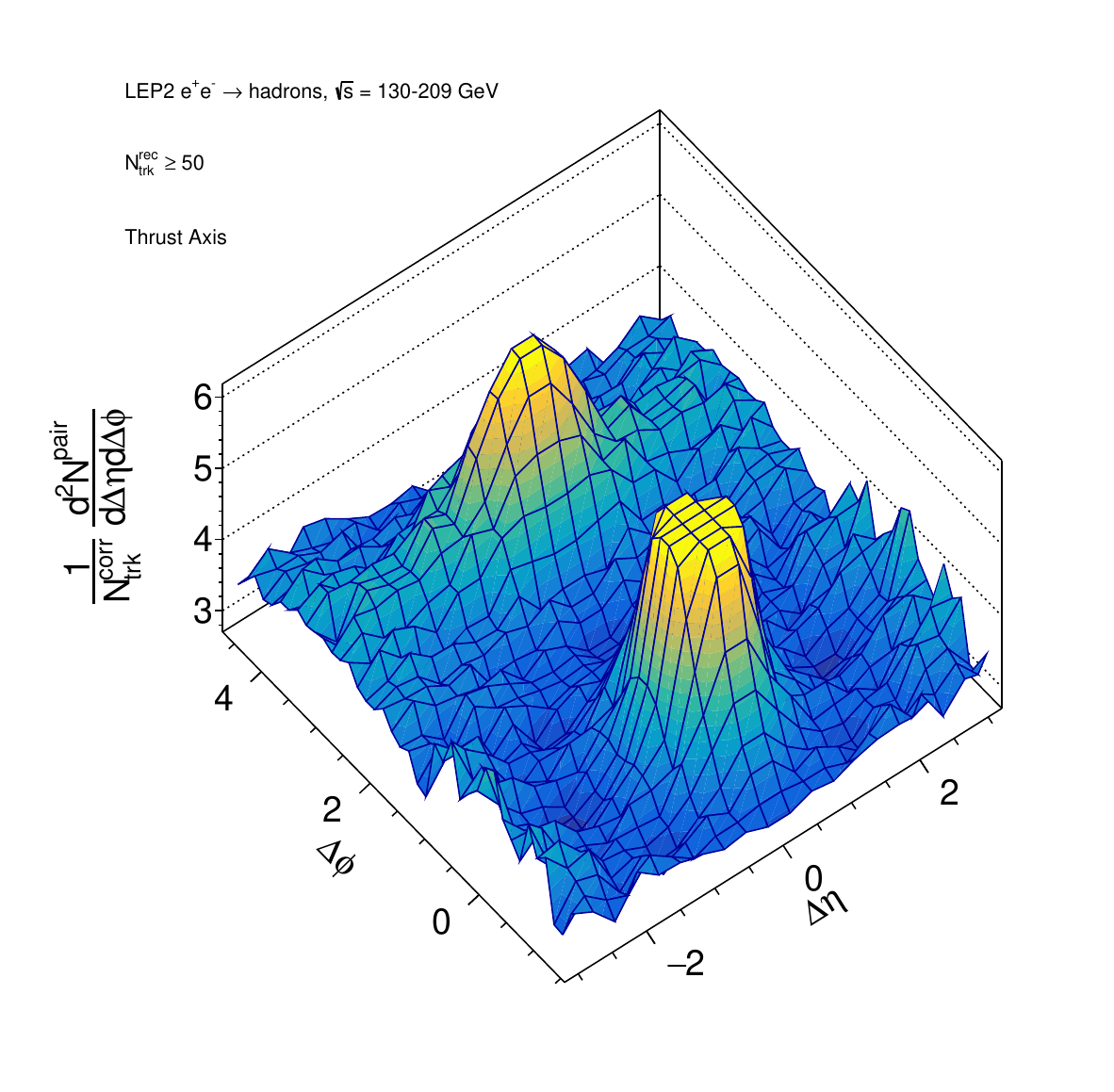}
\caption{[No BDT selection] Two-particle correlation function with respect to the
  thrust axis in the multiplicity intervals $[30,35)$, $[35,40)$, $[40,45)$,
  $[45,50)$, and $[50,\infty)$.}
\label{fig:rst_nobdt_40phibins_0908_final_2PC}
\end{figure}

\begin{figure}[ht]
\centering
    \includegraphics[width=0.45\textwidth]{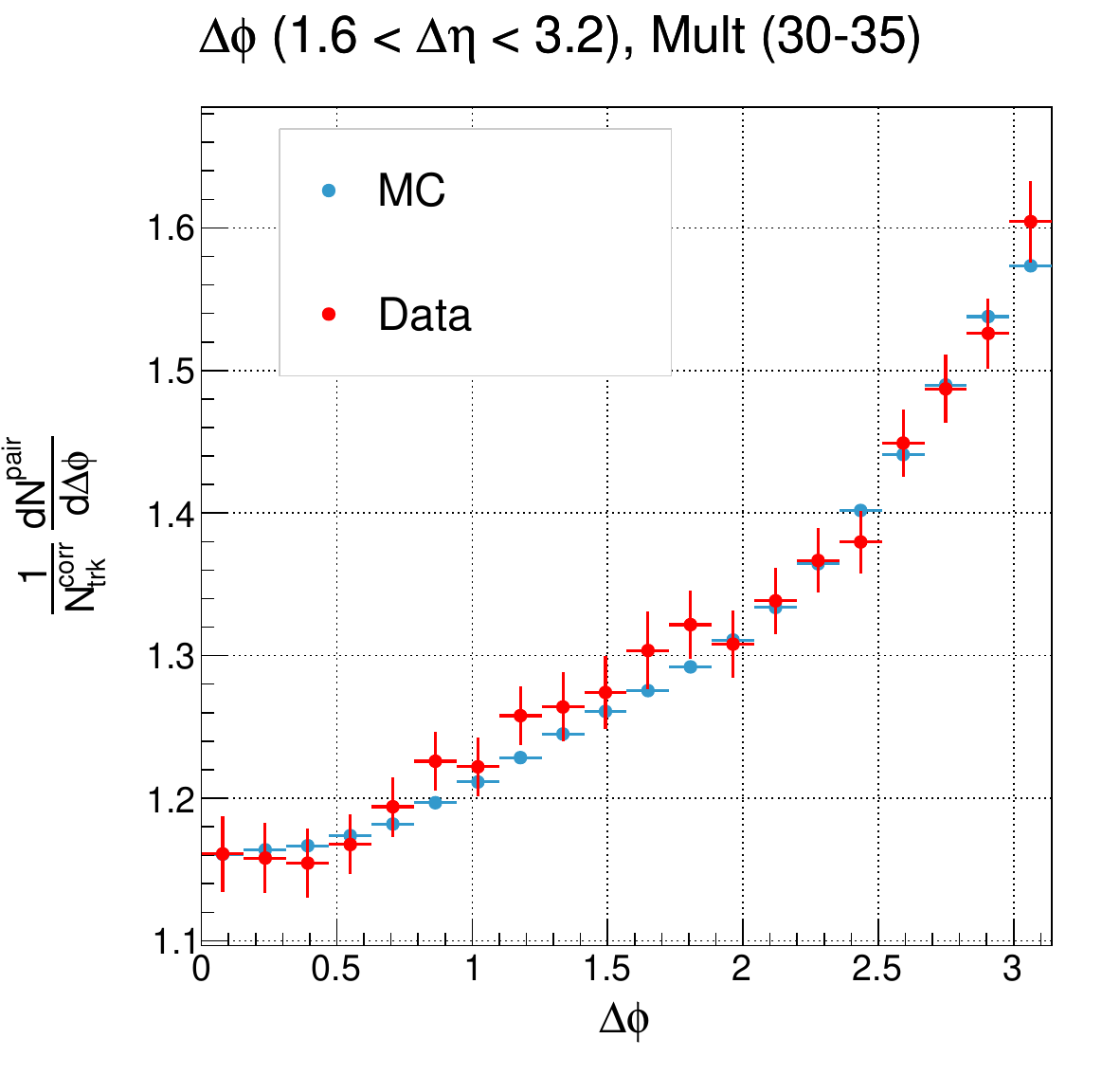}
    \includegraphics[width=0.45\textwidth]{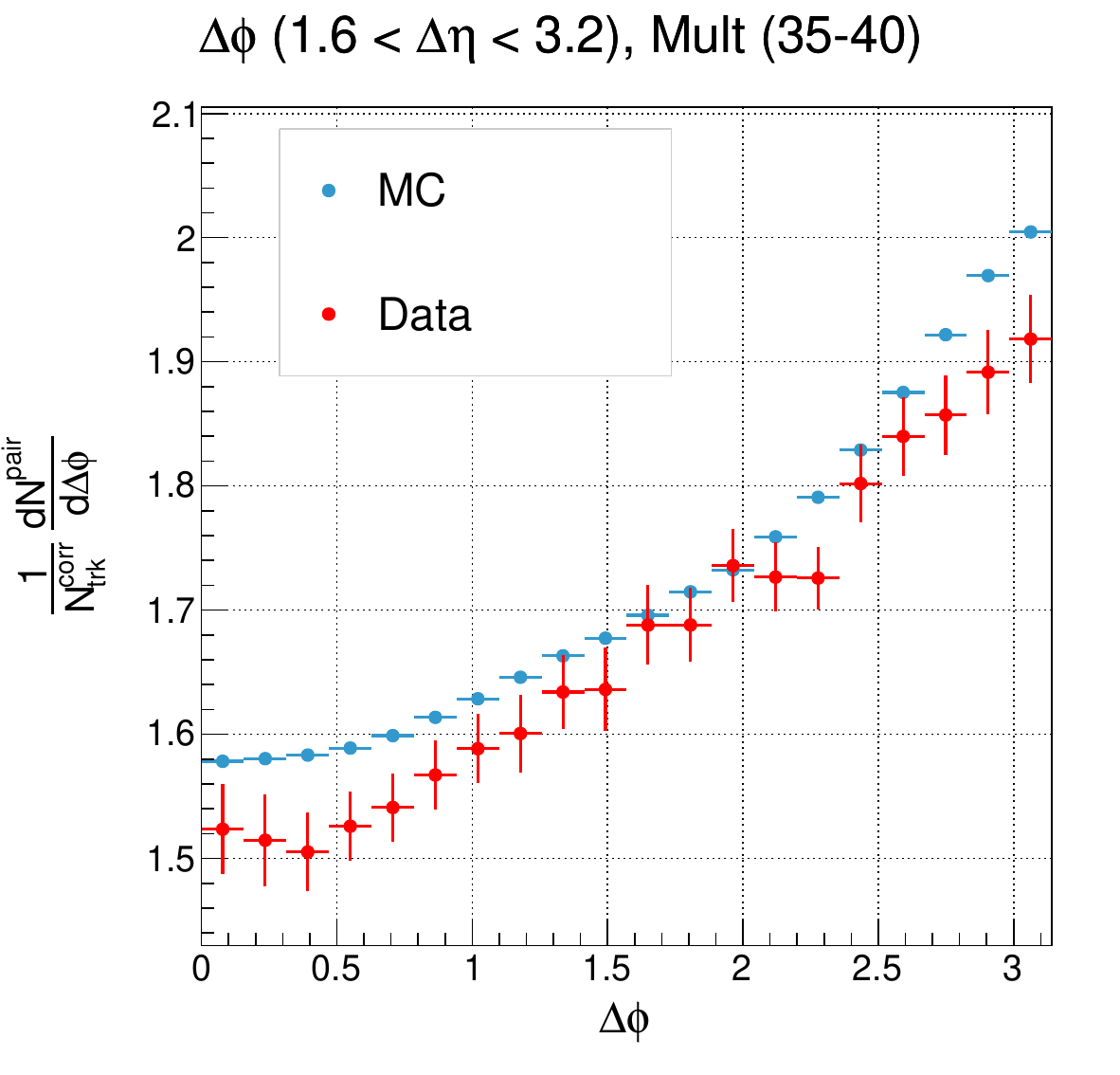}
    \includegraphics[width=0.45\textwidth]{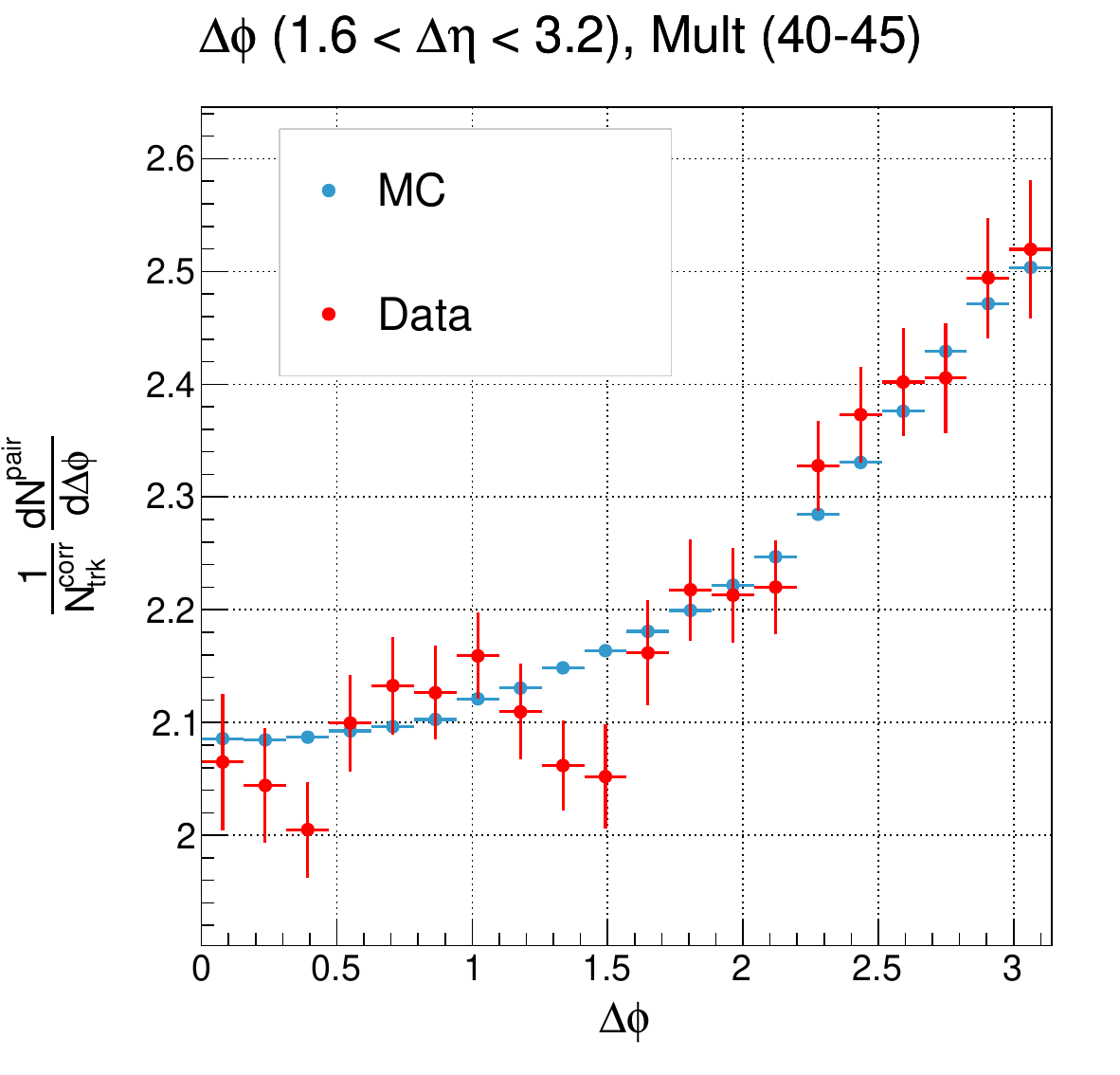}
    \includegraphics[width=0.45\textwidth]{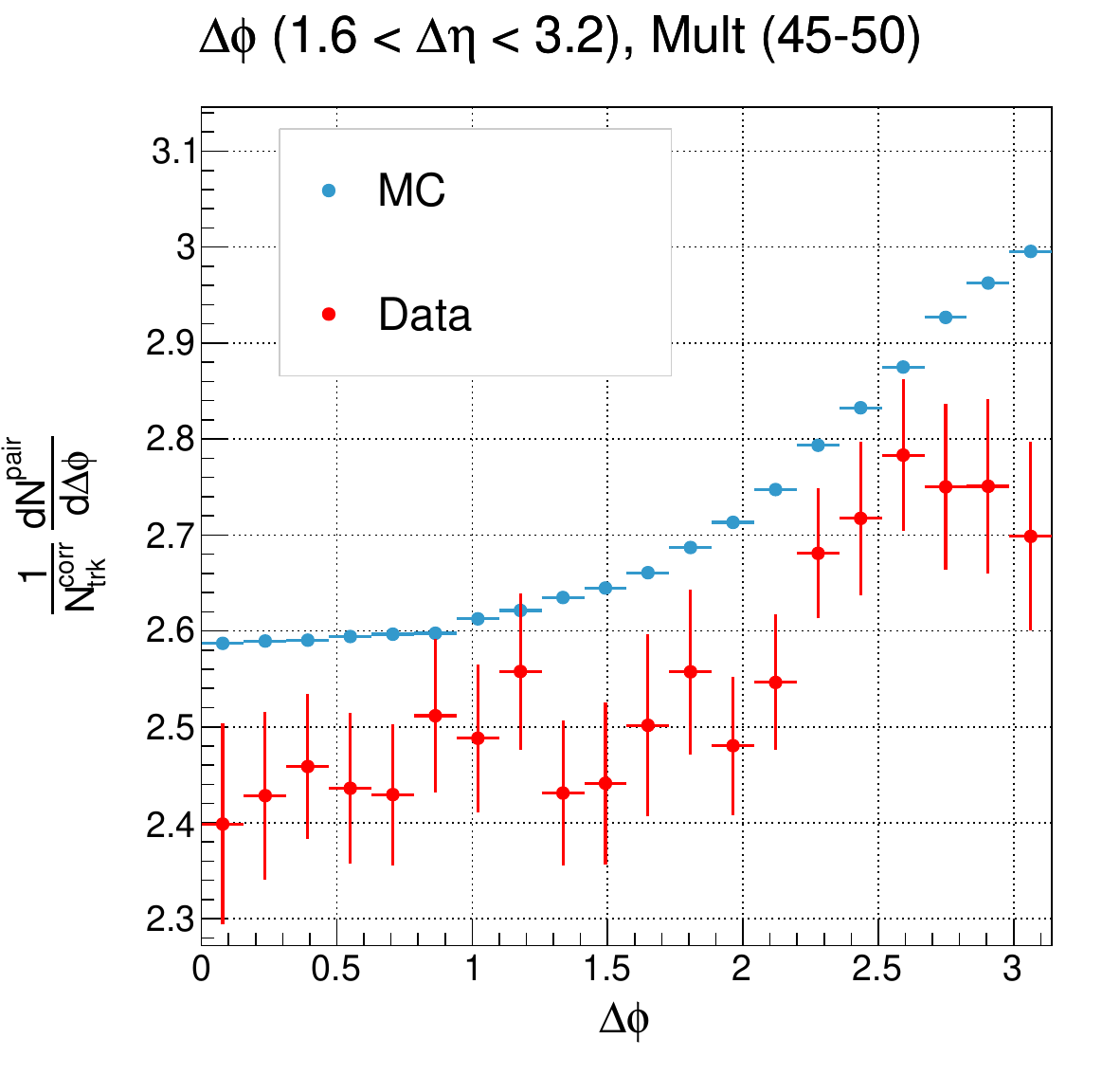}
    \includegraphics[width=0.45\textwidth]{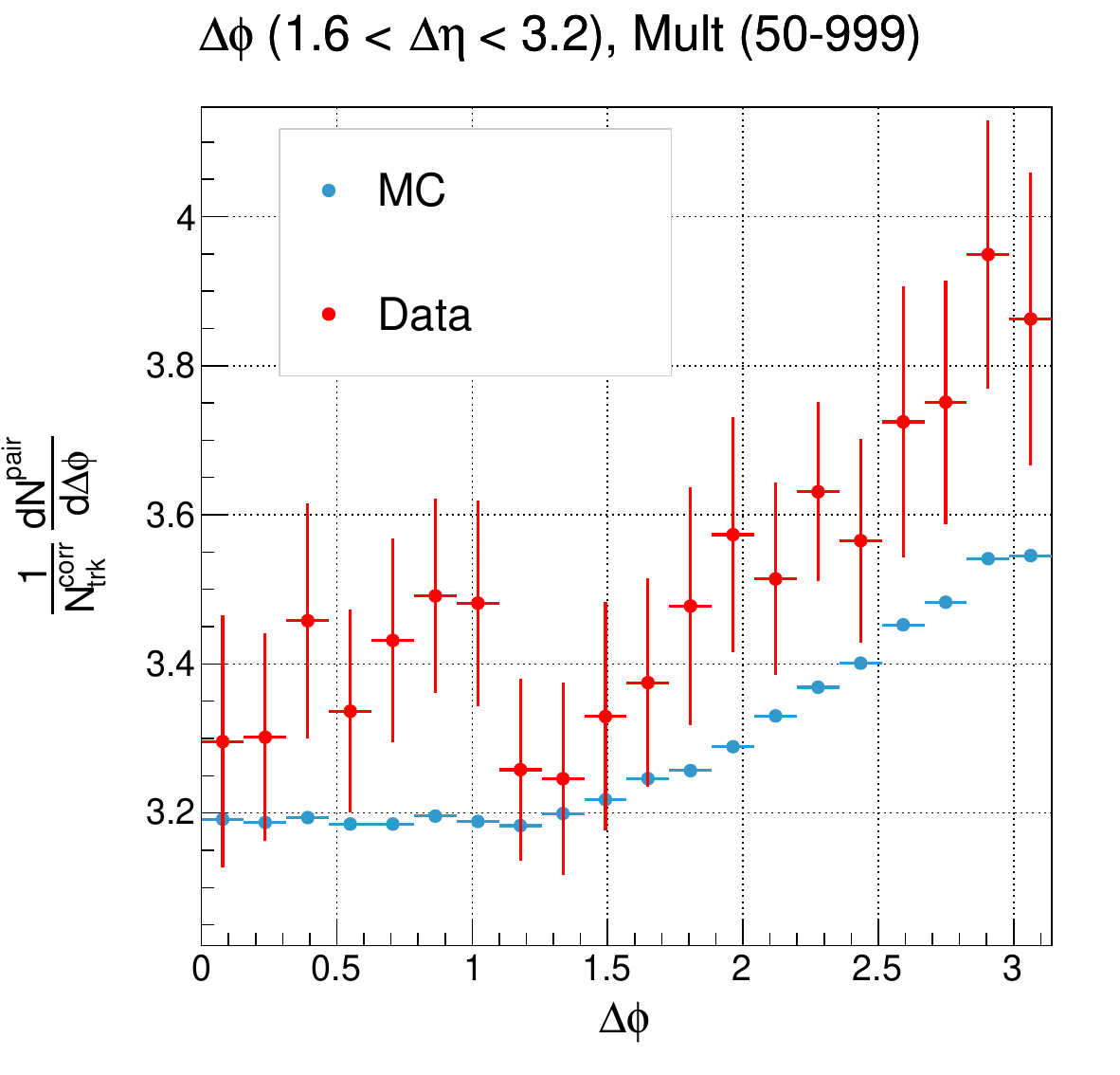}
\caption{[No BDT selection] Long-range ($1.6<\Delta\eta<3.2$) azimuthal differential
  yield $Y_l(\Delta\phi)$ with respect to the thrust axis in the multiplicity
  intervals $[30,35)$, $[35,40)$, $[40,45)$, $[45,50)$, and $[50,\infty)$.
  Data (red) carry bootstrap-derived statistical uncertainties
  (Sec.~\ref{sec:flow_bootstrap}); MC (blue) shows statistical errors.}
\label{fig:rst_nobdt_40phibins_0908_final_dNdphi}
\end{figure}

\begin{figure}[ht]
\centering
    \includegraphics[width=0.45\textwidth]{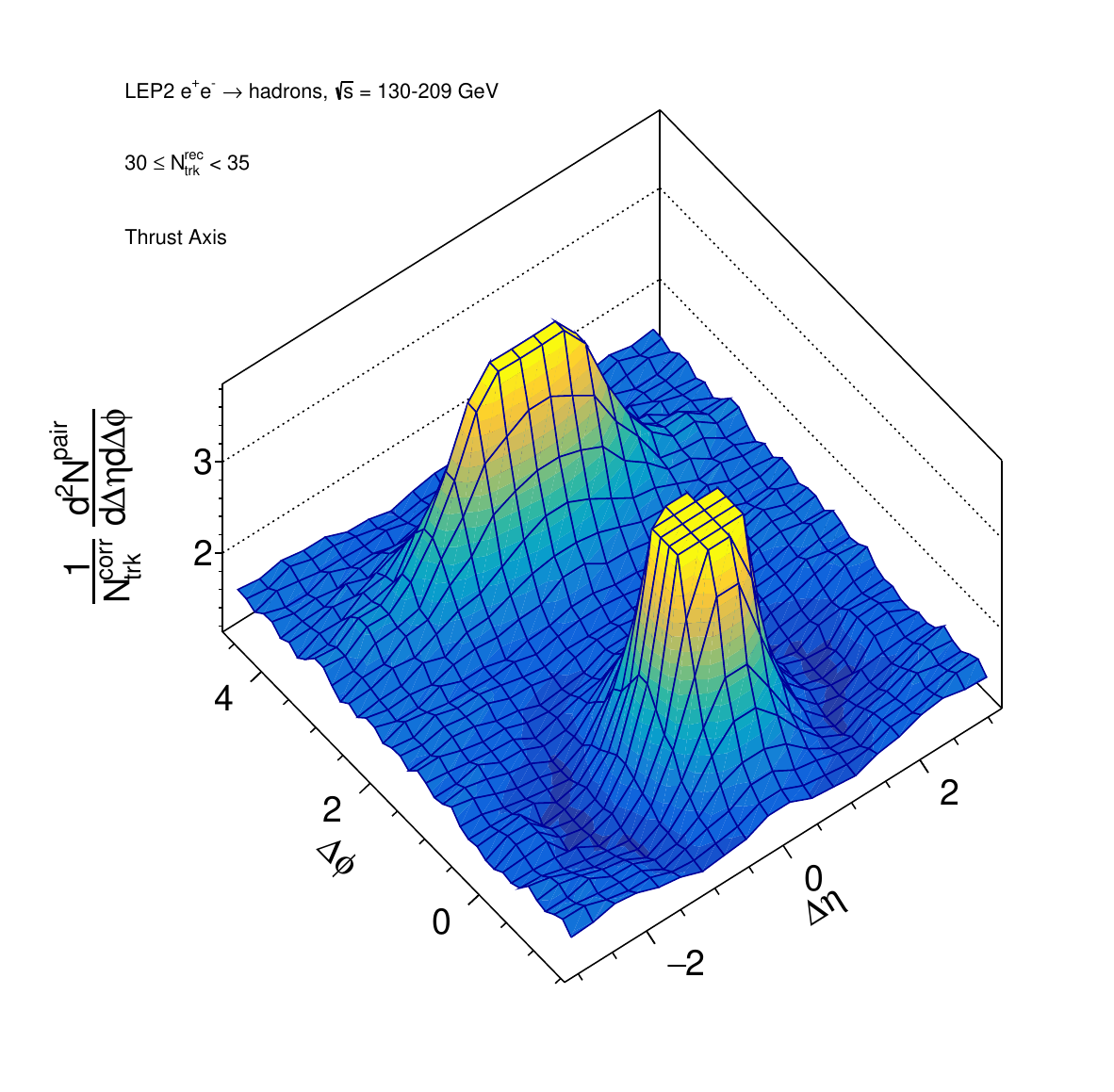}
    \includegraphics[width=0.45\textwidth]{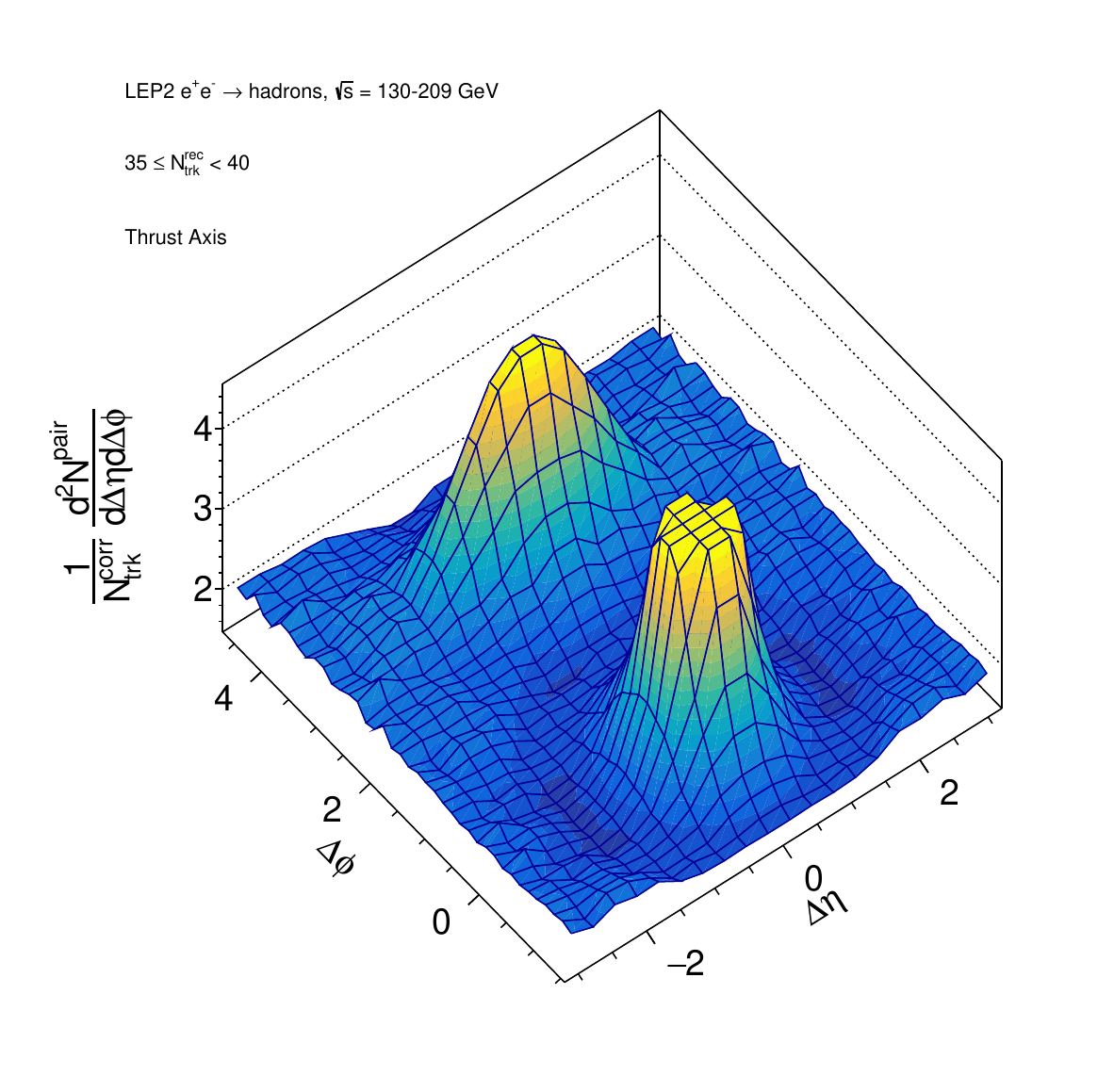}
    \includegraphics[width=0.45\textwidth]{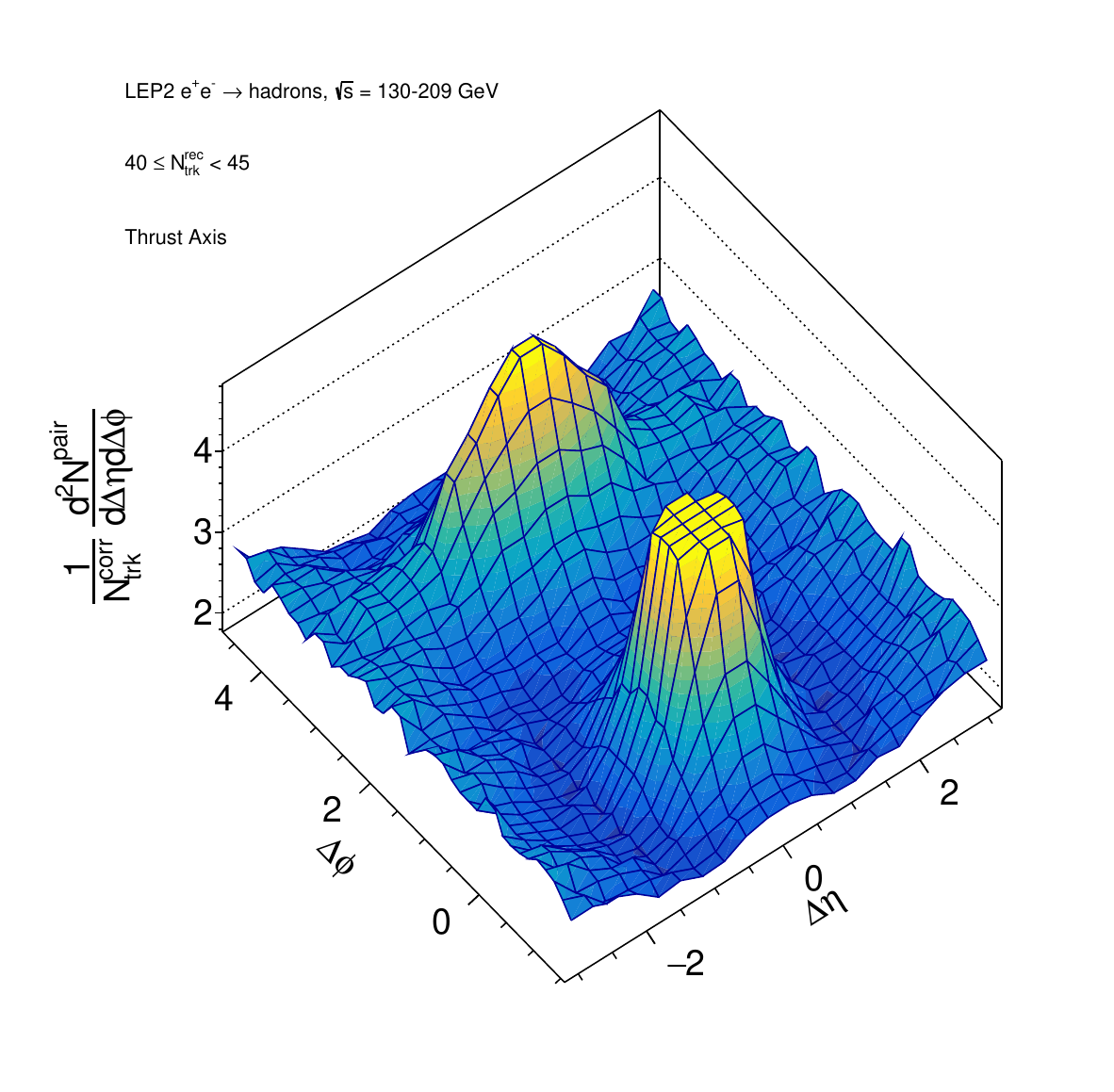}
    \includegraphics[width=0.45\textwidth]{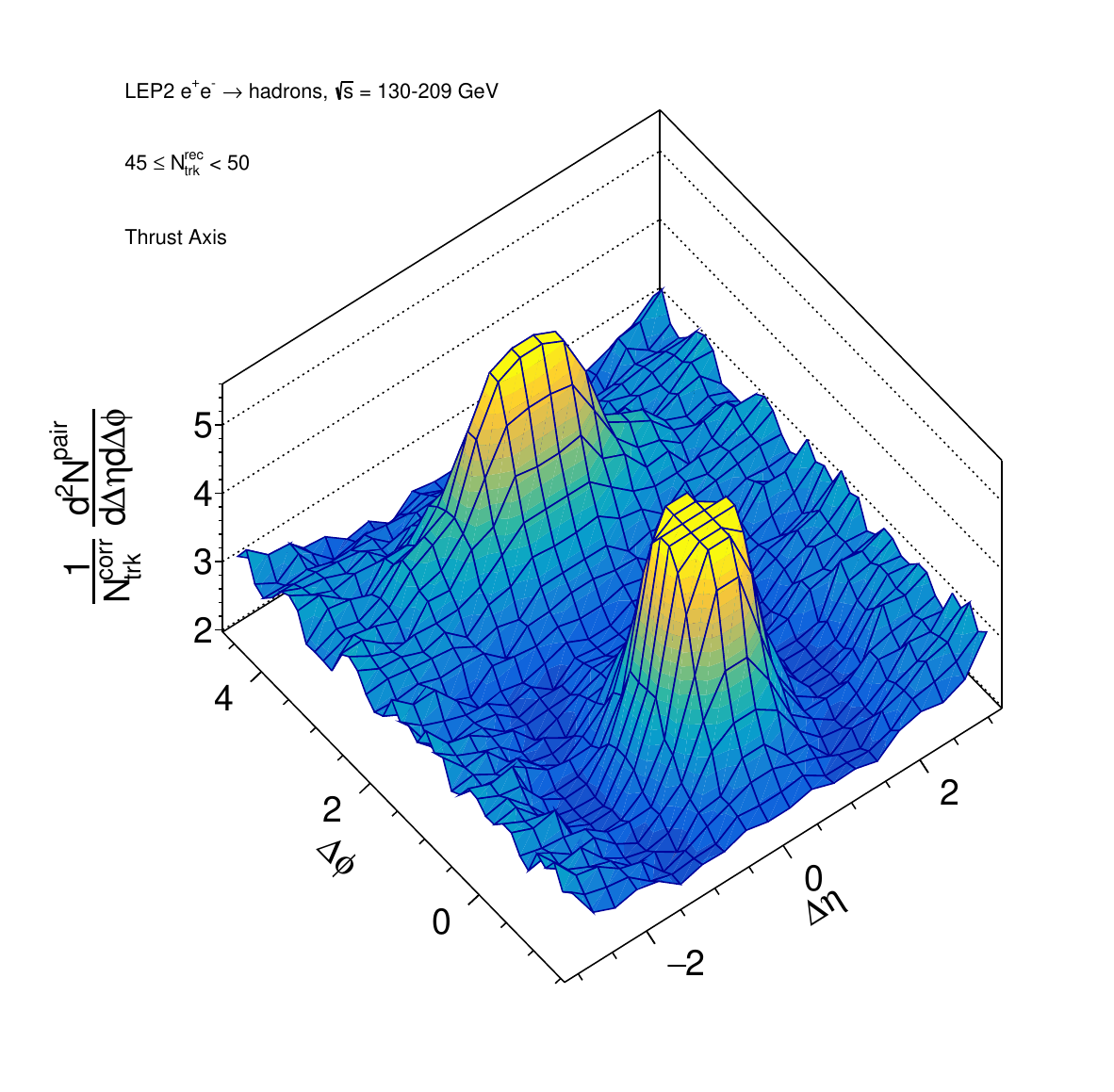}
    \includegraphics[width=0.45\textwidth]{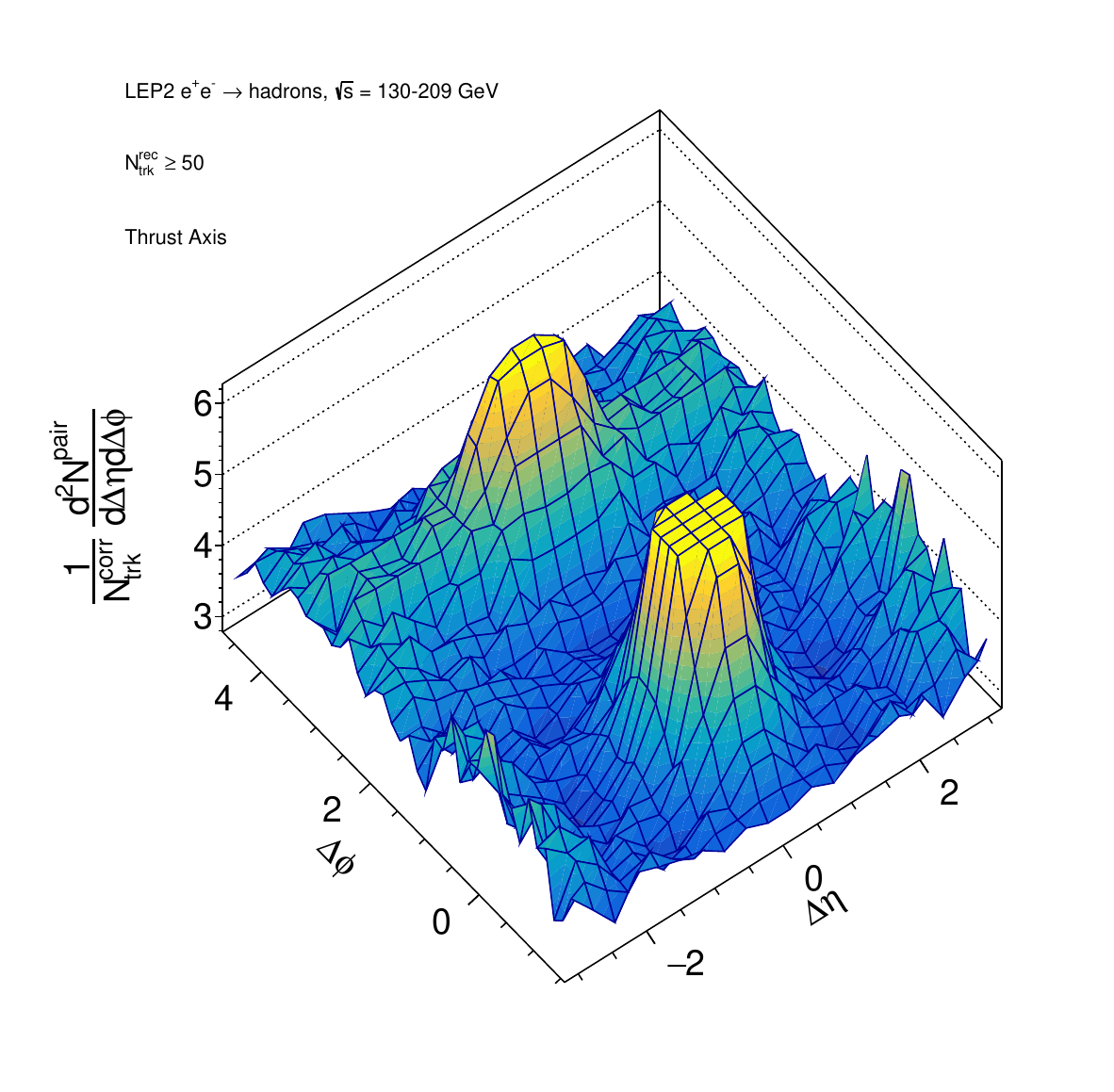}
\caption{[BDT at $\epsilon_{\rm sig}\approx 95\%$] Two-particle correlation function
  with respect to the thrust axis in the multiplicity intervals $[30,35)$,
  $[35,40)$, $[40,45)$, $[45,50)$, and $[50,\infty)$.}
\label{fig:rst_95pbdt_40phibins_0908_final_2PC}
\end{figure}

\begin{figure}[ht]
\centering
    \includegraphics[width=0.45\textwidth]{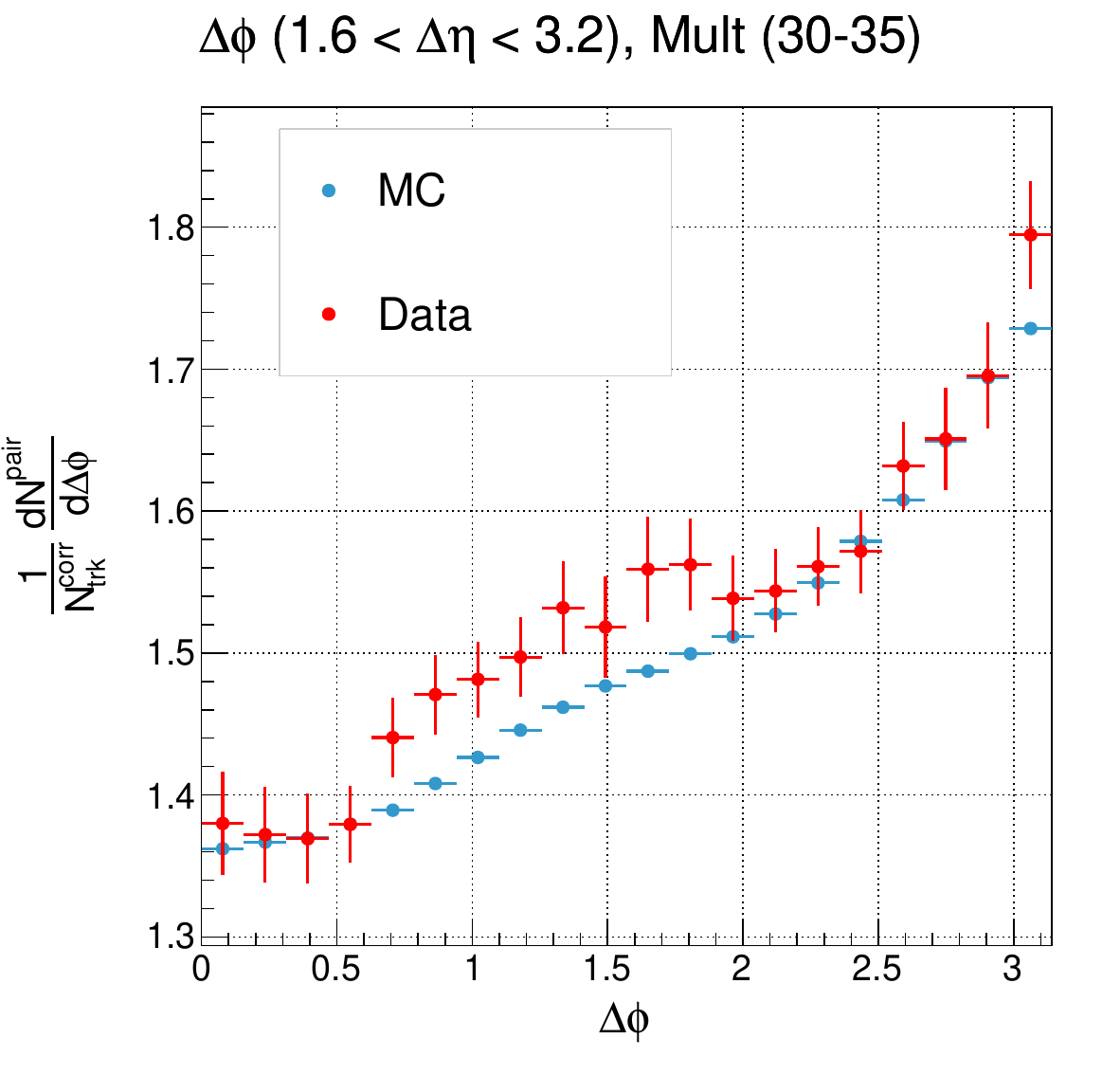}
    \includegraphics[width=0.45\textwidth]{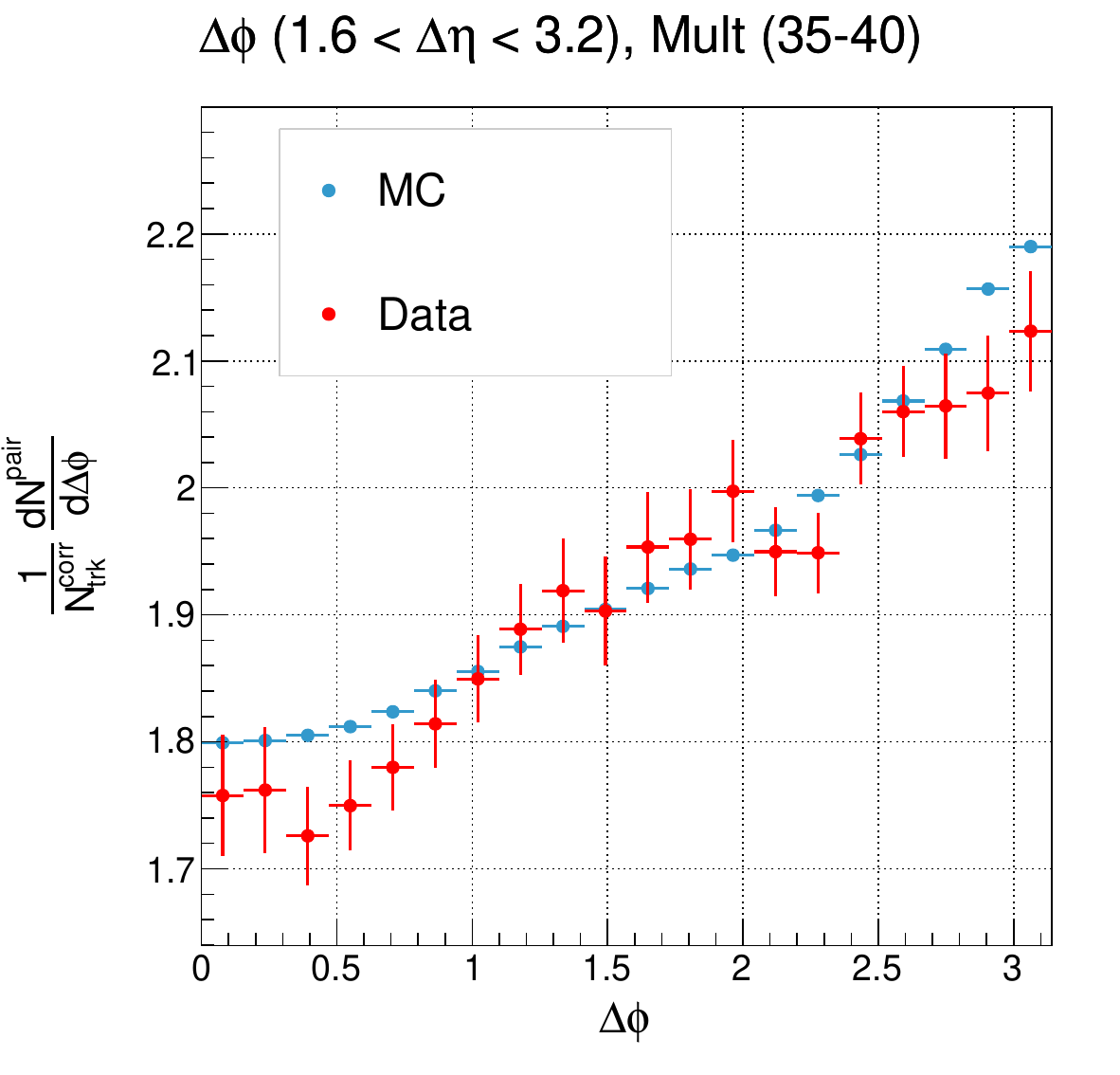}
    \includegraphics[width=0.45\textwidth]{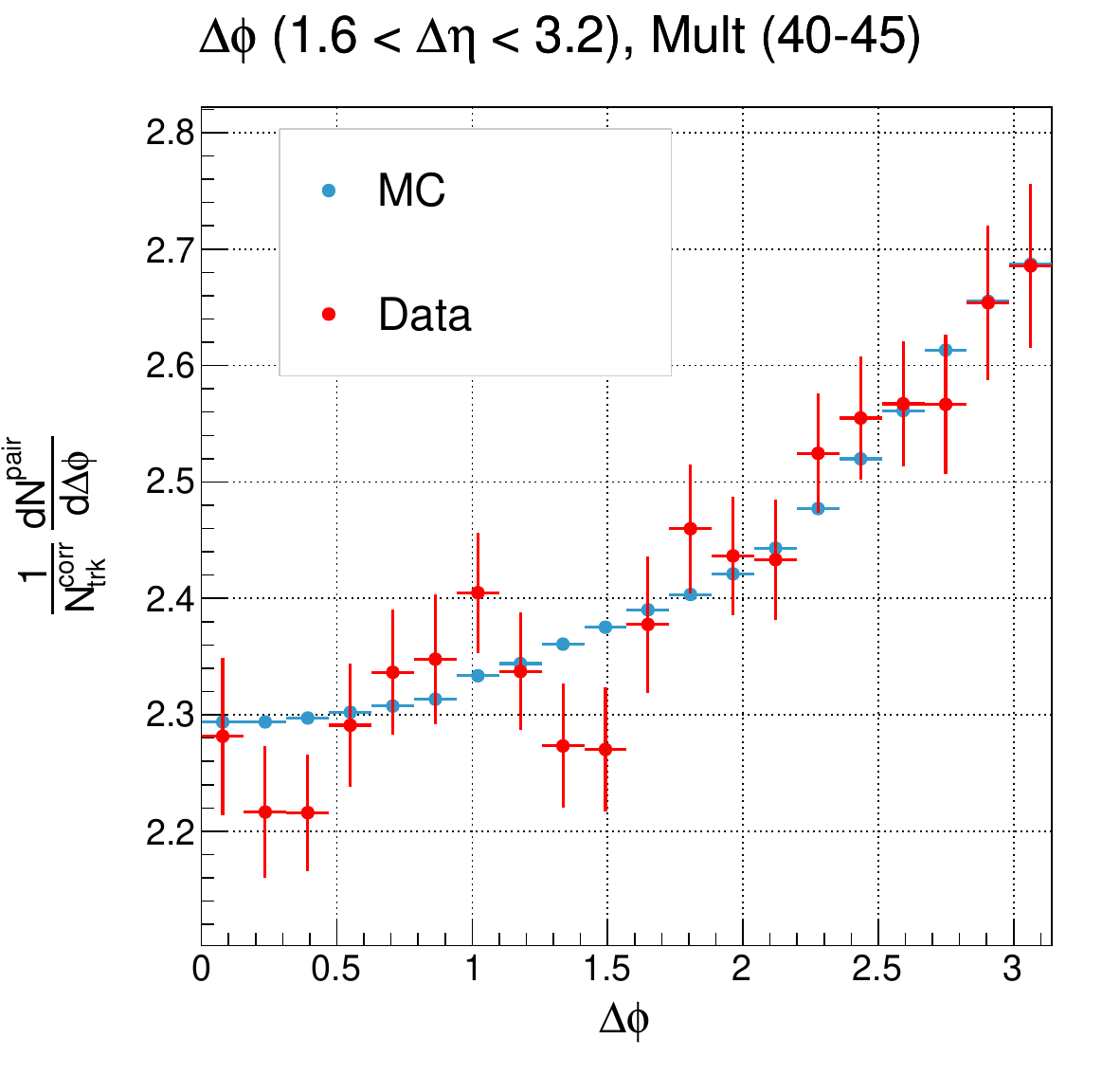}
    \includegraphics[width=0.45\textwidth]{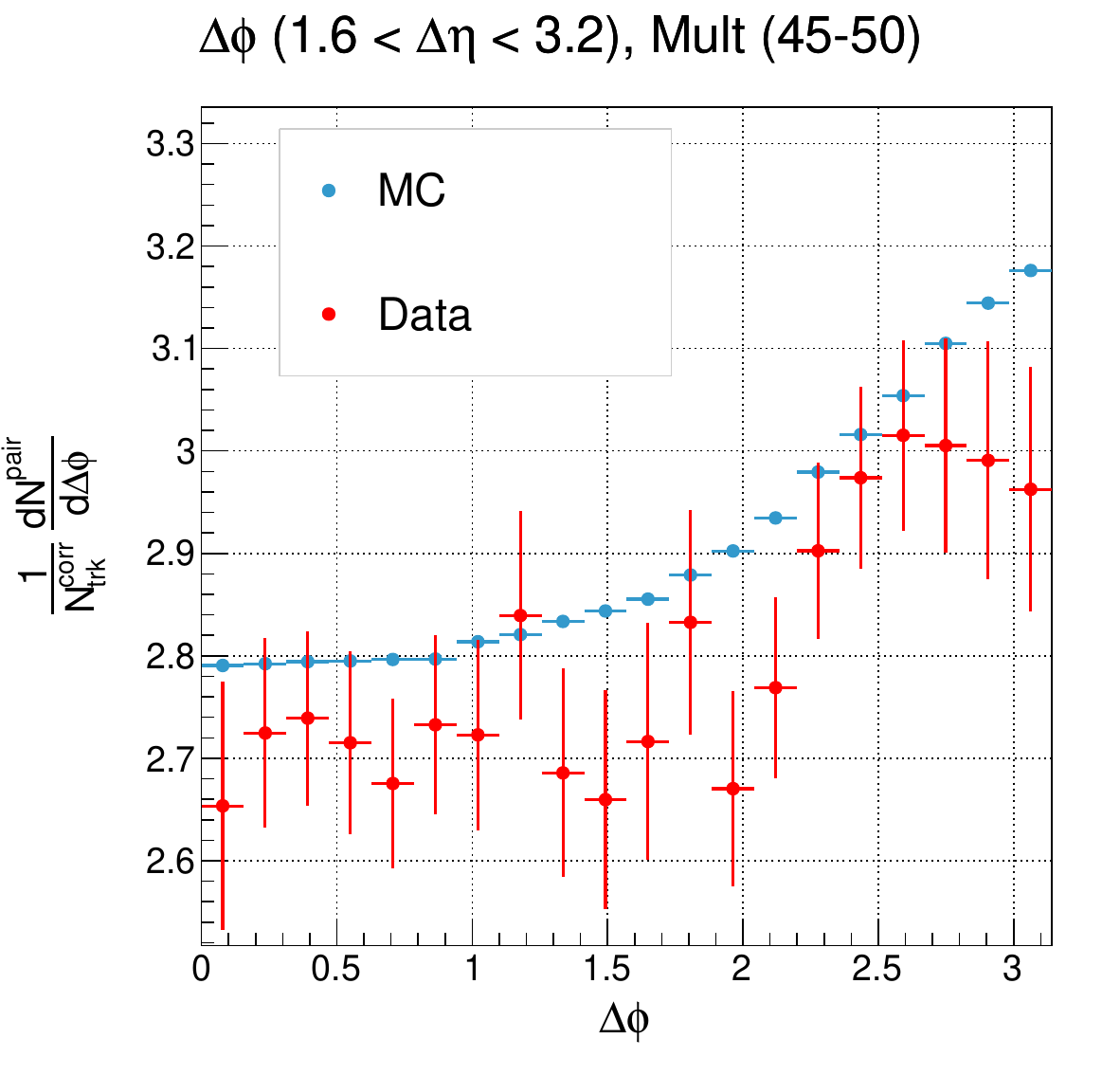}
    \includegraphics[width=0.45\textwidth]{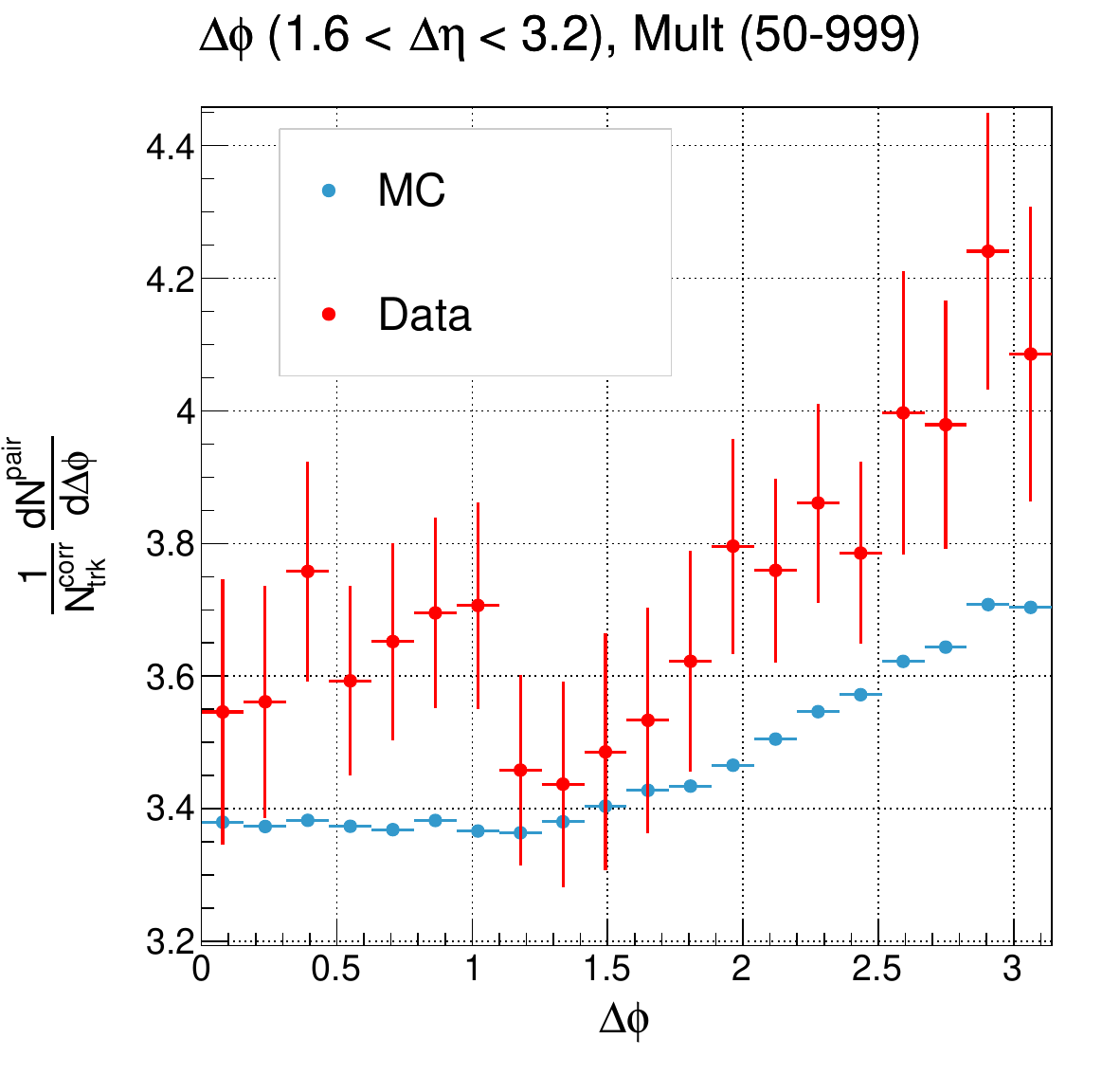}
\caption{[BDT at $\epsilon_{\rm sig}\approx 95\%$] Long-range
  ($1.6<\Delta\eta<3.2$) azimuthal differential yield $Y_l(\Delta\phi)$ with respect
  to the thrust axis in the multiplicity intervals $[30,35)$, $[35,40)$,
  $[40,45)$, $[45,50)$, and $[50,\infty)$.
  Data (red) carry bootstrap-derived statistical uncertainties
  (Sec.~\ref{sec:flow_bootstrap}); MC (blue) shows statistical errors.}
\label{fig:rst_95pbdt_40phibins_0908_final_dNdphi}
\end{figure}

\begin{figure}[ht]
\centering
    \includegraphics[width=0.45\textwidth]{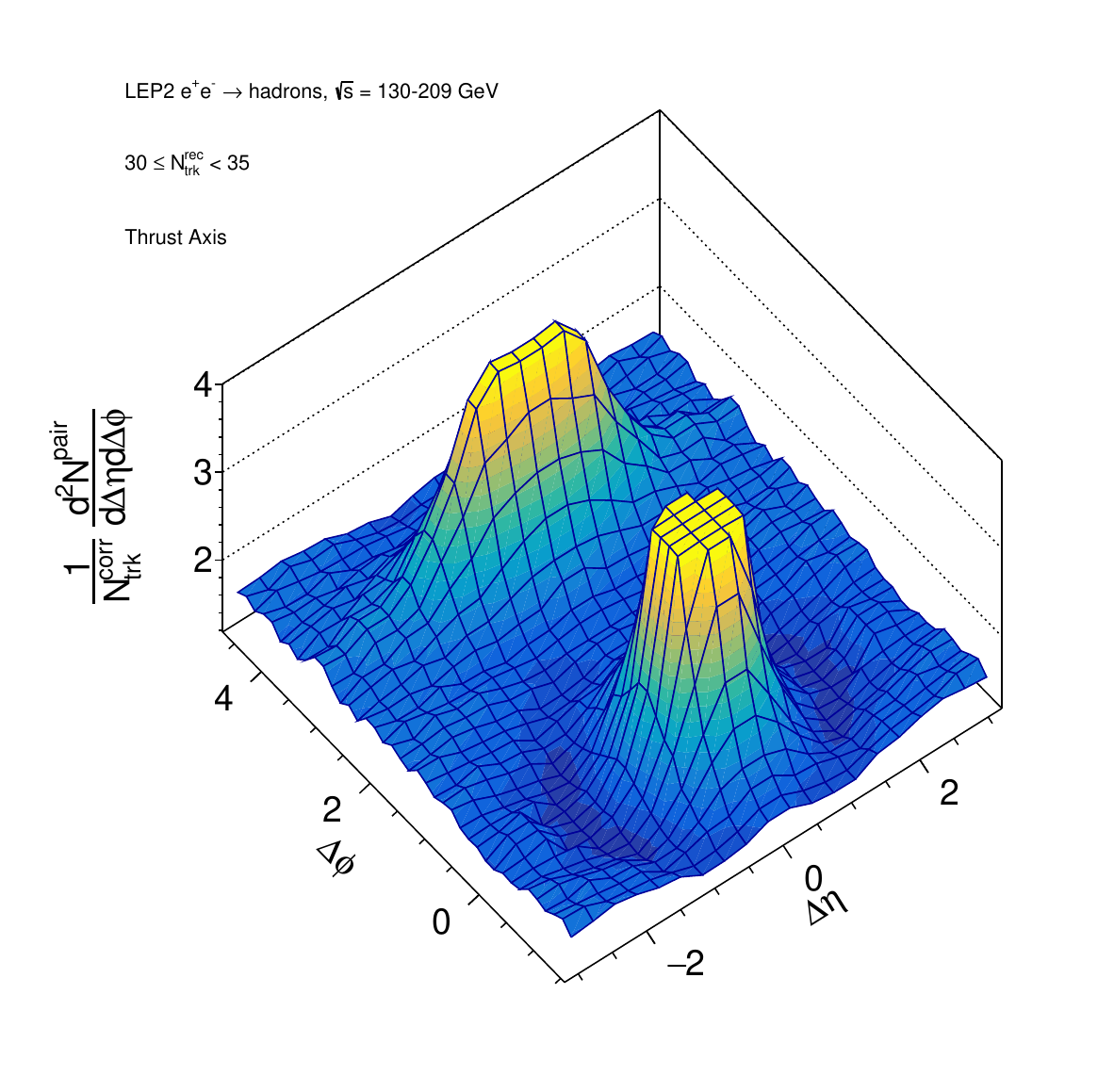}
    \includegraphics[width=0.45\textwidth]{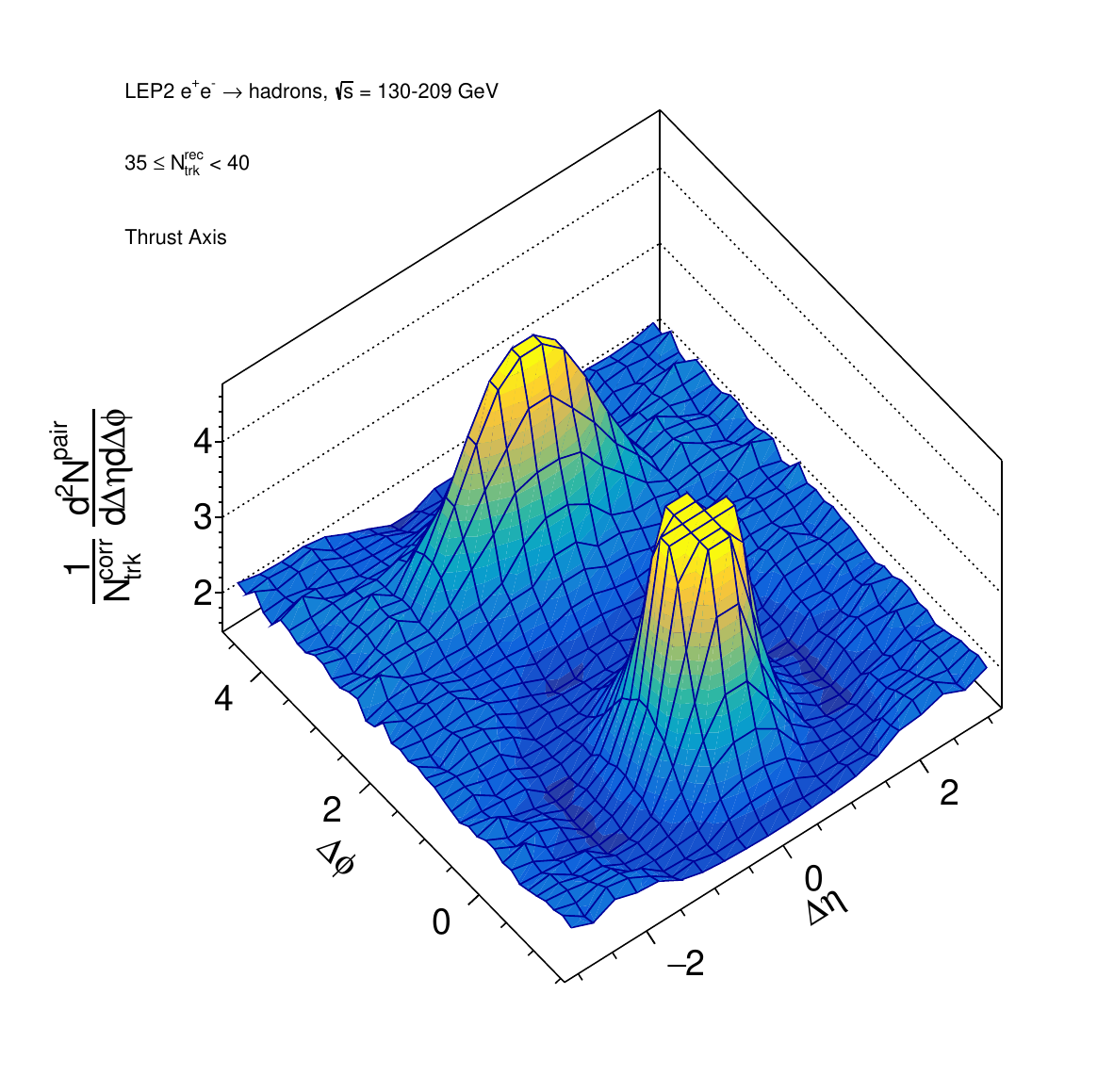}
    \includegraphics[width=0.45\textwidth]{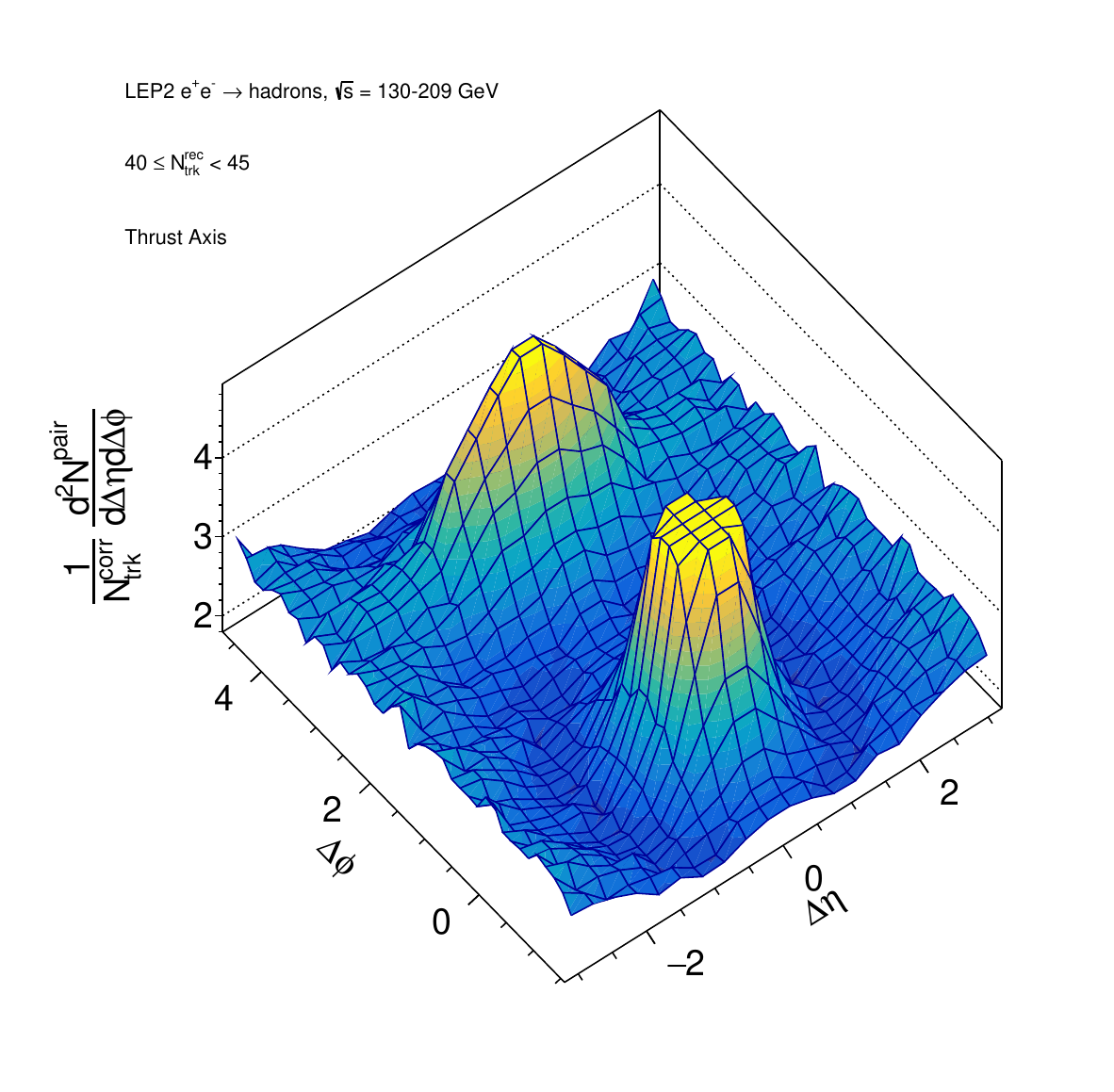}
    \includegraphics[width=0.45\textwidth]{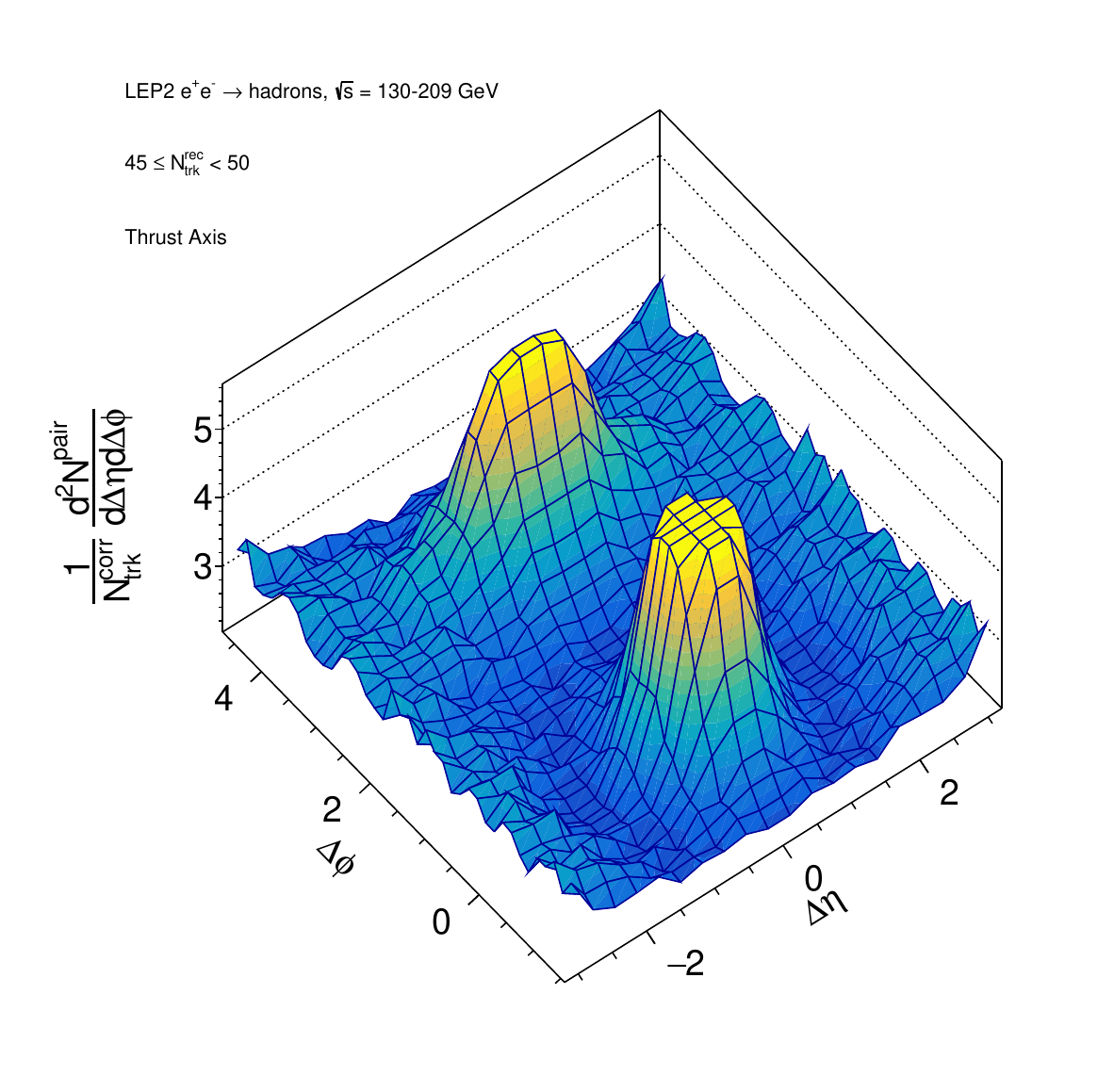}
    \includegraphics[width=0.45\textwidth]{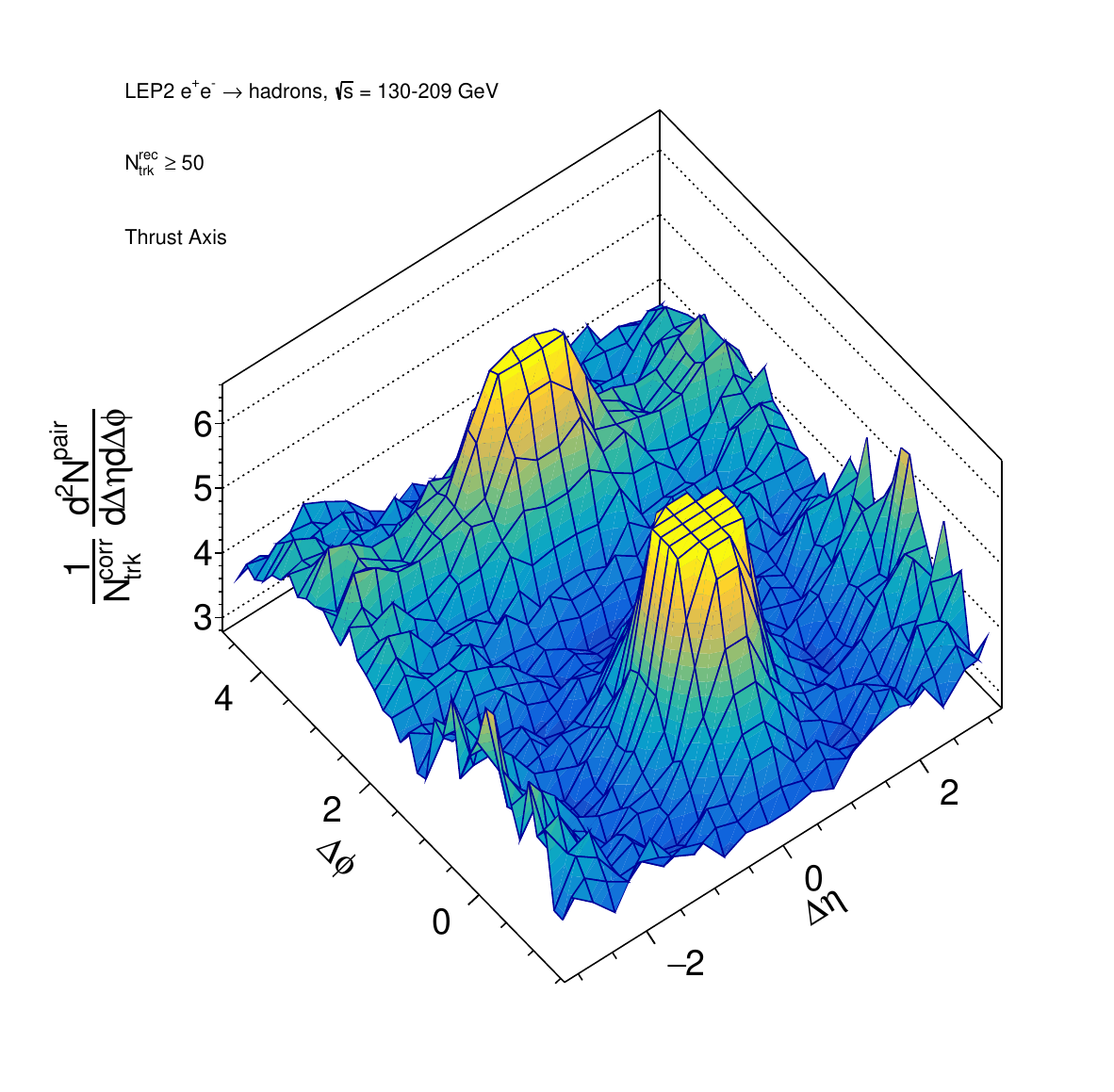}
\caption{[BDT at $\epsilon_{\rm sig}\approx 90\%$] Two-particle correlation function
  with respect to the thrust axis in the multiplicity intervals $[30,35)$,
  $[35,40)$, $[40,45)$, $[45,50)$, and $[50,\infty)$.}
\label{fig:rst_90pbdt_40phibins_0908_final_2PC}
\end{figure}

\begin{figure}[ht]
\centering
    \includegraphics[width=0.45\textwidth]{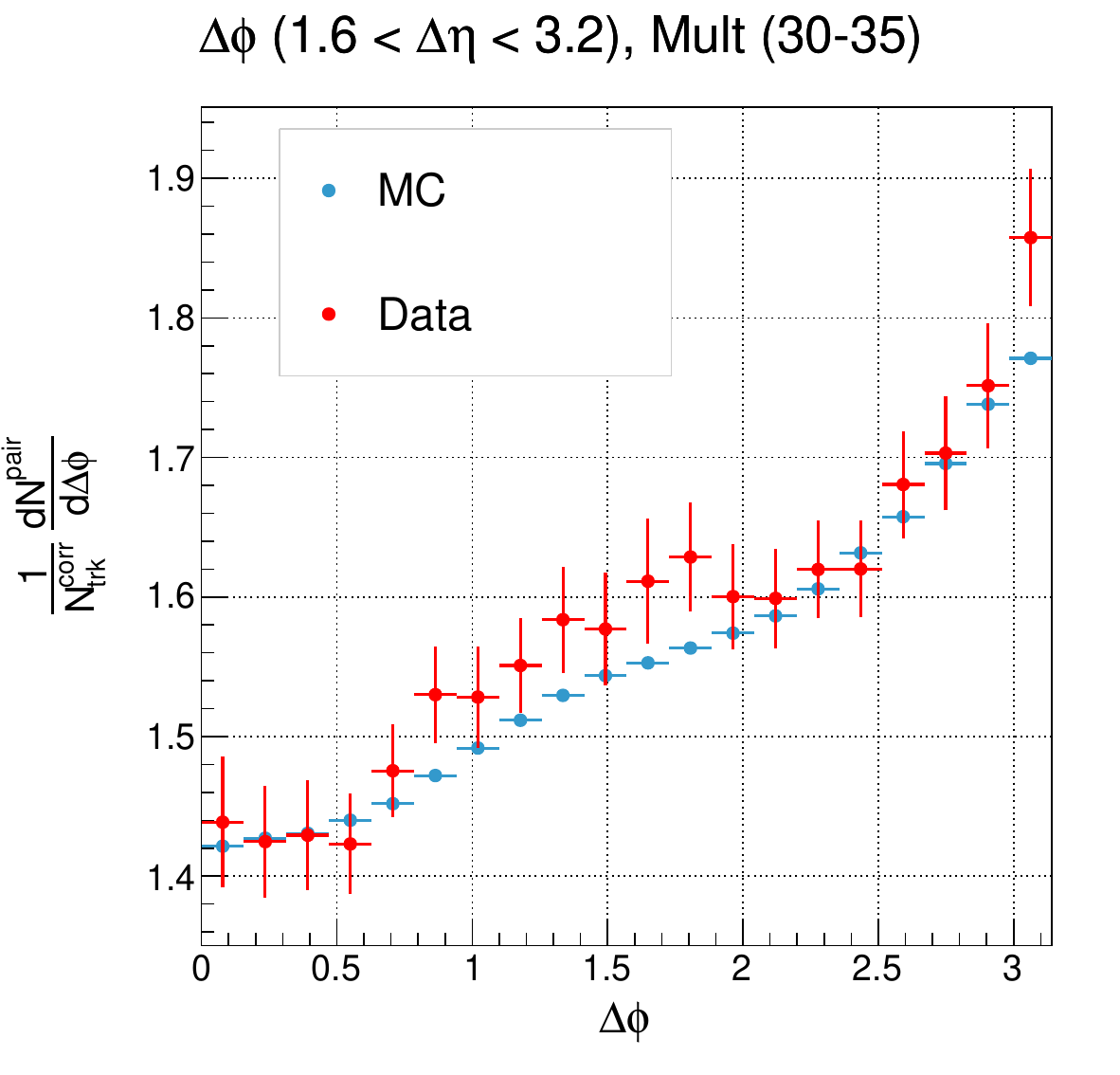}
    \includegraphics[width=0.45\textwidth]{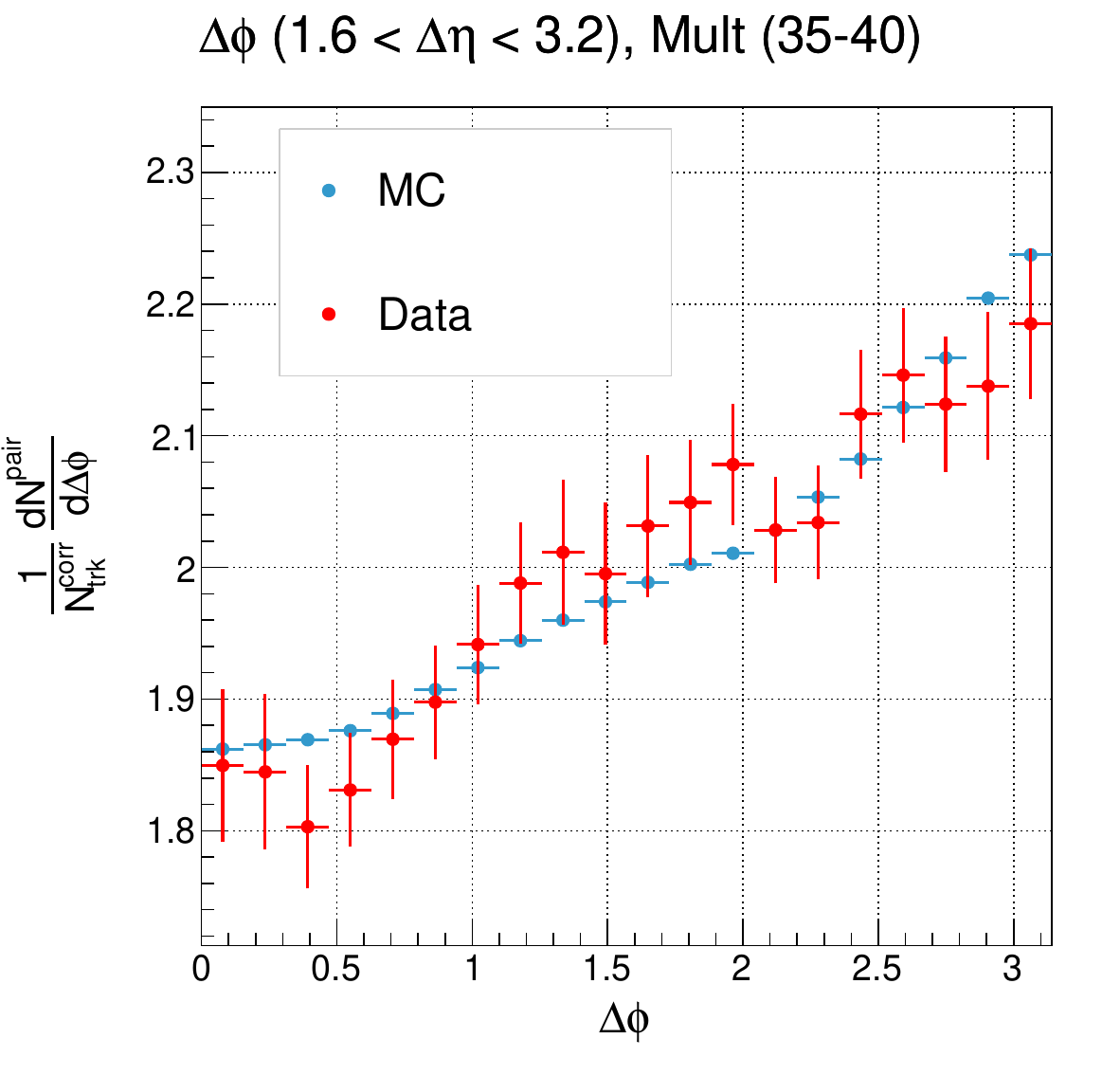}
    \includegraphics[width=0.45\textwidth]{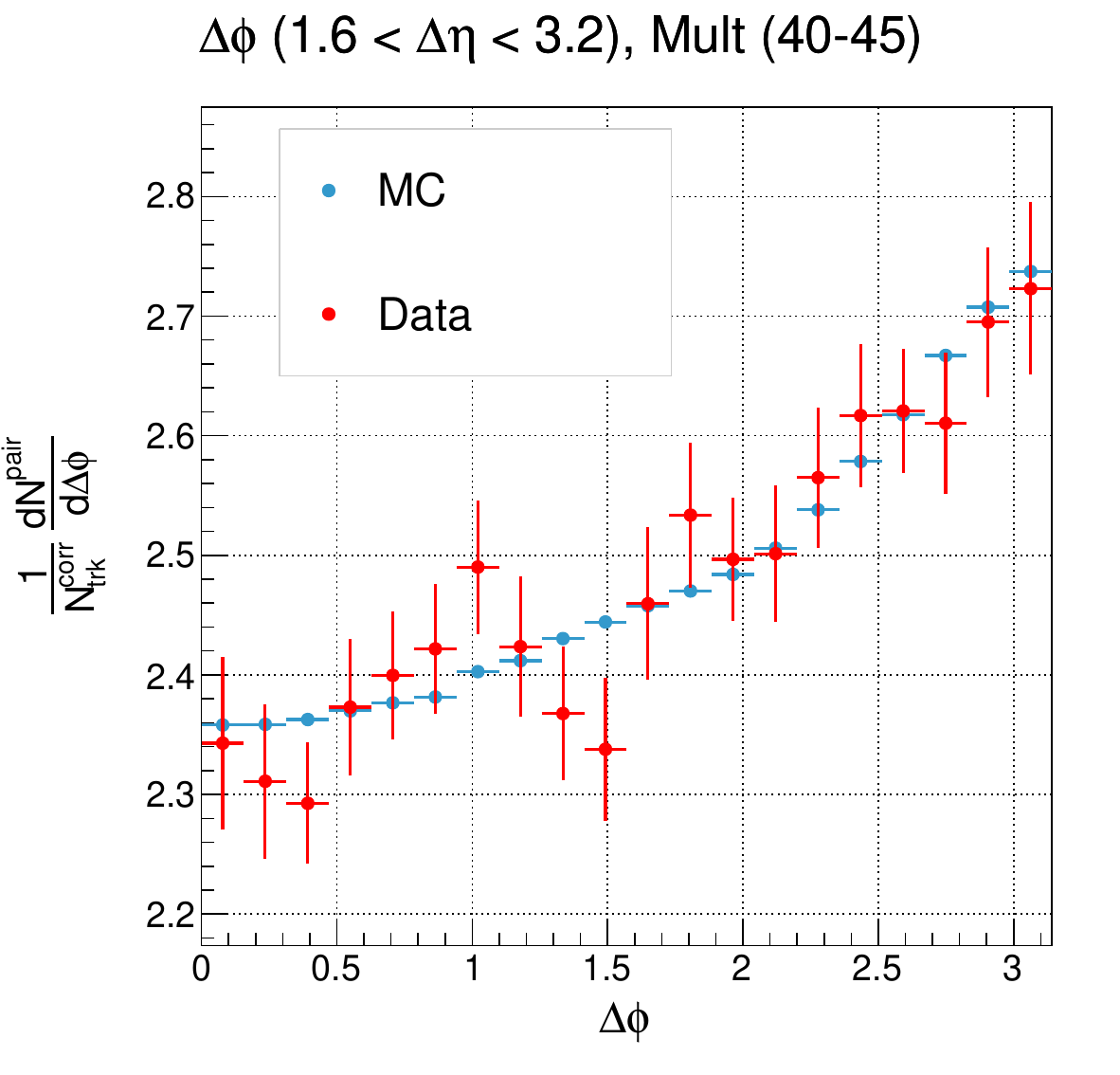}
    \includegraphics[width=0.45\textwidth]{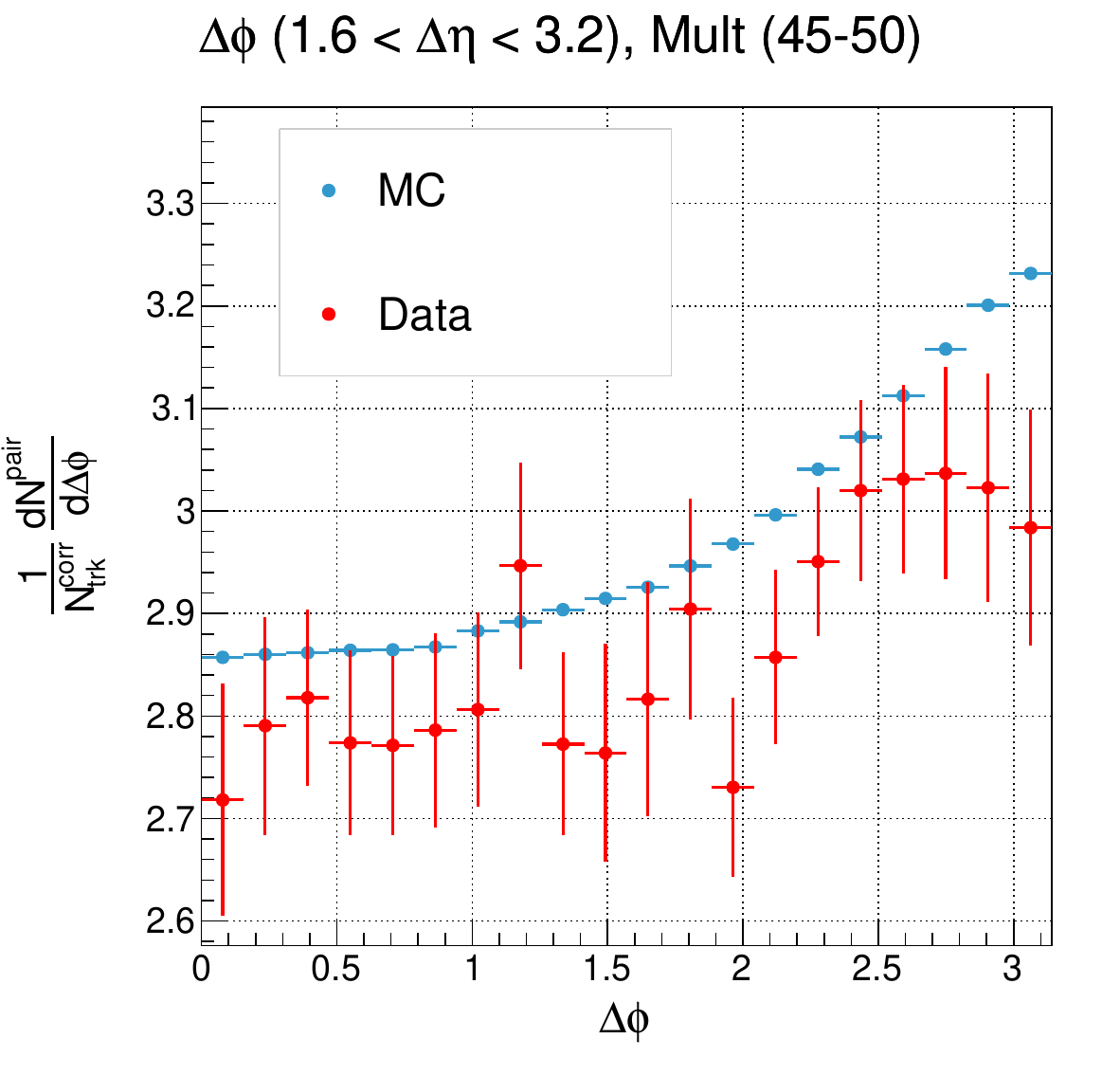}
    \includegraphics[width=0.45\textwidth]{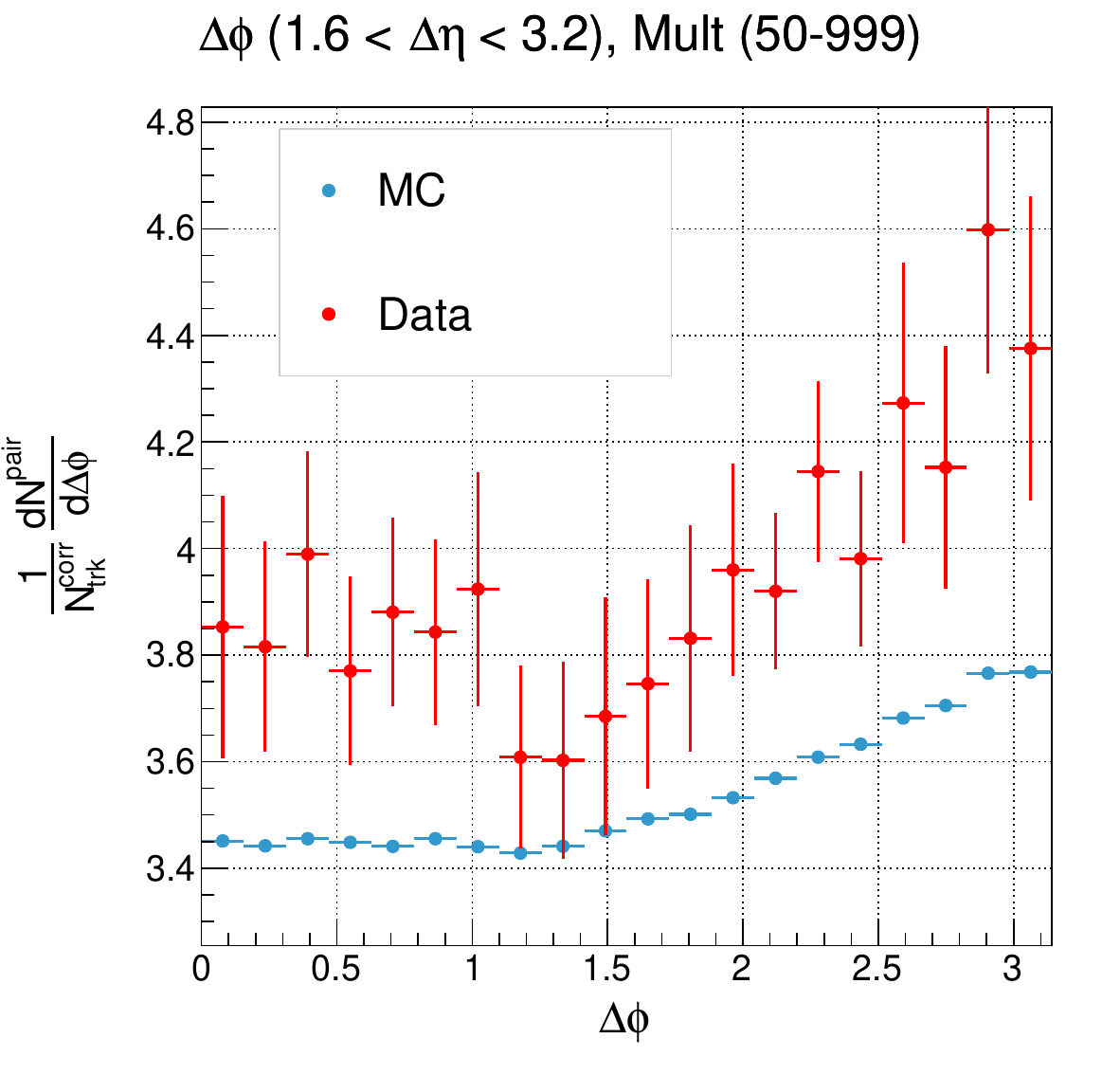}
\caption{[BDT at $\epsilon_{\rm sig}\approx 90\%$] Long-range
  ($1.6<\Delta\eta<3.2$) azimuthal differential yield $Y_l(\Delta\phi)$ in the
  multiplicity intervals $[30,35)$, $[35,40)$, $[40,45)$, $[45,50)$, and
  $[50,\infty)$. Data (red) carry bootstrap-derived statistical uncertainties
  (Sec.~\ref{sec:flow_bootstrap}); MC (blue) shows statistical errors.}
\label{fig:rst_90pbdt_40phibins_0908_final_dNdphi}
\end{figure}

\begin{figure}[ht]
\centering
    \includegraphics[width=0.45\textwidth]{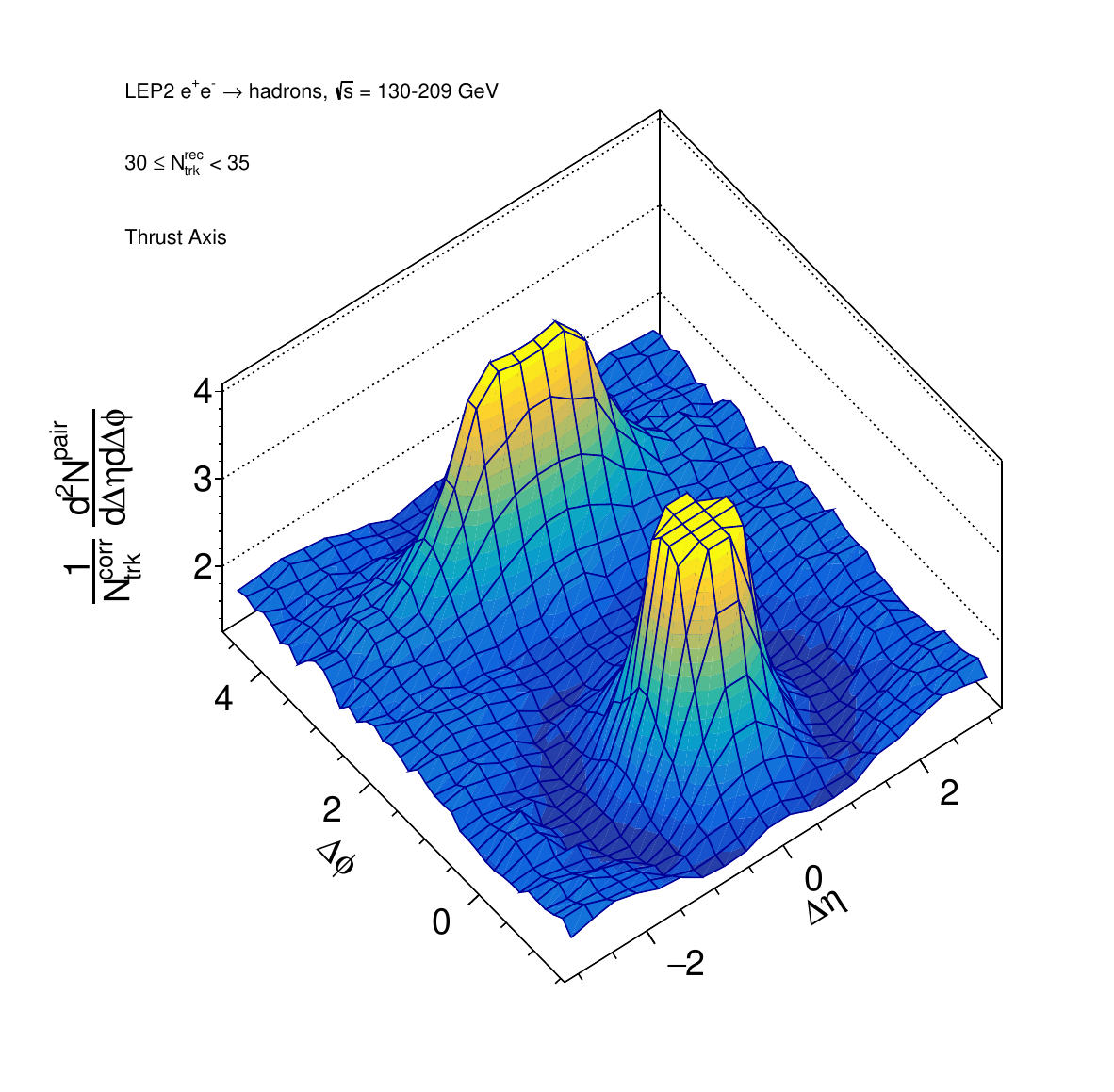}
    \includegraphics[width=0.45\textwidth]{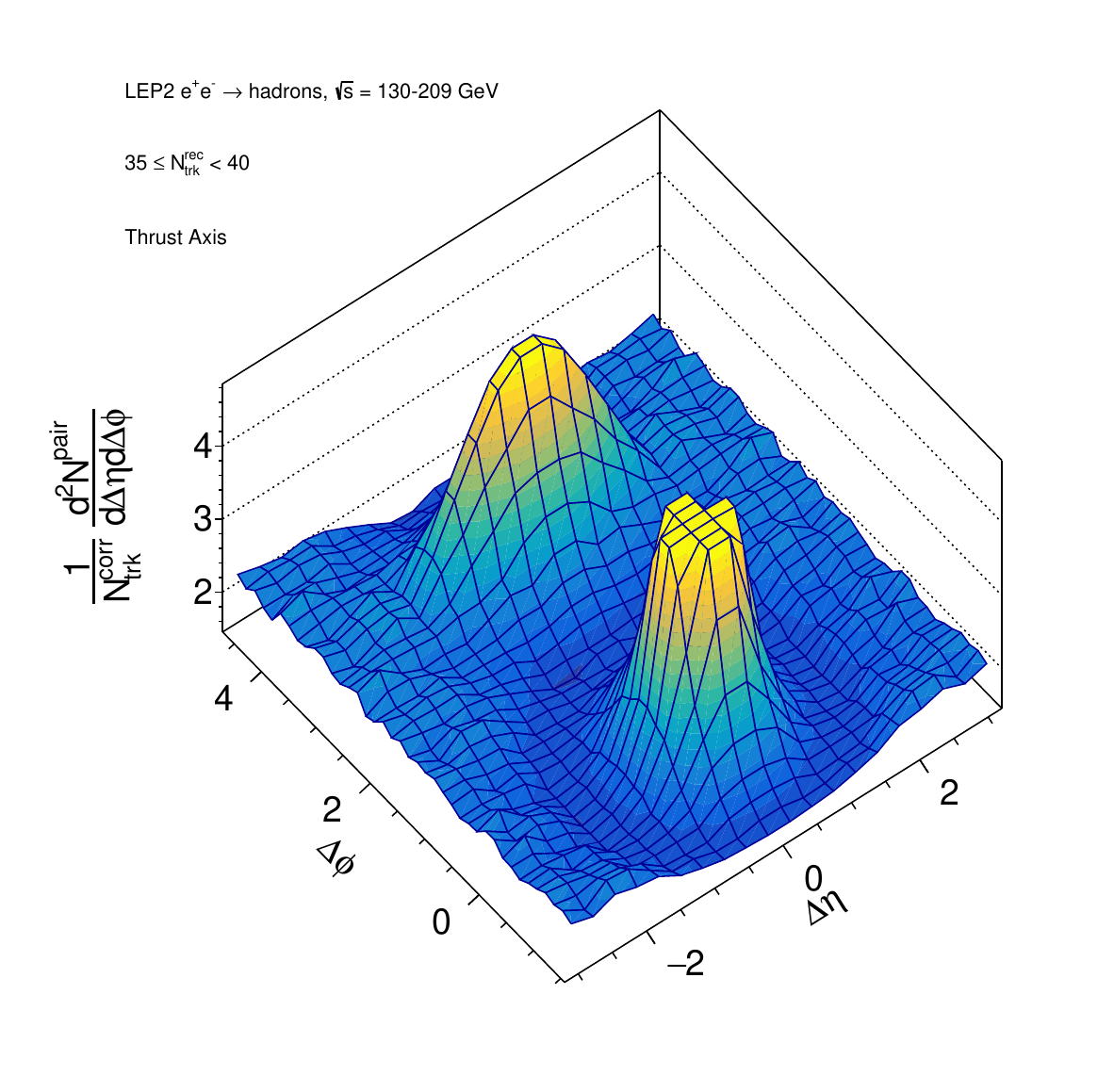}
    \includegraphics[width=0.45\textwidth]{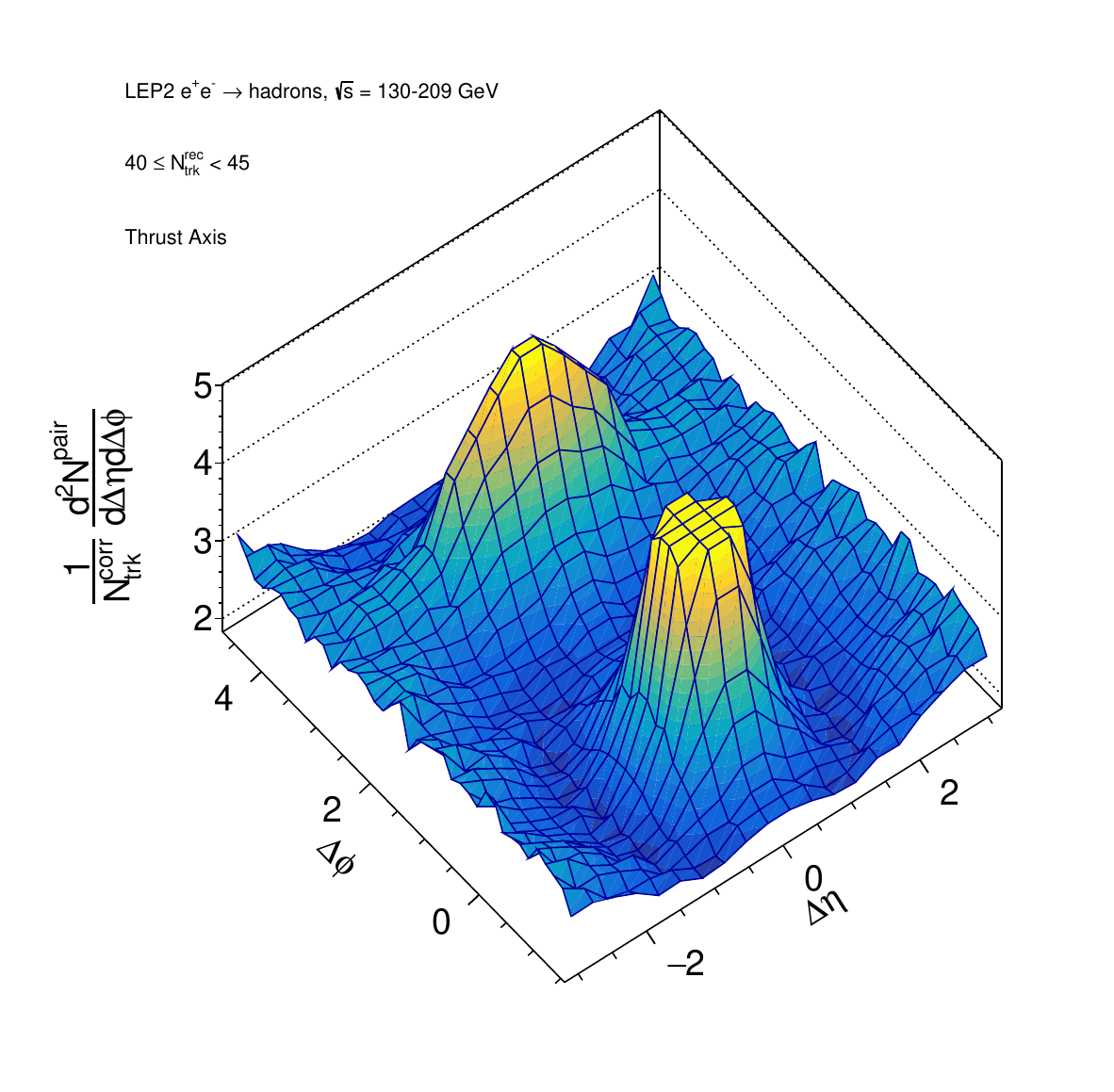}
    \includegraphics[width=0.45\textwidth]{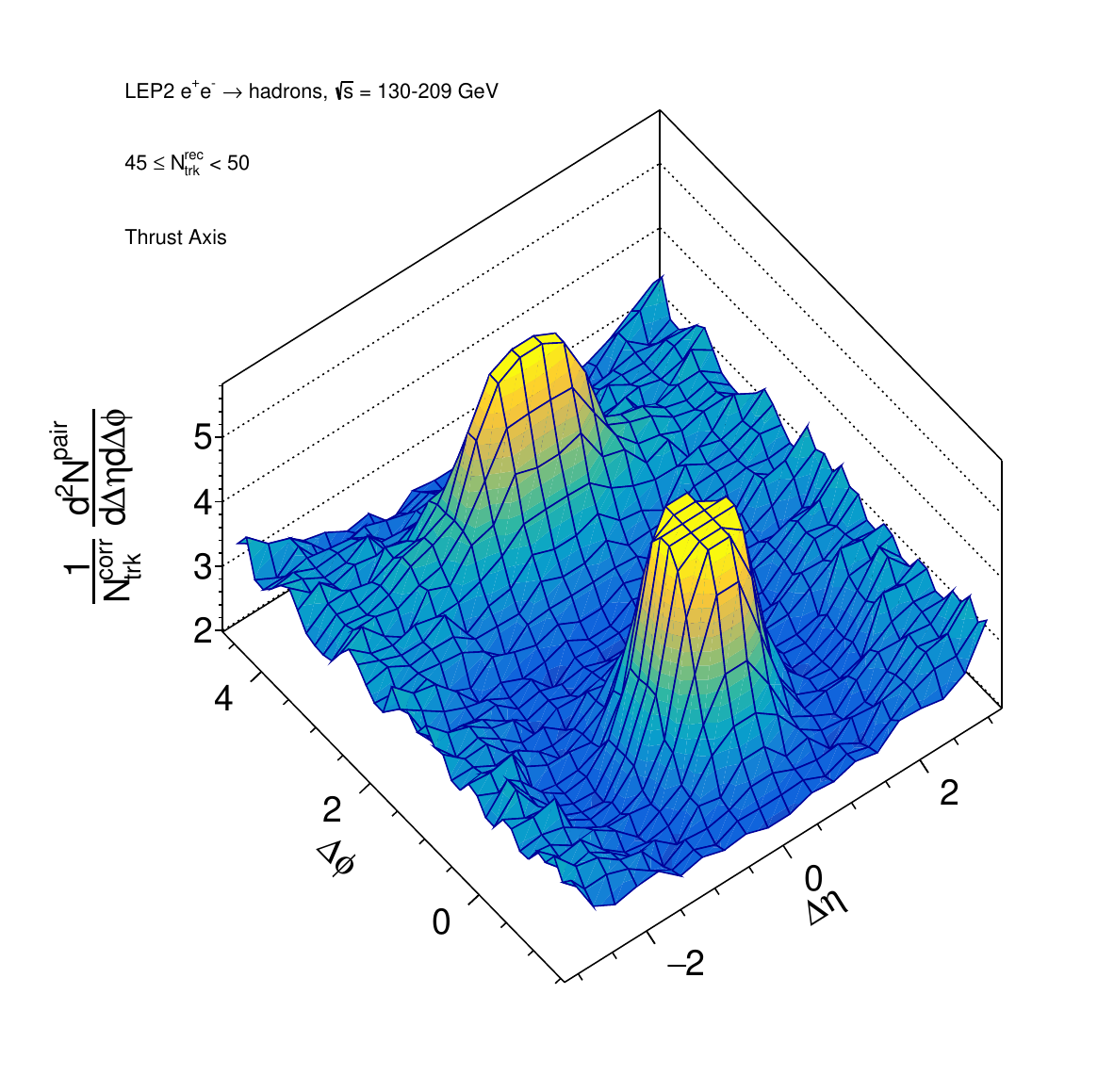}
    \includegraphics[width=0.45\textwidth]{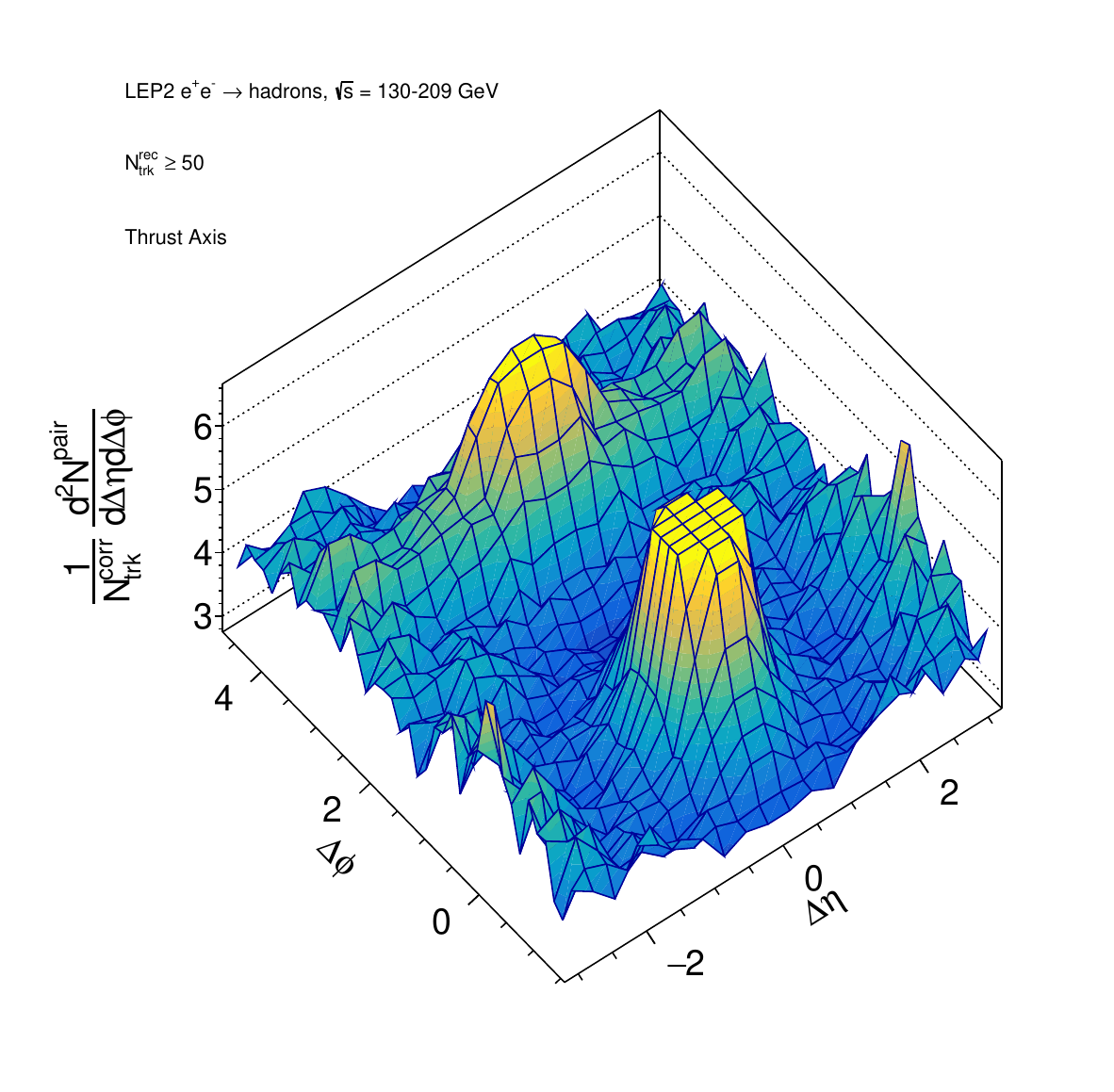}
\caption{[BDT at $\epsilon_{\rm sig}\approx 85\%$] Two-particle correlation function
  with respect to the thrust axis in the multiplicity intervals $[30,35)$,
  $[35,40)$, $[40,45)$, $[45,50)$, and $[50,\infty)$.}
\label{fig:rst_85pbdt_40phibins_0908_final_2PC}
\end{figure}

\begin{figure}[ht]
\centering
    \includegraphics[width=0.45\textwidth]{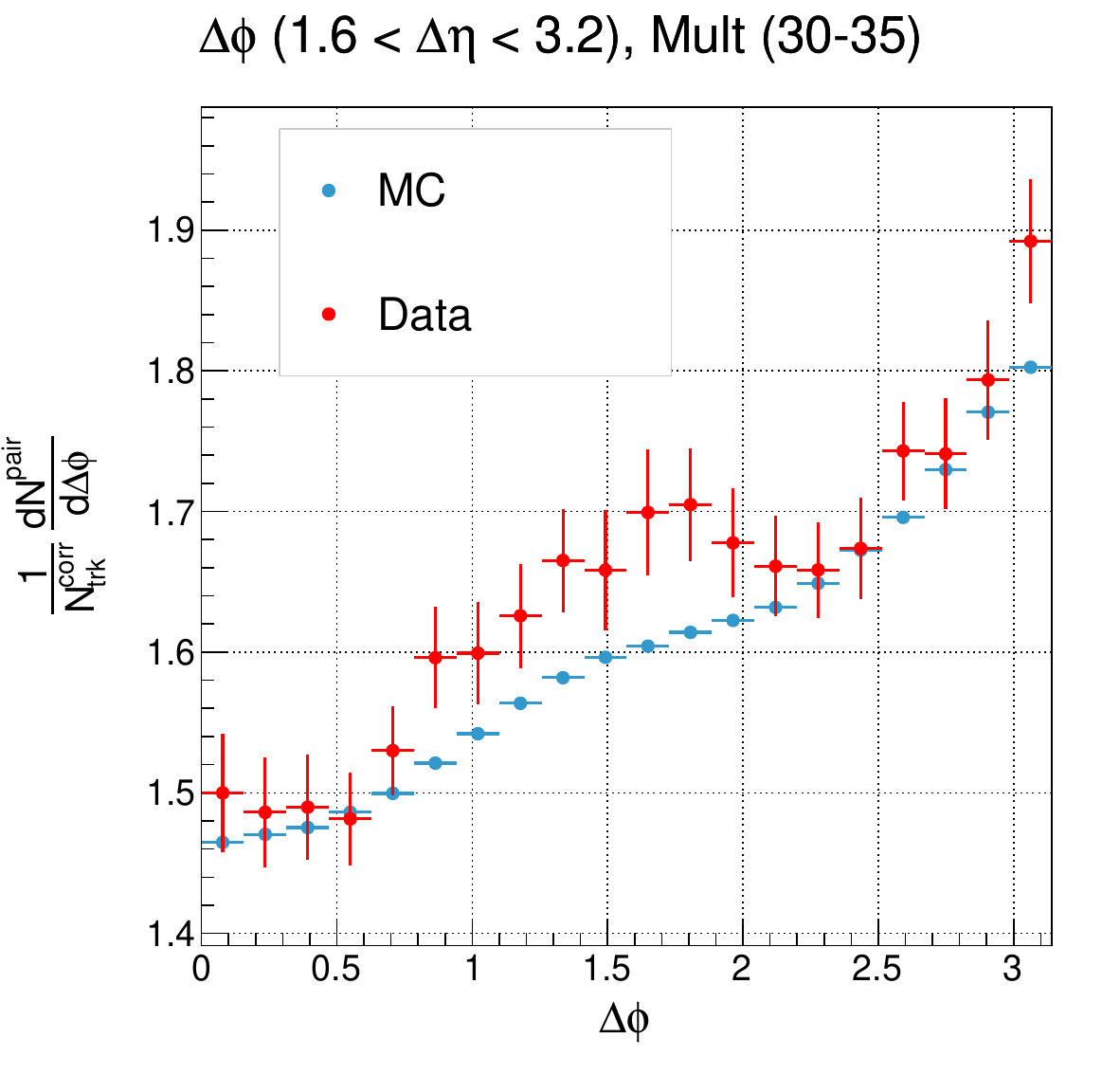}
    \includegraphics[width=0.45\textwidth]{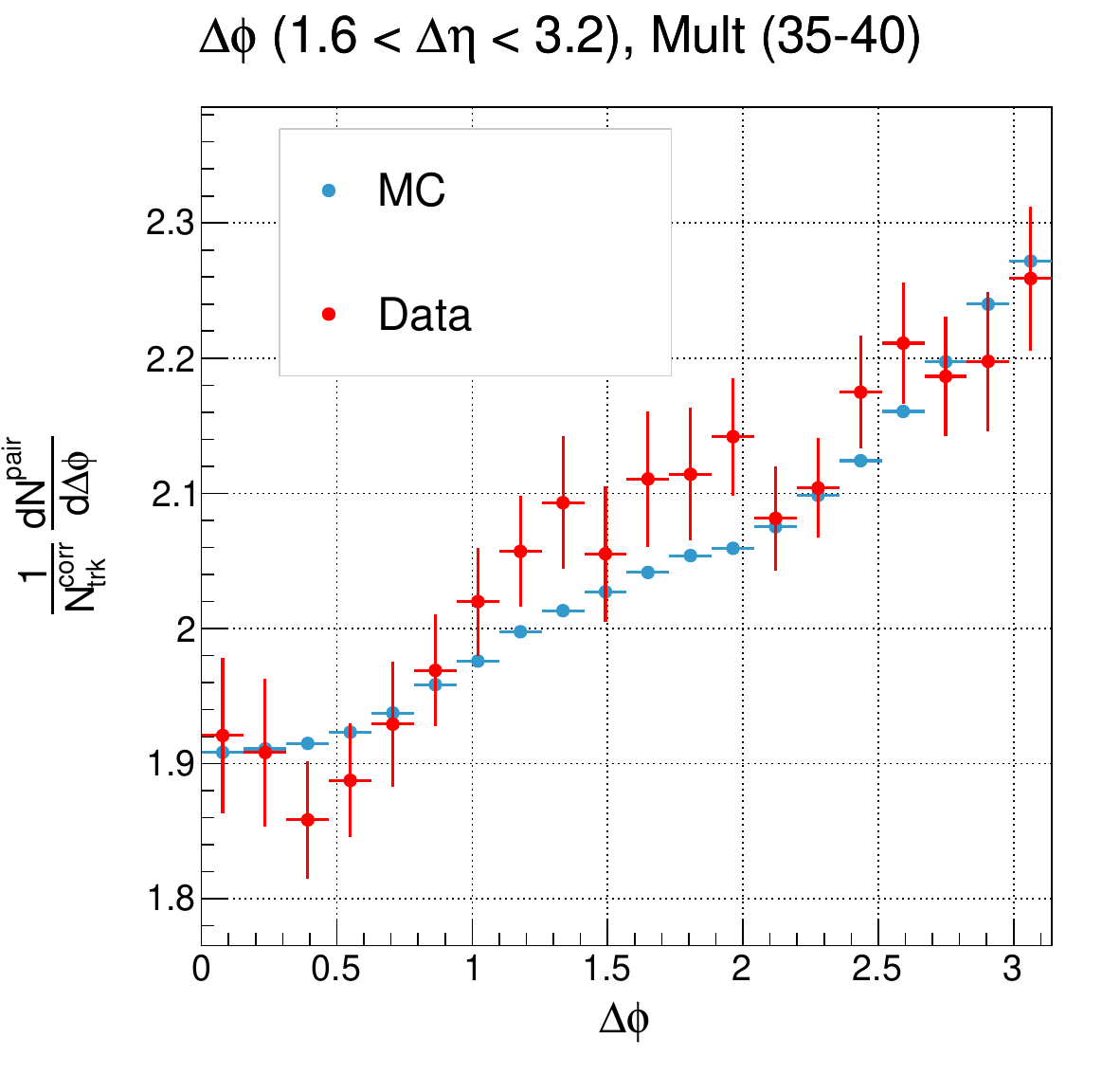}
    \includegraphics[width=0.45\textwidth]{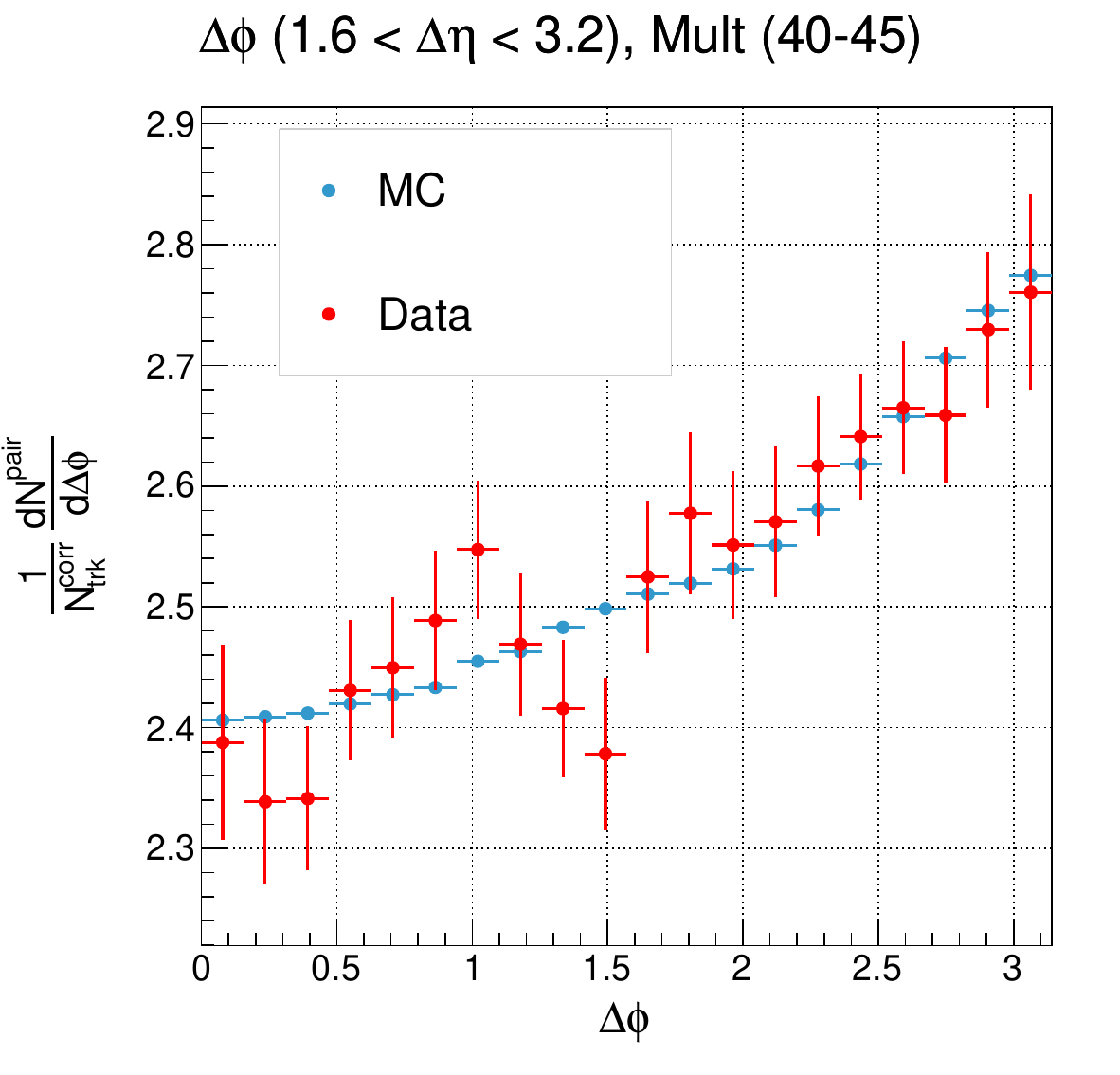}
    \includegraphics[width=0.45\textwidth]{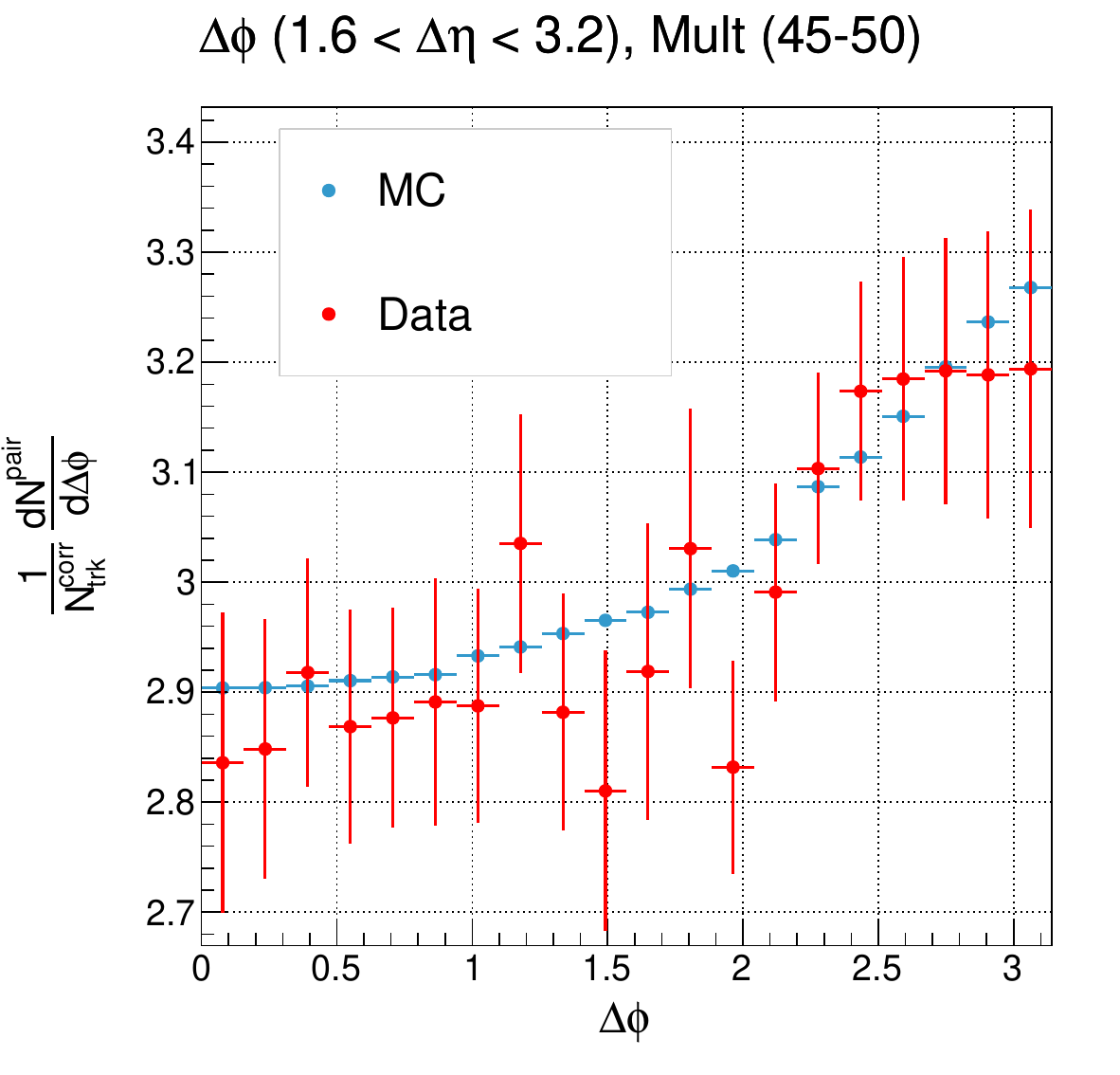}
    \includegraphics[width=0.45\textwidth]{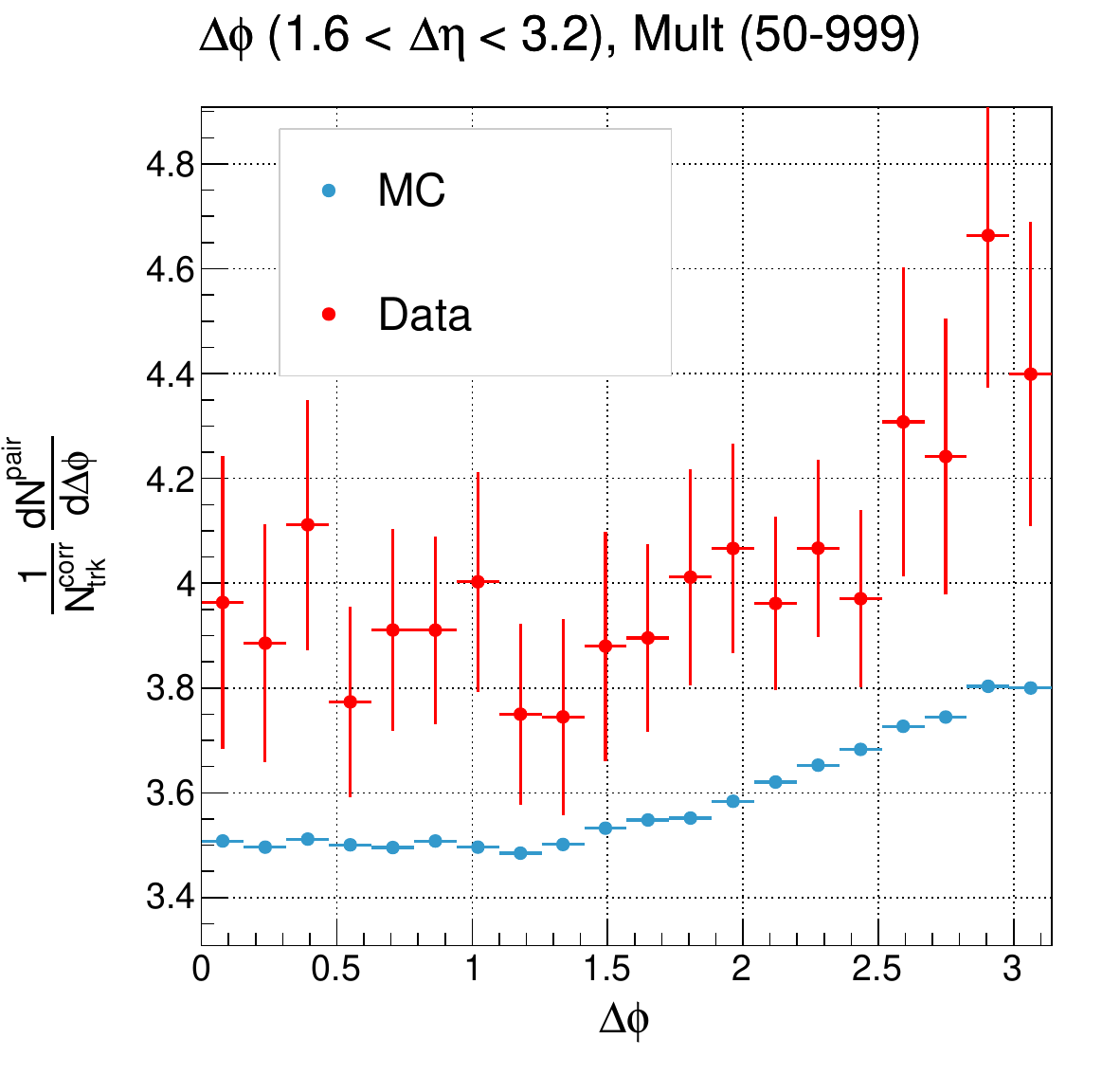}
\caption{[BDT at $\epsilon_{\rm sig}\approx 85\%$] Long-range
  ($1.6<\Delta\eta<3.2$) azimuthal differential yield $Y_l(\Delta\phi)$ in the
  multiplicity intervals $[30,35)$, $[35,40)$, $[40,45)$, $[45,50)$, and
  $[50,\infty)$. Data (red) carry bootstrap-derived statistical uncertainties
  (Sec.~\ref{sec:flow_bootstrap}); MC (blue) shows statistical errors.}
\label{fig:rst_85pbdt_40phibins_0908_final_dNdphi}
\end{figure}

\begin{figure}[ht]
\centering
    \includegraphics[width=0.45\textwidth]{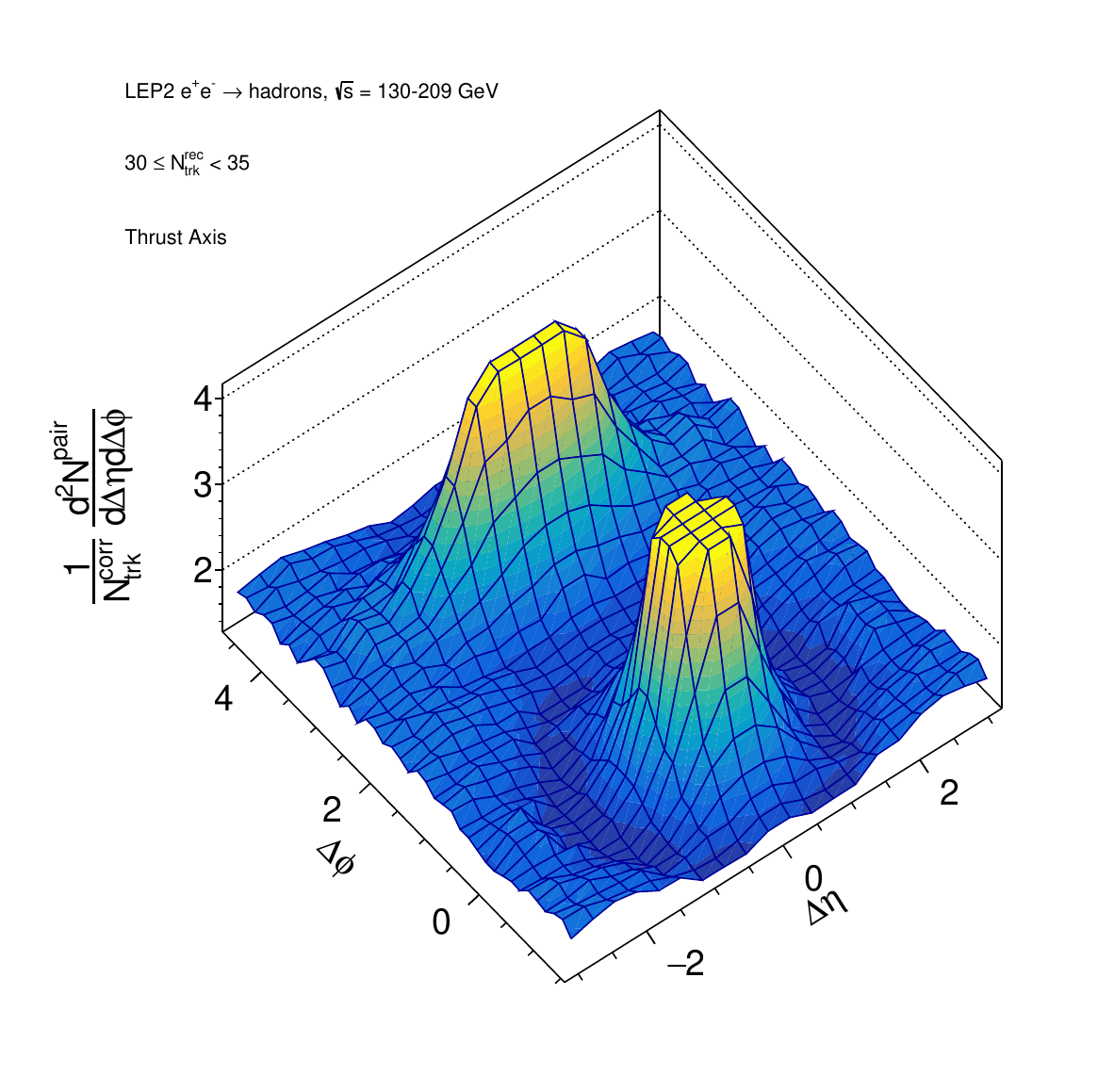}
    \includegraphics[width=0.45\textwidth]{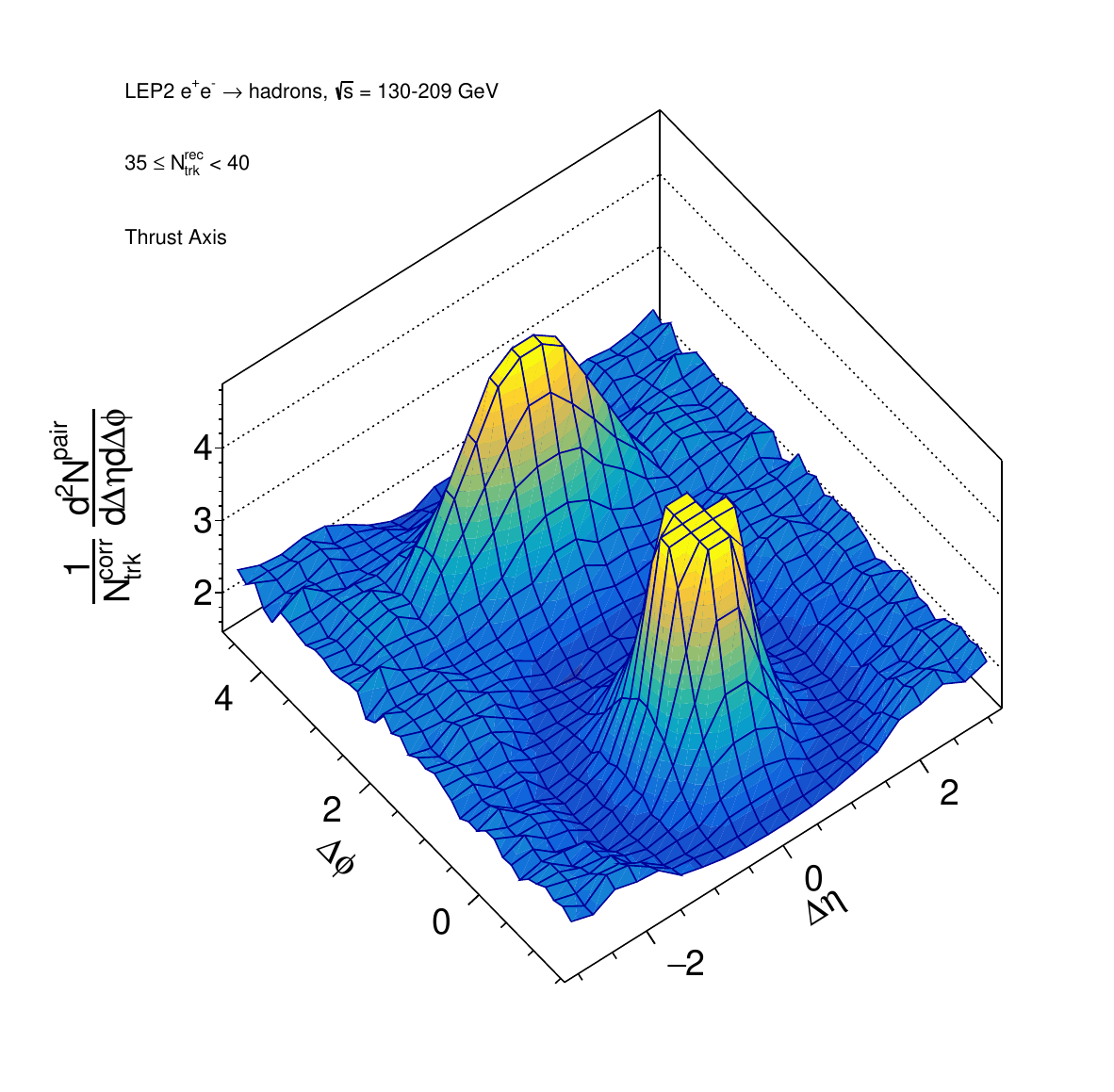}
    \includegraphics[width=0.45\textwidth]{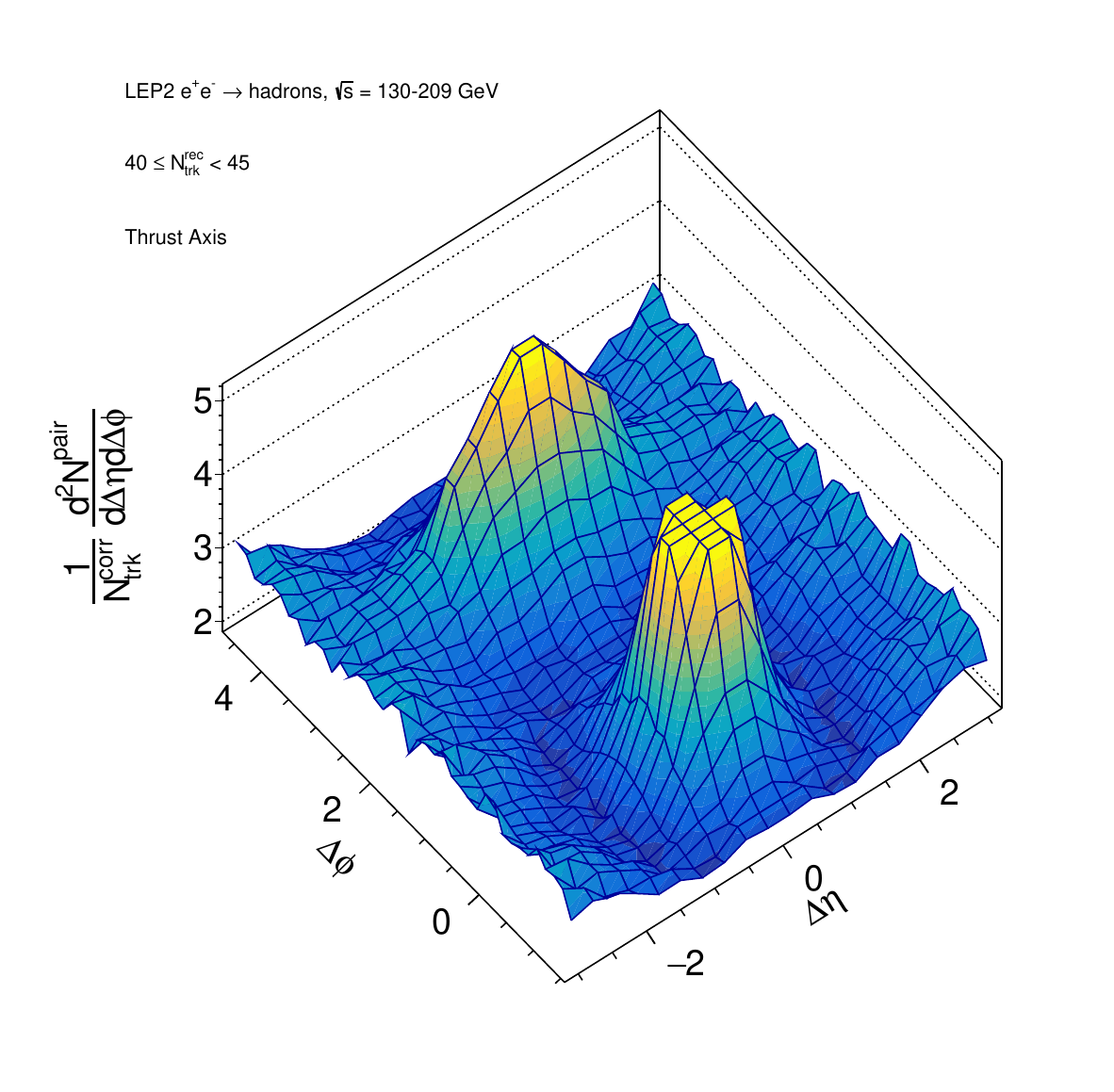}
    \includegraphics[width=0.45\textwidth]{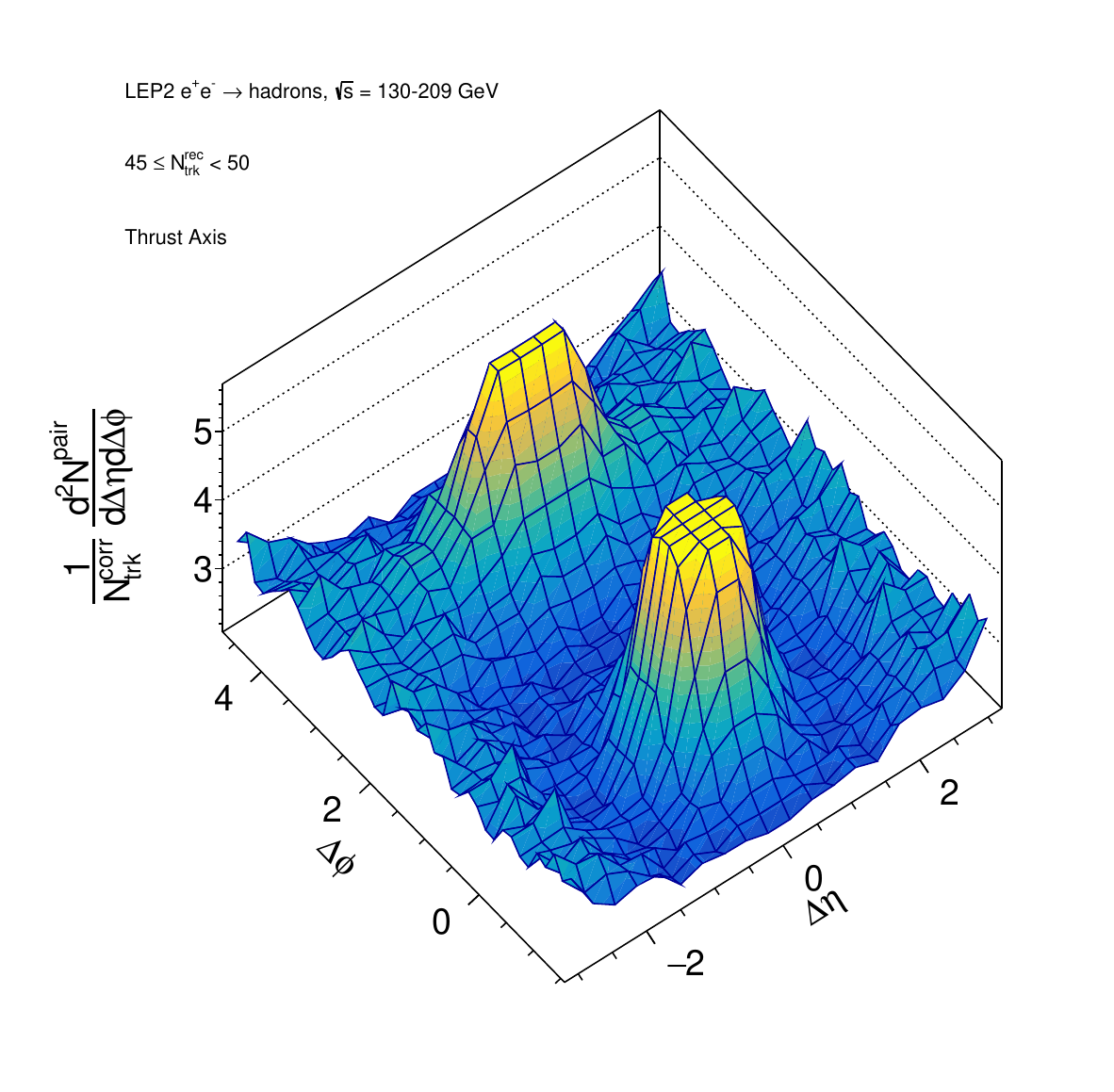}
    \includegraphics[width=0.45\textwidth]{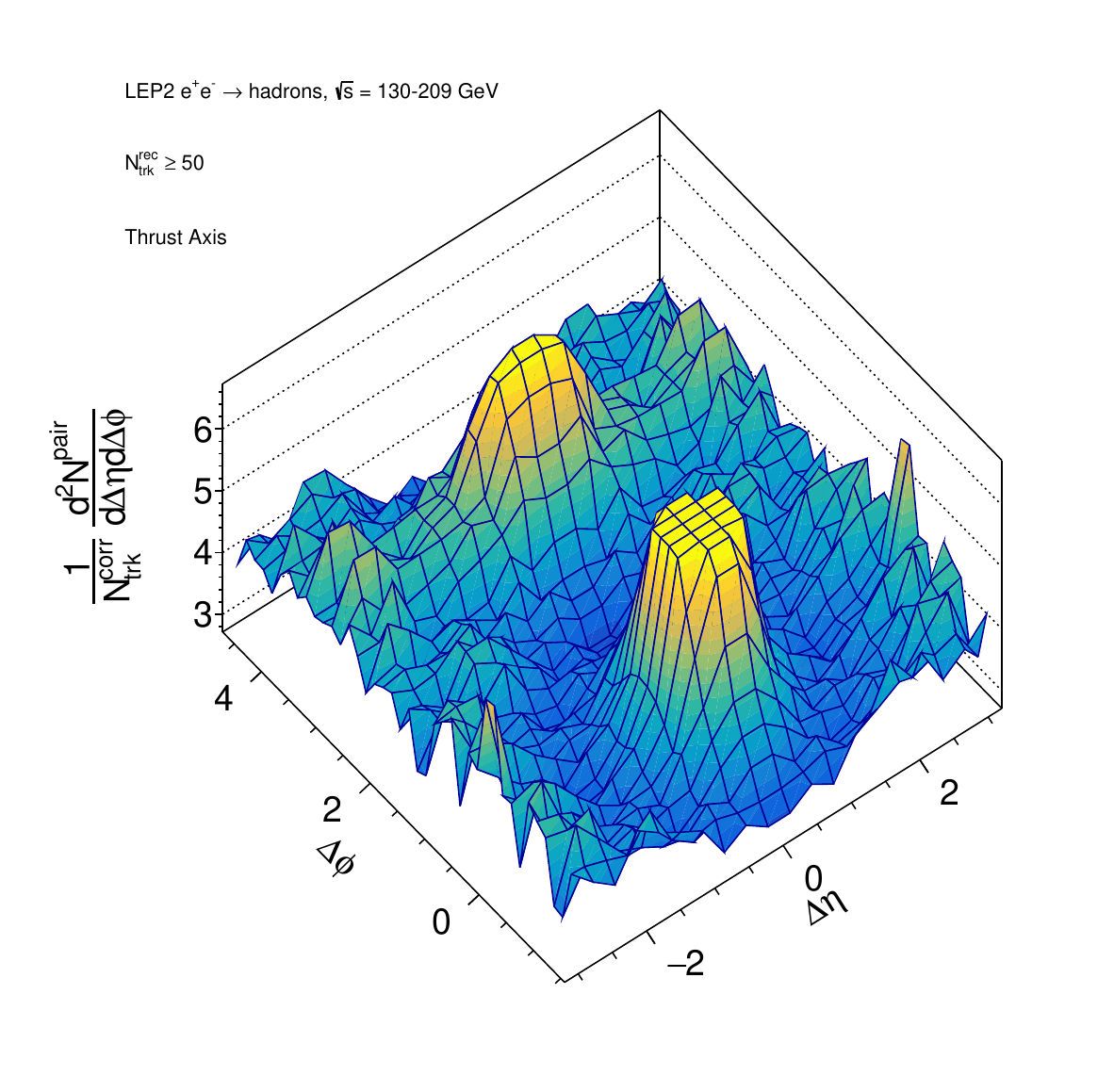}
\caption{[BDT at $\epsilon_{\rm sig}\approx 80\%$] Two-particle correlation function
  with respect to the thrust axis in the multiplicity intervals $[30,35)$,
  $[35,40)$, $[40,45)$, $[45,50)$, and $[50,\infty)$.}
\label{fig:rst_80pbdt_40phibins_0908_final_2PC}
\end{figure}

\begin{figure}[ht]
\centering
    \includegraphics[width=0.45\textwidth]{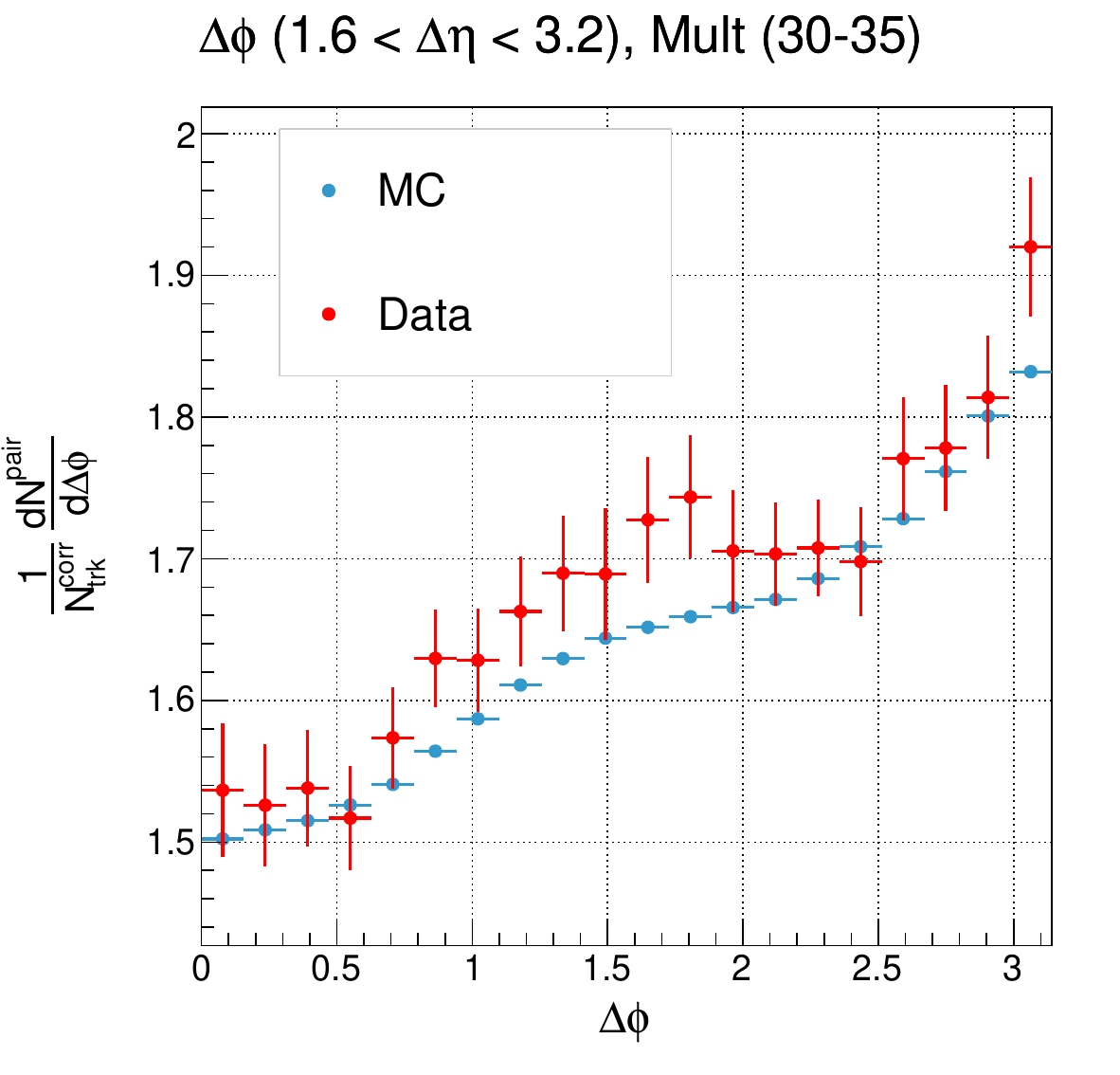}
    \includegraphics[width=0.45\textwidth]{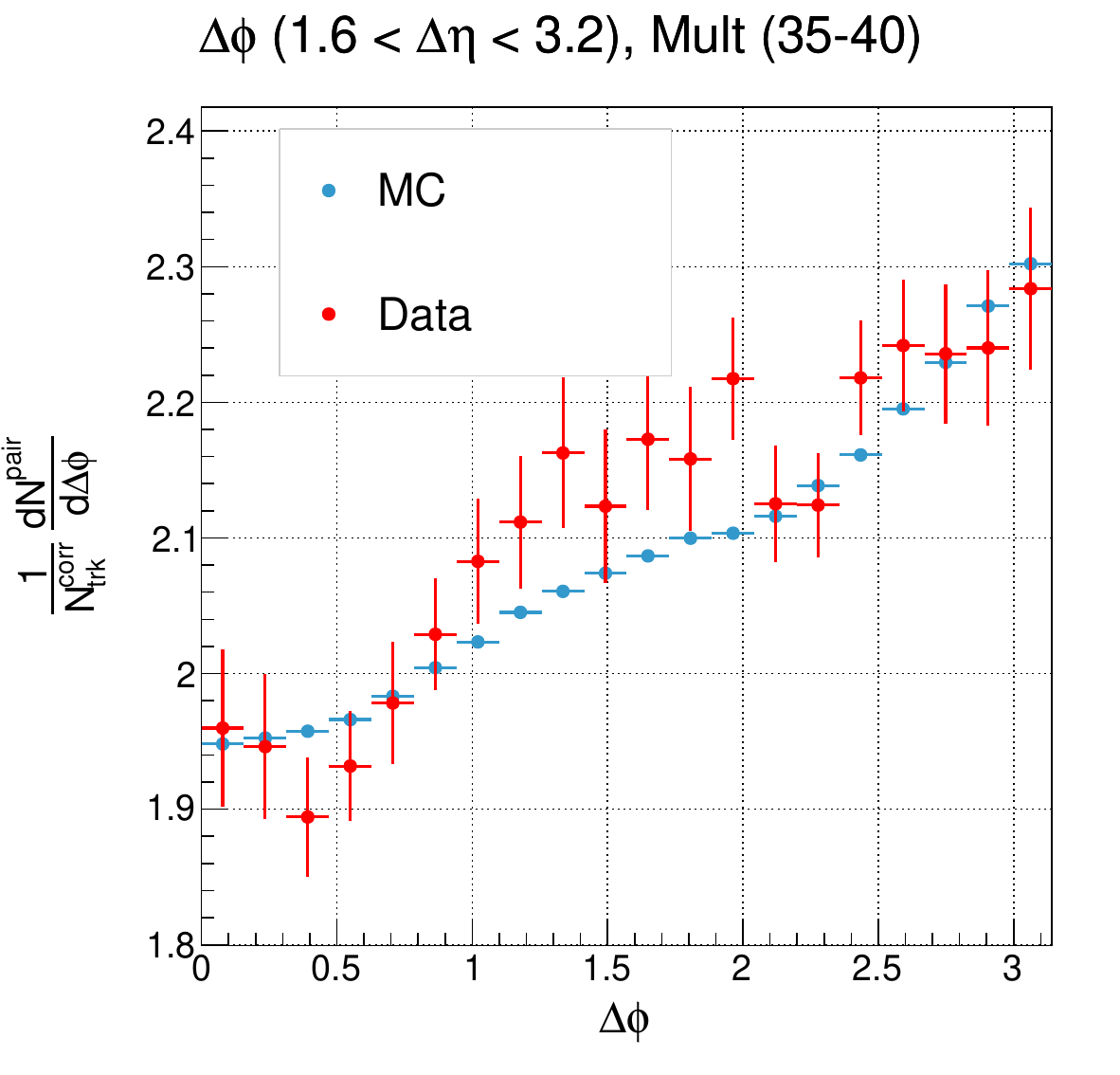}
    \includegraphics[width=0.45\textwidth]{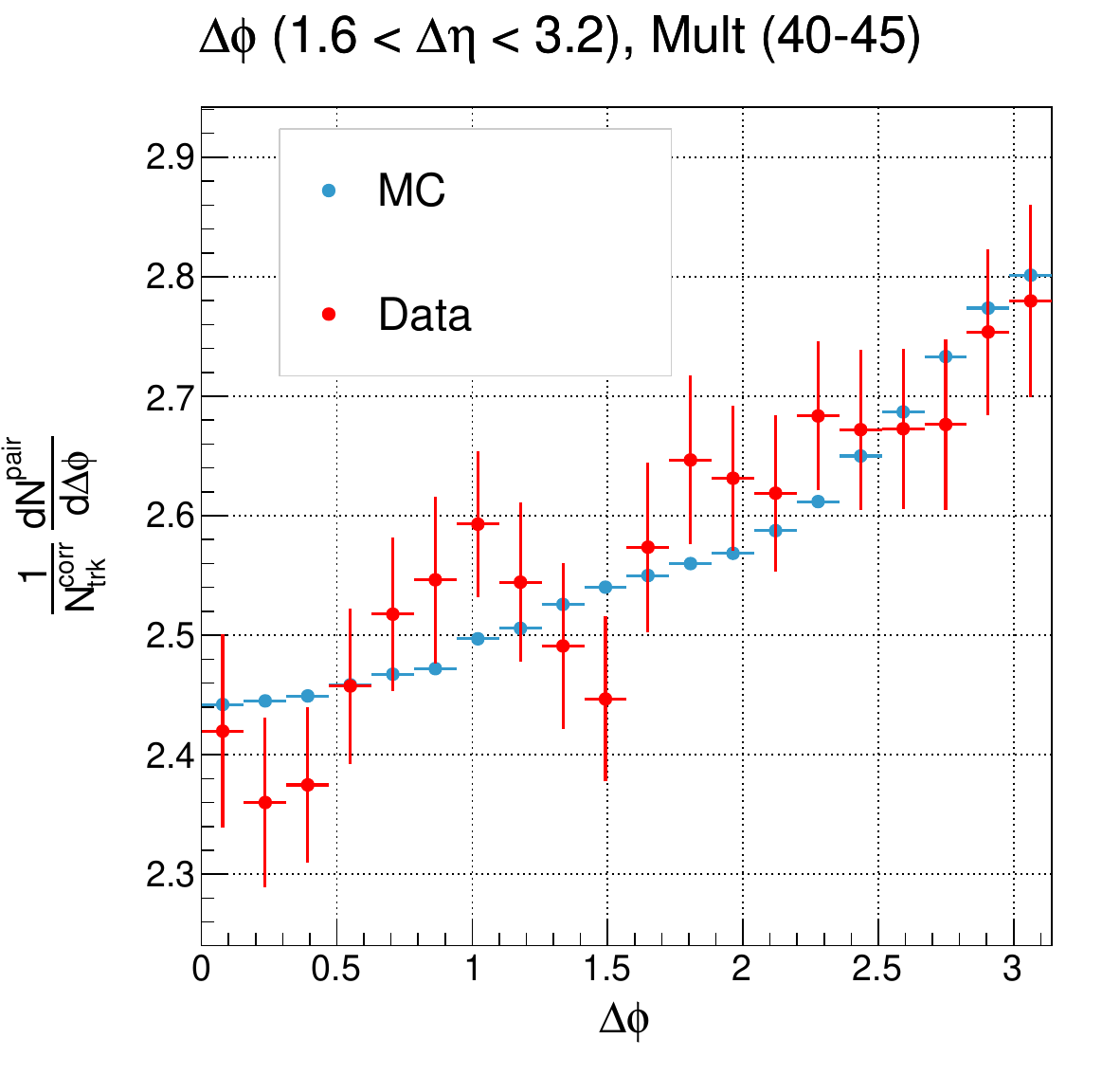}
    \includegraphics[width=0.45\textwidth]{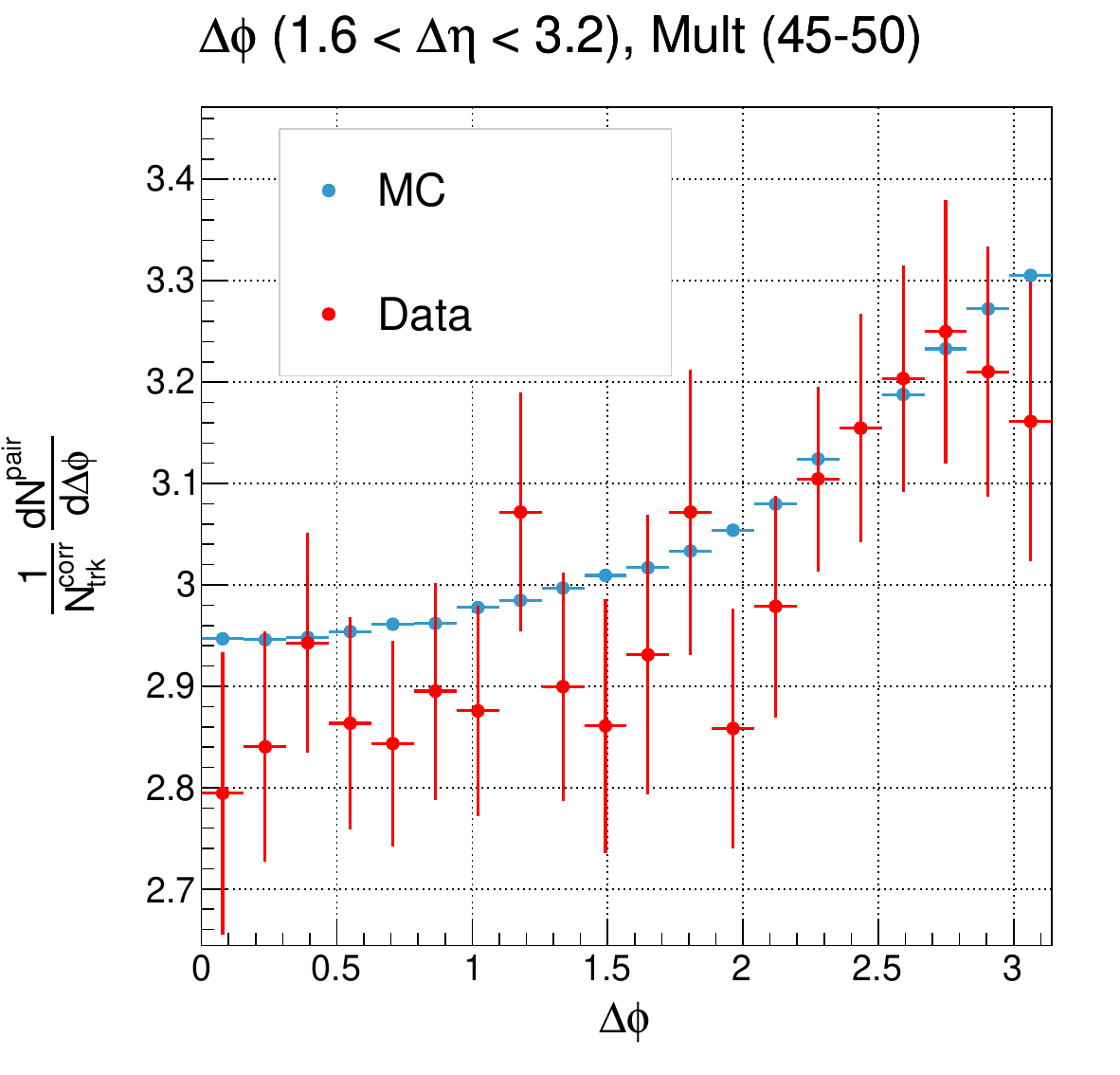}
    \includegraphics[width=0.45\textwidth]{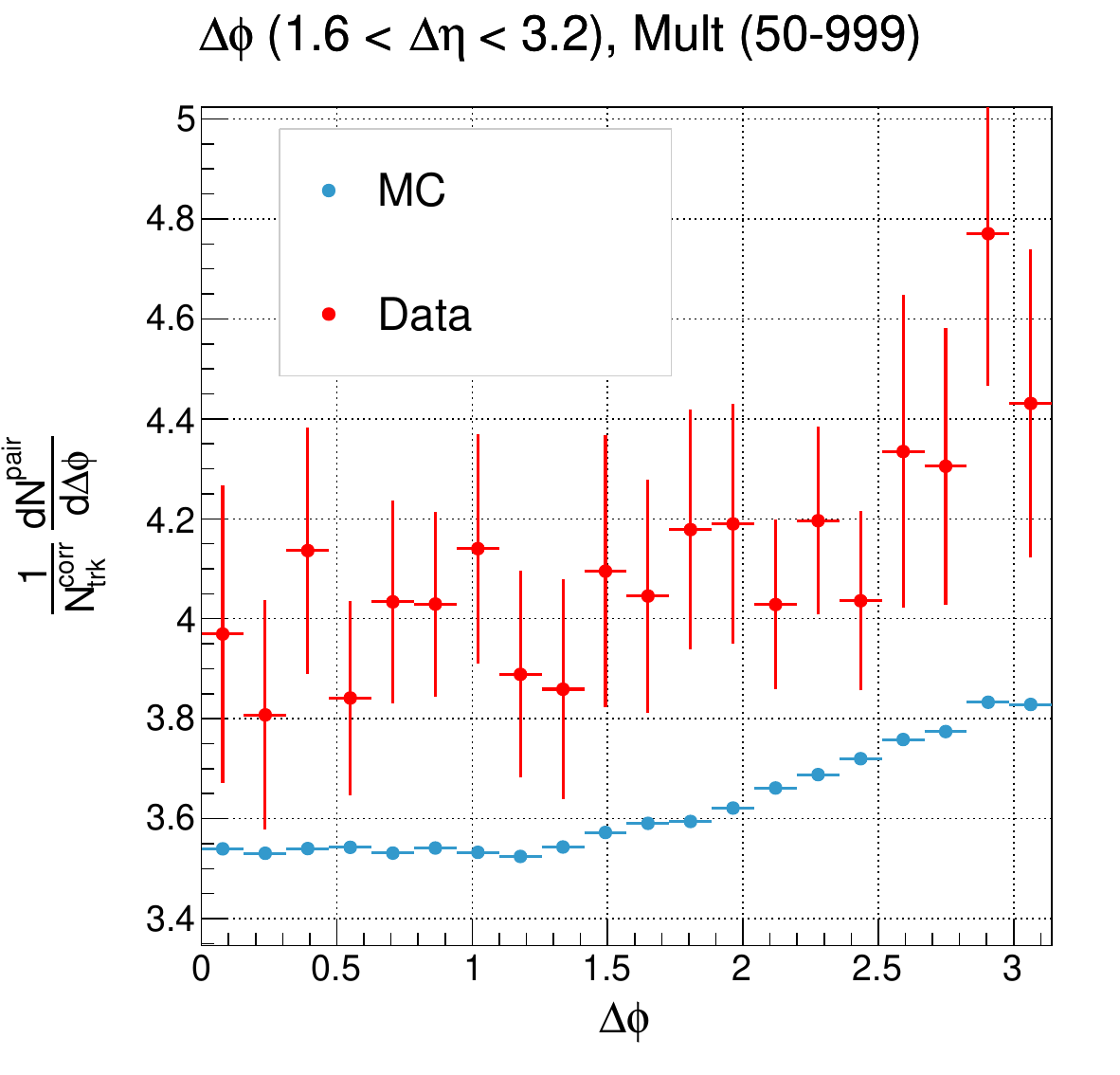}
\caption{[BDT at $\epsilon_{\rm sig}\approx 80\%$] Long-range
  ($1.6<\Delta\eta<3.2$) azimuthal differential yield $Y_l(\Delta\phi)$ in the
  multiplicity intervals $[30,35)$, $[35,40)$, $[40,45)$, $[45,50)$, and
  $[50,\infty)$. Data (red) carry bootstrap-derived statistical uncertainties
  (Sec.~\ref{sec:flow_bootstrap}); MC (blue) shows statistical errors.}
\label{fig:rst_80pbdt_40phibins_0908_final_dNdphi}
\end{figure}

In the appendix section, we show the detailed two-particle correlations and azimuthal differential yields for the multiplicity-inclusive result and as a function of offline multiplicity in the $[30,35)$, $[35,40)$, $[40,45)$, $[45,50)$, and $[50,\infty)$ intervals. The results of inclusive LEP-II events (without $WW$-enhanced BDT selection) are shown in Figure~\ref{fig:rst_nobdt_40phibins_0908_final_2PC} and~\ref{fig:rst_nobdt_40phibins_0908_final_dNdphi}. The two-particle correlations are also evaluated for selected events passing $WW$-enhanced BDT requirements.  Figures~\ref{fig:rst_95pbdt_40phibins_0908_final_2PC} to~\ref{fig:rst_80pbdt_40phibins_0908_final_dNdphi} show results with increasingly stringent BDT cuts, corresponding to 95\% to 80\% signal efficiency.

\begin{figure}[ht]
\centering
    \begin{subfigure}[b]{0.32\textwidth}
    \includegraphics[width=\textwidth]{figures/vnStudy/ntrkskim30_nobdt_thrust/Data_reco_183_207_v123.pdf}
    \caption{No BDT selection}
    \end{subfigure}
    \begin{subfigure}[b]{0.32\textwidth}
    \includegraphics[width=\textwidth]{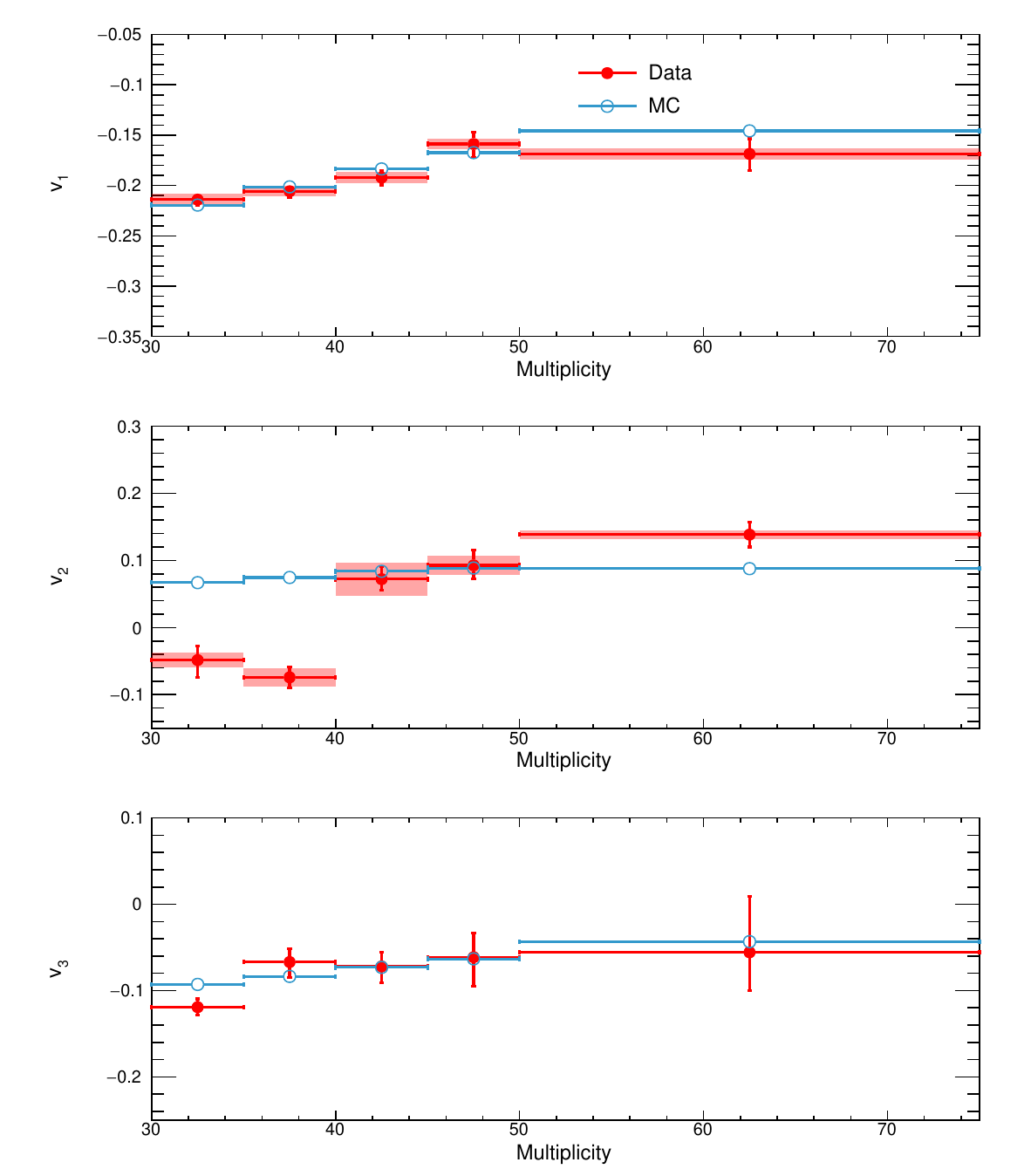}
    \caption{$\epsilon_{\rm sig} \approx 95\%$}
    \end{subfigure}
    \begin{subfigure}[b]{0.32\textwidth}
    \includegraphics[width=\textwidth]{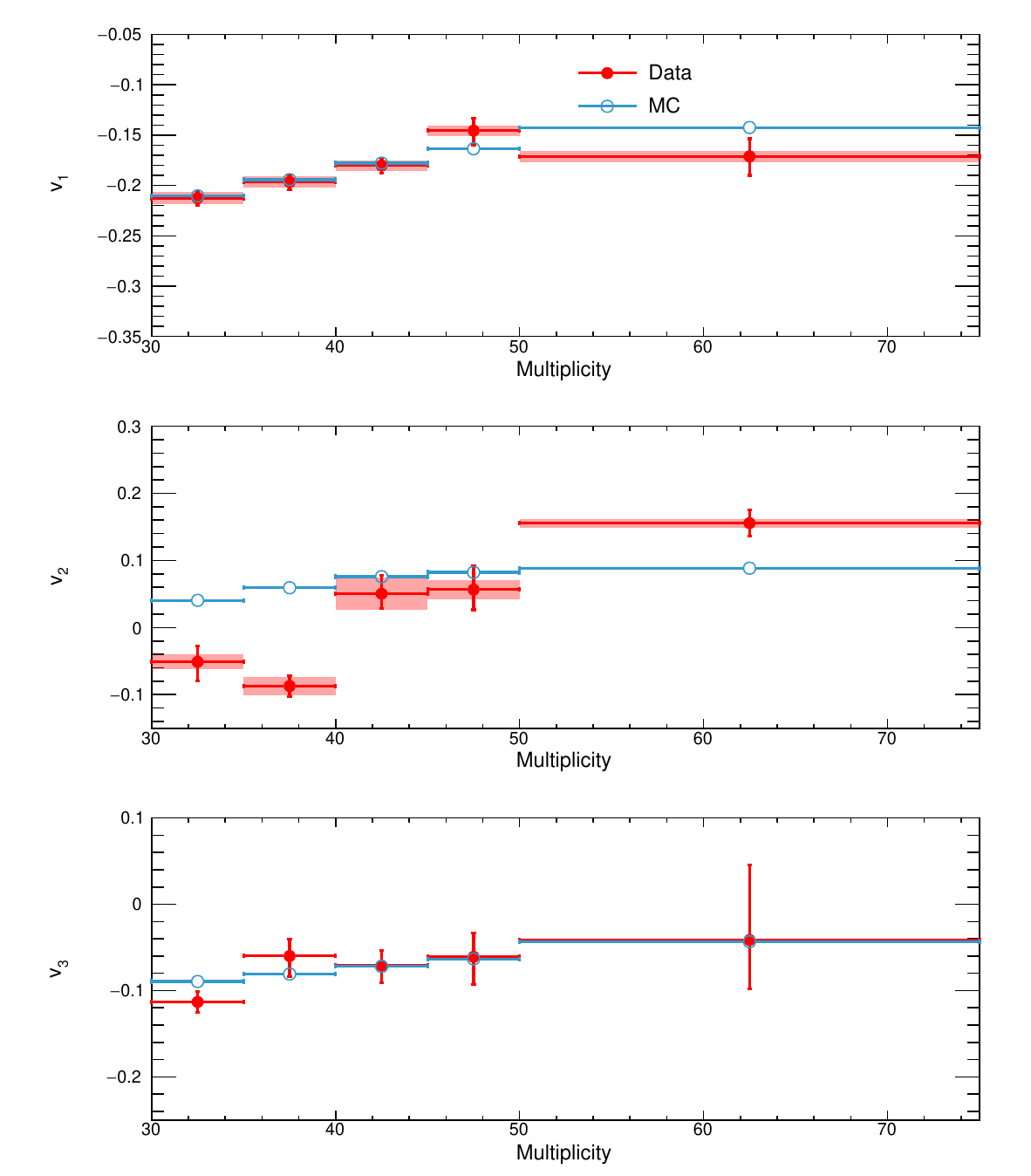}
    \caption{$\epsilon_{\rm sig} \approx 90\%$}
    \end{subfigure}
    \begin{subfigure}[b]{0.32\textwidth}
    \includegraphics[width=\textwidth]{figures/vnStudy/ntrkskim30_85pbdt_thrust/Data_reco_183_207_v123_with_bdt_syst.pdf}
    \caption{$\epsilon_{\rm sig} \approx 85\%$}
    \end{subfigure}
    \begin{subfigure}[b]{0.32\textwidth}
    \includegraphics[width=\textwidth]{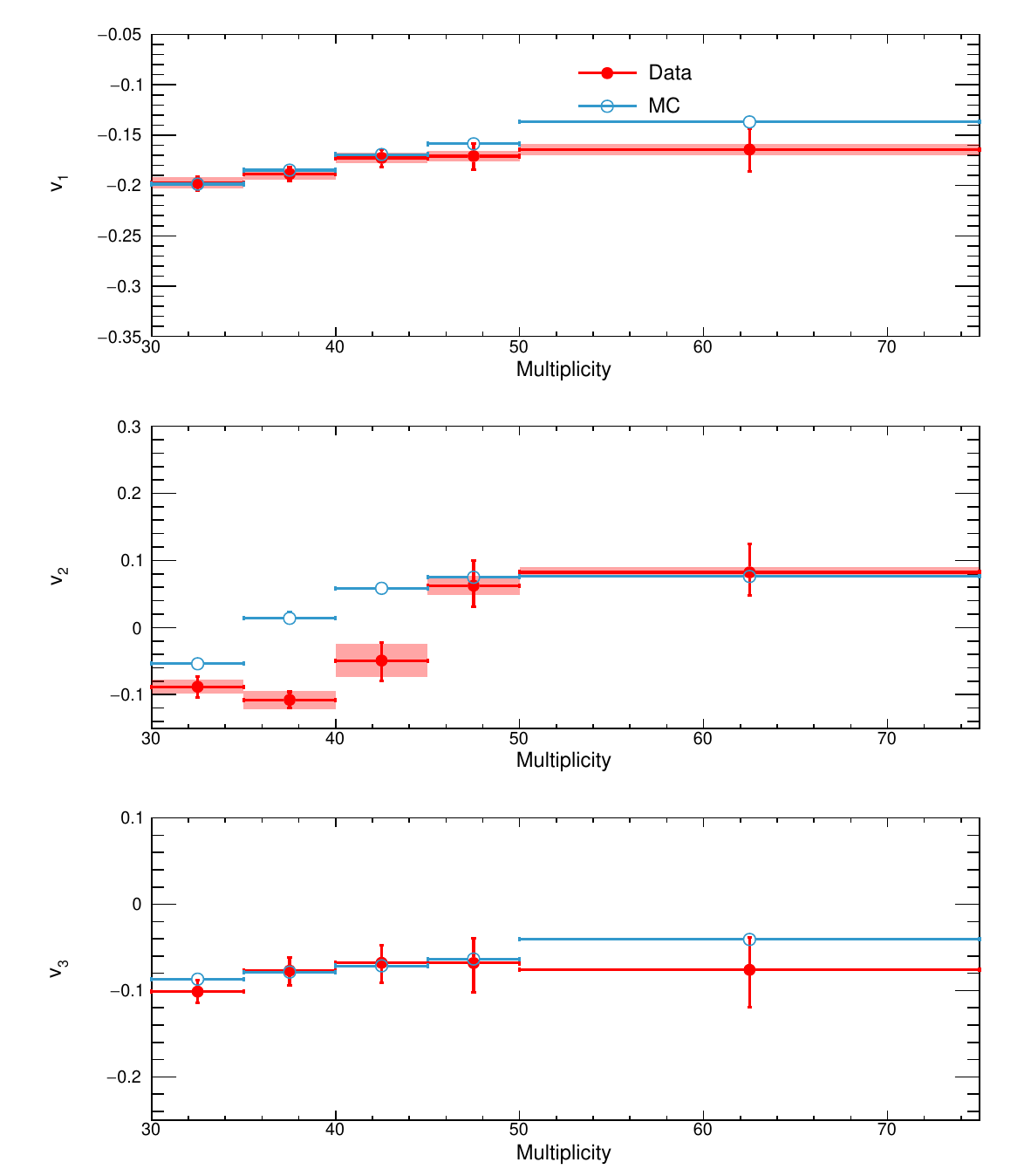}
    \caption{$\epsilon_{\rm sig} \approx 80\%$}
    \end{subfigure}
    \caption{Signed square-root proxies, labeled $v_{n}$, as a function of the event multiplicity interval for the thrust-axis analysis of the LEP-II high-energy sample without and with the \WW-enhanced BDT selections at $\epsilon_{\rm sig} \approx 95\%$, 90\%, 85\%, and 80\% working points, shown in (a) to (e), respectively. The data's $v_1$, $v_2$, and $v_3$ are shown as black, red, and purple error bars. MC results are open dots with corresponding colors.}
\label{fig:vnVsMult_thrust_details}
\end{figure}

\begin{figure}[ht]
\centering
    \begin{subfigure}[b]{0.40\textwidth}
    \includegraphics[width=\textwidth]{figures/Dv2/ntrkskim30_nobdt_thrust/Dv2_183_207_SquareShift.pdf}
    \caption{No BDT selection}
    \end{subfigure}
    \begin{subfigure}[b]{0.40\textwidth}
    \includegraphics[width=\textwidth]{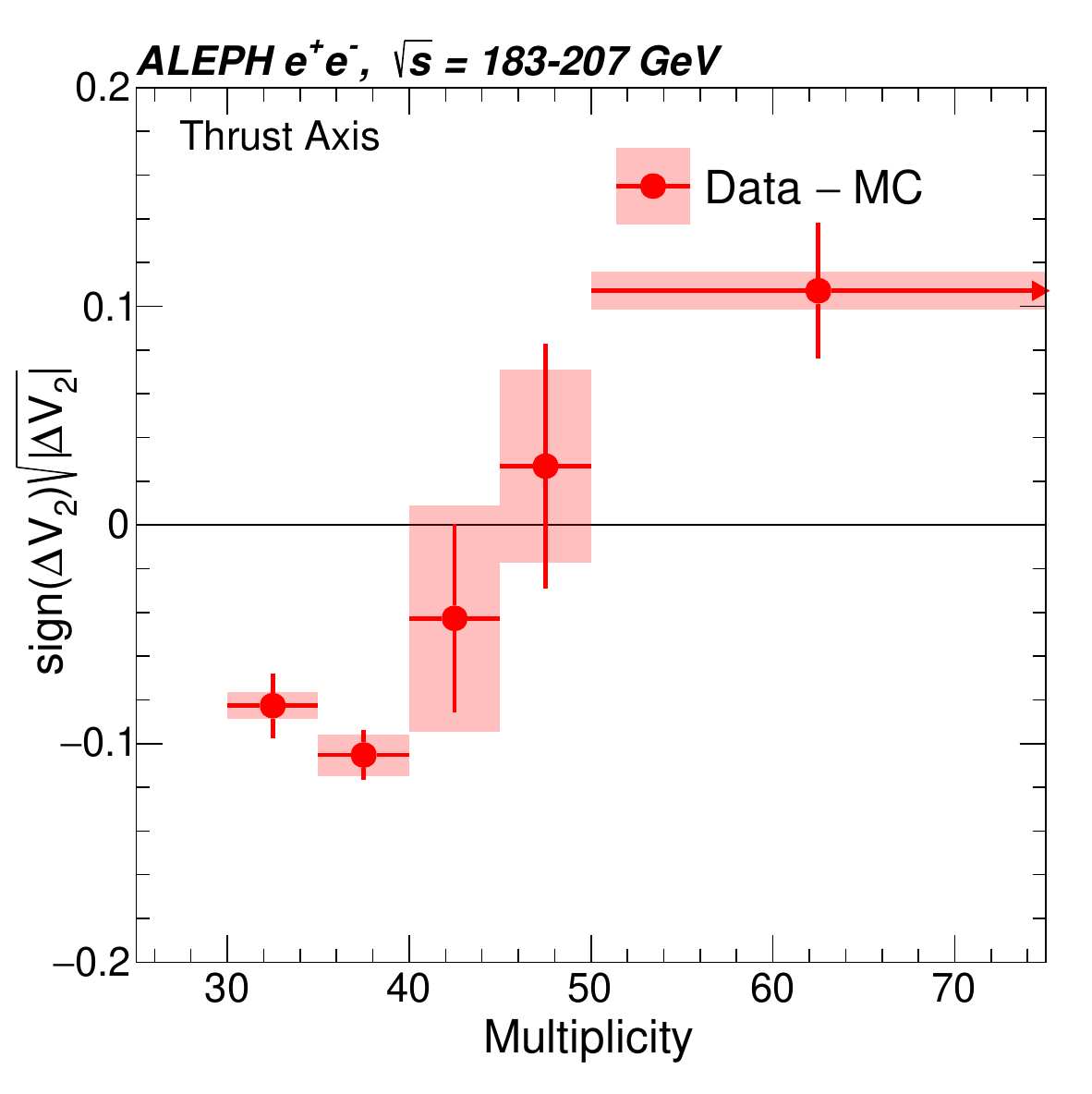}
    \caption{$\epsilon_{\rm sig} \approx 95\%$}
    \end{subfigure}
    \begin{subfigure}[b]{0.40\textwidth}
    \includegraphics[width=\textwidth]{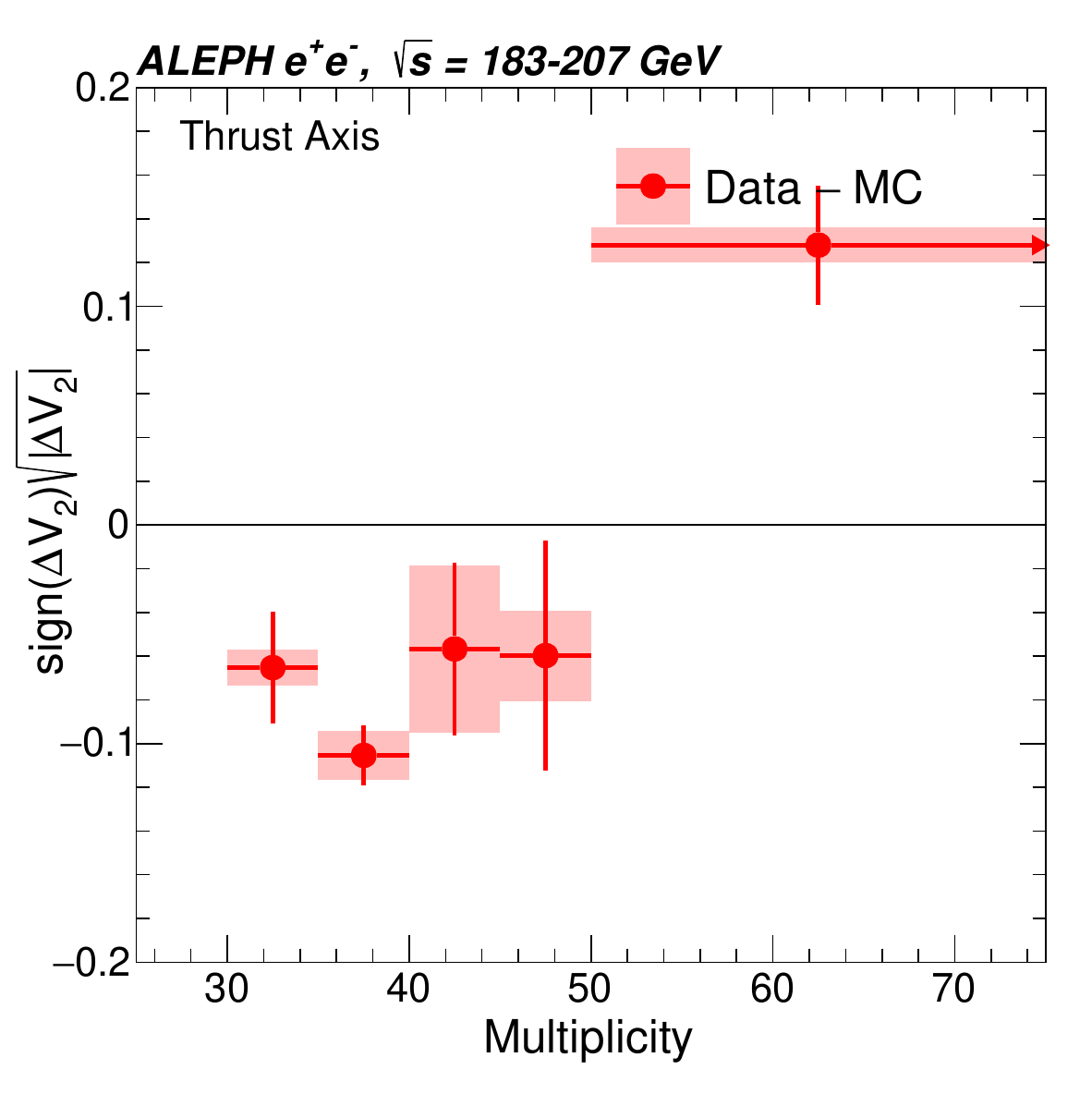}
    \caption{$\epsilon_{\rm sig} \approx 90\%$}
    \end{subfigure}
    \begin{subfigure}[b]{0.40\textwidth}
    \includegraphics[width=\textwidth]{figures/Dv2/ntrkskim30_85pbdt_thrust/Dv2_183_207_SquareShift_with_bdt_syst.pdf}
    \caption{$\epsilon_{\rm sig} \approx 85\%$}
    \end{subfigure}
    \begin{subfigure}[b]{0.40\textwidth}
    \includegraphics[width=\textwidth]{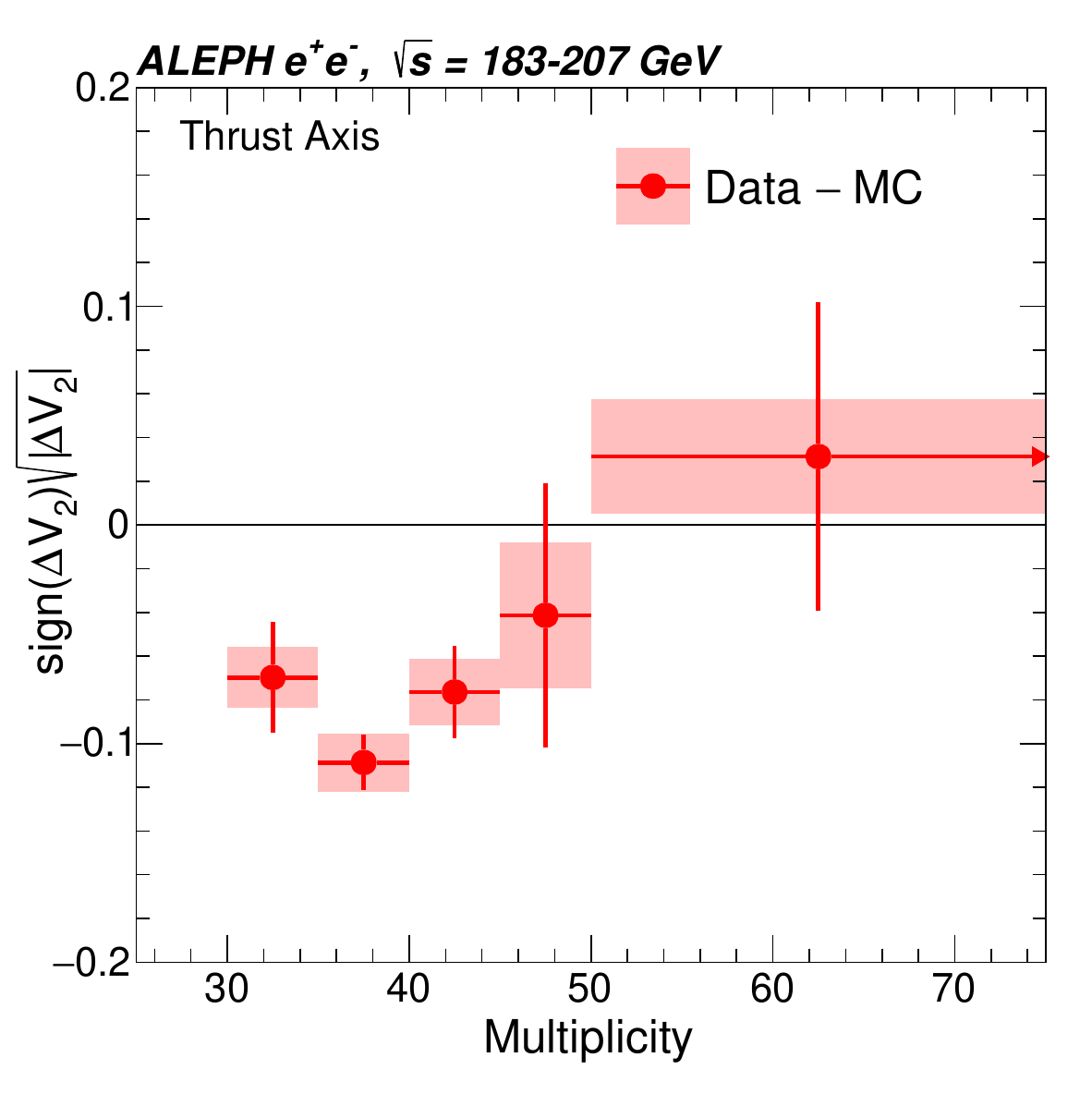}
    \caption{$\epsilon_{\rm sig} \approx 80\%$}
    \end{subfigure}
\caption{$\mathrm{sign}(\Delta V_{2\Delta})\sqrt{|\Delta V_{2\Delta}|}$ as a function of the
offline multiplicity for the thrust-axis analysis of the LEP-II high-energy
sample, without the BDT selection (a) and with the \WW-enhanced BDT selection at
$\epsilon_{\rm sig}\approx 95\%$, 90\%, 85\%, 80\% working points, shown in
(b)--(e), where $\Delta V_{2\Delta} \equiv V_{2\Delta}^{\rm data} - V_{2\Delta}^{\rm MC}$.
This is a signed square-root representation of the data--MC difference in the fitted two-particle Fourier coefficient, not a direct difference of $v_2$ values. Vertical bars indicate the
bootstrap-propagated statistical uncertainty; shaded red boxes show the
systematic uncertainty.}
\label{fig:Dv2VsMult_thrust_details}
\end{figure}

The signed square-root proxies, labeled $v_1$, $v_2$, and $v_3$ for compactness, extracted from the long-range azimuthal differential yields $Y_l(\Delta \phi)$ at all the working points are shown in Figure~\ref{fig:vnVsMult_thrust_details}, for both data (error bars) and MC (open dots). Figure~\ref{fig:Dv2VsMult_thrust_details} shows the transformed data--MC difference of the fitted two-particle coefficient, $\mathrm{sign}(\Delta V_{2\Delta}) \sqrt{|\Delta V_{2\Delta}|}$.

\clearpage

\clearpage

\clearpage

\clearpage

\clearpage

\clearpage

\begin{figure}[ht]
    \begin{center}
        \begin{subfigure}{1.0\textwidth}
            \includegraphics[width=1.0\textwidth]{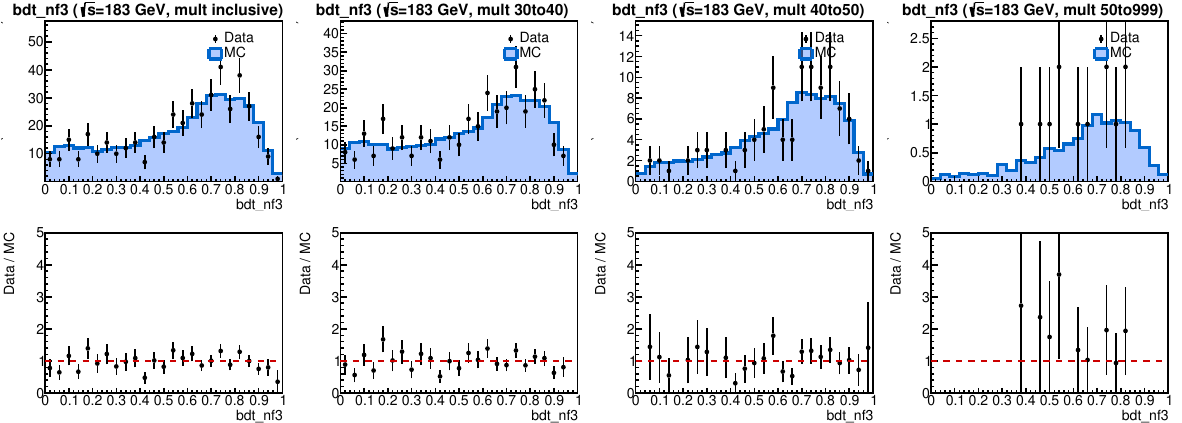}
            \vspace{-20pt}
            \caption{bdt\_nf3 with $\sqrt{s}=183$ GeV}
            \vspace{7pt}
        \end{subfigure}
        \begin{subfigure}{1.0\textwidth}
            \includegraphics[width=1.0\textwidth]{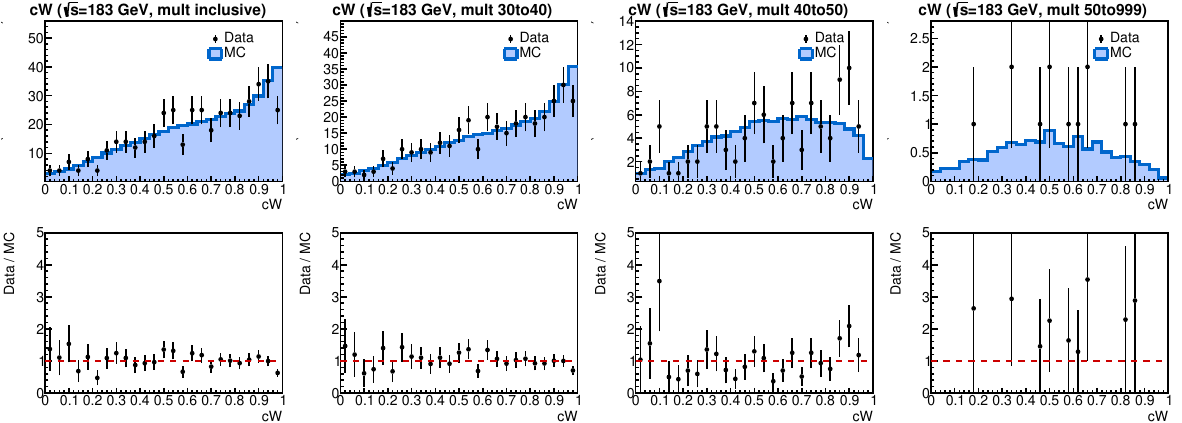}
            \vspace{-20pt}
            \caption{$c_W$ with $\sqrt{s}=183$ GeV}
            \vspace{7pt}
        \end{subfigure}
        
        \begin{subfigure}{1.0\textwidth}
            \includegraphics[width=1.0\textwidth]{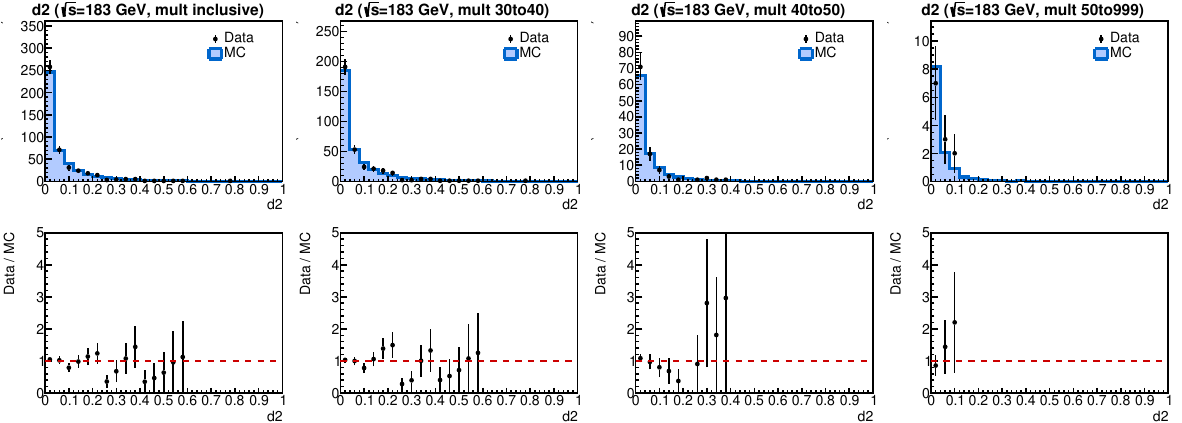}
            \vspace{-20pt}
            \caption{$d^2$ with $\sqrt{s}=183$ GeV}
        \end{subfigure}
    \end{center}
    \caption{Data and normalized Monte Carlo (MC) histograms \& Data/MC ratio plots for $\sqrt{s}=183$ GeV. Variables compared: bdt\_nf3, $c_W$, and $d^2$. For each set of histograms, the data (black markers) are compared to MC (blue histogram) across charged-particle multiplicity bins: inclusive, 30-40, 40-50, 50+, left to right. Below each histogram is a Data/MC ratio plot, where the red dashed line corresponds to unity.}
  \label{fig:DataVsMCVariables183A}
\end{figure}

\begin{figure}[h]
    \begin{center}
        \begin{subfigure}{1.0\textwidth}
            \includegraphics[width=1.0\textwidth]{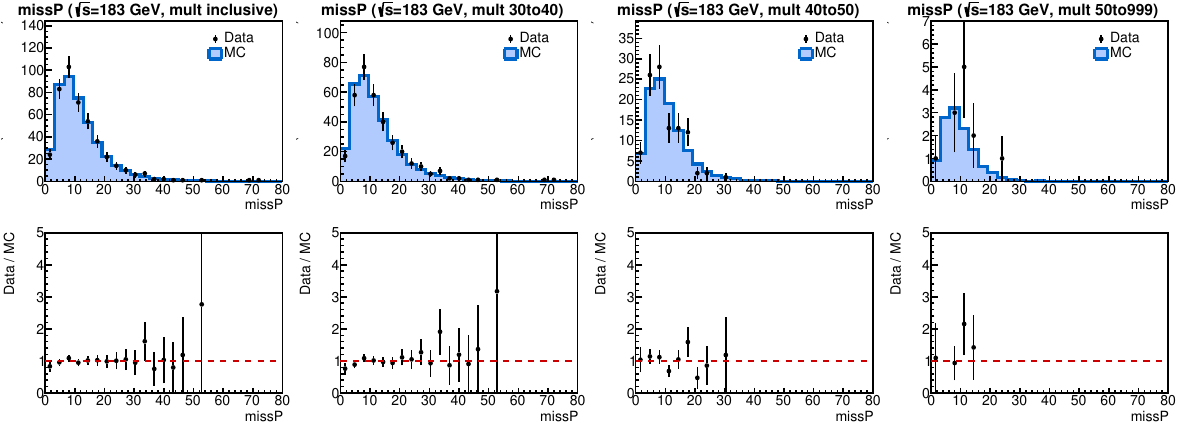}
            \vspace{-20pt}
            \caption{$\Vec{p}_{\rm miss}$ with $\sqrt{s}=183$ GeV}
            \vspace{7pt}
        \end{subfigure}
        \begin{subfigure}{1.0\textwidth}
            \includegraphics[width=1.0\textwidth]{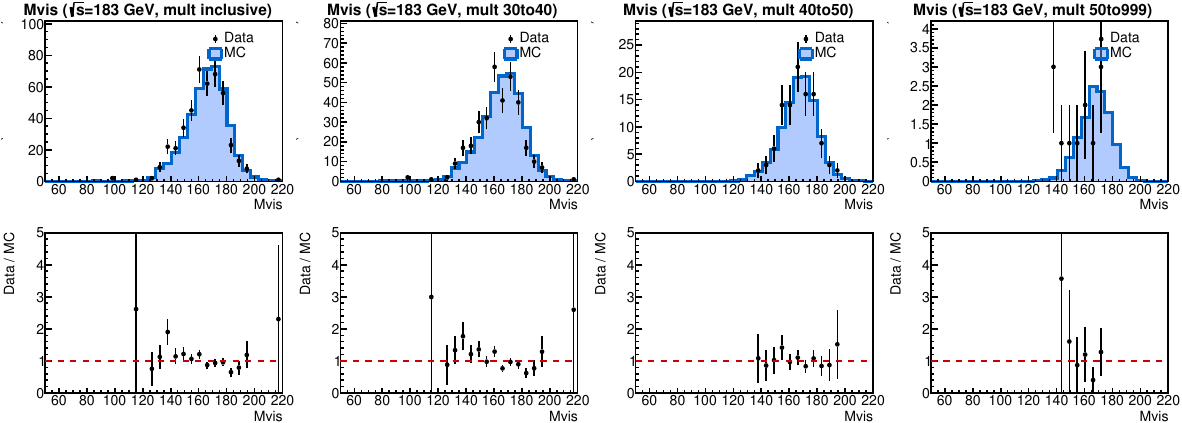}
            \vspace{-20pt}
            \caption{$M_{\rm vis}$ with $\sqrt{s}=183$ GeV}
            \vspace{7pt}
        \end{subfigure}
        \begin{subfigure}{1.0\textwidth}
            \includegraphics[width=1.0\textwidth]{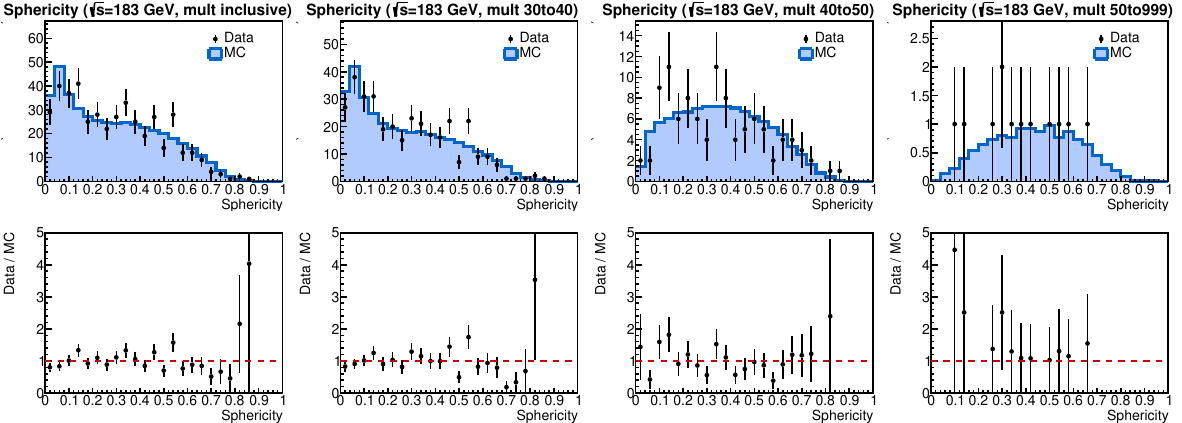}
            \vspace{-20pt}
            \caption{Sphericity with $\sqrt{s}=183$ GeV}
        \end{subfigure}
    \end{center}
    \caption{Same as Fig.~\ref{fig:DataVsMCVariables183A}, but with variables compared: $\Vec{p}_{\rm miss}$, $M_{\rm vis}$, and Sphericity. $\sqrt{s}=183$ GeV.}
  \label{fig:DataVsMCVariables183B}
\end{figure}

\begin{figure}[ht]
    \begin{center}
        \begin{subfigure}{1.0\textwidth}
            \includegraphics[width=1.0\textwidth]{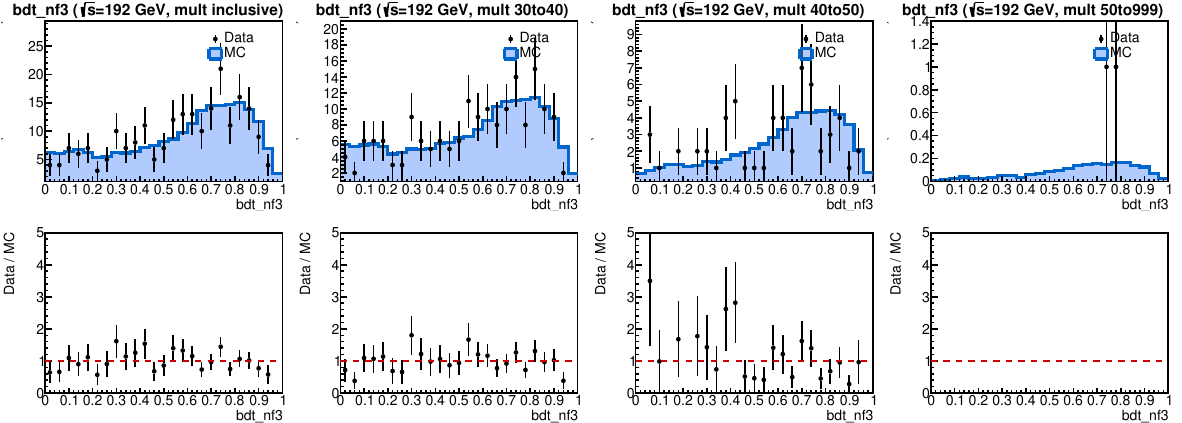}
            \vspace{-20pt}
            \caption{bdt\_nf3 with $\sqrt{s}=192$ GeV}
            \vspace{7pt}
        \end{subfigure}
        \begin{subfigure}{1.0\textwidth}
            \includegraphics[width=1.0\textwidth]{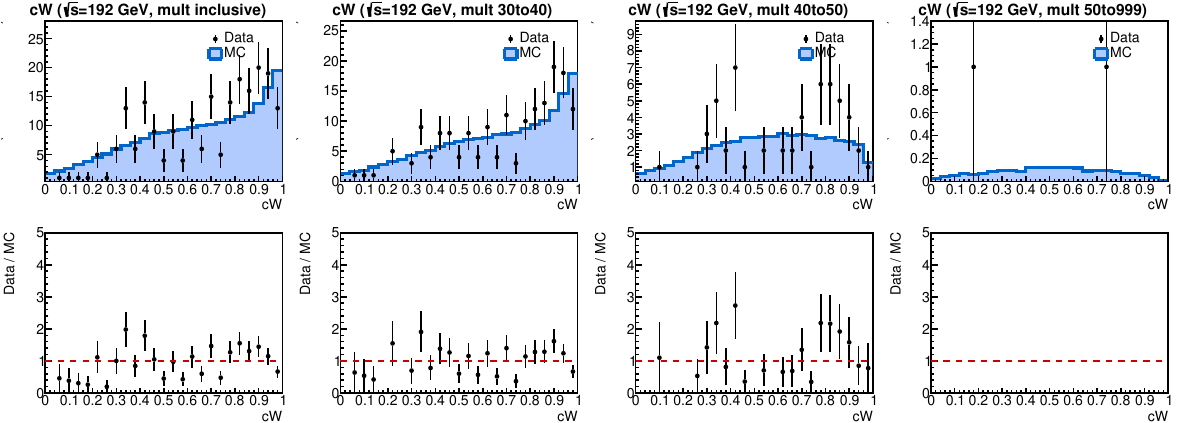}
            \vspace{-20pt}
            \caption{$c_W$ with $\sqrt{s}=192$ GeV}
            \vspace{7pt}
        \end{subfigure}
        
        \begin{subfigure}{1.0\textwidth}
            \includegraphics[width=1.0\textwidth]{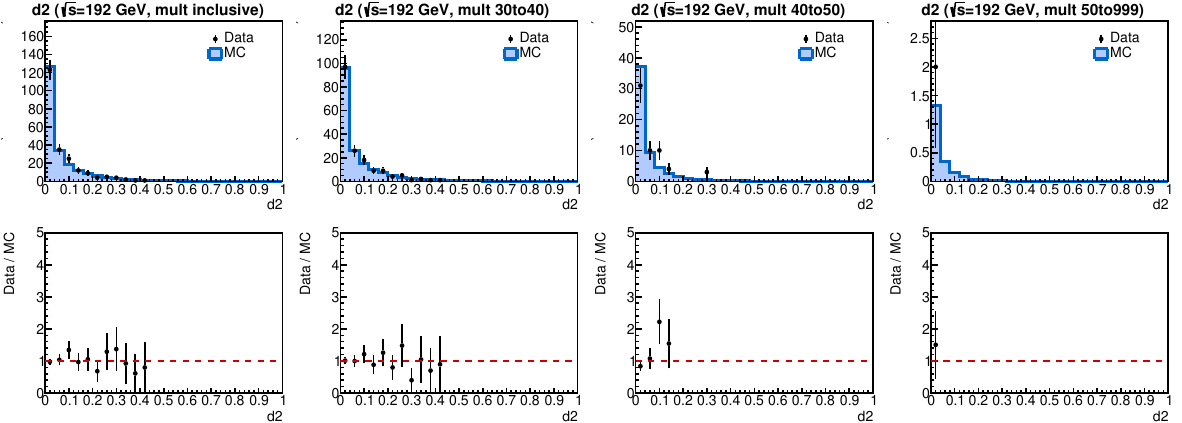}
            \vspace{-20pt}
            \caption{$d^2$ with $\sqrt{s}=192$ GeV}
        \end{subfigure}
    \end{center}
    \caption{Same as Fig.~\ref{fig:DataVsMCVariables183A}, but with variables compared: bdt\_nf3, $c_W$, and $d^2$. $\sqrt{s}=192$ GeV.} 
  \label{fig:DataVsMCVariables192A}
\end{figure}

\begin{figure}[h]
    \begin{center}
        \begin{subfigure}{1.0\textwidth}
            \includegraphics[width=1.0\textwidth]{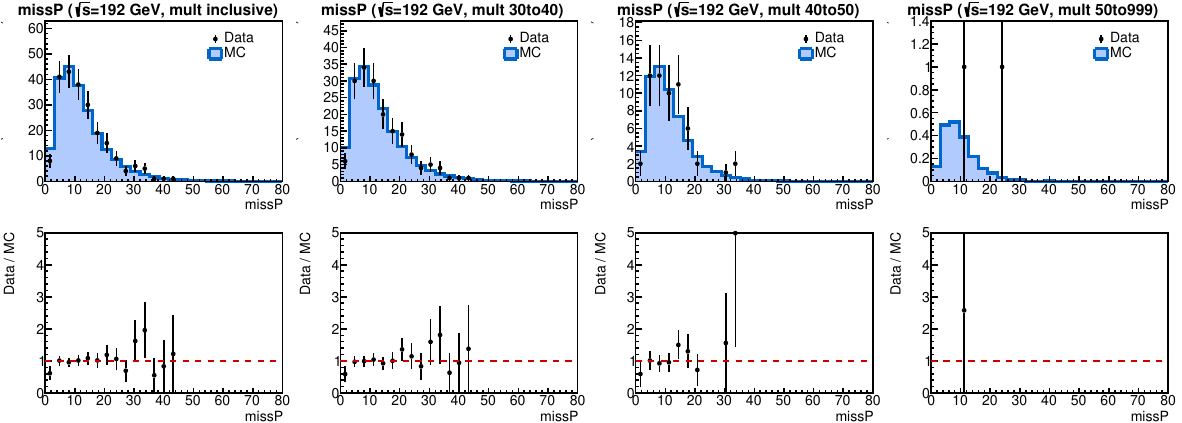}
            \vspace{-20pt}
            \caption{$\Vec{p}_{\rm miss}$ with $\sqrt{s}=192$ GeV}
            \vspace{7pt}
        \end{subfigure}
        \begin{subfigure}{1.0\textwidth}
            \includegraphics[width=1.0\textwidth]{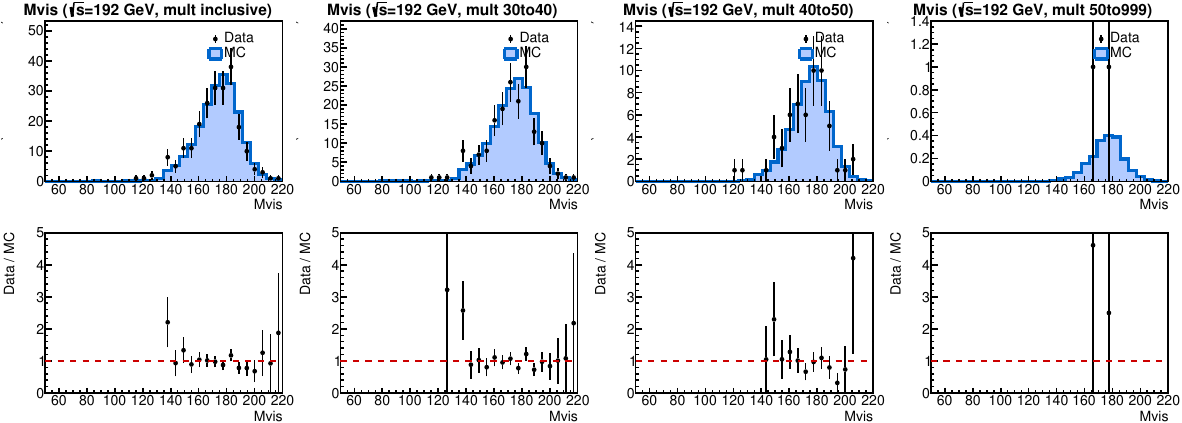}
            \vspace{-20pt}
            \caption{$M_{\rm vis}$ with $\sqrt{s}=192$ GeV}
            \vspace{7pt}
        \end{subfigure}
        \begin{subfigure}{1.0\textwidth}
            \includegraphics[width=1.0\textwidth]{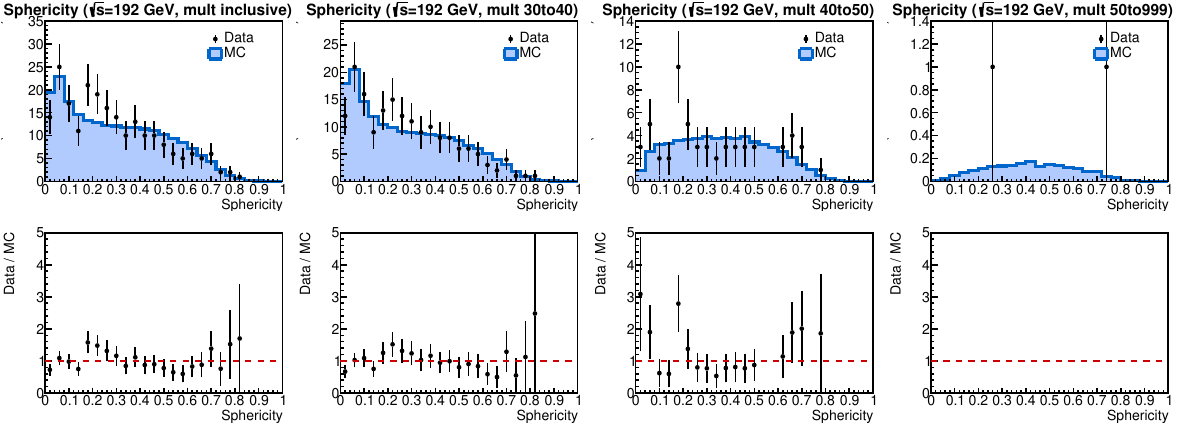}
            \vspace{-20pt}
            \caption{Sphericity with $\sqrt{s}=192$ GeV}
        \end{subfigure}
    \end{center}
    \caption{Same as Fig.~\ref{fig:DataVsMCVariables183A}, but with variables compared: $\Vec{p}_{\rm miss}$, $M_{\rm vis}$, and Sphericity. $\sqrt{s}=192$ GeV.}
  \label{fig:DataVsMCVariables192B}
\end{figure}

\begin{figure}[ht]
    \begin{center}
        \begin{subfigure}{1.0\textwidth}
            \includegraphics[width=1.0\textwidth]{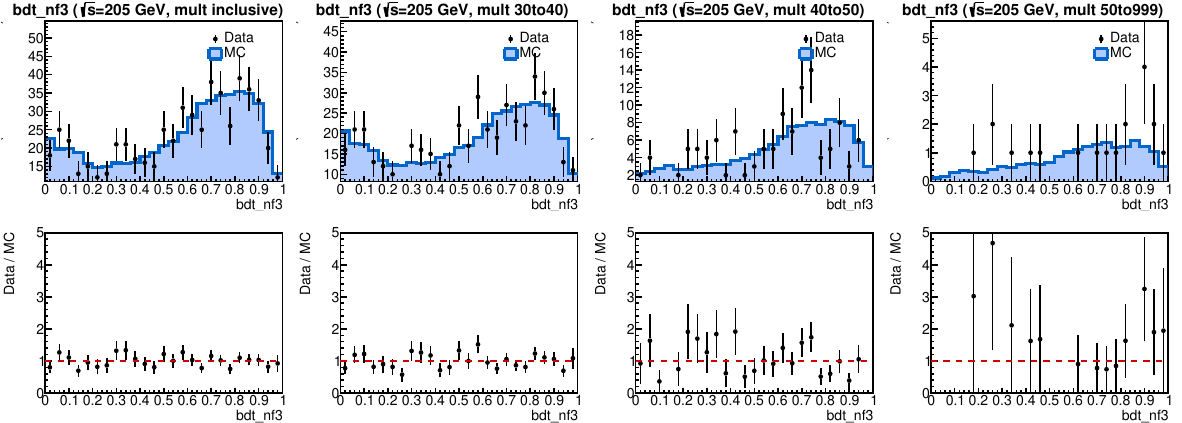}
            \vspace{-20pt}
            \caption{bdt\_nf3 with $\sqrt{s}=205$ GeV}
            \vspace{7pt}
        \end{subfigure}
        \begin{subfigure}{1.0\textwidth}
            \includegraphics[width=1.0\textwidth]{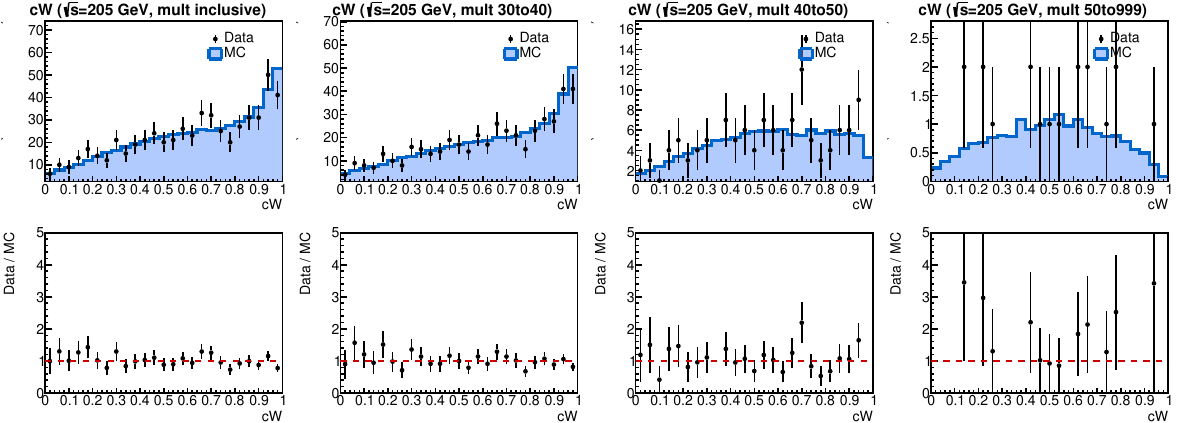}
            \vspace{-20pt}
            \caption{$c_W$ with $\sqrt{s}=205$ GeV}
            \vspace{7pt}
        \end{subfigure}
        
        \begin{subfigure}{1.0\textwidth}
            \includegraphics[width=1.0\textwidth]{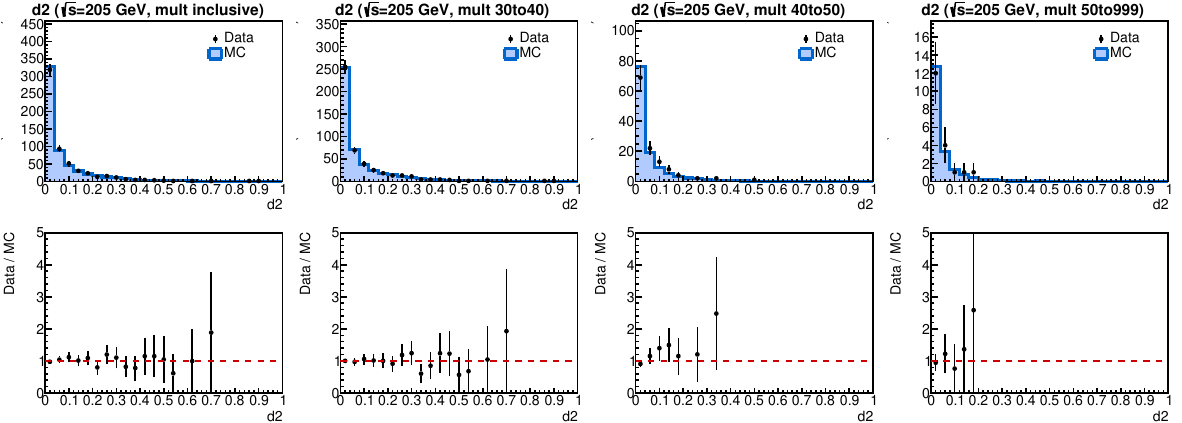}
            \vspace{-20pt}
            \caption{$d^2$ with $\sqrt{s}=205$ GeV}
        \end{subfigure}
    \end{center}
    \caption{Same as Fig.~\ref{fig:DataVsMCVariables183A}, but with variables compared: bdt\_nf3, $c_W$, and $d^2$. $\sqrt{s}=205$ GeV.}  
  \label{fig:DataVsMCVariables205A}
\end{figure}

\begin{figure}[h]
    \begin{center}
        \begin{subfigure}{1.0\textwidth}
            \includegraphics[width=1.0\textwidth]{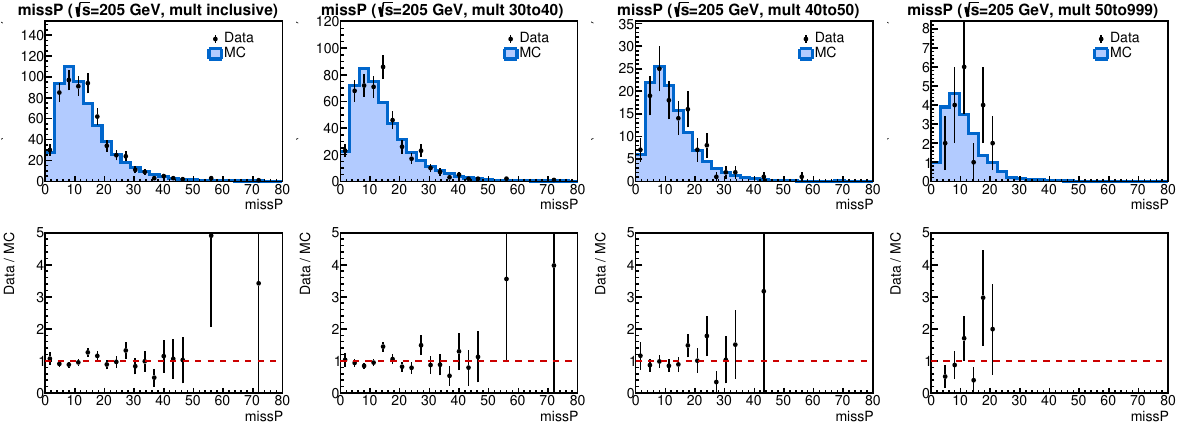}
            \vspace{-20pt}
            \caption{$\Vec{p}_{\rm miss}$ with $\sqrt{s}=205$ GeV}
            \vspace{7pt}
        \end{subfigure}
        \begin{subfigure}{1.0\textwidth}
            \includegraphics[width=1.0\textwidth]{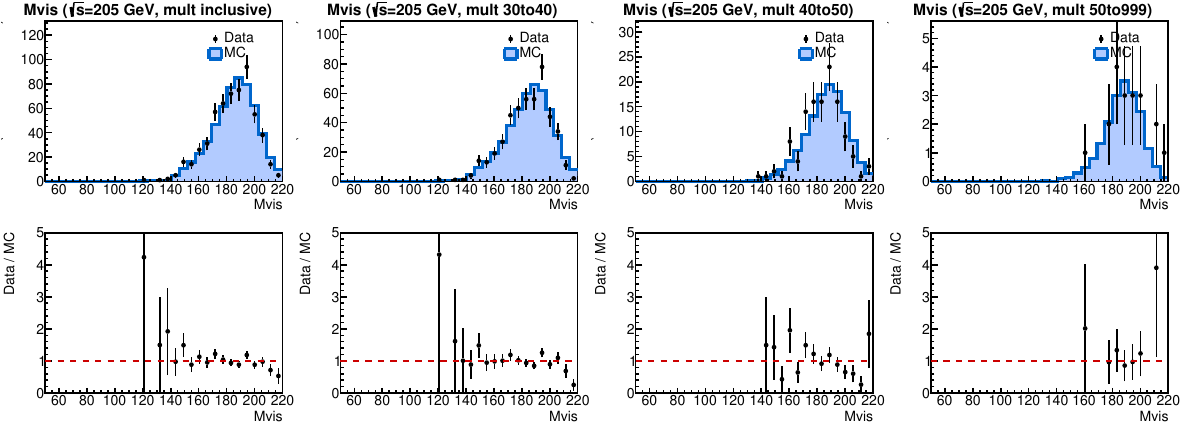}
            \vspace{-20pt}
            \caption{$M_{\rm vis}$ with $\sqrt{s}=205$ GeV}
            \vspace{7pt}
        \end{subfigure}
        \begin{subfigure}{1.0\textwidth}
            \includegraphics[width=1.0\textwidth]{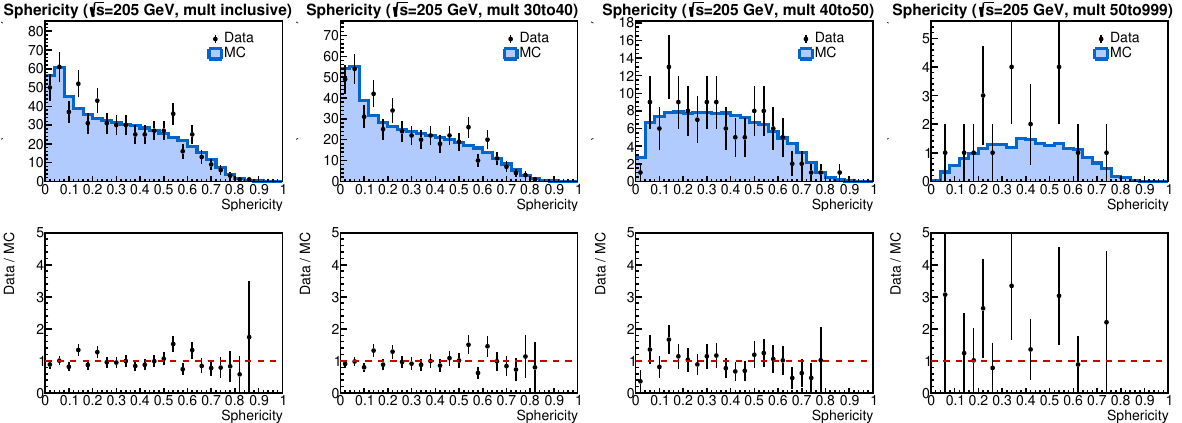}
            \vspace{-20pt}
            \caption{Sphericity with $\sqrt{s}=205$ GeV}
        \end{subfigure}
    \end{center}
    \caption{Same as Fig.~\ref{fig:DataVsMCVariables183A}, but with variables compared: $\Vec{p}_{\rm miss}$, $M_{\rm vis}$, and Sphericity. $\sqrt{s}=205$ GeV.}
  \label{fig:DataVsMCVariables205B}
\end{figure}


\bibliographystyle{JHEP}
\typeout{}
\bibliography{TPC}

@article{ALEPH:1994ayc,
    author = "Buskulic, D. and others",
    collaboration = "ALEPH",
    title = "{Performance of the ALEPH detector at LEP}",
    reportNumber = "CERN-PPE-94-170, FSU-SCRI-95-70",
    doi = "10.1016/0168-9002(95)00138-7",
    journal = "Nucl. Instrum. Meth. A",
    volume = "360",
    pages = "481--506",
    year = "1995"
}

@article{Creanza:1998nim,
    author = "Creanza, D. and others",
    collaboration = "ALEPH",
    title = "{The new ALEPH Silicon Vertex Detector}",
    journal = "Nucl. Instrum. Meth. A",
    volume = "409",
    pages = "157--160",
    year = "1998",
    doi = "10.1016/S0168-9002(97)91255-9"
}

@article{Khachatryan:2010gv,
      author         = "Khachatryan, Vardan and others",
      title          = "{Observation of Long-Range Near-Side Angular Correlations
                        in Proton-Proton Collisions at the LHC}",
      collaboration  = "CMS",
      journal        = "JHEP",
      volume         = "09",
      year           = "2010",
      pages          = "091",
      doi            = "10.1007/JHEP09(2010)091",
      eprint         = "1009.4122",
      archivePrefix  = "arXiv",
      primaryClass   = "hep-ex",
      reportNumber   = "CMS-QCD-10-002, CERN-PH-EP-2010-031",
      SLACcitation   = "%%CITATION = ARXIV:1009.4122;%%"
}

@article{Aad:2015gqa,
      author         = "Aad, Georges and others",
      title          = "{Observation of Long-Range Elliptic Azimuthal
                        Anisotropies in $\sqrt{s}=$13 and 2.76 TeV $pp$
                        Collisions with the ATLAS Detector}",
      collaboration  = "ATLAS",
      journal        = "Phys. Rev. Lett.",
      volume         = "116",
      year           = "2016",
      number         = "17",
      pages          = "172301",
      doi            = "10.1103/PhysRevLett.116.172301",
      eprint         = "1509.04776",
      archivePrefix  = "arXiv",
      primaryClass   = "hep-ex",
      reportNumber   = "CERN-PH-EP-2015-251",
      SLACcitation   = "%%CITATION = ARXIV:1509.04776;%%"
}

@article{CMS:2012qk,
      author         = "Chatrchyan, Serguei and others",
      title          = "{Observation of long-range near-side angular correlations
                        in proton-lead collisions at the LHC}",
      collaboration  = "CMS",
      journal        = "Phys. Lett.",
      volume         = "B718",
      year           = "2013",
      pages          = "795-814",
      doi            = "10.1016/j.physletb.2012.11.025",
      eprint         = "1210.5482",
      archivePrefix  = "arXiv",
      primaryClass   = "nucl-ex",
      reportNumber   = "CMS-HIN-12-005, CERN-PH-EP-2012-320, CMS-HIN-12-015",
      SLACcitation   = "%%CITATION = ARXIV:1210.5482;%%"
}

@article{Chatrchyan:2012wg,
      author         = "Chatrchyan, Serguei and others",
      title          = "{Centrality dependence of dihadron correlations and
                        azimuthal anisotropy harmonics in $\mathrm{PbPb}$ collisions at
                        $\sqrt{s_{NN}}=2.76\text{ }\mathrm{TeV}$}",
      collaboration  = "CMS",
      journal        = "Eur. Phys. J.",
      volume         = "C72",
      year           = "2012",
      pages          = "2012",
      doi            = "10.1140/epjc/s10052-012-2012-3",
      eprint         = "1201.3158",
      archivePrefix  = "arXiv",
      primaryClass   = "nucl-ex",
      reportNumber   = "CMS-HIN-11-006, CERN-PH-EP-2011-222",
      SLACcitation   = "%%CITATION = ARXIV:1201.3158;%%"
}

@article{Aaij:2015qcq,
      author         = "Aaij, Roel and others",
      title          = "{Measurements of long-range near-side angular
                        correlations in $\sqrt{s_{\text{NN}}}=5\text{ }\mathrm{TeV}$ proton-lead
                        collisions in the forward region}",
      collaboration  = "LHCb",
      journal        = "Phys. Lett.",
      volume         = "B762",
      year           = "2016",
      pages          = "473-483",
      doi            = "10.1016/j.physletb.2016.09.064",
      eprint         = "1512.00439",
      archivePrefix  = "arXiv",
      primaryClass   = "nucl-ex",
      reportNumber   = "LHCB-PAPER-2015-040, CERN-PH-EP-2015-308",
      SLACcitation   = "%%CITATION = ARXIV:1512.00439;%%"
}

@article{Nagle:2018nvi,
      author         = "Nagle, James L. and Zajc, William A.",
      title          = "{Small System Collectivity in Relativistic Hadronic and
                        Nuclear Collisions}",
      journal        = "Ann. Rev. Nucl. Part. Sci.",
      volume         = "68",
      year           = "2018",
      pages          = "211-235",
      doi            = "10.1146/annurev-nucl-101916-123209",
      eprint         = "1801.03477",
      archivePrefix  = "arXiv",
      primaryClass   = "nucl-ex",
      SLACcitation   = "%%CITATION = ARXIV:1801.03477;%%"
}

@article{Alver:2010gr,
      author         = "Alver, B. and Roland, G.",
      title          = "{Collision geometry fluctuations and triangular flow in
                        heavy-ion collisions}",
      journal        = "Phys. Rev.",
      volume         = "C81",
      year           = "2010",
      pages          = "054905",
      doi            = "10.1103/PhysRevC.82.039903, 10.1103/PhysRevC.81.054905",
      note           = "[Erratum: Phys. Rev.C82,039903(2010)]",
      eprint         = "1003.0194",
      archivePrefix  = "arXiv",
      primaryClass   = "nucl-th",
      SLACcitation   = "%%CITATION = ARXIV:1003.0194;%%"
}

@article{Dusling:2013qoz,
      author         = "Dusling, Kevin and Venugopalan, Raju",
      title          = "{Comparison of the color glass condensate to dihadron
                        correlations in proton-proton and proton-nucleus
                        collisions}",
      journal        = "Phys. Rev.",
      volume         = "D87",
      year           = "2013",
      number         = "9",
      pages          = "094034",
      doi            = "10.1103/PhysRevD.87.094034",
      eprint         = "1302.7018",
      archivePrefix  = "arXiv",
      primaryClass   = "hep-ph",
      SLACcitation   = "%%CITATION = ARXIV:1302.7018;%%"
}

@article{Bozek:2011if,
      author         = "Bozek, Piotr",
      title          = "{Collective flow in p-Pb and d-Pd collisions at TeV
                        energies}",
      journal        = "Phys. Rev.",
      volume         = "C85",
      year           = "2012",
      pages          = "014911",
      doi            = "10.1103/PhysRevC.85.014911",
      eprint         = "1112.0915",
      archivePrefix  = "arXiv",
      primaryClass   = "hep-ph",
      SLACcitation   = "%%CITATION = ARXIV:1112.0915;%%"
}

@article{Nagle:2017sjv,
      author         = "Nagle, J. L. and Belmont, Ron and Hill, Kurt and Orjuela
                        Koop, Javier and Perepelitsa, Dennis V. and Yin, Pengqi
                        and Lin, Zi-Wei and McGlinchey, Darren",
      title          = "{Minimal conditions for collectivity in $e^+e^-$ and
                        $p+p$ collisions}",
      journal        = "Phys. Rev.",
      volume         = "C97",
      year           = "2018",
      number         = "2",
      pages          = "024909",
      doi            = "10.1103/PhysRevC.97.024909",
      eprint         = "1707.02307",
      archivePrefix  = "arXiv",
      primaryClass   = "nucl-th",
      SLACcitation   = "%%CITATION = ARXIV:1707.02307;%%"
}

@article{Bierlich:2024lmb,
    author = {Bierlich, Christian and Christiansen, Peter and Gustafson, G{\"o}sta and L{\"o}nnblad, Leif and T{\"o}rnkvist, Robin and Zapp, Korinna},
    title = "{Going against the flow: Revealing the QCD degrees of freedom in hadronic collisions}",
    eprint = "2409.16093",
    archivePrefix = "arXiv",
    primaryClass = "hep-ph",
    month = "9",
    year = "2024"
}

@article{He:2015hfa,
      author         = "He, Liang and Edmonds, Terrence and Lin, Zi-Wei and Liu,
                        Feng and Molnar, Denes and Wang, Fuqiang",
      title          = "{Anisotropic parton escape is the dominant source of
                        azimuthal anisotropy in transport models}",
      journal        = "Phys. Lett.",
      volume         = "B753",
      year           = "2016",
      pages          = "506-510",
      doi            = "10.1016/j.physletb.2015.12.051",
      eprint         = "1502.05572",
      archivePrefix  = "arXiv",
      primaryClass   = "nucl-th",
      SLACcitation   = "%%CITATION = ARXIV:1502.05572;%%"
}

@article{Badea:2019vey,
    author = "Badea, Anthony and Baty, Austin and Chang, Paoti and Innocenti, Gian Michele and Maggi, Marcello and Mcginn, Christopher and Peters, Michael and Sheng, Tzu-An and Thaler, Jesse and Lee, Yen-Jie",
    title = "{Measurements of two-particle correlations in $e^+e^-$ collisions at 91 GeV with ALEPH archived data}",
    eprint = "1906.00489",
    archivePrefix = "arXiv",
    primaryClass = "hep-ex",
    reportNumber = "MITHIG-MOD-19-001",
    doi = "10.1103/PhysRevLett.123.212002",
    journal = "Phys. Rev. Lett.",
    volume = "123",
    number = "21",
    pages = "212002",
    year = "2019"
}

@article{Sjostrand:1993yb,
    author = "Sjostrand, Torbjorn",
    title = "{High-energy physics event generation with PYTHIA 5.7 and JETSET 7.4}",
    reportNumber = "CERN-TH-7111-93",
    doi = "10.1016/0010-4655(94)90132-5",
    journal = "Comput. Phys. Commun.",
    volume = "82",
    pages = "74--90",
    year = "1994"
}

@article{STAR:2005ryu,
    author = "Adams, J. and others",
    collaboration = "STAR",
    title = {Distributions of Charged Hadrons Associated with High Transverse Momentum Particles in $pp$ and $\mathrm{Au}+\mathrm{Au}$ Collisions at $\sqrt{{s}_{NN}}=200\text{ }\mathrm{GeV}$},
    eprint = "nucl-ex/0501016",
    archivePrefix = "arXiv",
    doi = "10.1103/PhysRevLett.95.152301",
    journal = "Phys. Rev. Lett.",
    volume = "95",
    pages = "152301",
    year = "2005"
}

@article{STAR:2009ngv,
    author = "Abelev, B. I. and others",
    collaboration = "STAR",
    title = "{Long range rapidity correlations and jet production in high energy nuclear collisions}",
    eprint = "0909.0191",
    archivePrefix = "arXiv",
    primaryClass = "nucl-ex",
    doi = "10.1103/PhysRevC.80.064912",
    journal = "Phys. Rev. C",
    volume = "80",
    pages = "064912",
    year = "2009"
}

@article{PHOBOS:2009sau,
    author = "Alver, B. and others",
    collaboration = "PHOBOS",
    title = {High Transverse Momentum Triggered Correlations over a Large Pseudorapidity Acceptance in $\mathrm{Au}+\mathrm{Au}$ Collisions at $\sqrt{{s}_{NN}}=200\text{ }\mathrm{GeV}$},
    eprint = "0903.2811",
    archivePrefix = "arXiv",
    primaryClass = "nucl-ex",
    doi = "10.1103/PhysRevLett.104.062301",
    journal = "Phys. Rev. Lett.",
    volume = "104",
    pages = "062301",
    year = "2010"
}

@article{Aamodt:2011by,
    author = "Aamodt, K. and others",
    collaboration = "ALICE",
    title = "{Harmonic decomposition of two-particle angular correlations in Pb-Pb collisions at $\sqrt{s_{NN}}= 2.76\text{ }\mathrm{TeV}$}",
    eprint = "1109.2501",
    archivePrefix = "arXiv",
    primaryClass = "nucl-ex",
    reportNumber = "CERN-PH-EP-2011-152",
    doi = "10.1016/j.physletb.2012.01.060",
    journal = "Phys. Lett. B",
    volume = "708",
    pages = "249--264",
    year = "2012"
}

@article{Adam:2019woz,
    author = "Adam, Jaroslav and others",
    collaboration = "STAR",
    title = "{Azimuthal Harmonics in Small and Large Collision Systems at RHIC Top Energies}",
    eprint = "1901.08155",
    archivePrefix = "arXiv",
    primaryClass = "nucl-ex",
    doi = "10.1103/PhysRevLett.122.172301",
    journal = "Phys. Rev. Lett.",
    volume = "122",
    number = "17",
    pages = "172301",
    year = "2019"
}

@article{ZEUS:2019jya,
    author = "Abt, I. and others",
    collaboration = "ZEUS",
    title = "{Two-particle azimuthal correlations as a probe of collective behaviour in deep inelastic $ep$ scattering at HERA}",
    eprint = "1912.07431",
    archivePrefix = "arXiv",
    primaryClass = "hep-ex",
    reportNumber = "DESY-19-174",
    doi = "10.1007/JHEP04(2020)070",
    journal = "JHEP",
    volume = "04",
    pages = "070",
    year = "2020"
}

@article{PhysRevLett.39.1587,
  title = {Quantum Chromodynamics Test for Jets},
  author = {Farhi, Edward},
  journal = {Phys. Rev. Lett.},
  volume = {39},
  issue = {25},
  pages = {1587--1588},
  numpages = {0},
  year = {1977},
  month = {Dec},
  publisher = {American Physical Society},
  doi = {10.1103/PhysRevLett.39.1587},
  url = {https://link.aps.org/doi/10.1103/PhysRevLett.39.1587}
}

@article{Dumitru:2010iy,
    author = "Dumitru, Adrian and Dusling, Kevin and Gelis, Francois and Jalilian-Marian, Jamal and Lappi, Tuomas and Venugopalan, Raju",
    title = "{The Ridge in proton-proton collisions at the LHC}",
    eprint = "1009.5295",
    archivePrefix = "arXiv",
    primaryClass = "hep-ph",
    reportNumber = "INT-PUB-10-051, BCCUNY-HEP-10-03, BNL-94103-2010-JA, RBRC-858",
    doi = "10.1016/j.physletb.2011.01.024",
    journal = "Phys. Lett. B",
    volume = "697",
    pages = "21--25",
    year = "2011"
}

@article{ATLAS:2012cix,
    author = "Aad, Georges and others",
    collaboration = "ATLAS",
    title = "{Observation of Associated Near-Side and Away-Side Long-Range Correlations in $\sqrt{s_{NN}}$=5.02  TeV Proton-Lead Collisions with the ATLAS Detector}",
    eprint = "1212.5198",
    archivePrefix = "arXiv",
    primaryClass = "hep-ex",
    reportNumber = "CERN-PH-EP-2012-366",
    doi = "10.1103/PhysRevLett.110.182302",
    journal = "Phys. Rev. Lett.",
    volume = "110",
    number = "18",
    pages = "182302",
    year = "2013"
}

@article{PHENIX:2013ktj,
    author = "Adare, A. and others",
    collaboration = "PHENIX",
    title = "{Quadrupole Anisotropy in Dihadron Azimuthal Correlations in Central $d+$Au Collisions at $\sqrt{s_{_{NN}}}$=200 GeV}",
    eprint = "1303.1794",
    archivePrefix = "arXiv",
    primaryClass = "nucl-ex",
    doi = "10.1103/PhysRevLett.111.212301",
    journal = "Phys. Rev. Lett.",
    volume = "111",
    number = "21",
    pages = "212301",
    year = "2013"
}

@article{Busza:2018rrf,
    author = "Busza, Wit and Rajagopal, Krishna and van der Schee, Wilke",
    title = "{Heavy Ion Collisions: The Big Picture, and the Big Questions}",
    eprint = "1802.04801",
    archivePrefix = "arXiv",
    primaryClass = "hep-ph",
    reportNumber = "MIT-CTP-4892, MIT-CTP/4892",
    doi = "10.1146/annurev-nucl-101917-020852",
    journal = "Ann. Rev. Nucl. Part. Sci.",
    volume = "68",
    pages = "339--376",
    year = "2018"
}

@article{Bierlich:2020naj,
    author = {Bierlich, Christian and Chakraborty, Smita and Gustafson, G\"osta and L\"onnblad, Leif},
    title = "{Setting the string shoving picture in a new frame}",
    eprint = "2010.07595",
    archivePrefix = "arXiv",
    primaryClass = "hep-ph",
    reportNumber = "LU-TP 20-48, MCnet-20-22",
    doi = "10.1007/JHEP03(2021)270",
    journal = "JHEP",
    volume = "03",
    pages = "270",
    year = "2021"
}

@article{Castorina:2020iia,
    author = "Castorina, P. and Lanteri, D. and Satz, H.",
    title = "{Strangeness enhancement and flow-like effects in $e^+e^-$ annihilation at high parton density}",
    eprint = "2011.06966",
    archivePrefix = "arXiv",
    primaryClass = "hep-ph",
    doi = "10.1140/epja/s10050-021-00393-z",
    journal = "Eur. Phys. J. A",
    volume = "57",
    number = "3",
    pages = "111",
    year = "2021"
}

@article{Agostini:2021xca,
    author = "Agostini, Pedro and Altinoluk, Tolga and Armesto, N\'estor",
    title = "{Multi-particle production in proton\textendash{}nucleus collisions in the color glass condensate}",
    eprint = "2103.08485",
    archivePrefix = "arXiv",
    primaryClass = "hep-ph",
    doi = "10.1140/epjc/s10052-021-09475-0",
    journal = "Eur. Phys. J. C",
    volume = "81",
    number = "8",
    pages = "760",
    year = "2021"
}

@article{Belle:2022fvl,
    author = "Chen, Y. -C. and others",
    collaboration = "Belle",
    title = "{Measurement of Two-Particle Correlations of Hadrons in $e^{+}e^{-}$ Collisions at Belle}",
    eprint = "2201.01694",
    archivePrefix = "arXiv",
    primaryClass = "hep-ex",
    reportNumber = "Belle Preprint 2021-32; KEK Preprint 2021-57",
    month = "1",
    year = "2022"
}

@article{Larkoski:2021hee,
    author = "Larkoski, Andrew J. and Melia, Tom",
    title = "{A large-N expansion for minimum bias}",
    eprint = "2107.04041",
    archivePrefix = "arXiv",
    primaryClass = "hep-ph",
    doi = "10.1007/JHEP10(2021)094",
    journal = "JHEP",
    volume = "10",
    pages = "094",
    year = "2021"
}

@article{ATLAS:2021jhn,
    author = "Aad, Georges and others",
    collaboration = "ATLAS",
    title = "{Two-particle azimuthal correlations in photonuclear ultraperipheral Pb+Pb collisions at 5.02 TeV with ATLAS}",
    eprint = "2101.10771",
    archivePrefix = "arXiv",
    primaryClass = "nucl-ex",
    reportNumber = "CERN-EP-2020-246",
    doi = "10.1103/PhysRevC.104.014903",
    journal = "Phys. Rev. C",
    volume = "104",
    number = "1",
    pages = "014903",
    year = "2021"
}

@article{CMS:2022doq,
    collaboration = "CMS",
    title = "{Two-particle azimuthal correlations in $\gamma$p interactions using pPb collisions at $\sqrt{s_\mathrm{NN}}$ = 8.16 TeV}",
    eprint = "2204.13486",
    archivePrefix = "arXiv",
    primaryClass = "nucl-ex",
    reportNumber = "CMS-HIN-18-008, CERN-EP-2021-100",
    month = "4",
    year = "2022"
}

@article{The:2022lun,
    author = "The, Belle and Collaboration",
    collaboration = "Belle",
    title = "{Two-particle angular correlations in $e^+ e^-$ collisions to hadronic final states in two reference coordinates at Belle}",
    eprint = "2206.09440",
    archivePrefix = "arXiv",
    primaryClass = "hep-ex",
    reportNumber = "Belle Preprint 2022-11; KEK Preprint 2022-10",
    month = "6",
    year = "2022"
}

@article{Baty:2021ugw,
    author = "Baty, Austin and Gardner, Parker and Li, Wei",
    title = "{Collective evolution of a parton in the vacuum: the ultimate partonic ''droplet'', non-perturbative QCD and quantum entanglement}",
    eprint = "2104.11735",
    archivePrefix = "arXiv",
    primaryClass = "hep-ph",
    month = "4",
    year = "2021"
}

@article{Decamp:1990jra,
      author         = "Decamp, D. and others",
      title          = "{ALEPH: A detector for electron-positron annnihilations
                        at LEP}",
      collaboration  = "ALEPH",
      journal        = "Nucl. Instrum. Meth.",
      volume         = "A294",
      year           = "1990",
      pages          = "121-178",
      doi            = "10.1016/0168-9002(90)91831-U",
      note           = "[Erratum: Nucl. Instrum. Meth.A303,393(1991)]",
      reportNumber   = "CERN-EP-90-25",
      SLACcitation   = "%%CITATION = NUIMA,A294,121;%%"
}

@article{Sjostrand:2000wi,
      author         = "Sjostrand, Torbjorn and Eden, Patrik and Friberg,
                        Christer and Lonnblad, Leif and Miu, Gabriela and Mrenna,
                        Stephen and Norrbin, Emanuel",
      title          = "{High-energy physics event generation with PYTHIA 6.1}",
      journal        = "Comput. Phys. Commun.",
      volume         = "135",
      year           = "2001",
      pages          = "238-259",
      doi            = "10.1016/S0010-4655(00)00236-8",
      eprint         = "hep-ph/0010017",
      archivePrefix  = "arXiv",
      primaryClass   = "hep-ph",
      reportNumber   = "LU-TP-00-30",
      SLACcitation   = "%%CITATION = HEP-PH/0010017;%%"
}

@article{ALEPH:2003obs,
    author = "Heister, A. and others",
    collaboration = "ALEPH",
    title = "{Studies of QCD at e+ e- centre-of-mass energies between 91-GeV and 209-GeV}",
    reportNumber = "CERN-EP-2003-084",
    doi = "10.1140/epjc/s2004-01891-4",
    journal = "Eur. Phys. J. C",
    volume = "35",
    pages = "457--486",
    year = "2004"
}

@article{Cacciari:2011ma,
    author = "Cacciari, Matteo and Salam, Gavin P. and Soyez, Gregory",
    title = "{FastJet User Manual}",
    eprint = "1111.6097",
    archivePrefix = "arXiv",
    primaryClass = "hep-ph",
    doi = "10.1140/epjc/s10052-013-1896-2",
    journal = "Eur. Phys. J. C",
    volume = "72",
    pages = "1896",
    year = "2012"
}

@article{ALEPH:2006cdc,
    author = "Schael, S. and others",
    collaboration = "ALEPH",
    title = "{Measurement of the $W$ boson mass and width in $e^{+} e^{-}$ collisions at LEP}",
    eprint = "hep-ex/0605011",
    archivePrefix = "arXiv",
    reportNumber = "CERN-PH-EP-2006-004",
    doi = "10.1140/epjc/s2006-02576-8",
    journal = "Eur. Phys. J. C",
    volume = "47",
    pages = "309--335",
    year = "2006"
}

@article{CMS:2011cqy,
    author = "Chatrchyan, Serguei and others",
    collaboration = "CMS",
    title = "{Long-range and short-range dihadron angular correlations in central PbPb collisions at a nucleon-nucleon center of mass energy of 2.76 TeV}",
    eprint = "1105.2438",
    archivePrefix = "arXiv",
    primaryClass = "nucl-ex",
    reportNumber = "CERN-PH-EP-2011-056, CMS-HIN-11-001",
    doi = "10.1007/JHEP07(2011)076",
    journal = "JHEP",
    volume = "07",
    pages = "076",
    year = "2011"
}

@article{ALICE:2011svq,
    author = "Aamodt, K. and others",
    collaboration = "ALICE",
    title = "{Harmonic decomposition of two-particle angular correlations in Pb-Pb collisions at $\sqrt{s_{NN}}=$ 2.76 TeV}",
    eprint = "1109.2501",
    archivePrefix = "arXiv",
    primaryClass = "nucl-ex",
    reportNumber = "CERN-PH-EP-2011-152",
    doi = "10.1016/j.physletb.2012.01.060",
    journal = "Phys. Lett. B",
    volume = "708",
    pages = "249--264",
    year = "2012"
}

@article{ATLAS:2012at,
    author = "Aad, Georges and others",
    collaboration = "ATLAS",
    title = "{Measurement of the azimuthal anisotropy for charged particle production in $\sqrt{s_{NN}}=2.76$ TeV lead-lead collisions with the ATLAS detector}",
    eprint = "1203.3087",
    archivePrefix = "arXiv",
    primaryClass = "hep-ex",
    reportNumber = "CERN-PH-EP-2012-035",
    doi = "10.1103/PhysRevC.86.014907",
    journal = "Phys. Rev. C",
    volume = "86",
    pages = "014907",
    year = "2012"
}

@article{Bierlich:2019wld,
    author = "Bierlich, Christian and Rasmussen, Christine O.",
    title = "{Dipole evolution: perspectives for collectivity and $\gamma^*$A collisions}",
    eprint = "1907.12871",
    archivePrefix = "arXiv",
    primaryClass = "hep-ph",
    reportNumber = "LU TP 19-32, MCnet-19-17",
    doi = "10.1007/JHEP10(2019)026",
    journal = "JHEP",
    volume = "10",
    pages = "026",
    year = "2019"
}

@article{ALICE:2012eyl,
    author = "Abelev, Betty and others",
    collaboration = "ALICE",
    title = "{Long-range angular correlations on the near and away side in $p$-Pb collisions at $\sqrt{s_{NN}}=5.02$ TeV}",
    eprint = "1212.2001",
    archivePrefix = "arXiv",
    primaryClass = "nucl-ex",
    reportNumber = "CERN-PH-EP-2012-359",
    doi = "10.1016/j.physletb.2013.01.012",
    journal = "Phys. Lett. B",
    volume = "719",
    pages = "29--41",
    year = "2013"
}

@article{ALICE:2013snk,
    author = "Abelev, Betty Bezverkhny and others",
    collaboration = "ALICE",
    title = "{Long-range angular correlations of $\rm \pi$, K and p in p-Pb collisions at $\sqrt{s_{\rm NN}}$ = 5.02 TeV}",
    eprint = "1307.3237",
    archivePrefix = "arXiv",
    primaryClass = "nucl-ex",
    reportNumber = "CERN-PH-EP-2013-115",
    doi = "10.1016/j.physletb.2013.08.024",
    journal = "Phys. Lett. B",
    volume = "726",
    pages = "164--177",
    year = "2013"
}

@article{Chen:2021uws,
    author = "Chen, Yi and others",
    title = "{Jet energy spectrum and substructure in e$^{+}$e$^{-}$ collisions at 91.2 GeV with ALEPH Archived Data}",
    eprint = "2111.09914",
    archivePrefix = "arXiv",
    primaryClass = "hep-ex",
    reportNumber = "MITHIG-MOD-21-001",
    doi = "10.1007/JHEP06(2022)008",
    journal = "JHEP",
    volume = "06",
    pages = "008",
    year = "2022"
}

@article{Chen:2021uwsNote,
    author = "Chen, Yi and Lee, Yen-Jie and Maggi, Marcello and Chang, Paoti and Chien, Yang-Ting and McGinn, Christopher and Perepelitsa, Dennis",
    title = "{Analysis note: jet reconstruction, energy spectra, and substructure analyses with archived ALEPH data}",
    eprint = "2108.04877",
    archivePrefix = "arXiv",
    primaryClass = "hep-ex",
    month = "8",
    year = "2021"
}

@article{ALICE:2023ulm,
    author = "Acharya, Shreyasi and others",
    collaboration = "ALICE",
    title = "{Emergence of Long-Range Angular Correlations in Low-Multiplicity Proton-Proton Collisions}",
    eprint = "2311.14357",
    archivePrefix = "arXiv",
    primaryClass = "nucl-ex",
    reportNumber = "CERN-EP-2023-265",
    doi = "10.1103/PhysRevLett.132.172302",
    journal = "Phys. Rev. Lett.",
    volume = "132",
    number = "17",
    pages = "172302",
    year = "2024"
}

@article{CMS:2023iam,
    author = "Hayrapetyan, Aram and others",
    collaboration = "CMS",
    title = "{Observation of Enhanced Long-Range Elliptic Anisotropies Inside High-Multiplicity Jets in pp Collisions at s=13{\,}{\,}TeV}",
    eprint = "2312.17103",
    archivePrefix = "arXiv",
    primaryClass = "hep-ex",
    reportNumber = "CMS-HIN-21-013, CERN-EP-2023-281",
    doi = "10.1103/PhysRevLett.133.142301",
    journal = "Phys. Rev. Lett.",
    volume = "133",
    number = "14",
    pages = "142301",
    year = "2024"
}

@article{Chen:2023njr,
    author = "Chen, Yu-Chen and others",
    title = "{Long-range near-side correlation in e+e{\ensuremath{-}} collisions at 183-209 GeV with ALEPH archived data}",
    eprint = "2312.05084",
    archivePrefix = "arXiv",
    primaryClass = "hep-ex",
    reportNumber = "MITHIG-MOD-23-001",
    doi = "10.1016/j.physletb.2024.138957",
    journal = "Phys. Lett. B",
    volume = "856",
    pages = "138957",
    year = "2024"
}

@article{Chen:2023nsi,
    author = "Chen, Yu-Chen and Lee, Yen-Jie and Chen, Yi and Chang, Paoti and McGinn, Chris and Sheng, Tzu-An and Innocenti, Gian Michele and Maggi, Marcello",
    title = "{Analysis note: two-particle correlation in $e^+e^-$ collisions at 91-209 GeV with archived ALEPH data}",
    eprint = "2309.09874",
    archivePrefix = "arXiv",
    primaryClass = "hep-ex",
    reportNumber = "MITHIG-MOD-NOTE-23-001",
    month = "9",
    year = "2023"
}

@article{Bossi:2025eecNote,
    author = "Bossi, Hannah and Chen, Yu-Chen and Chen, Yi and Zhang, Jingyu and Innocenti, Gian Michele and Badea, Anthony and Baty, Austin and Maggi, Marcello and McGinn, Chris and Lee, Yen-Jie",
    title = "{Analysis note: measurement of energy-energy correlator in $e^{+}e^{-}$ collisions at $91$ GeV with archived ALEPH data}",
    eprint = "2505.11828",
    archivePrefix = "arXiv",
    primaryClass = "hep-ex",
    month = "5",
    year = "2025"
}

@article{Badea:2025thrustNote,
    author = "Badea, Anthony and Baty, Austin and Bossi, Hannah and Chen, Yu-Chen and Chen, Yi and Zhang, Jingyu and Innocenti, Gian Michele and Maggi, Marcello and McGinn, Chris and Peters, Michael and Sheng, Tzu-An and Mikuni, Vinicius and Avaylon, Matthew and Komiske, Patrick and Metodiev, Eric and Thaler, Jesse and Nachman, Benjamin and Lee, Yen-Jie",
    title = "{Analysis note: measurement of thrust in $e^{+}e^{-}$ collisions at $\sqrt{s}=91$ GeV with archived ALEPH data}",
    eprint = "2507.14349",
    archivePrefix = "arXiv",
    primaryClass = "hep-ex",
    month = "7",
    year = "2025"
}

@article{Jadach:2001mp,
    author = "Jadach, S. and Placzek, W. and Skrzypek, M. and Ward, B. F. L. and Was, Z.",
    title = "{The Monte Carlo program KoralW version 1.51 and the concurrent Monte Carlo KoralW and YFSWW3 with all background graphs and first order corrections to W pair production}",
    eprint = "hep-ph/0104049",
    archivePrefix = "arXiv",
    reportNumber = "CERN-TH-2001-040, UTHEP-01-0102",
    doi = "10.1016/S0010-4655(01)00296-X",
    journal = "Comput. Phys. Commun.",
    volume = "140",
    pages = "475--512",
    year = "2001"
}

@article{Jadach:1999vf,
    author = "Jadach, S. and Ward, B. F. L. and Was, Z.",
    title = "{The Precision Monte Carlo event generator K K for two fermion final states in e+ e- collisions}",
    eprint = "hep-ph/9912214",
    archivePrefix = "arXiv",
    reportNumber = "DESY-99-106, CERN-TH-99-235, UTHEP-99-08-01",
    doi = "10.1016/S0010-4655(00)00048-5",
    journal = "Comput. Phys. Commun.",
    volume = "130",
    pages = "260--325",
    year = "2000"
}

@inproceedings{chen_xgboost_2016,
	address = {New York, NY, USA},
	series = {{KDD} '16},
	title = {{XGBoost}: {A} {Scalable} {Tree} {Boosting} {System}},
	isbn = {978-1-4503-4232-2},
	shorttitle = {{XGBoost}},
	url = {https://dl.acm.org/doi/10.1145/2939672.2939785},
	doi = {10.1145/2939672.2939785},
	urldate = {2025-11-02},
	booktitle = {Proceedings of the 22nd {ACM} {SIGKDD} {International} {Conference} on {Knowledge} {Discovery} and {Data} {Mining}},
	publisher = {Association for Computing Machinery},
	author = {Chen, Tianqi and Guestrin, Carlos},
	month = aug,
	year = {2016},
	pages = {785--794},
}

@article{Grosse-Oetringhaus:2024bwr,
    author = "Grosse-Oetringhaus, Jan Fiete and Wiedemann, Urs Achim",
    title = "{A Decade of Collectivity in Small Systems}",
    eprint = "2407.07484",
    archivePrefix = "arXiv",
    primaryClass = "hep-ex",
    reportNumber = "CERN-TH-2024-110",
    month = "7",
    year = "2024"
}

@article{2013119,
title = {Electroweak measurements in electron–positron collisions at W-boson-pair energies at LEP},
journal = {Physics Reports},
volume = {532},
number = {4},
pages = {119-244},
year = {2013},
note = {Electroweak Measurements in Electron-Positron Collisions at W-Boson-Pair Energies at LEP},
issn = {0370-1573},
doi = {https://doi.org/10.1016/j.physrep.2013.07.004},
url = {https://www.sciencedirect.com/science/article/pii/S0370157313002706}
}

@article{ATLAS:2025OOFlow,
  author        = "{ATLAS Collaboration}",
  title         = "{Measurement of the azimuthal anisotropy of charged particles in $\sqrt{s_{\mathrm{NN}}}=5.36$ TeV $^{16}$O$+^{16}$O and $^{20}$Ne$+^{20}$Ne collisions with the ATLAS detector}",
  eprint        = "2509.05171",
  archivePrefix = "arXiv",
  primaryClass  = "nucl-ex",
  reportNumber  = "CERN-EP-2025-200",
  year          = "2025",
  note          = "Submitted to Phys. Rev. C"
}

@article{ALICE:2025OOFlow,
  author        = "{ALICE Collaboration}",
  title         = "{Evidence of nuclear geometry-driven anisotropic flow in OO and Ne$-$Ne collisions at $\sqrt{s_{\rm NN}}=5.36$ TeV}",
  eprint        = "2509.06428",
  archivePrefix = "arXiv",
  primaryClass  = "nucl-ex",
  reportNumber  = "CERN-EP-2025-203",
  year          = "2025",
  note          = "Submitted to Phys. Rev. Lett."
}

@article{CMS:2025OOFlow,
  author        = "{CMS Collaboration}",
  title         = "{Observation of long-range collective flow in OO and NeNe collisions and implications for nuclear structure studies}",
  eprint        = "2510.02580",
  archivePrefix = "arXiv",
  primaryClass  = "nucl-ex",
  reportNumber  = "CERN-EP-2025-222, CMS-HIN-25-009",
  year          = "2025",
  note          = "Submitted to Phys. Rev. Lett."
}
\end{document}